\documentclass[pdflatex,sn-mathphys-num,iicol]{sn-jnl}% iicol = two-column layout

\usepackage{graphicx}      % \includegraphics, \rotatebox
\usepackage{amsmath}       % align, cases
\usepackage{amssymb}
\usepackage{multirow}      % \multirow in tables
\usepackage{booktabs}      % \toprule \midrule \bottomrule
\usepackage{tabularx}      % tabularx X-columns
\usepackage{threeparttable}% \begin{threeparttable}, \tablenotes
\usepackage{siunitx}       % \sisetup, \num
\usepackage{algorithm}     % algorithm* float
\usepackage{algpseudocode} % algorithmic environment
\usepackage{subcaption}    % \subcaption / subfigure
\usepackage[capitalise]{cleveref} % \cref / \Cref  (load after hyperref, which sn-jnl loads)

\newlength{\plotaxesheighttwo}
\newlength{\plotaxesheightthree}
\newlength{\plotaxesheightfour}
\newlength{\plotwidthtwo}
\newlength{\plotwidththree}
\newlength{\plotwidthfour}
\newlength{\plotwidthgrid}
\newlength{\cbarwidthtwo}
\newlength{\cbarwidththree}
\newlength{\cbarwidthfour}
\newlength{\cbarwidthgrid}
\newlength{\gridpagewidth}
\newlength{\totalgridheight}
\newcommand{\interspace}{\hfill} % horizontal filler between side-by-side panels

\begin{document}

\title[Human-Agent Interaction in Synthetic Social Networks]{Human-Agent
Interaction in Synthetic Social Networks: A Framework for Studying Online
Polarization}

\author*[1]{\fnm{Tim} \sur{Donkers}}\email{tim.donkers@uni-due.de}
\author[1]{\fnm{J\"urgen} \sur{Ziegler}}\email{juergen.ziegler@uni-due.de}

\affil[1]{\orgdiv{Interactive Systems}, \orgname{University of
Duisburg-Essen}, \orgaddress{\street{Forsthausweg 2}, \city{Duisburg},
\postcode{47057}, \state{North Rhine-Westphalia}, \country{Germany}}}

%% "Google Scholar profiles" block reproduced via sn-jnl's \presentaddress hook.
\presentaddresstxt{Google Scholar profiles}
\presentaddress{Tim Donkers:
\url{https://scholar.google.com/citations?user=QX3ITdwAAAAJ}\\
J\"urgen Ziegler:
\url{https://scholar.google.com/citations?user=UD-ha3YAAAAJ}}

%% NOTE: abstract transcribed from reference/Formatting_Paper.pdf -- please verify.
\abstract{Online political polarization operates across interconnected
dimensions---beliefs radicalize, networks fragment, discourse
transforms---yet we lack understanding of how these processes causally
interact and whether interventions targeting one dimension inadvertently
affect others. This paper introduces a framework enabling controlled causal
investigation by integrating formal opinion dynamics models with
LLM-generated communication, allowing computational agents and humans to
interact in shared social environments while maintaining experimental control
over psychological, structural, and algorithmic factors. Systematic
computational experiments reveal counterintuitive dynamics that challenge
conventional theories: complete network segregation buffers against extreme
polarization while moderate integration creates optimal conditions for
extremism, contradicting echo chamber explanations. Opinion radicalization
and network segregation can decouple, with algorithmic interventions showing
sharp threshold effects where identical approaches succeed preventatively but
fail remedially. Polarization manifests as a systematic communicative
syndrome---simultaneously increased emotionality, heightened group identity
salience, and reduced uncertainty. Human validation with 122 participants
confirms computational environments produce psychologically meaningful effects
with large effect sizes while revealing that discourse climate influences
opinion change through ambient processing that bypasses deliberate evaluation.
This framework provides infrastructure for evidence-based polarization theory
and intervention design that neither purely computational nor purely empirical
approaches could achieve alone.}


\maketitle
\clearpage  % start the body on a fresh page (\newpage only breaks the column in two-column mode)

%% =========================== MAIN TEXT ===========================
\section{Introduction}

Online political polarization operates across multiple interconnected dimensions—beliefs radicalize, networks fragment, and discourse transforms—yet we lack systematic understanding of how these processes causally interact. Do echo chambers cause belief extremization, or does belief polarization drive network sorting \cite{sasahara_social_2021}? Can algorithmic interventions reduce opinion polarization while inadvertently increasing structural segregation? Does discourse toxicity reflect or drive attitude change? Understanding these causal relationships requires frameworks that can isolate specific mechanisms while maintaining communicative realism and validating computational predictions against actual human behavior \cite{lazer_computational_2009, epstein_generative_1999}. Without such frameworks, polarization theory rests on correlational observational patterns and untested computational assumptions, constraining both theoretical understanding and evidence-based intervention design.

This validation gap stems from fundamental trade-offs across existing research approaches. Observational studies leverage large-scale social media data to identify macro-level patterns \cite{karjus_evolving_2024,bond_political_2022,al_amin_unveiling_2017}, providing invaluable ecological insights, but cannot manipulate key variables or establish causal relationships between specific mechanisms and polarization outcomes. Traditional opinion dynamics models enable controlled experimentation with mathematically precise belief update mechanisms \cite{hegselmann_opinion_2002,degroot_reaching_1974,sasahara_social_2021}, yet rely on simplified interaction rules that sacrifice communicative realism and prevent direct validation against human responses. Recent LLM-based simulations achieve unprecedented communicative authenticity \cite{chuang_simulating_2024,breum_persuasive_2024,ohagi_polarization_2024}, but operate as "black boxes" lacking the explicit, formal mechanisms necessary for systematic theoretical testing. Each approach has distinct strengths, yet none achieves the combination of formal mechanisms, realistic communication, and human validation necessary for causal investigation of polarization dynamics.

This paper introduces a framework that bridges this gap by enabling seamless interaction between humans and computational agents in shared social environments. Our approach combines the analytical rigor of formal opinion dynamics models with the communicative authenticity of LLM-based agents: agents' beliefs evolve according to mathematical equations grounded in Social Judgment Theory \cite{sherif_attitude_1965}, while their communication occurs through natural language, allowing direct interaction with human participants. Agents and humans interact in simulated social media environments where they encounter posts, react with likes or comments, and experience algorithmically-curated content streams. The framework provides systematic control over psychological, structural, and algorithmic factors, enabling investigation of how these dimensions interact to drive polarization.

Our work makes three primary contributions to polarization research.

First, we provide novel \emph{theoretical insights} that reveal patterns challenging structural echo chamber explanations \cite{sunstein_republic_2017} while aligning with theoretical perspectives conceptualizing polarization as an inherent communicative order rather than a pathological deviation from democratic norms \cite{kumkar_polarisierung_2025}. Our computational experiments demonstrate that complete network segregation actually buffers against extreme polarization while moderately integrated networks create optimal conditions for extremism—consistent with political identity formation requiring friction with opposing viewpoints and recent computational work showing that digital media drive affective polarization through partisan sorting rather than isolation \cite{tornberg_how_2022, tornberg_modeling_2021}. These findings suggest polarization may function as the inherent logic of political discourse, with profound implications for intervention design: structural fixes such as breaking echo chambers or maximizing cross-cutting exposure may intensify rather than reduce polarization by increasing the friction necessary for identity crystallization. We reveal an "engagement valley" where moderate disagreement produces the lowest interaction levels, suggesting fundamental instability in the measured discourse often considered essential for democracy. Our co-evolutionary analysis demonstrates that structural network segregation can occur with only moderate belief polarization, indicating that network-based studies may systematically overestimate ideological extremization by conflating social sorting with radical opinion change. By analyzing LLM-generated discourse, we show that polarization manifests as a systematic "communicative syndrome"—discourse with simultaneously higher emotionality, greater group salience, and reduced uncertainty—with message interpretation shaped by individual personality traits and interaction history in ways that create path dependencies extending beyond initial polarization triggers.

Second, our \emph{empirical validation study} with 122 participants demonstrates that computational environments produce psychologically meaningful effects while revealing critical gaps in current models. Participants accurately perceived the distinct communicative differences our computational analysis identified, with large effect sizes (emotionality $g=1.53$, group salience $g=0.73$, uncertainty $g=-1.26$). Critically, human validation revealed that environmental polarization influenced opinion change through pathways our computational model did not capture—discourse climate affects attitudes through rapid, ambient contextual processing \cite{petty_elaboration_1986} that bypasses conscious evaluation of specific message features. This divergence between computational predictions and human mechanisms points toward refinements needed in opinion dynamics models, demonstrating the value of systematic human-in-the-loop validation for advancing theoretical understanding.

Third, we introduce a \emph{methodological framework} that enables systematic validation of computational polarization models against human behavior. By integrating computational modeling and human experimentation \cite{lazer_computational_2009}, we provide replicable infrastructure for evidence-based model development that neither approach could achieve alone, enabling systematic investigation of social dynamics at scales and levels of control impossible with traditional approaches.

The remainder of this paper is structured as follows. Section~2 reviews related work establishing our theoretical foundation. Section~3 details our hybrid framework integrating LLM agents with formal opinion dynamics. Section~4 presents systematic computational experiments examining how individual psychology, network structure, and algorithmic design interact to drive polarization across opinion, network, and communicative dimensions. Section~5 validates computational predictions through human-subject experimentation investigating how individuals perceive and respond to simulated environments. Section~6 synthesizes computational and empirical findings to advance polarization theory, addressing the echo chamber paradox, algorithmic intervention dynamics, and communicative transformation while discussing methodological contributions and limitations. Section~7 concludes with contributions and future directions.

Given the extensive technical detail required for reproducibility, we provide streamlined exposition of key methods and findings in the main text to maintain narrative clarity, while \cref{app:simulation-model,app:computational_experiments,app:message-analysis,app:user-study} contain complete experimental protocols, mathematical specifications, and detailed results to ensure full reproducibility and facilitate framework adoption by other researchers.

\section{Related Work}
Research on social polarization in online networks spans multiple disciplines and methodological approaches. The existing literature reveals three main streams: (1) observational studies analyzing large-scale empirical data, (2) theoretical models simulating opinion dynamics, and (3) experimental studies with human participants. While observational and theoretical approaches are well-developed, human experiments providing direct causal insights into individual-level effects remain comparatively scarce.

\paragraph{Observational and Theoretical Approaches}
Observational studies leveraging social media data employ sophisticated analytical techniques to identify macro-level polarization patterns. Sentiment analysis approaches track emotional polarization and linguistic divergence over time \cite{karjus_evolving_2024,alsinet_measuring_2021,buder_does_2021}, while network clustering methods reveal ideological segregation and echo chamber formation \cite{treuillier_gaining_2024,bond_political_2022,al_amin_unveiling_2017}, driven by user choice and algorithmic curation \cite{conover_political_2011,bakshy_exposure_2015,xing_research_2022,yarchi_political_2021}. Topic modeling techniques analyze how polarization manifests across different subjects and contexts \cite{kim_polarized_2019,chen_modeling_2021}, and specialized approaches explore the roles of specific users in shaping information flow \cite{jiang_political_2020,recuero_using_2019}.

Complementing these empirical findings, theoretical research has developed increasingly sophisticated mathematical models of opinion dynamics. Classical models established foundational mechanisms like consensus through averaging \cite{degroot_reaching_1974,friedkin_social_1990}, clustering via bounded confidence thresholds \cite{deffuant_mixing_2000,hegselmann_opinion_2002}, imitation \cite{holley_ergodic_1975}, social validation \cite{sznajd-weron_opinion_2000}, or cultural trait dissemination \cite{axelrod_agent-based_2006}. Modern extensions enhance realism by incorporating cognitive biases like confirmation bias \cite{del_vicario_modeling_2017,allahverdyan_opinion_2014} and reactance \cite{flache_models_2017,cornacchia_polarization_2020}, modeling network co-evolution \cite{baumann_modeling_2020,sasahara_social_2021}, accounting for varying influence \cite{amelkin_polar_2017,chen_characteristics_2016}, and including algorithmic amplification effects \cite{donkers_-sounding_2023,donkers_dual_2021,geschke_triple-filter_2019}. However, these mathematical models, while analytically powerful, typically abstract complex human communication into simplified numerical or rule-based interactions, often sacrificing linguistic richness and contextual nuance vital to understanding online discourse.

Recent advances in Large Language Models (LLMs) offer a different simulation paradigm focused on communicative realism. LLM-based agents can simulate polarization dynamics \cite{ohagi_polarization_2024}, replicate emergent network phenomena like diffusion through text \cite{gao_social_2024}, and allow exploration of linguistic influence strategies \cite{breum_persuasive_2024} and agent biases \cite{velarde_principles_2024}, building on foundational work in generative agent-based modeling \cite{ghaffarzadegan_generative_2024,gurcan_llm_2024,junprung_exploring_2024}. Surveys highlight the growing potential and challenges of these methods \cite{gao_large_2024}. While capturing linguistic interaction effectively, these LLM simulations often lack the explicit, interpretable belief update mechanisms and formal grounding of traditional opinion dynamics models \cite{chuang_simulating_2024}, creating a methodological disconnect between communicative realism and analytical precision.

\paragraph{Experimental Human Studies}

Experimental studies with human participants, though scarce \cite{gao_large_2024}, provide crucial causal insights unavailable from observation or pure simulation alone, drawing on foundations in social psychology regarding social influence and group dynamics \cite{sherif_psychology_1936,asch_effects_1951,festinger_theory_1957,tajfel_social_1971,janis_groupthink_1982,kunda_case_1990}. Contemporary online experiments typically fall into two categories. \textit{Targeted intervention studies} examine specific causal drivers like content effects \cite{banks_polarizedfeeds_2021}, platform affordances \cite{wuestenenk_influence_2023}, affective dynamics \cite{schieferdecker_affective_2024}, identity cues \cite{wuestenenk_influence_2023}, or network influence \cite{stewart_information_2019}, but often use simplified stimuli or abstract away from complex platform environments. \textit{Model calibration studies} use controlled tasks, often involving numeric judgments \cite{chacoma_opinion_2015,vande_kerckhove_modelling_2016,das_modeling_2014}, to parameterize formal opinion dynamics models \cite[complementing large-scale fitting, e.g.,][]{phillips_high-dimensional_2023,valensise_drivers_2023,peralta_multidimensional_2024}, but these constrained interactions rarely capture the richness of online discourse.

Summarized, a significant methodological challenge remains in creating an experimental framework that integrates the analytical rigor of formal models with the ecological validity of realistic, LLM-driven communication and direct human interaction. The following section details our approach to addressing this challenge.

\section{Simulation Framework}
\label{sec:method-overview}

To investigate online polarization, we developed a computational framework that integrates opinion dynamics principles with the communicative realism afforded by Large Language Model (LLM)-based artificial agents. This hybrid approach bridges the analytical precision of formal models with the ecological validity of natural language simulations. The framework comprises four core components: an adaptive agent architecture, a psychologically-grounded model of opinion dynamics, an explicit model of communicative actions, and a co-evolving social network. This section outlines these components, while the full mathematical specification is provided in \cref{app:simulation-model}.

\subsection{Adaptive Agent Architecture}

At the core of our framework are adaptive agents ($A_i$) that possess rich, evolving internal states: a continuous \emph{opinion score} ($o_i(t) \in [-1, 1]$), a static \emph{persona} ($P_i$) defining personality traits, and a dynamic \emph{interaction history} ($H_i(t)$) that serves as memory. A fundamental principle of our architecture is the distinction between these hidden internal states and the \emph{manifest communication} they produce. We model a realistic social process where individuals cannot directly access others' private thoughts---they only observe public expressions, i.e., the messages people post.

A key aspect of our framework is how these messages are generated: an LLM dynamically crafts text by conditioning on the agent's \emph{entire} internal state. This produces nuanced communication where message content reflects not just the agent's core opinion, but also their personality and recent interactions (\cref{tab:prompt-components,tab:message-intensity}). Equally important is the framework's \emph{interpretive layer}. Unlike traditional models where an author's opinion score is directly transmitted, an LLM-driven evaluation function, $\pi$, simulates subjective perception---it reads the textual content of an encountered message ($m$) to estimate its perceived stance ($o_m$). Crucially, all subsequent opinion updates and reactions are based on this \emph{perception} of the message, not on privileged access to the author's true beliefs. This explicit interpretation step introduces realistic potential for misunderstanding, context-dependent perception, and bias.

Importantly, this design choice enables agents to interact seamlessly with human users without requiring explicit opinion measurements. While traditional opinion dynamics models depend on direct access to other agents' opinion values (e.g., to apply averaging mechanisms), our agents derive perceived opinions purely from textual content evaluation. This eliminates the need to elicit numerical opinion scores from humans, who may themselves be uncertain about their precise positions, and allows agents to update their beliefs based on natural language communication just as humans do in real-world interactions.

\begin{table*}[h!]
\centering
\small
\caption{Message Generation Prompting Components}
\label{tab:prompt-components}
\begin{tabularx}{\textwidth}{lX}
\toprule
\textbf{Component} & \textbf{Description and Format} \\
\midrule
Opinion Value & Numerical stance (-1 to 1) determining message position and intensity. Format: "You hold a \{strong/moderate/weak\} \{positive/negative\} opinion (value: X) on this topic" \\
\midrule
Topic Description & Core topic information and contextual framework. Format: "The topic is Universal Basic Income, focusing on economic, social and ethical implications" \\
\midrule
Personality Profile & Agent's character traits and communication patterns. Format: "You are a \{thoughtful/passionate/analytical\} person who tends to \{communication style\}" \\
\midrule
Interaction History & Recent interactions and response context. Format: "You recently \{agreed/disagreed\} with user X about \{topic aspect\}" \\
\midrule
Intensity Instructions & Guidelines for emotional expression and assertion strength based on opinion intensity (see Table $2$). Format: For $|o_i| > 0.7$: "Express strong conviction and emotional investment" \\
\bottomrule
\end{tabularx}
\end{table*}

\begin{table*}[h!]
\centering
\small
\caption{Message Generation Characteristics by Opinion Intensity}
\label{tab:message-intensity}
\begin{tabularx}{\textwidth}{lX}
\toprule
\textbf{Intensity} & \textbf{Message Characteristics} \\
\midrule
Low ($|o_i| < 0.3$) & Balanced argumentation; acknowledges multiple perspectives; uses conditional statements; minimal group identification; emphasizes uncertainty. \\
Moderate ($0.3 \leq |o_i| \leq 0.7$) & Clear directional bias; in-group preference; moderate skepticism of opposing views; emotional undertones while maintaining reasoned discourse. \\
High ($|o_i| > 0.7$) & Strong emotional language; pronounced group identification; dehumanization of opponents; hyperbolic terminology; portrays opposing views as threats. \\
\bottomrule
\end{tabularx}
\end{table*}

\subsection{Opinion Dynamics}
\begin{figure*}[ht] % Use [ht] for better placement flexibility
    \centering % Center everything within the figure environment

    \begin{minipage}[b]{0.48\textwidth}
        \centering
        \includegraphics[width=\textwidth]{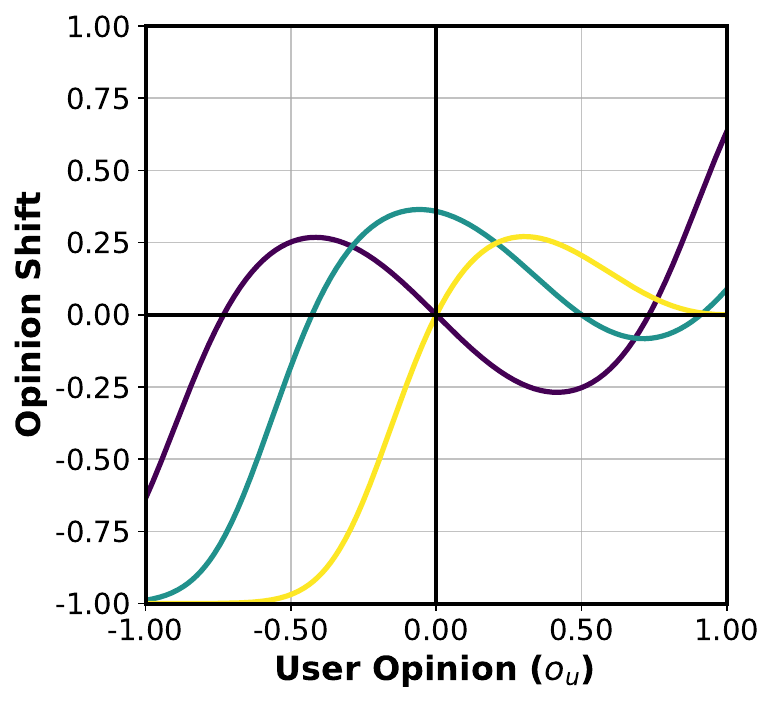}
        \centerline{(a) Opinion shift} % Manual label and caption
    \end{minipage}
    \hfill % Creates space between the minipages
    \begin{minipage}[b]{0.48\textwidth}
        \centering
        \includegraphics[width=\textwidth]{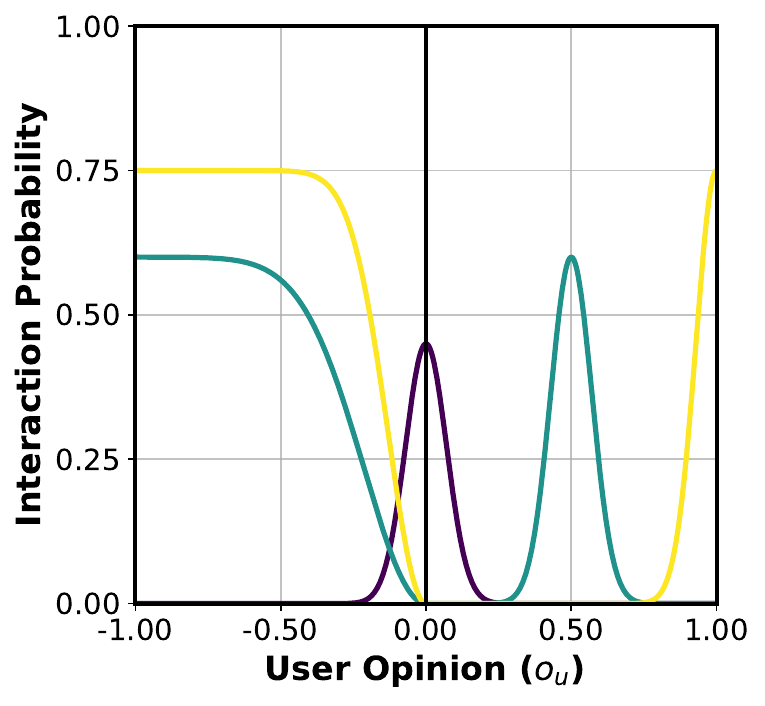}
        \centerline{(b) Interaction probability} % Manual label and caption
    \end{minipage}

    \vspace{1ex} % Add some vertical space before the legend
    \begin{minipage}{\textwidth}
        \centering
        \includegraphics[width=0.5\textwidth]{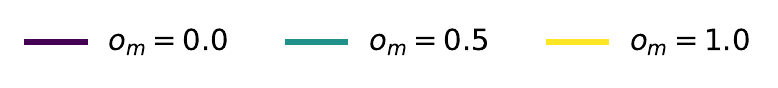}
    \end{minipage}
    \vspace{1ex} % Optional: Add space between legend and main caption

    \caption{Agent response functions versus user opinion ($o_u$) for varying message opinions ($o_m$).
    \textbf{(a)} Opinion shift illustrates assimilation (positive shift towards $o_m$) and repulsion (negative shift away from $o_m$); zero crossings represent stability points.
    \textbf{(b)} Interaction probability peaks show preference for aligned opinions (homophily). Stronger effects and asymmetry for extreme messages ($o_m=1.0$) capture varied interaction dynamics.}
    \label{fig:opinion-interaction-functions} % One label for the entire figure
\end{figure*}

Our framework models opinion evolution based on principles from Social Judgment Theory \cite{sherif_attitude_1965}, which suggests that individuals evaluate new information through the lens of their existing attitudes. Agent opinions evolve through a process of assimilation and repulsion, governed by their personal latitudes of acceptance and rejection, as visualized in \cref{fig:opinion-interaction-functions}a.

When an agent encounters a message, the opinion update process begins with the message evaluation function $\pi$, which determines the perceived opinion $o_m$ expressed in the message content. The agent then calculates the opinion difference $d = o_i(t) - o_m$ and compares this difference to their current "latitude of acceptance"—the range of opinion differences they can tolerate before shifting from attraction to repulsion.

The core opinion shift is governed by a cubic polynomial function:
\begin{align}
    \omega_i(m) = d \cdot (d^2 - \sigma_{\text{eff}}^2(o_i, o_m)) \label{eq:polynomial_shift_main}
\end{align}
where $d$ is the opinion difference and $\sigma_{\text{eff}}^2$ represents the effective squared width of the attraction zone. This function creates distinct behavioral regimes: when the opinion difference falls within the effective width ($|d|^2 < \sigma_{\text{eff}}^2$), the agent is attracted toward the message; when outside this range ($|d|^2 > \sigma_{\text{eff}}^2$), the agent is repelled away from the message. The final opinion shift is bounded using the hyperbolic tangent function: $\Delta o_i(m) = \tanh(\omega_i(m))$.

A distinctive feature of our model is that the attraction width is not fixed but dynamically adjusted based on the agent's current conviction and the relationship to the encountered message. This models motivated reasoning \cite{kunda_case_1990}, where an agent's existing beliefs influence how they process new information.

The effective attraction width is calculated as:
\begin{align}
    \sigma_{\text{eff}}^2(o_i, o_m) = \left( \sigma_{\text{base}} \cdot f_{\text{scale}}(o_i, o_m) \right)^2 \label{eq:effective_width_sq_main}
\end{align}
where $\sigma_{\text{base}}$ represents an agent's \emph{tolerance for disagreement}—technically, the baseline width of their "latitude of acceptance" \cite{sherif_attitude_1965}. Since agents operate on opinions in $[-1, 1]$, this parameter ranges from 0 to 2: $\sigma_{\text{base}} = 0$ creates agents who reject any deviation from their exact position, while $\sigma_{\text{base}} = 2$ creates agents who assimilate across the entire opinion spectrum. Opinions falling within this tolerance range are assimilated; those outside trigger reactance. $f_{\text{scale}}$ is a scaling factor that modulates this width based on the agent's conviction and the concordance between agent and message opinions.

The scaling factor significantly narrows the attraction width for highly convicted agents when they encounter ideologically aligned but more moderate messages. This mechanism models psychological resistance to moderating influences from one's own side—a highly convicted liberal, for instance, might reject a more moderate liberal message as insufficiently committed. By shrinking the effective attraction width in these scenarios, the model makes repulsion more likely even for relatively similar opinions, preventing agents with extreme views from being easily moderated by their own side.

When opinions are discordant (opposite sides of the spectrum), the effective width remains close to the baseline, allowing normal assimilation-repulsion dynamics to operate. This asymmetric treatment captures the psychological reality that people are often more resistant to moderation from allies than to opposition from adversaries. The full mathematical derivation is available in \cref{app:opinion-dynamics}.

\subsection{Communicative Actions}

Distinct from internal opinion evolution, our framework explicitly models observable communicative behaviors. Drawing on theories of social interaction, we assume engagement is driven by two competing motivations: the desire to affirm in-group identity (\emph{homophily}) and the desire to challenge out-group views (\emph{antagonism}).

These motivations are captured by a \emph{concordance trigger} ($P_{\text{con}}$) that activates on opinion alignment, and a \emph{discordance trigger} whose base probability ($P_{\text{dis}}$) is scaled by an agent's intrinsic \emph{propensity for discordant reaction ($p_{\text{dis}}$)}. This dual-trigger model proves highly flexible (see \cref{fig:opinion-interaction-functions}b): \emph{likes} can be modeled as low-effort agreement with $p_{\text{dis}} \approx 0$ (making the discordance term vanish and leaving only the homophilic trigger), while argumentative \emph{comments} involve both concordance and antagonism components with non-zero $p_{\text{dis}}$.

Assuming these triggers are independent, the total probability of interaction is based on their union. This is calculated by summing the two trigger probabilities and then subtracting their joint probability to avoid double-counting the (extremely rare) cases where both triggers might activate simultaneously, yielding the final formula:

\begin{align}
    P_{\text{int}}(A_i, m) = p_{\text{base}} \left( P_{\text{con}} + p_{\text{dis}} \cdot P_{\text{dis}} \right. \nonumber \\
    \left. - p_{\text{dis}} \cdot P_{\text{con}} \cdot P_{\text{dis}} \right)
    \label{eq:interaction_probability_main} 
\end{align}

This allows us to simulate not just whether agents engage, but the nature and function of that engagement (see \cref{app:communicative-actions} for the full derivation of $P_{\text{con}}$ and $P_{\text{dis}}$).

\subsection{Network Structure and Information Flow}

Agent interactions unfold within a co-evolving directed social network, $G(t)$, that governs information flow. We initialize the network using a hybrid model that blends principles from \emph{Stochastic Block Models (SBMs)} and \emph{Preferential Attachment}. This creates realistic initial conditions with both homophilous community clustering (based on opinion) and heterogeneous "hub" structures, reflecting the "rich-get-richer" dynamics common in online networks. 

The network then evolves as agents rewire their connections based on opinion similarity, modeling the process of social sorting that can lead to echo chambers. Information propagation is managed by a recommendation system that simulates algorithmic curation, balancing exposure to content from an agent's existing network (the "filter bubble") with content discovered from outside (controlled by a discovery rate, $\delta_{\text{rec}}$). This key parameter allows systematic exploration of how algorithmic design choices affect polarization dynamics (see \cref{app:network-evolution}).

\subsection{Simulation Workflow}

The simulation progresses through an iterative workflow over discrete time steps, creating interconnected feedback loops (see \cref{alg:social-simulation} in \cref{app:simulation-workflow} for detailed pseudocode). Each time step represents a full cycle of social interaction. The process begins with \emph{Content Generation}, where a subset of agents post new messages reflecting their current states. Next, the \emph{Recommendation System} curates a personalized feed for every agent, determining their information exposure for that step. Agents then \emph{Interact} with this content, performing actions like liking or commenting based on the probabilistic models. Following these interactions, agents undergo \emph{Opinion Updates}, where their internal beliefs shift in response to the messages they perceived. Finally, agents may perform \emph{Network Updates}, rewiring their follow connections based on recent encounters.

This complete cycle—where opinions shape communication, which in turn shapes opinions and network ties—repeats, allowing the framework to model the emergent, multi-level dynamics of phenomena like polarization.

\begin{figure*}[htbp]
    \centering

    \setlength{\plotaxesheightthree}{6cm} % Use the specific length for 3-plot figures

    \begin{minipage}[b]{\plotwidththree} % Use specific width for 3-plot figures
        \centering
        \includegraphics[width=\linewidth]
            {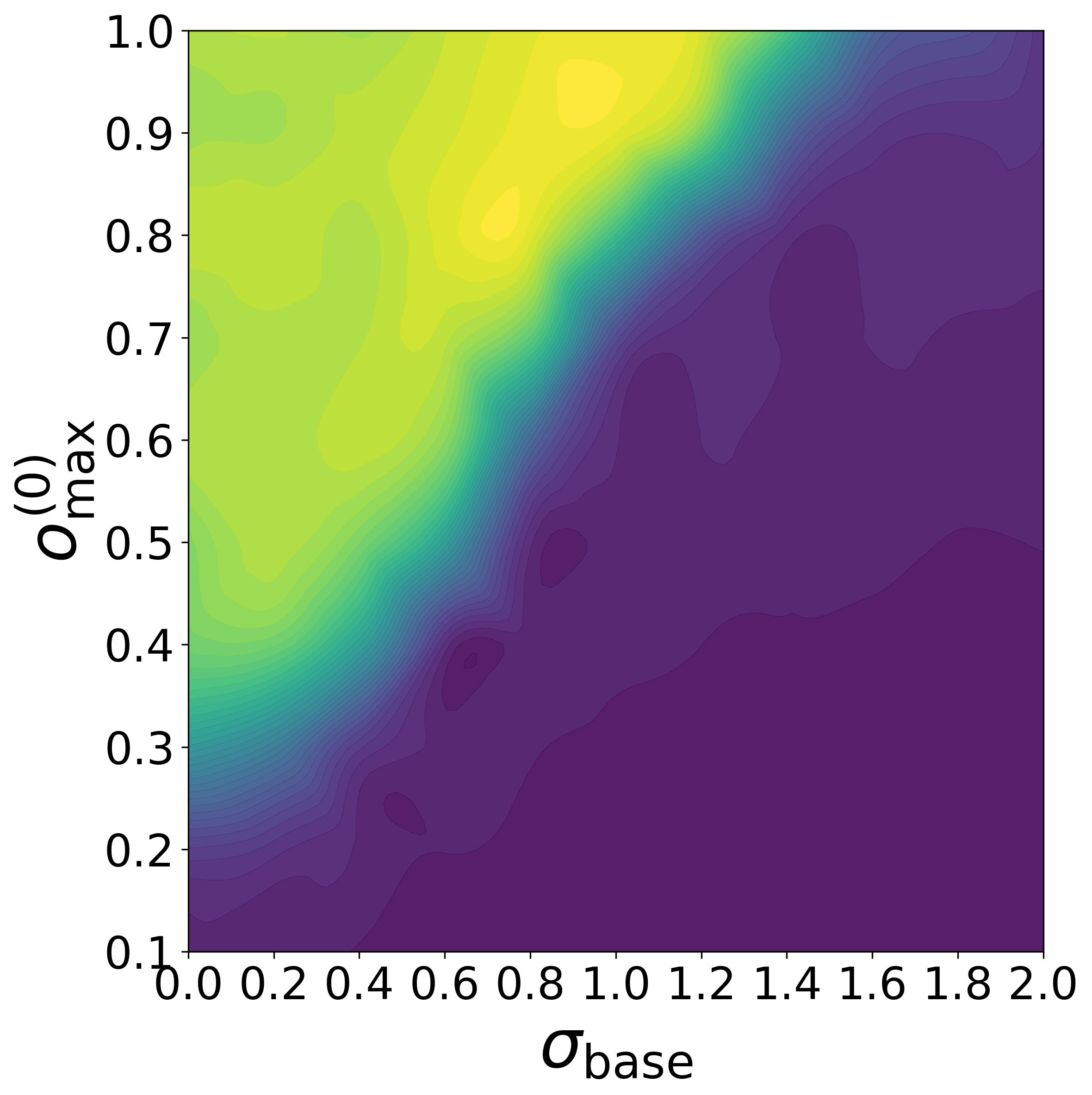}
        \centerline{(a) $\sigma_{\text{base}}$ vs $o_{\text{max}}^{(0)}$}
    \end{minipage}%
    \interspace
    \begin{minipage}[b]{\plotwidththree}
        \centering
        \includegraphics[width=\linewidth]
            {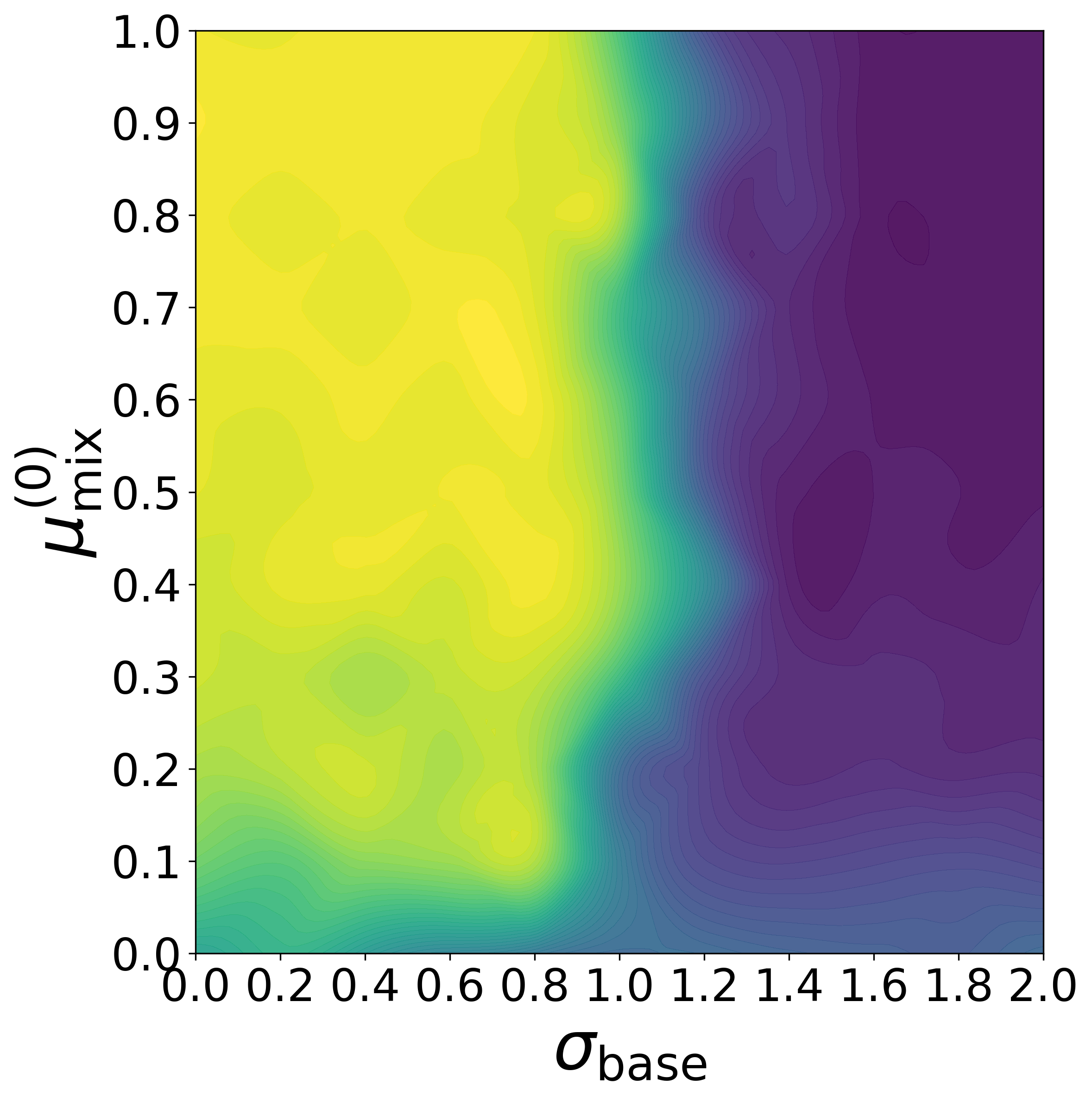}
        \centerline{(b) $\sigma_{\text{base}}$ vs $\mu_{\text{mix}}^{(0)}$}
    \end{minipage}%
    \interspace
    \begin{minipage}[b]{\plotwidththree}
        \centering
        \includegraphics[width=\linewidth]
            {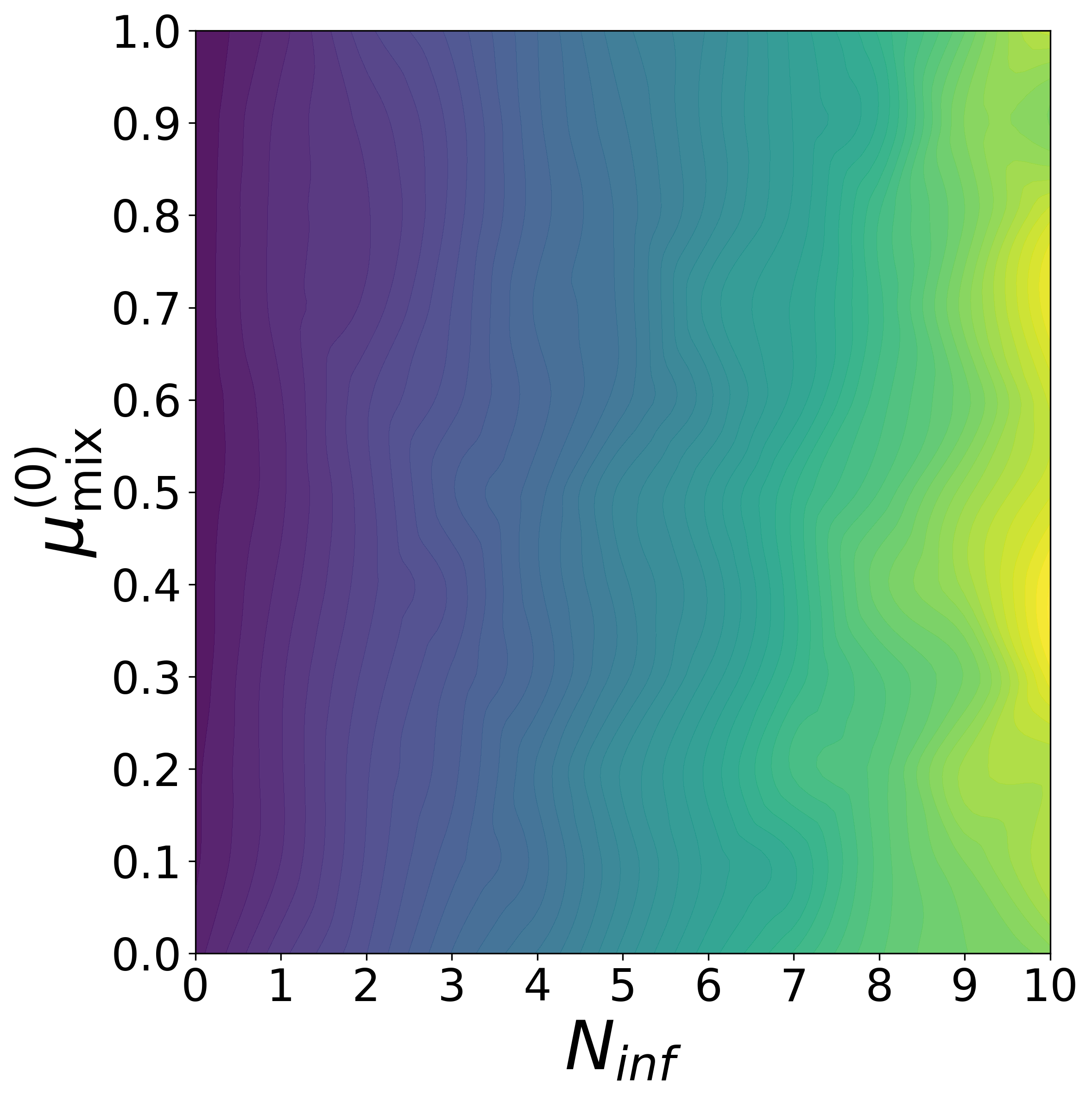}
        \centerline{(c) $N_{\text{inf}}$ vs $\mu_{\text{mix}}^{(0)}$}
    \end{minipage}%
    \interspace
    \begin{minipage}[b]{\cbarwidththree}
        \centering
        \includegraphics[height=\plotaxesheightthree, width=\linewidth, keepaspectratio]
            {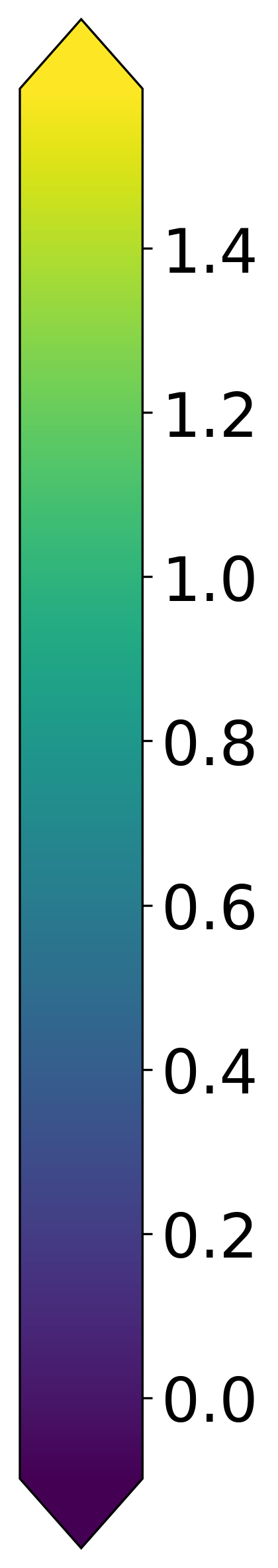}
        \vspace{4.5ex} % align colorbar top with panel tops (compensate for panel labels)
    \end{minipage}

    \caption{Final mean polarization (Esteban-Ray index; yellow=high, purple=low) across parameter spaces. Plots show polarization varying with: 
    \textbf{(a)} assimilation-repulsion balance ($\sigma_{\text{base}}$) vs. initial opinion spread ($o_{\text{max}}^{(0)}$); 
    \textbf{(b)} $\sigma_{\text{base}}$ vs. initial network mixing ($\mu_{\text{mix}}^{(0)}$); and 
    \textbf{(c)} number of extremists ($N_{\text{inf}}$) vs. $\mu_{\text{mix}}^{(0)}$. 
    Key drivers increasing polarization include strong repulsion (low $\sigma_{\text{base}}$), high initial diversity ($o_{\text{max}}^{(0)}$), moderate network mixing ($\mu_{\text{mix}}^{(0)}$), and a high number of influencers.}
    \label{fig:polarization-heatmaps}
\end{figure*} % Include relevant heatmaps, e.g., Fig 2b, 2c

\section{Computational Experiments}
\label{sec:evaluation}

Having established our simulation framework, we now demonstrate its capabilities through systematic computational experiments investigating the drivers of online polarization. Each experiment involves 50 agents interacting over 100 time steps, with specific parameters varied to isolate different mechanisms. All results represent averages across 5 simulation runs per parameter constellation to ensure statistical robustness, with each run employing freshly generated network structures and randomized initial conditions to ensure observed patterns reflect genuine dynamics rather than artifacts of specific realizations.

Polarization is measured using the Esteban-Ray polarization index: $P = K \sum_{i,j} \frac{1}{n^{2+\alpha}} |o_i - o_j|$, where $o_i$ and $o_j$ are agent opinions, $n$ is population size, $\alpha = 1.6$, and $K = 1$ is a normalization constant. This index captures opinion distance between agents weighted by population frequency, appropriately measuring polarization in contexts with multimodal distributions including the trimodal outcomes observed in our experiments \cite{esteban_measurement_1994}.

Our experimental design progresses systematically from structural to communicative dimensions. We first examine how individual psychology, network topology, and algorithmic design interact to shape opinion and network dynamics under both static and co-evolutionary conditions (\cref{subsec:eval_structure,subsec:eval_coevolution}). We then leverage our framework's LLM capabilities to analyze how polarization transforms communication content and how individual differences influence message perception (\cref{subsec:eval_communication_content_and_perception}). Finally, we investigate how polarized environments shape different forms of social engagement, from supportive interactions to antagonistic exchanges (\cref{subsec:eval_engagement_dynamics}). Detailed experimental setups and additional results are provided in \cref{app:computational_experiments}.

\begin{figure*}[htbp]
    \centering

    \setlength{\plotaxesheightthree}{6cm} % Use the specific length for 3-plot figures

    \begin{minipage}[b]{\plotwidththree} % Use specific width for 3-plot figures
        \centering
        \includegraphics[width=\linewidth]
            {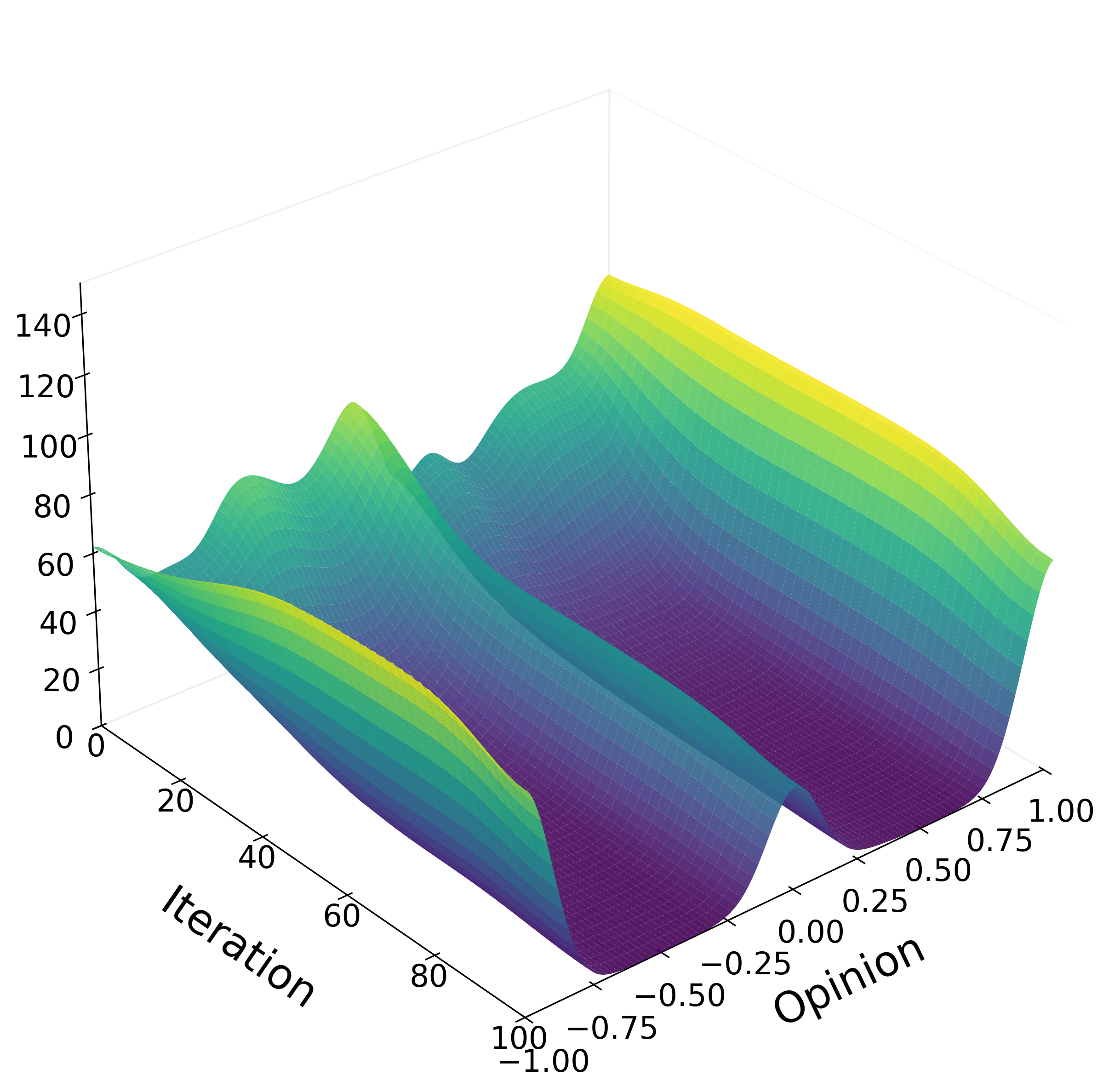}
        \centerline{(a) $\sigma_{\text{base}} = 0.0$}
    \end{minipage}%
    \interspace
    \begin{minipage}[b]{\plotwidththree}
        \centering
        \includegraphics[width=\linewidth]
            {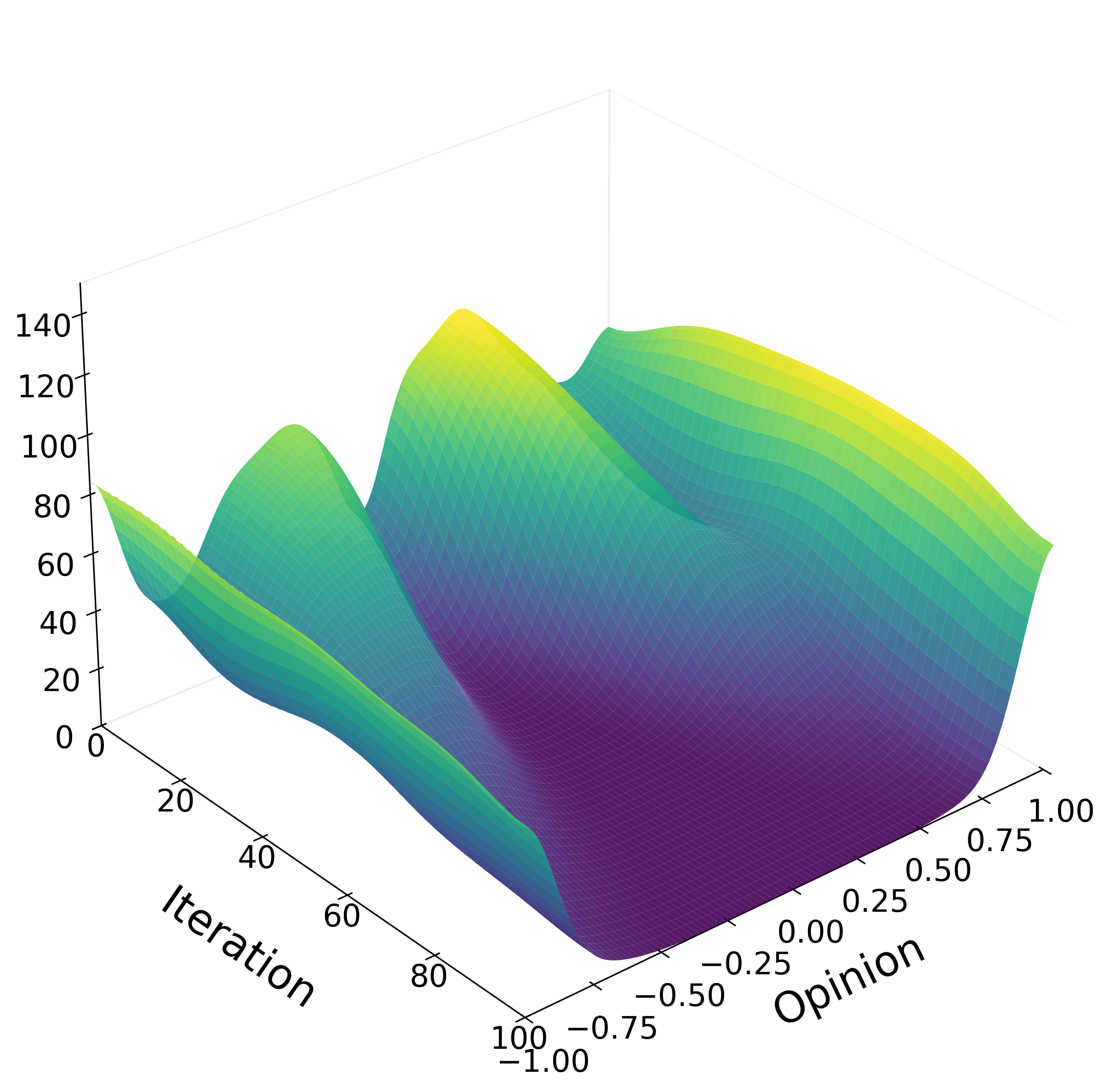}
        \centerline{(b) $\sigma_{\text{base}} = 1.0$}
    \end{minipage}%
    \interspace
    \begin{minipage}[b]{\plotwidththree}
        \centering
        \includegraphics[width=\linewidth]
            {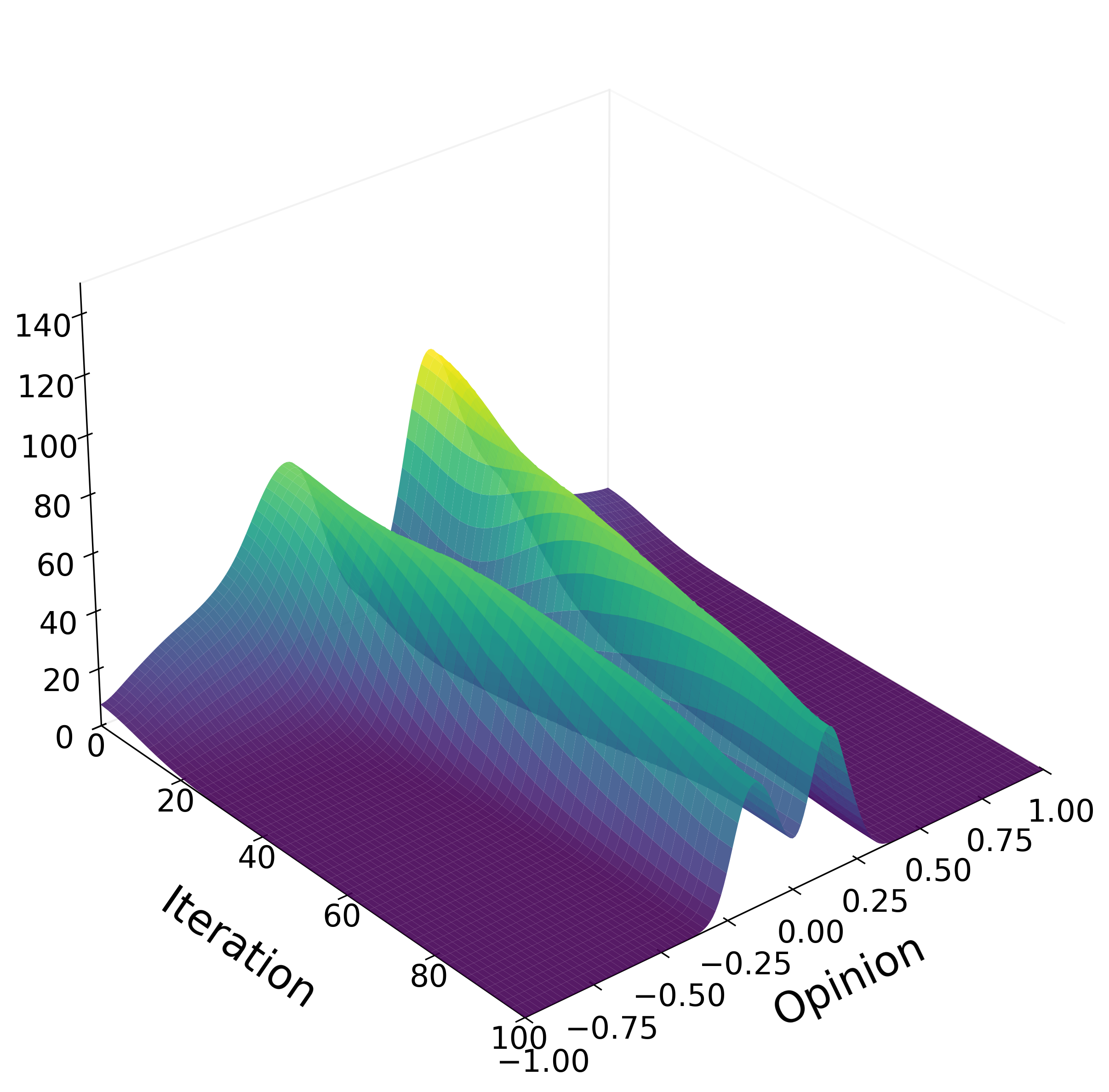}
        \centerline{(c) $\sigma_{\text{base}} = 2.0$}
    \end{minipage}%
    \interspace
    \begin{minipage}[b]{\cbarwidththree}
        \centering
        \includegraphics[height=\plotaxesheightthree, width=\linewidth, keepaspectratio]
            {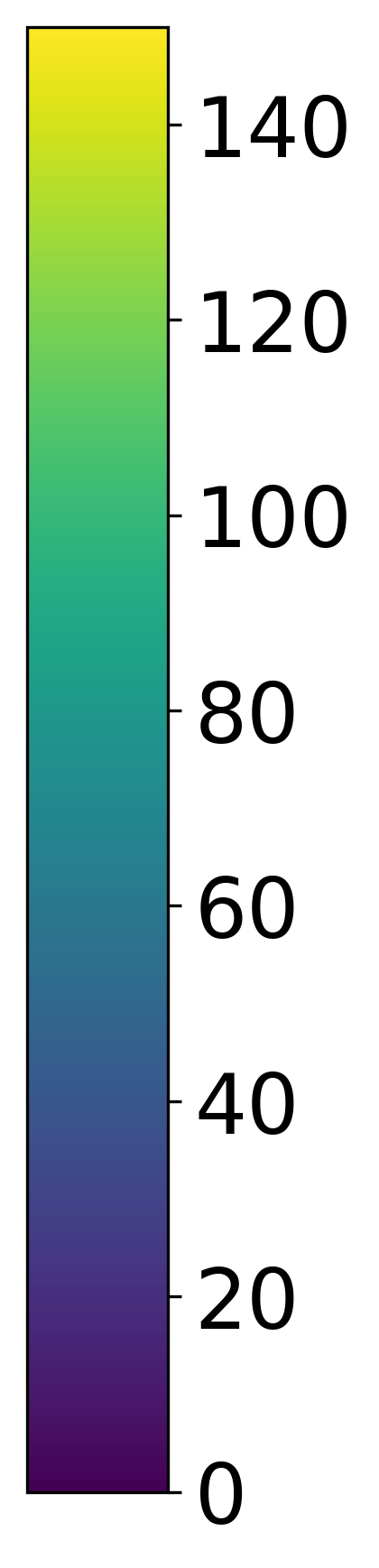}
        \vspace{4.5ex} % align colorbar top with panel tops (compensate for panel labels)
    \end{minipage}

    \caption{Agent opinion distribution (Kernel Density Estimates; height/color = density) evolution over simulation time (y-axis) vs. opinion (x-axis), for fixed initial spread ($o_{\text{max}}^{(0)}=1.0$) and varying assimilation-repulsion balance ($\sigma_{\text{base}}$). 
    \textbf{(a)} Pure repulsion ($\sigma_{\text{base}}=0.0$) leads to a trimodal distribution. 
    \textbf{(b)} Balanced assimilation/repulsion ($\sigma_{\text{base}}=1.0$) produces strong bipolarization. 
    \textbf{(c)} Strong assimilation ($\sigma_{\text{base}}=2.0$) leads to consensus (unimodal distribution).}
    \label{fig:kdes-center-limit}
\end{figure*}

\subsection{Polarization Drivers: Psychology, Diversity, and Network Structure}
\label{subsec:eval_structure}

Understanding when and how societies become polarized requires examining the complex interplay between individual psychology, social structure, and initial conditions. Our simulations reveal that collective opinion outcomes depend critically on three specific factors: agents' tolerance for disagreement ($\sigma_{\text{base}}$), initial opinion diversity, and network connectivity between opposing camps. Timeline analyses with confidence intervals for all experiments are provided in \cref{app:computational_experiments}.

\subsubsection{Initial Opinion Diversity}
\label{subsubsec:balance}

We first explored how agents' psychological tolerance for disagreement—their "latitude of acceptance" \cite{sherif_attitude_1965}—interacts with the initial spread of opinions in the population. The parameter $\sigma_{\text{base}}$ governs this fundamental balance: low values create agents who reject nearly any differing opinion, while high values create agents who assimilate broadly. To isolate these psychological dynamics, we held network structure constant at moderate integration ($\mu_{\text{mix}}^{(0)} = 0.5$), representing a realistic balance between community clustering and cross-cutting ties that allows both within-group and between-group interactions to influence opinion dynamics. The systematic effects of varying this network parameter are examined in \cref{subsubsec:structure}.

The relationship between tolerance and polarization proves highly nonlinear and depends critically on initial diversity (\cref{fig:polarization-heatmaps}a). To understand this complex landscape, we examine three prototypical cases representing the natural parameter range (0-2) established in our methods that capture distinct edge conditions. Under pure repulsion ($\sigma_{\text{base}}=0$), polarization occurs only when sufficient initial diversity exists ($o_{\text{max}}^{(0)} \geq 0.4$, confirmed across multiple simulation runs), producing characteristic trimodal distributions—two extreme peaks flanking a persistent moderate center (\cref{fig:kdes-center-limit}a). This trimodal structure emerges from the interaction between pure repulsion and biased network connectivity: moderate agents are predominantly exposed to extreme content from their own side, causing repulsion toward the center. However, when they cross the neutral line (opinion = 0), they become exposed to extremists from the opposite side, again causing repulsion back toward the center. This creates stable oscillations around neutrality, sustaining a substantial moderate camp. This center-repulsion mechanism aligns with theoretical accounts of polarization as an inherent communicative order where moderates may be alienated by the dogmatic discourse of their own ideological camp \cite{kumkar_polarisierung_2025}.

\begin{figure*}[h]
    \centering

    \setlength{\totalgridheight}{10.5cm} % <--- *** ADJUST THIS CRITICAL VALUE ***

    \begin{minipage}[t]{\gridpagewidth} % Uses command from preamble
        \centering

        \begin{minipage}[b]{\plotwidthgrid} % Uses command from preamble
            \centering
            \includegraphics[width=\linewidth]
                {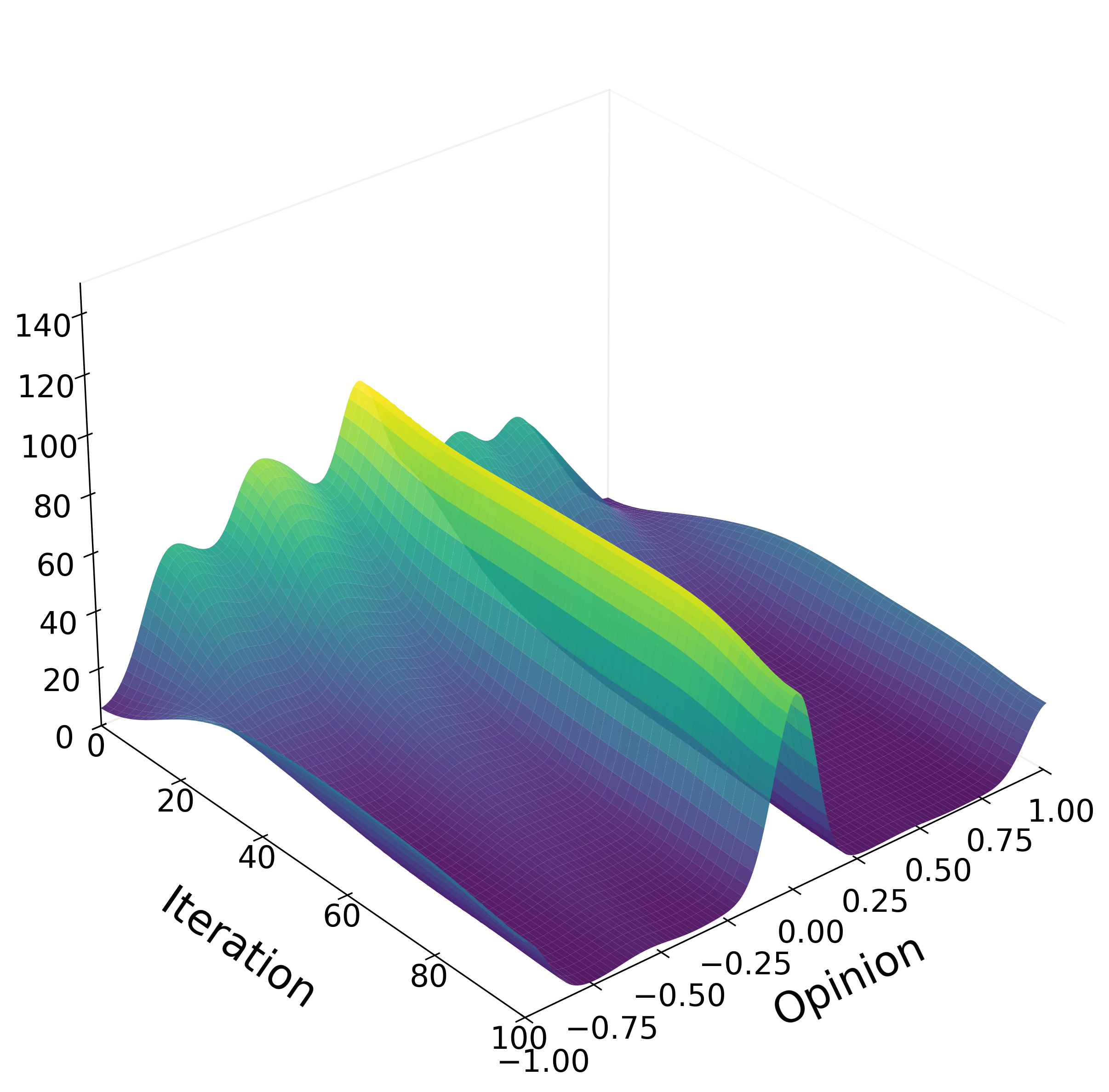}
            \centerline{(a) $\mu_c=0.0, \sigma_{base}=0.0$}
        \end{minipage}%
        \hfill
        \begin{minipage}[b]{\plotwidthgrid}
            \centering
            \includegraphics[width=\linewidth]
                {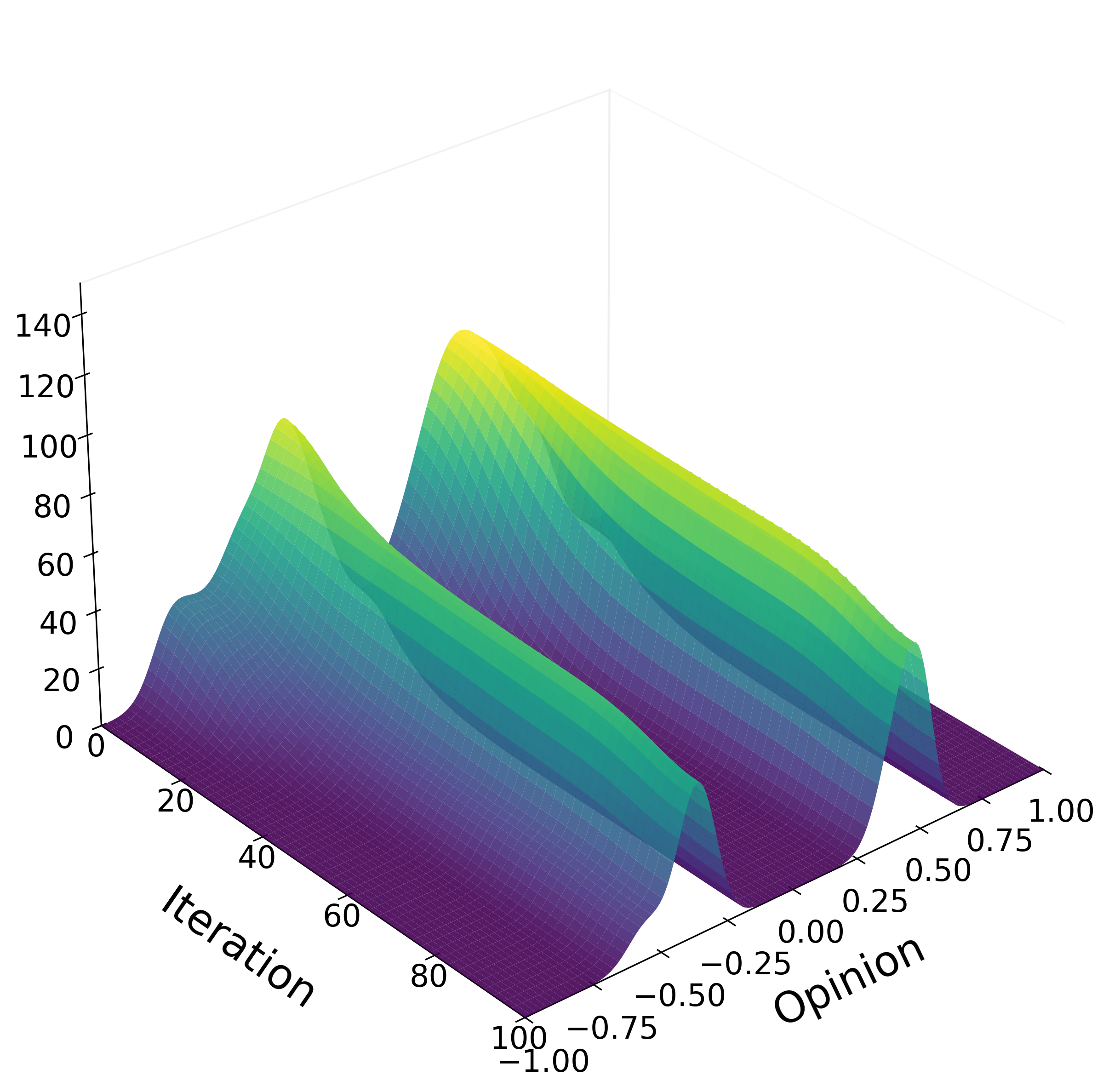}
            \centerline{(b) $\mu_c=0.0, \sigma_{base}=1.0$}
        \end{minipage}%
        \hfill
        \begin{minipage}[b]{\plotwidthgrid}
            \centering
            \includegraphics[width=\linewidth]
                {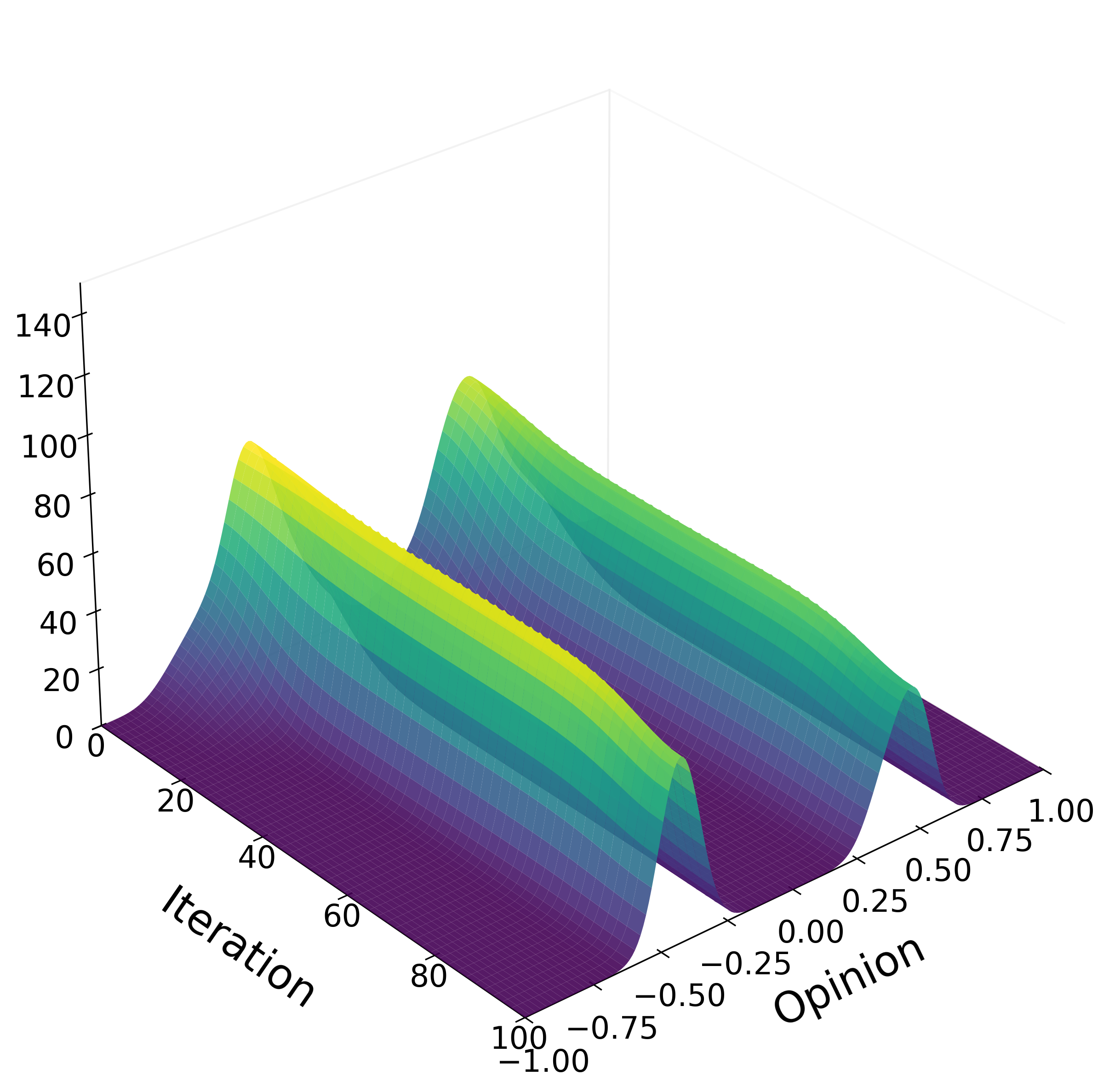}
            \centerline{(c) $\mu_c=0.0, \sigma_{base}=2.0$}
        \end{minipage}
        
        \vspace{2ex} % Vertical space between rows

        \begin{minipage}[b]{\plotwidthgrid}
            \centering
            \includegraphics[width=\linewidth]
                {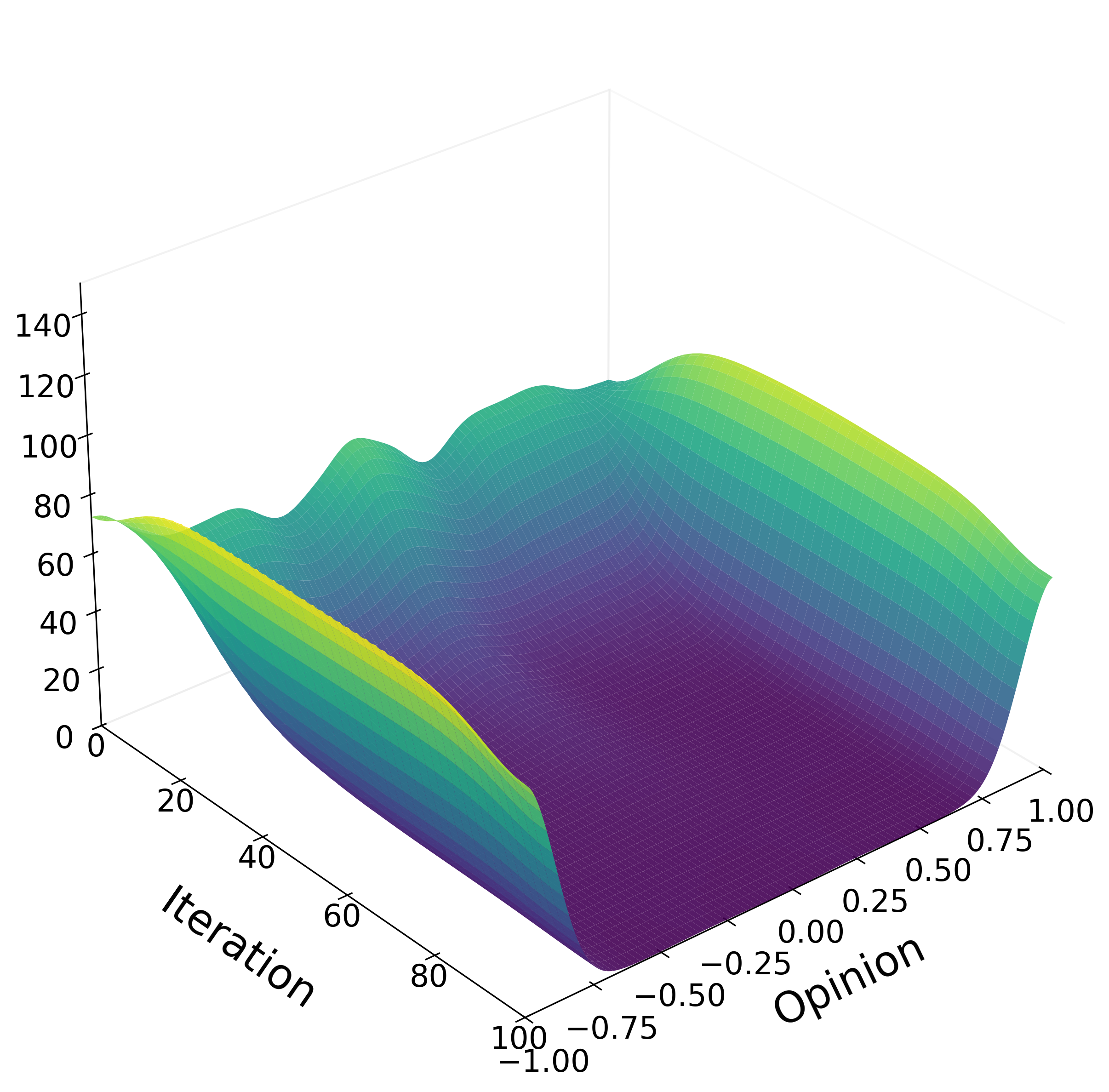}
            \centerline{(d) $\mu_c=1.0, \sigma_{base}=0.0$}
        \end{minipage}%
        \hfill
        \begin{minipage}[b]{\plotwidthgrid}
            \centering
            \includegraphics[width=\linewidth]
                {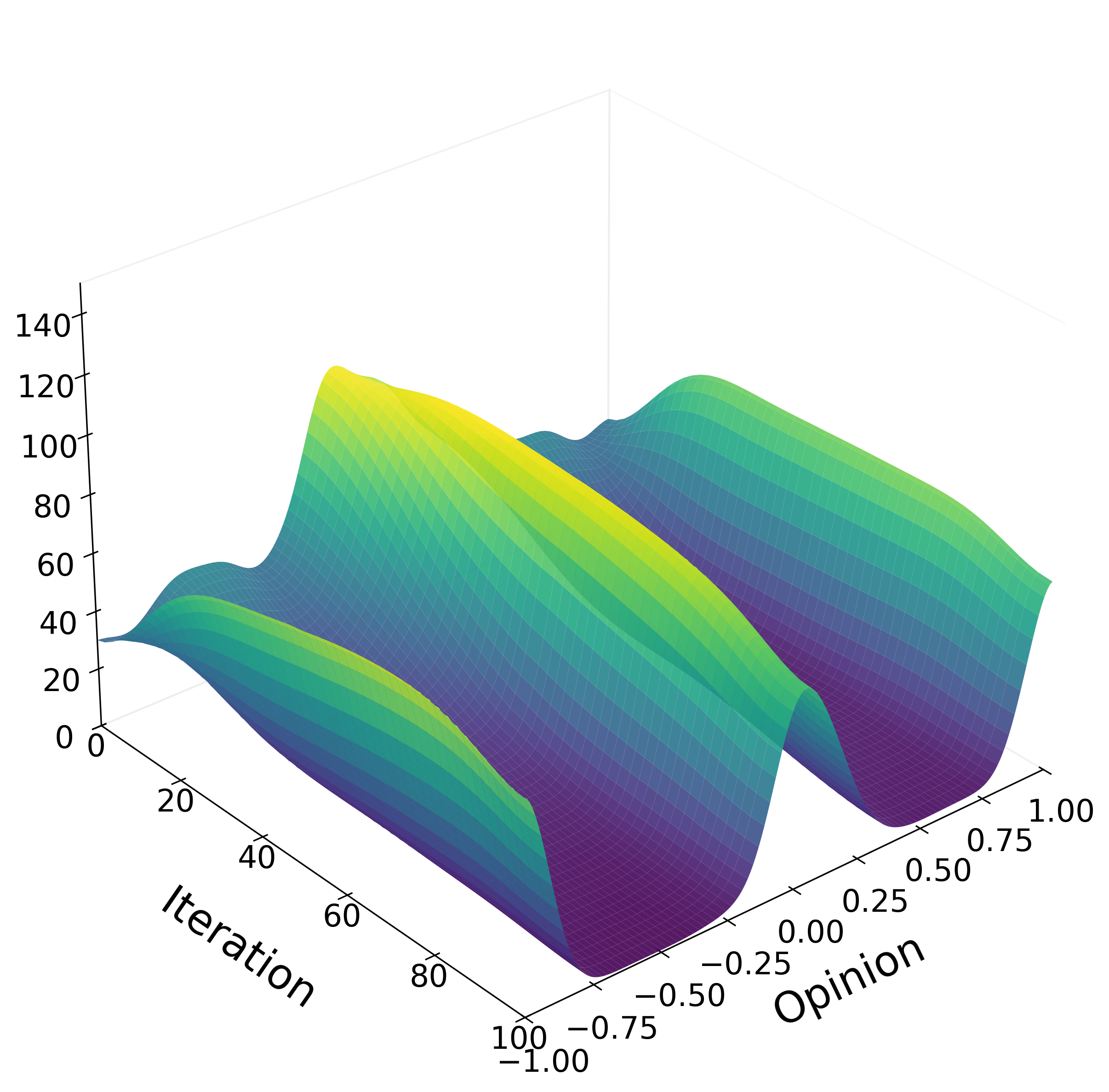}
            \centerline{(e) $\mu_c=1.0, \sigma_{base}=1.0$}
        \end{minipage}%
        \hfill
        \begin{minipage}[b]{\plotwidthgrid}
            \centering
            \includegraphics[width=\linewidth]
                {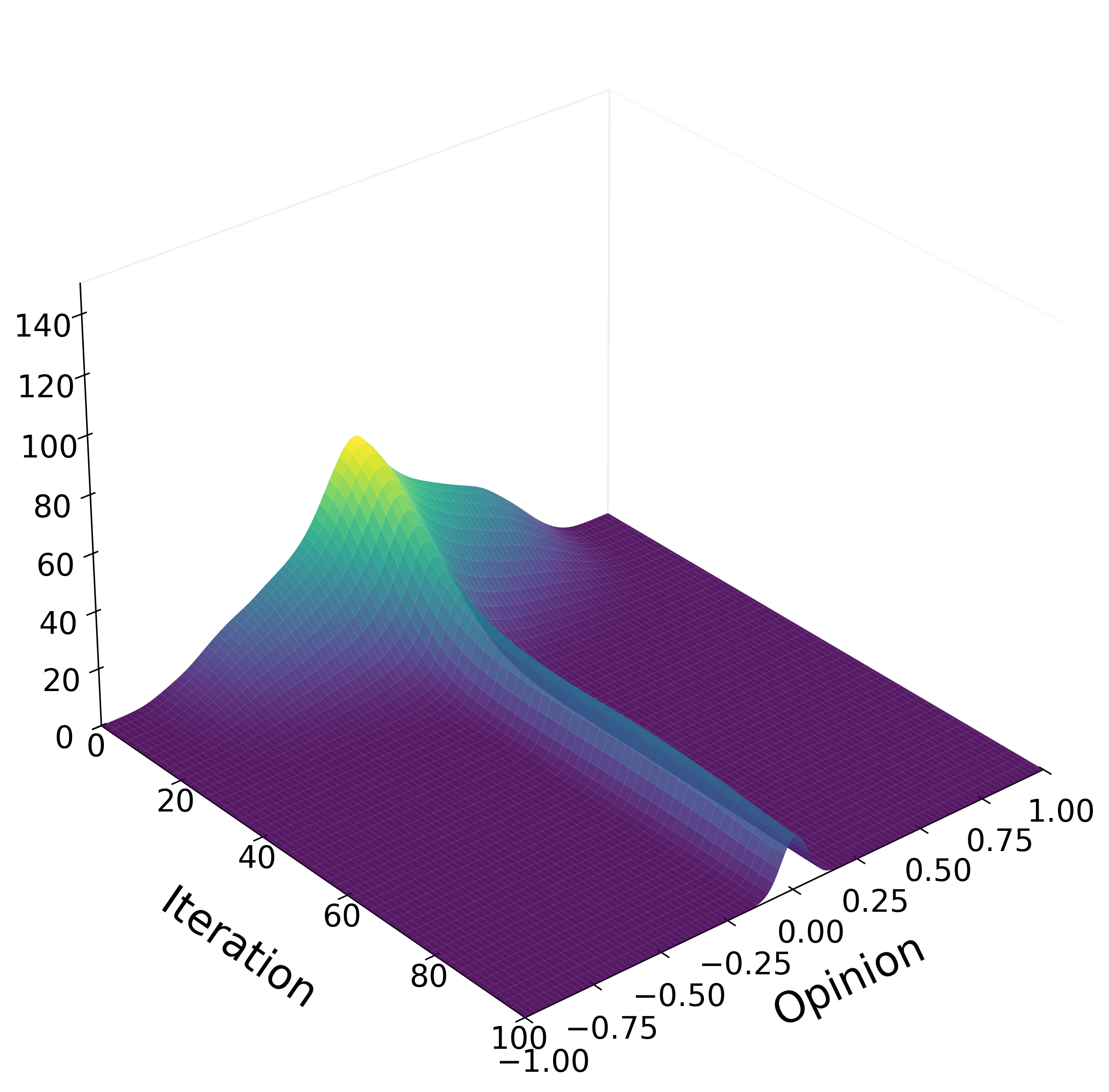}
            \centerline{(f) $\mu_c=1.0, \sigma_{base}=2.0$}
        \end{minipage}

    \end{minipage}%
    \hfill
    \begin{minipage}[t]{\cbarwidthgrid} % Uses command from preamble
        \vspace{0pt}
        \includegraphics[height=\totalgridheight, width=\linewidth, keepaspectratio]
            {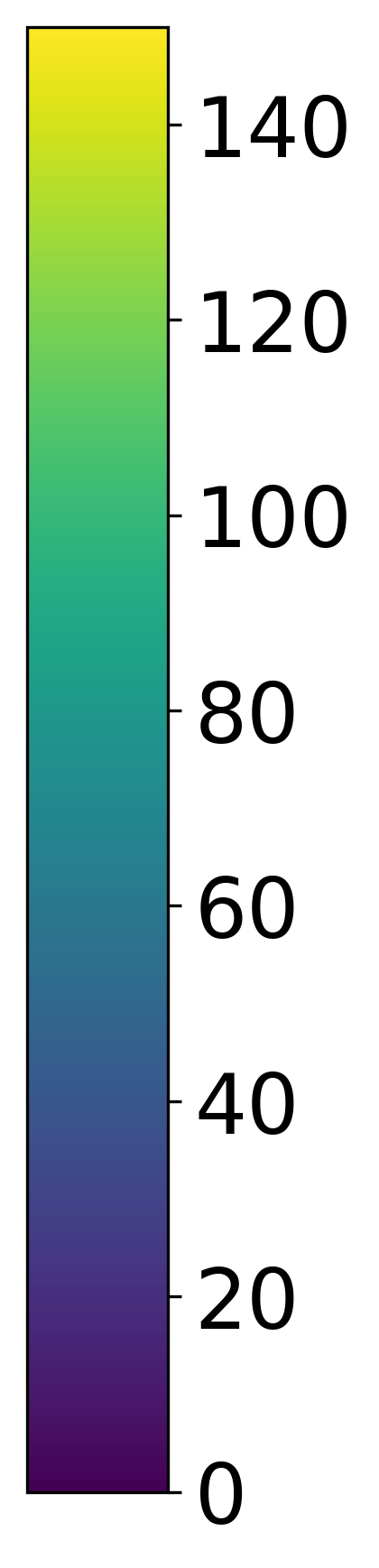}
    \end{minipage}

    \caption{Agent opinion distribution (KDE) evolution comparing effects of assimilation-repulsion ($\sigma_{base}$) and initial network mixing ($\mu_{\text{mix}}^{(0)}$). 
    Top row: No mixing ($\mu_{\text{mix}}^{(0)}=0.0$, echo chambers) shows a trimodal shape for pure repulsion \textbf{(a)} and moderate bimodal distributions for both balanced assimilation/repulsion \textbf{(b)} and strong assimilation \textbf{(c)}. 
    Bottom row: Full mixing ($\mu_{\text{mix}}^{(0)}=1.0$) allows $\sigma_{base}$ to drive outcomes: pure repulsion \textbf{(d)} leads to strong bipolarization, balanced assimilation/repulsion \textbf{(e)} yields a trimodal distribution, and strong assimilation \textbf{(f)} forces consensus.}
    \label{fig:kdes-center-mixing}
\end{figure*}

However, the \emph{highest polarization scores} emerge under intermediate tolerance ($\sigma_{\text{base}} \approx 0.8$-$1.2$), but only when combined with high initial diversity ($o_{\text{max}}^{(0)} \ge 0.9$). This "sweet spot" produces clean bipolarization (\cref{fig:kdes-center-limit}b), where the political center completely empties as agents sort into two opposing camps. Under these conditions, pre-existing extreme factions provide stable attractors: intermediate tolerance enables selective attraction to similar agents within the same broad ideological camp while still repelling opposing extremists, allowing agents to consolidate around opposing poles rather than exhibiting the center-oscillation dynamics observed under pure repulsion.

Crucially, the same intermediate tolerance parameters produce fundamentally different outcomes under lower initial diversity conditions. With moderate or low initial diversity ($o_{\text{max}}^{(0)} \le 0.7$), the system converges to consensus rather than polarization. Without sufficient pre-existing extreme factions, the same tolerance parameters drive toward consensus as moderate agents find common ground. This bifurcation reveals the conditional nature of intermediate tolerance: outcomes depend critically on whether the initial distribution contains extreme factions that can serve as ideological attractors.

Conversely, high tolerance ($\sigma_{\text{base}} \geq 1.4$) consistently drives consensus across all initial diversity conditions (\cref{fig:kdes-center-limit}c), as agents' broad latitude of acceptance enables them to assimilate diverse viewpoints. The tolerance-diversity interaction thus reveals that extreme polarization requires a precise balance: sufficient psychological flexibility to enable both in-group attraction and out-group repulsion \cite{dandekar_biased_2013}, combined with pre-existing extreme factions that can serve as ideological attractors for moderate agents.

\subsubsection{Network Structure: Integration Versus Segregation}
\label{subsubsec:structure}

We examined how network connectivity between opposing communities mediates polarization dynamics by varying the inter-community mixing parameter $\mu_{\text{mix}}^{(0)}$ from complete segregation (0.0) to full integration (1.0). To isolate network effects, we held initial opinion spread at a moderate level ($o_{\text{max}}^{(0)}=0.7$), representing societies with substantial but not extreme initial disagreement where pre-existing factions exist but most agents hold moderate positions.

\begin{figure*}[htbp] % Use placement specifiers like h, t, b, p
    \centering

    \setlength{\plotaxesheighttwo}{7.0cm} % <--- *** ADJUST THIS CRITICAL VALUE ***

    \begin{minipage}[b]{\plotwidthtwo} % Align by BOTTOM [b]
        \centering
        \includegraphics[width=\linewidth]
            {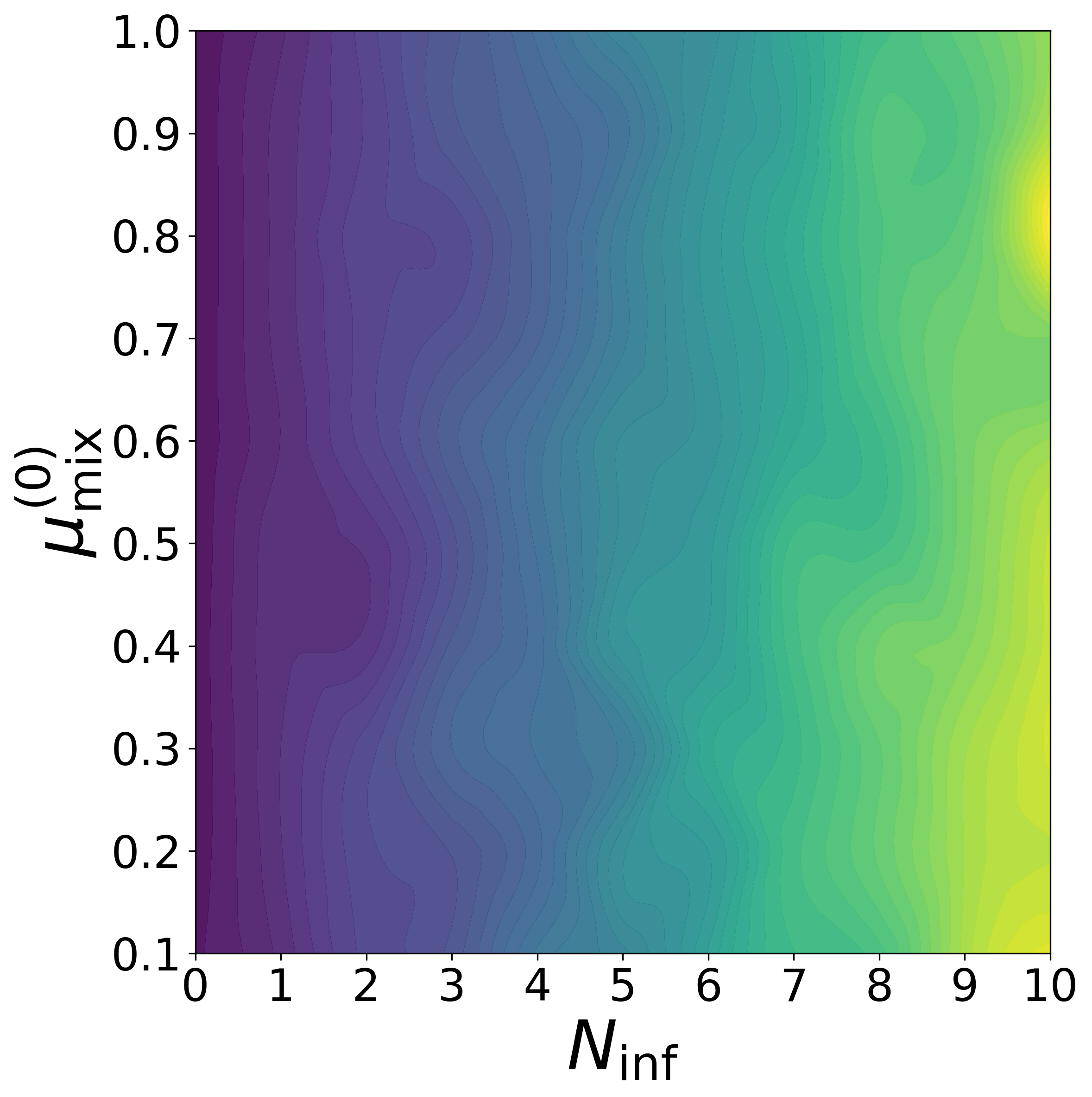}
        \centerline{(a) $\mu_{\text{mix}}^{(0)} vs N_{\text{inf}}$} % REPLACED \subcaption
    \end{minipage}% <--- No space
    \hfill
    \begin{minipage}[b]{\plotwidthtwo} % Align by BOTTOM [b]
        \centering
        \includegraphics[width=\linewidth]
             {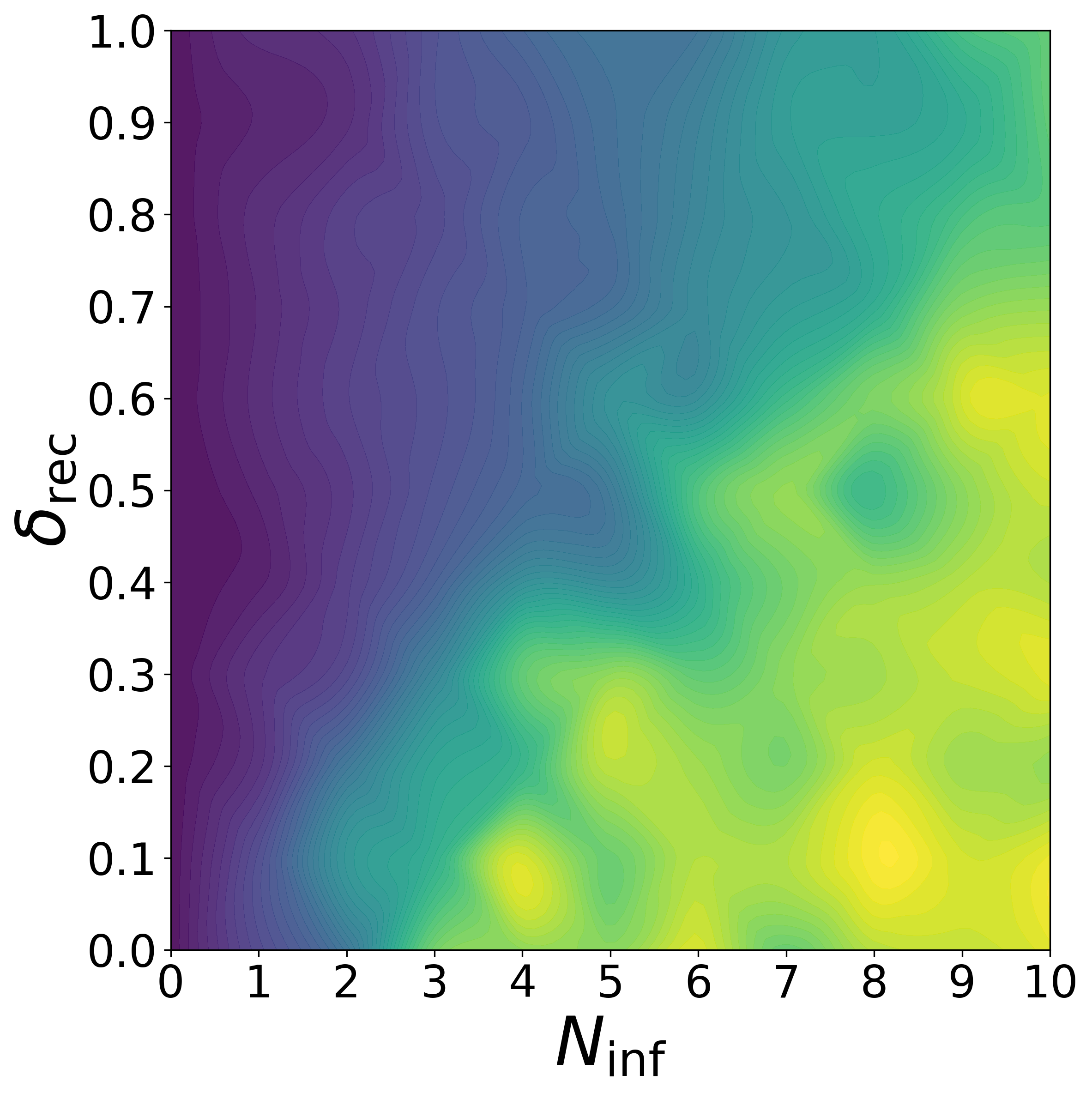}
        \centerline{(b) $\delta_{\text{rec}} vs N_{\text{inf}}$} % REPLACED \subcaption
    \end{minipage}% <--- No space
    \hfill
    \begin{minipage}[b]{\cbarwidthtwo} % Align by BOTTOM [b]
        \centering
        \includegraphics[height=\plotaxesheighttwo, width=\linewidth, keepaspectratio]
            {figures/heatmaps/polarization/colorbar.png}
        \vspace{4.5ex} % Adjust if bottoms don't quite align perfectly
    \end{minipage}

    \caption{Final mean polarization (yellow=high) landscapes under co-evolution. Plots show polarization varying with: 
    \textbf{(a)} initial network mixing ($\mu_{\text{mix}}^{(0)}$) vs. number of influencers ($N_{\text{inf}}$); and 
    \textbf{(b)} discovery rate ($\delta_{\text{rec}}$) vs. $N_{\text{inf}}$. 
    Polarization increases strongly with $N_{\text{inf}}$. Notably, polarization peaks under moderate initial mixing ($\mu_{\text{mix}}^{(0)} \approx 0.2-0.7$) and is highest when discovery rates ($\delta_{\text{rec}}$) are low ($\le 0.1$), contrasting with modularity findings (cf. Figure~\ref{fig:bimodal-unimodal-modularity-heatmaps}).}
    \label{fig:coevolution-polarization-heatmaps} % Original label is kept
\end{figure*}

Our findings reveal a key theoretical insight that challenges conventional echo chamber theories: complete segregation actually \emph{buffers against} extreme polarization (\cref{fig:polarization-heatmaps}b). This counter-intuitive result provides computational evidence for theoretical arguments about polarization as communicative order (detailed in \cref{sec:general_discussion}). The heatmap shows a predominantly linear relationship between tolerance and network connectivity, with tolerance playing the dominant role—except at the extreme of complete segregation ($\mu_{\text{mix}}^{(0)}=0$), where archetypical echo chambers produce only moderate polarization regardless of tolerance settings. The temporal dynamics explain this result: under segregation with pure repulsion ($\sigma_{\text{base}}=0$), agents develop trimodal distributions (\cref{fig:kdes-center-mixing}a) because moderate agents within each community encounter only their own side's extremists, causing them to be repelled toward the center rather than toward opposing extremes. Under wide attraction ($\sigma_{\text{base}} \ge 1.4$), segregated agents converge toward consensus within their respective communities (\cref{fig:kdes-center-mixing}c), creating moderate bifurcation since there is no cross-community contact to drive full consensus.

The highest polarization emerges under full integration ($\mu_{\text{mix}}^{(0)}=1.0$), where cross-community contact enables tolerance settings to drive system-wide outcomes. Under pure repulsion with full integration, agents develop strong bipolarization (\cref{fig:kdes-center-mixing}d) because repulsion strength increases with opinion distance: agents with positive leanings exposed equally to extremists from both camps are more strongly repelled by negative extremists than by positive ones, and vice versa, gradually pushing them toward their respective poles. Under intermediate tolerance with full integration, this produces trimodal rather than bimodal distributions (\cref{fig:kdes-center-mixing}e) because balanced assimilation-repulsion creates complex multi-modal outcomes where some agents consolidate around extremes while others remain moderate. High tolerance predictably drives consensus (\cref{fig:kdes-center-mixing}f) as agents' broad latitude of acceptance overcomes initial differences.

\begin{figure*}[htbp] % Placement specifiers are good
    \centering

    \setlength{\plotaxesheighttwo}{7.0cm} % <--- *** ADJUST THIS CRITICAL VALUE ***

    \begin{minipage}[b]{\plotwidthtwo} % Align by BOTTOM [b]
        \centering
        \includegraphics[width=\linewidth]
            {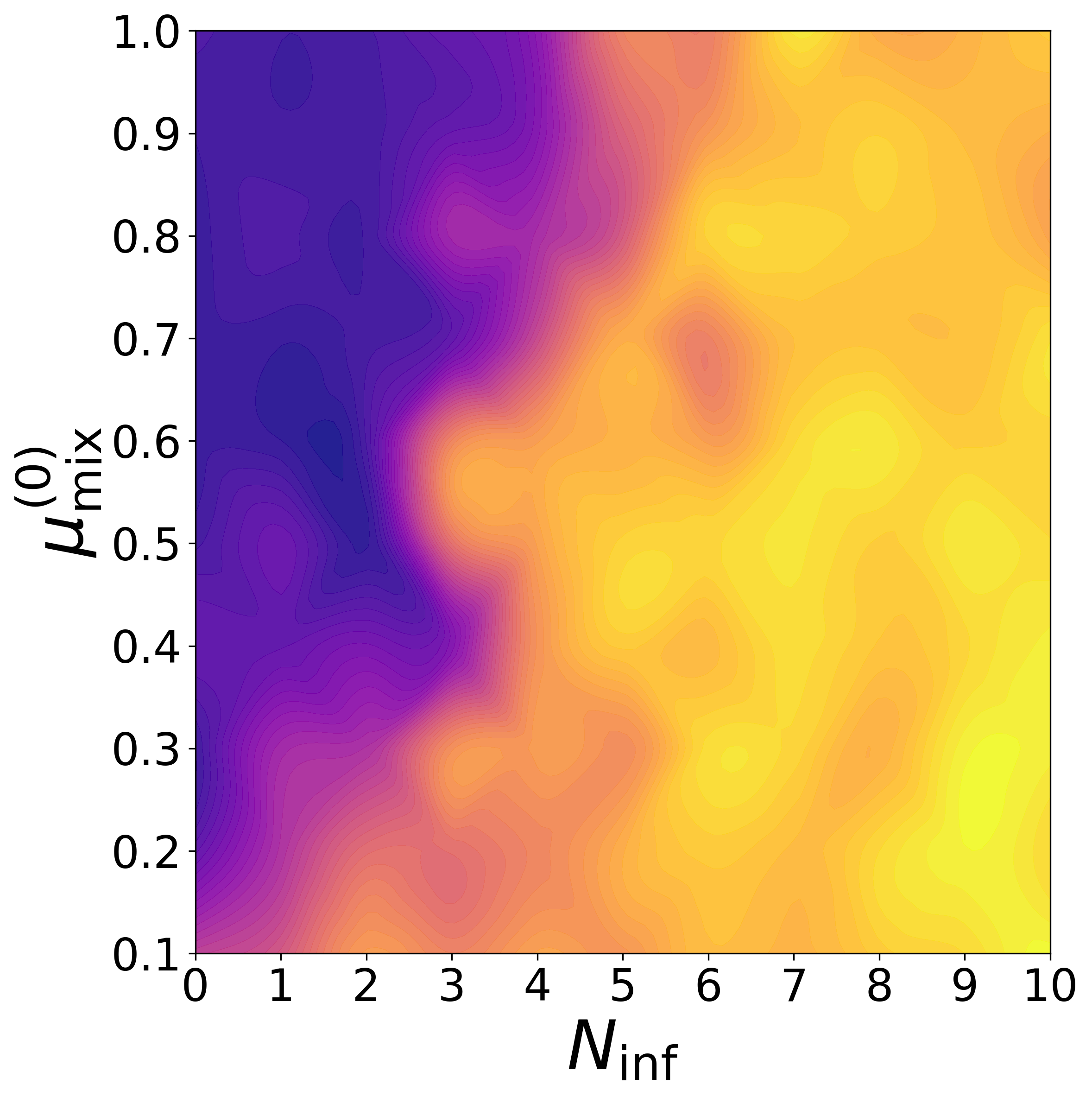}
        \centerline{(a) $\mu_{\text{mix}}^{(0)} vs N_{\text{inf}}$} % REPLACED \subcaption
    \end{minipage}% <--- No space
    \hfill
    \begin{minipage}[b]{\plotwidthtwo} % Align by BOTTOM [b]
        \centering
        \includegraphics[width=\linewidth]
             {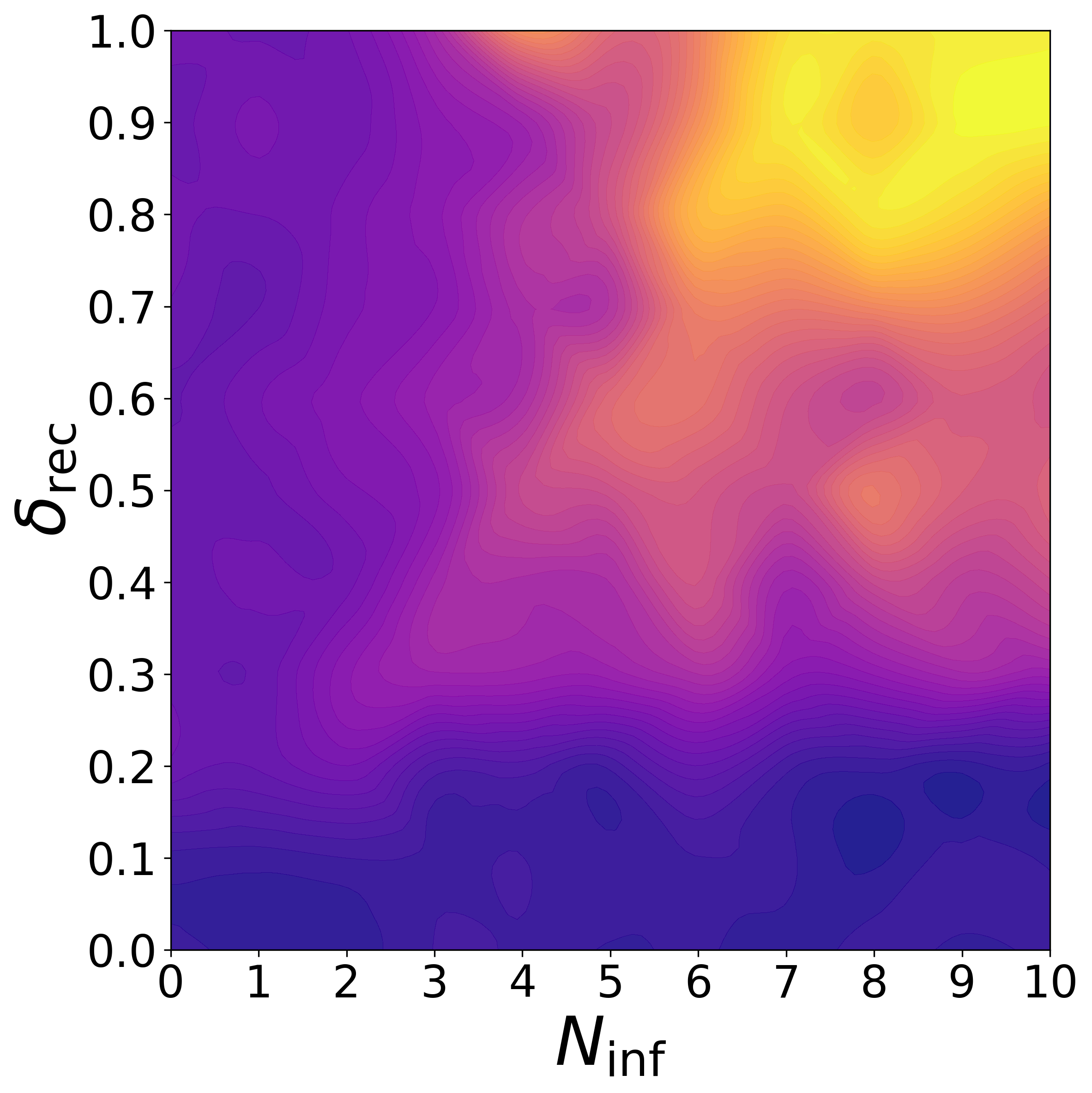}
        \centerline{(b) $\delta_{\text{rec}} vs N_{\text{inf}}$} % REPLACED \subcaption
    \end{minipage}% <--- No space
    \hfill
    \begin{minipage}[b]{\cbarwidthtwo} % Align by BOTTOM [b]
        \centering
        \includegraphics[height=\plotaxesheighttwo, width=\linewidth, keepaspectratio]
            {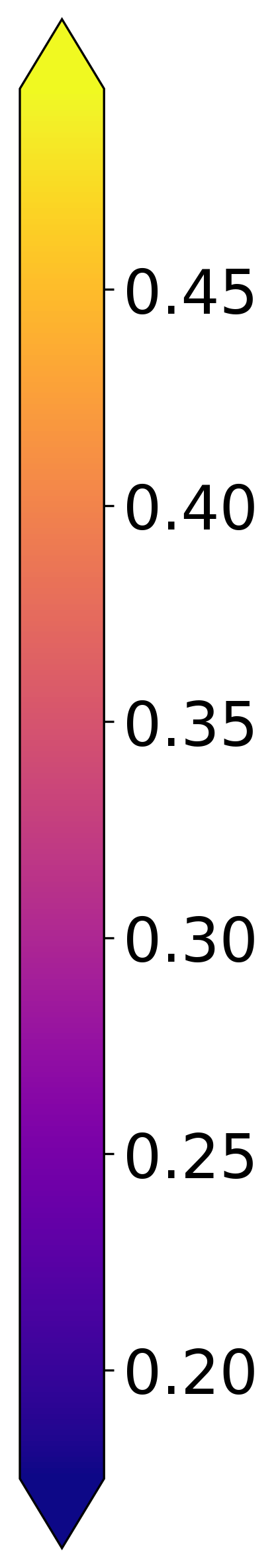}
        \vspace{4.5ex} % Adjust if bottoms don't quite align perfectly
    \end{minipage}

    \caption{Final network modularity (higher values = stronger community structure) landscapes. Plots show modularity varying with: 
    \textbf{(a)} initial network mixing ($\mu_{\text{mix}}^{(0)}$) vs. number of influencers ($N_{\text{inf}}$); and 
    \textbf{(b)} discovery rate ($\delta_{\text{rec}}$) vs. $N_{\text{inf}}$. 
    Modularity primarily increases with $N_{\text{inf}}$ and, significantly, with higher discovery rates ($\delta_{\text{rec}}$), indicating broad exposure promotes structural segregation when influencers are present.}
    \label{fig:coevolution-modularity-heatmaps} % Original label is kept
\end{figure*}

This reveals that network structure acts as a psychological amplifier: segregation constrains the scope of psychological processes by limiting encounter diversity, while integration unleashes the full potential of individual tolerance settings by ensuring agents encounter the complete spectrum of opinions and can respond according to their psychological predispositions.

Importantly, aside from the extreme of complete segregation, the relationship in the heatmap is predominantly influenced by tolerance rather than network connectivity, suggesting that the importance of network structure should not be overstated when some connections between camps exist—which is empirically almost always the case \cite{guess_avoiding_2018, barbera_tweeting_2015}. Network structure appears to have more of a mediating effect under these fixed network settings, while tolerance emerges as the substantial factor driving polarization outcomes. For a detailed synthesis of how tolerance, initial opinion diversity, and network structure interact to produce different polarization patterns, see \cref{app:synthesis-tolerance-structure}.

\subsubsection{The Role of Extremist Influencers}

The presence of extremist actors further demonstrates how structural context shapes polarization dynamics. We examined scenarios where extremist influencers ($\pm0.8$ opinions) attempt to influence a predominantly moderate population ($\pm0.1$ opinions) across different network mixing conditions. Our parameter ranges reflect empirical patterns: we varied influencer presence from 0-10 per faction (0-40\% of the population), encompassing estimates that extreme political activists comprise 10-15\% of politically engaged populations \cite{pew_political_typology_2021}, while dedicated political extremists represent 5-10\% \cite{enders_american_2020}. This range covers both minimal extremist presence observed in most democratic societies and higher concentrations found in polarized contexts where extremist voices can be algorithmically amplified beyond their demographic proportions \cite{tucker_social_2018}.

Extremist influencers achieve maximum impact in a "Goldilocks zone" of moderate network mixing ($\mu_{\text{mix}}^{(0)} \approx 0.3-0.7$) that provides optimal conditions for both faction consolidation and cross-cutting friction (\cref{fig:polarization-heatmaps}c). The heatmap reveals that polarization peaks at intermediate mixing levels, particularly when extremist presence is substantial, declining toward both complete segregation and full integration. Under complete segregation ($\mu_{\text{mix}}^{(0)}=0$), influencers' reach is constrained to their own communities, limiting their ability to create society-wide polarization. Under full integration ($\mu_{\text{mix}}^{(0)}=1.0$), extremist voices become diluted within diverse information streams, reducing their relative influence on moderate agents.

The moderate mixing zone represents the structural configuration most conducive to polarization. Here, influencers retain sufficient connectivity to reach across community boundaries while maintaining enough clustering to build cohesive opposing factions. This enables them to simultaneously leverage in-group dynamics (through community consolidation) and inter-group dynamics (through cross-cutting antagonism), maximizing their polarizing potential. The "Goldilocks zone" effect intensifies with the number of influencers, demonstrating that moderate network structures show highest susceptibility to extremist influence.

This finding reveals nonlinear relationships between network structure and polarization outcomes. Neither complete fragmentation nor total integration produces the highest extremist effectiveness; instead, moderately integrated networks—typically presumed to promote balanced discourse—demonstrate greatest susceptibility to polarization dynamics. This pattern suggests that structural parameters may have counterintuitive effects on polarization, where intermediate connectivity levels optimize conditions for extremist influence.

\begin{figure*}[h!]
    \centering
    \includegraphics[width=\textwidth]{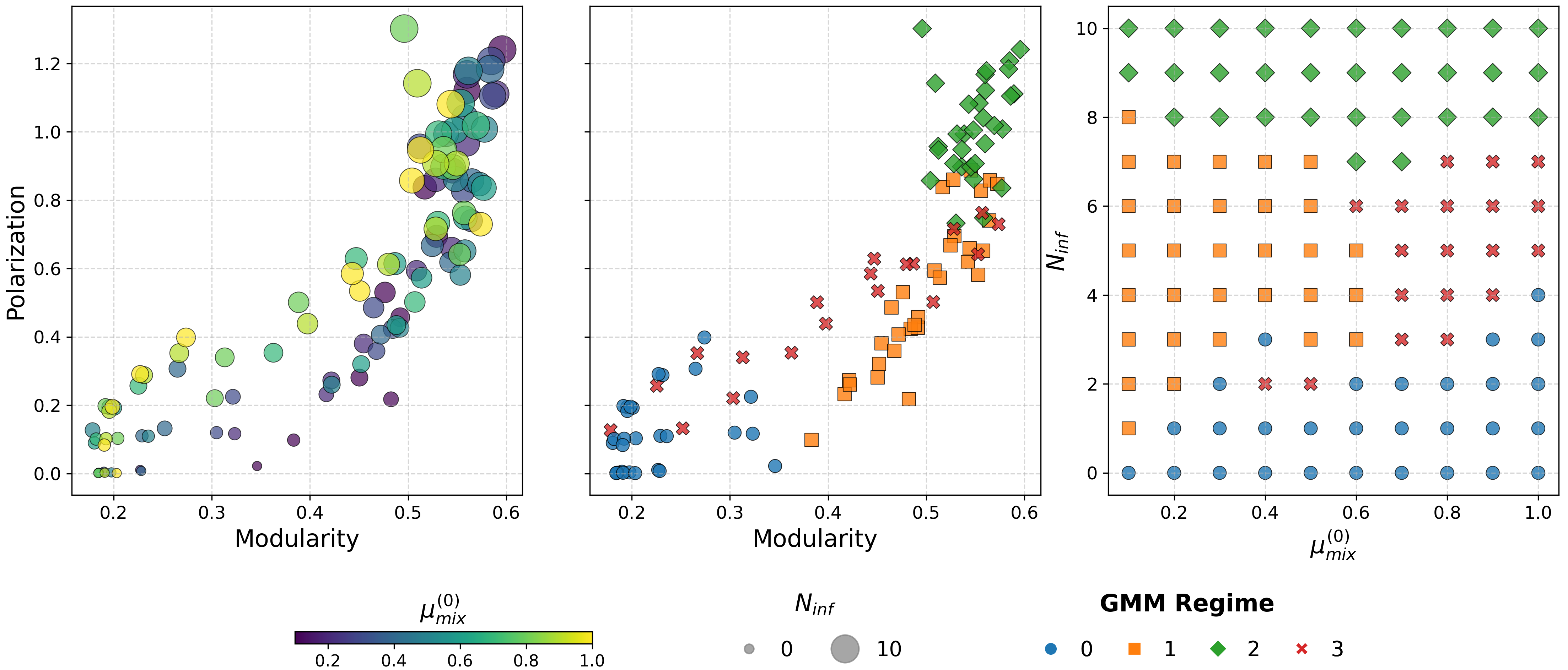}
    \caption{Co-evolution of modularity and polarization varying initial mixing ($\mu_{\text{mix}}^{(0)}$) and influencer count ($N_{\text{inf}}$). (a) Outcomes in Modularity-Polarization space (color=$\mu_{\text{mix}}^{(0)}$, size=$N_{\text{inf}}$) show positive correlation, driven mainly by $N_{\text{inf}}$. (b) Same space colored by GMM-identified behavioral regimes. (c) Regimes mapped onto input parameters ($\mu_{\text{mix}}^{(0)}$ vs $N_{\text{inf}}$). GMM effectively partitions parameter space into distinct outcomes (e.g., low influence yields low polarization/modularity; high influence yields high polarization/modularity), with intermediate mixing sometimes facilitating higher values.}
    \label{fig:mixing-influencers-scatter}
\end{figure*}

\subsection{Co-evolution of Opinions and Network Structure}
\label{subsec:eval_coevolution}

Our findings thus far assume static network structures, but real-world polarization involves dynamic feedback between opinions and relationships. We now activate network evolution to examine how opinion dynamics and structural adaptation co-evolve, revealing whether polarization of beliefs necessarily couples with social fragmentation. Both experiments employed varying numbers of extremist influencers ($\pm0.8$ opinions) attempting to influence a predominantly moderate population ($\pm0.1$ opinions).

\subsubsection{Opinion-Network Coupling Dynamics}

We first examine how initial network structure affects co-evolutionary dynamics by varying the inter-community mixing parameter ($\mu_{\text{mix}}$), which controls the relative likelihood of cross-community versus within-community connections (detailed in \cref{app:exp5_coevolution_mixing}). These experiments confirm a strong general tendency for coupling: influencer-driven opinion divergence often fuels structural segregation through homophily \cite{friedkin_structural_1998,centola_homophily_2007}, leading toward states of simultaneously high polarization (\cref{fig:coevolution-polarization-heatmaps}a) and high modularity (\cref{fig:coevolution-modularity-heatmaps}a). The specific trajectory, however, depends on initial conditions, with different starting network structures leading to distinct coupling patterns between structural and opinion dynamics \cite[cf.][]{lelkes_mass_2016}. A Gaussian Mixture Model analysis with four components (selected based on information criteria and interpretability; detailed methodology in \cref{app:exp5_coevolution_mixing}) identifies distinct co-evolutionary regimes (\cref{fig:mixing-influencers-scatter}), including conditions where moderate numbers of influencers create structural segregation with only moderate opinion polarization (Regime 1). This decoupling reveals that moderate influencer presence can drive community formation—where agents sort into distinct network clusters—without necessarily producing extreme belief polarization. The finding suggests that structural sorting and belief extremization may operate as partially independent processes \cite{dellaposta_why_2015}, with different sensitivity thresholds that enable early-stage community formation to precede ideological radicalization.

\begin{figure*}[h!]
    \centering
    \includegraphics[width=\textwidth]{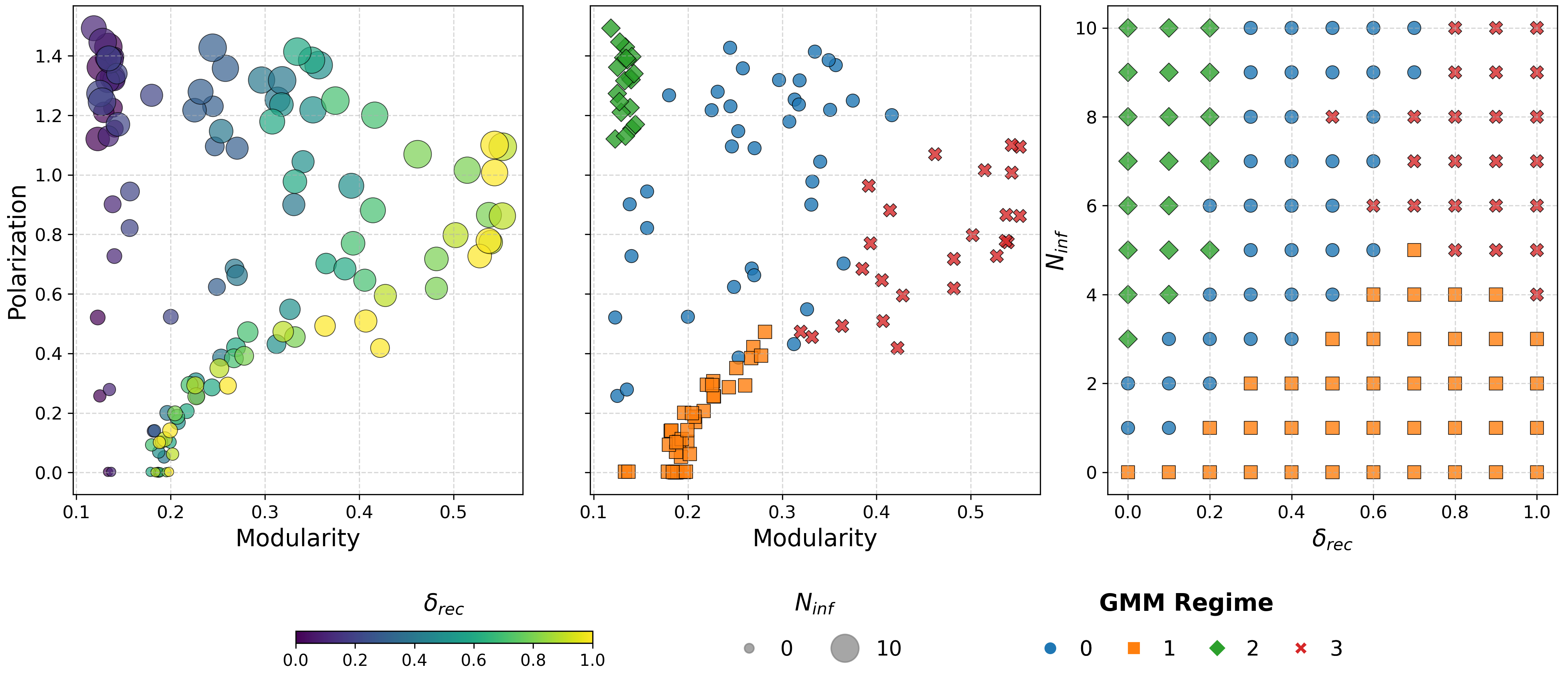}
    \caption{Co-evolution of modularity and polarization varying discovery rate ($\delta_{\text{rec}}$) and influencer count ($N_{\text{inf}}$). (a) Outcomes in Modularity-Polarization space (color=$\delta_{\text{rec}}$, size=$N_{\text{inf}}$) reveal the algorithmic trade-off: higher discovery rates (lighter colors) increase modularity but decrease polarization. (b) GMM-identified regimes in outcome space. (c) Regimes mapped onto input parameters ($\delta_{\text{rec}}$ vs $N_{\text{inf}}$). GMM highlights distinct outcomes, e.g., high influence/low discovery yields high polarization but lower modularity, while high influence/high discovery yields high modularity but lower polarization.}
    \label{fig:discovery-influencers-scatter}
\end{figure*}

\subsubsection{Algorithmic Discovery Effects}

\begin{figure*}[h]
    \centering
    \begin{minipage}[b]{0.49\textwidth}
        \centering
        \includegraphics[width=\textwidth]{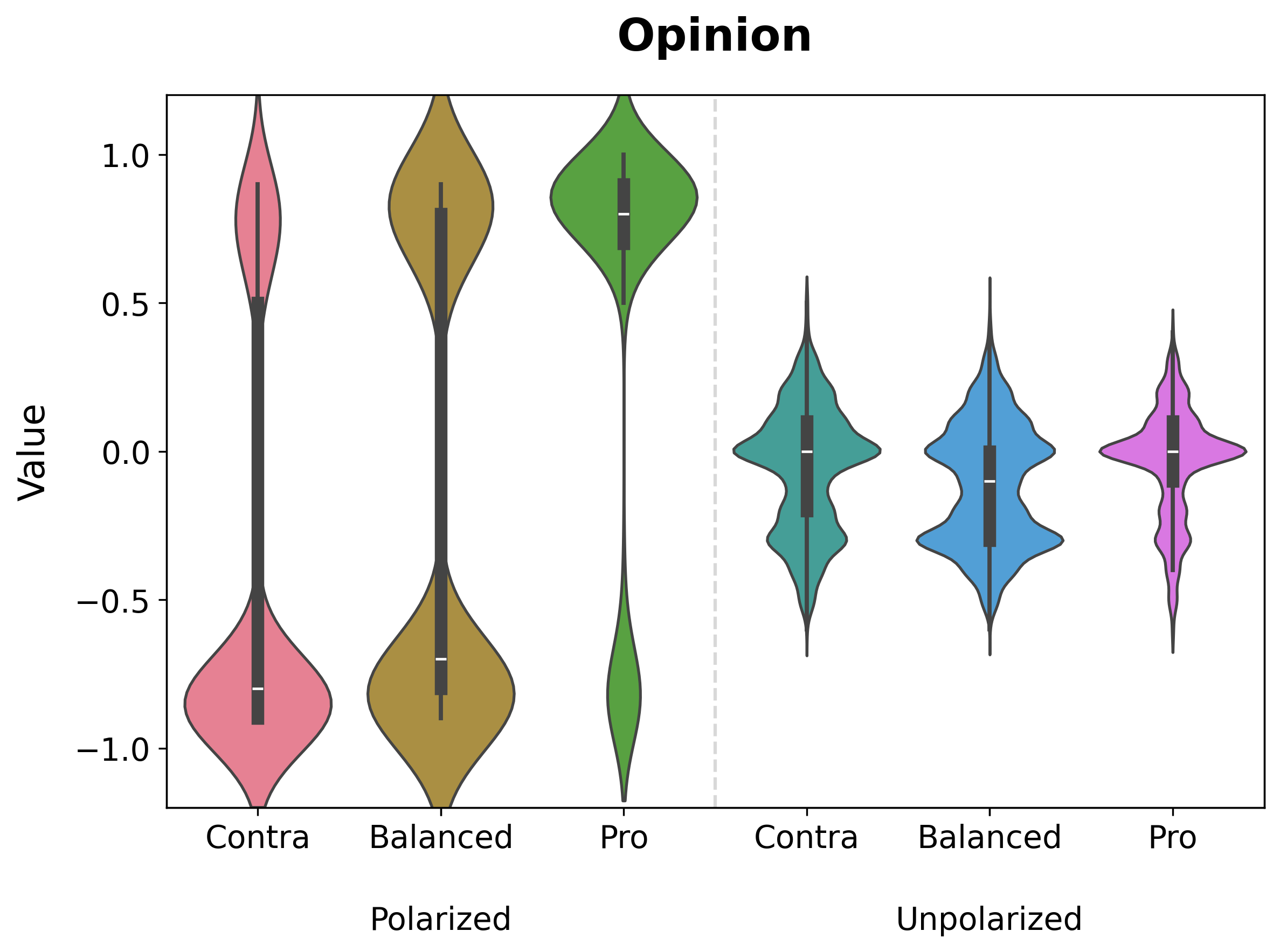}
        \centerline{(a) Message Opinion} % REPLACED \subcaption
    \end{minipage}
    \hfill
    \begin{minipage}[b]{0.49\textwidth}
        \centering
        \includegraphics[width=\textwidth]{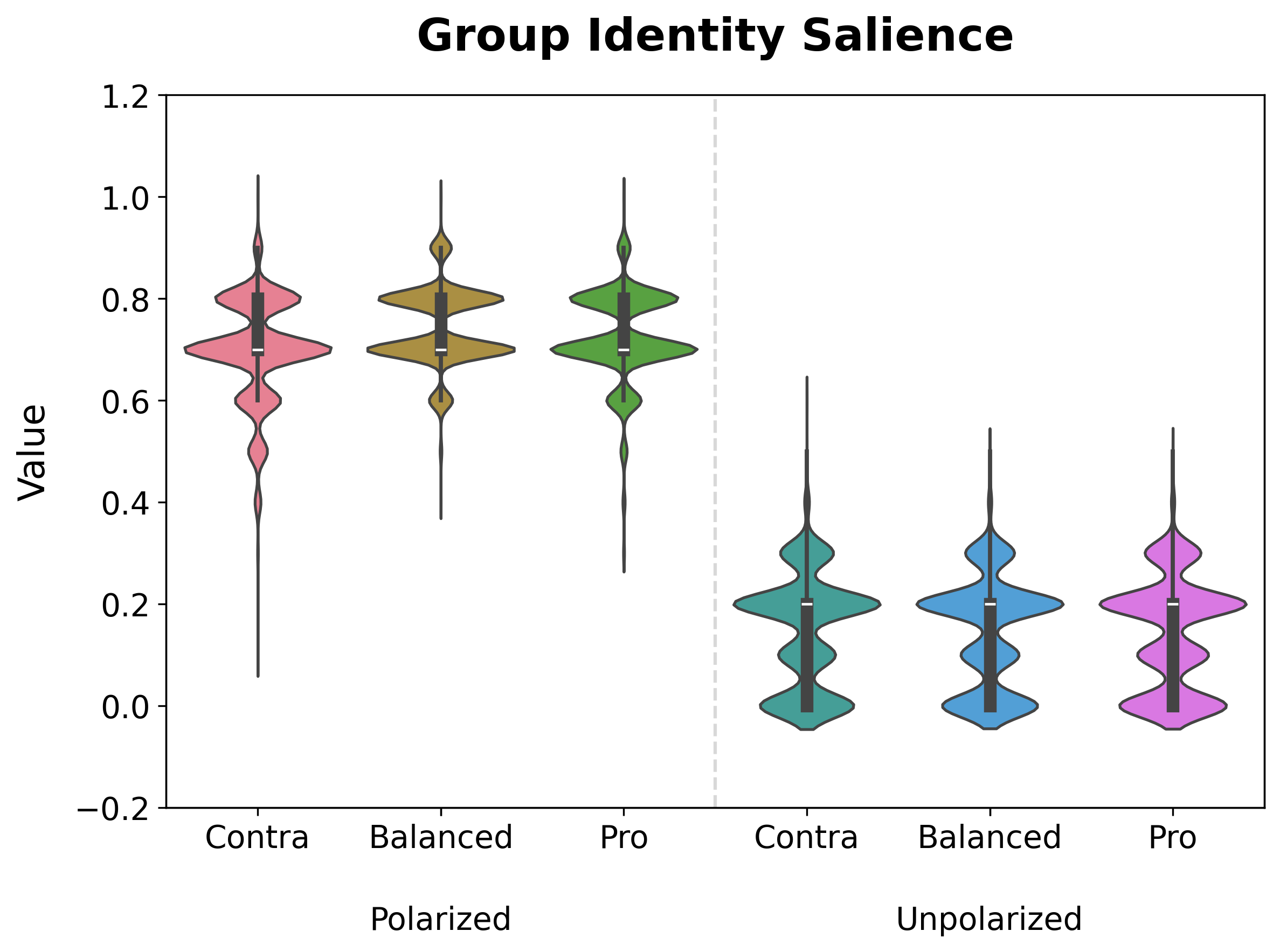}
        \centerline{(b) Group Identity Salience} % REPLACED \subcaption
    \end{minipage}
    
    \vspace{1em} % This provides the vertical spacing between the rows.
    
    \begin{minipage}[b]{0.49\textwidth}
        \centering
        \includegraphics[width=\textwidth]{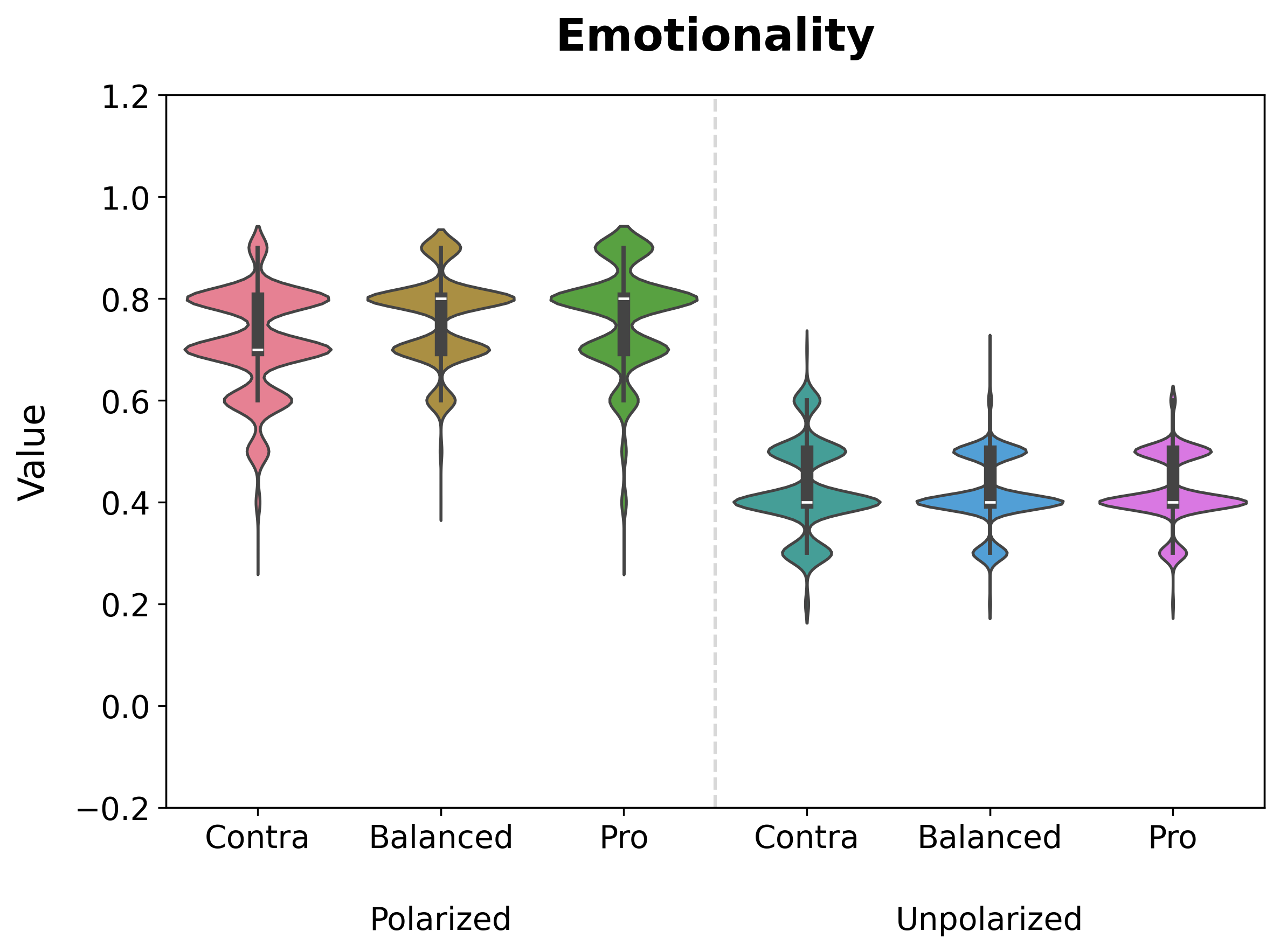}
        \centerline{(c) Emotionality} % REPLACED \subcaption
    \end{minipage}
    \hfill
    \begin{minipage}[b]{0.49\textwidth}
        \centering
        \includegraphics[width=\textwidth]{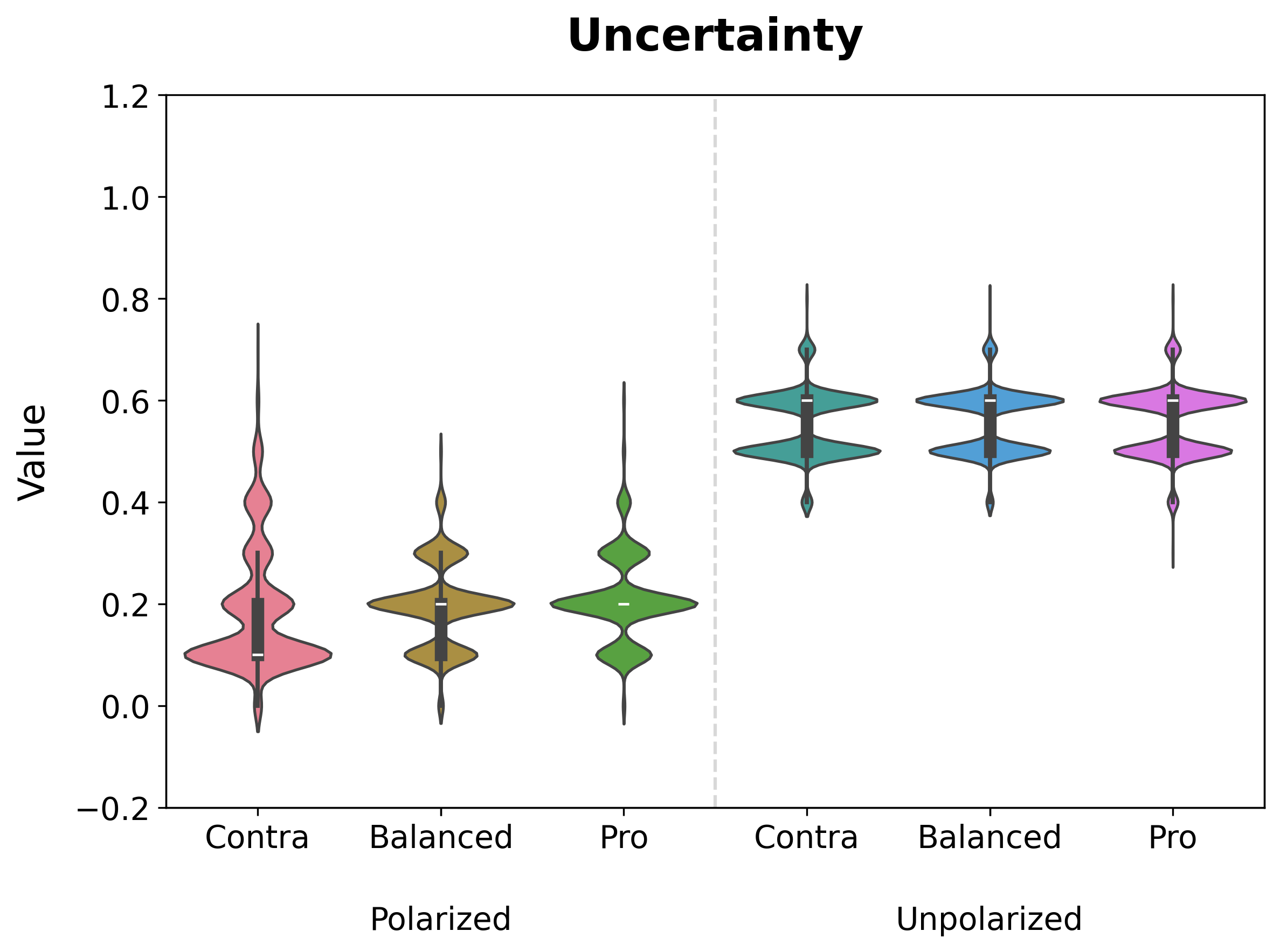}
        \centerline{(d) Uncertainty} % REPLACED \subcaption
    \end{minipage}

    \caption{The violin plots illustrate the distribution of values obtained from the LLM with respect to varying content dimensions. A differentiation is made between polarized and non-polarized populations, and the bias in the distribution of messages (pro, contra, balanced) is also considered.}
    \label{fig:message-analysis-violin-plots} % Original label is kept
\end{figure*}

We next examine algorithmic influence by varying the recommendation system's discovery rate ($\delta_{\text{rec}}$), revealing how the balance between extremist presence and information diversity determines algorithmic intervention effectiveness (\cref{fig:coevolution-polarization-heatmaps}b, \cref{fig:coevolution-modularity-heatmaps}b, detailed in \cref{app:exp6_coevolution_discovery}). The discovery rate parameter ranges from 0.0 (purely network-based recommendations) to 1.0 (purely discovery-based recommendations), reflecting real-world platform configurations. Major social media platforms typically employ mixed recommendation strategies: Facebook's News Feed balances friend connections with algorithmic discovery \cite{devito_algorithms_2017}, while Twitter's timeline varies between chronological friend networks and algorithmically-curated content \cite{eslami_i_2015}. Discovery-heavy approaches characterize platforms like TikTok, where recommendation algorithms prioritize content discovery over social connections \cite{simpson_understanding_2022}.

\begin{figure*}[h!]
    \centering
    \includegraphics[width=0.9\textwidth]{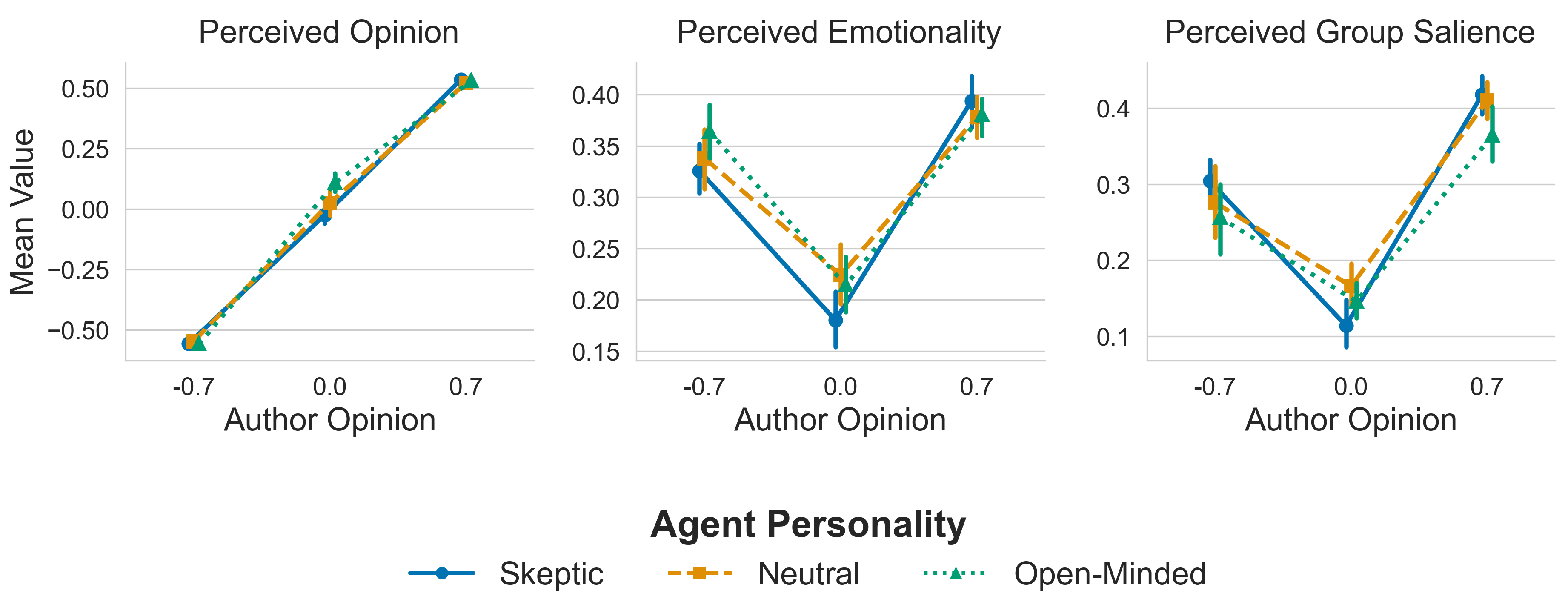}
    \caption{Mean perceived message characteristics (columns: opinion, emotionality, group salience) vs. actual author opinion (x-axis) for different agent personalities (lines) on Universal Basic Income topic. Error bars show SEM. Perceived opinion accurately tracks author opinion. Perceived emotionality and group salience show V-shaped patterns (minimal for neutral authors, maximal for extremists), largely consistent across personality types, indicating neutral messages are perceived as less emotional and group-focused.}
    \label{fig:message-analysis-interactions-personality}
\end{figure*}

\begin{figure*}[h!]
    \centering
    \includegraphics[width=0.9\textwidth]{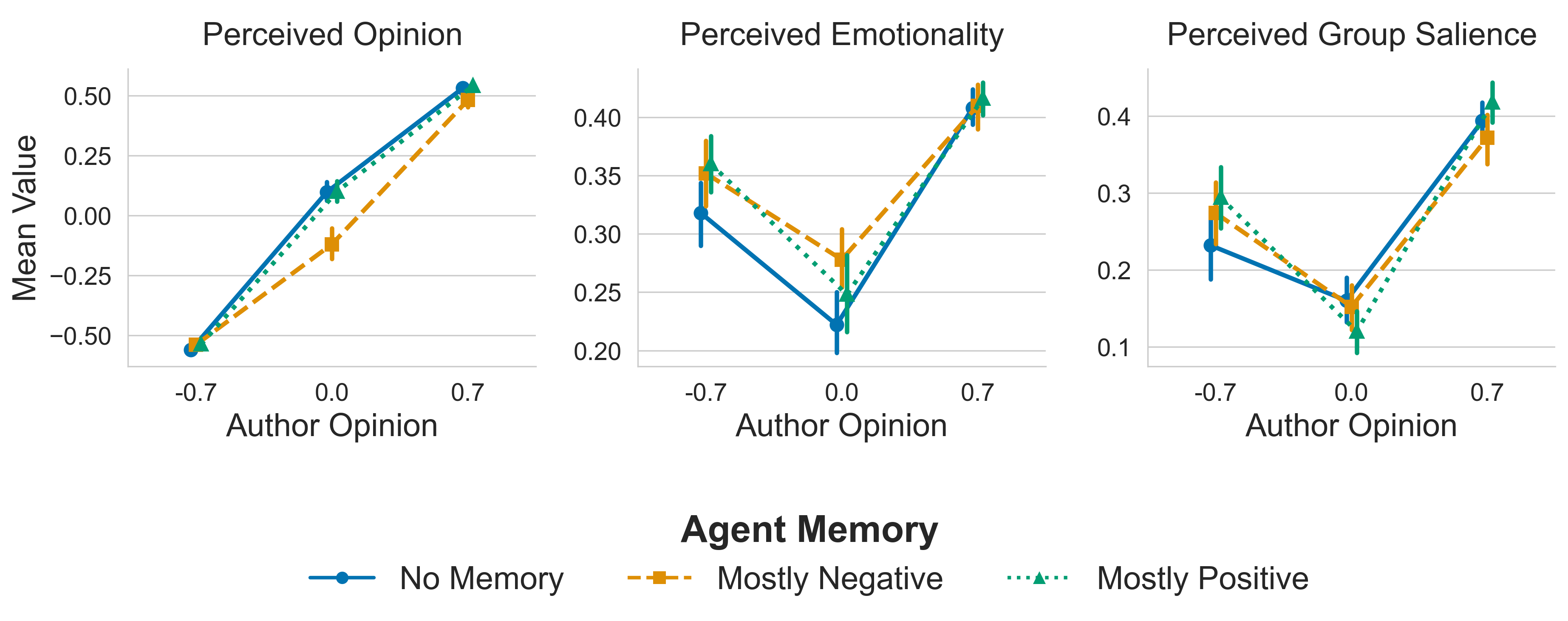}
    \caption{Mean perceived message characteristics (columns: opinion, emotionality, group salience) vs. actual author opinion (x-axis) for different agent memory states (lines: No Memory, Mostly Negative, Mostly Positive) on Universal Basic Income topic. Perceived opinion consistently tracks author opinion. Perceived emotionality and group salience generally show V-shaped patterns (minimal for neutral authors). Notably, agents with mostly positive memories perceive extreme messages (especially positive ones) as less emotional and group-salient than other agents.}
    \label{fig:message-analysis-interactions-memory}
\end{figure*}

The relationship between discovery rates, influencer numbers, and outcomes proves highly nonlinear, creating distinct operational regimes (timeline analyses with confidence intervals provided in \cref{app:exp6_coevolution_discovery}). When extremist presence is minimal (0-2 influencers), polarization remains low regardless of discovery rate, suggesting that isolated extremist voices cannot significantly influence moderate populations under any algorithmic configuration. The critical threshold emerges when extremist presence reaches intermediate levels (3-6 influencers), where discovery rates begin to strongly mediate outcomes.

Low discovery rates ($\delta_{\text{rec}} \le 0.1$) enable disproportionate extremist influence once sufficient influencers are present, creating filter bubble-like conditions \cite{pariser_filter_2011} where moderate agents become trapped within their initial network connections (\cref{fig:discovery-influencers-scatter}). Under these conditions, even moderate extremist presence can achieve substantial polarization by preventing moderates from discovering the broader moderate majority.

\begin{figure*}[htbp]
    \centering
    \setlength{\plotaxesheightfour}{7.2cm} % Adjusted for 4-panel layout

    % First row - Total interactions
    \begin{minipage}[b]{\plotwidthfour}
        \centering
        \includegraphics[width=\linewidth]{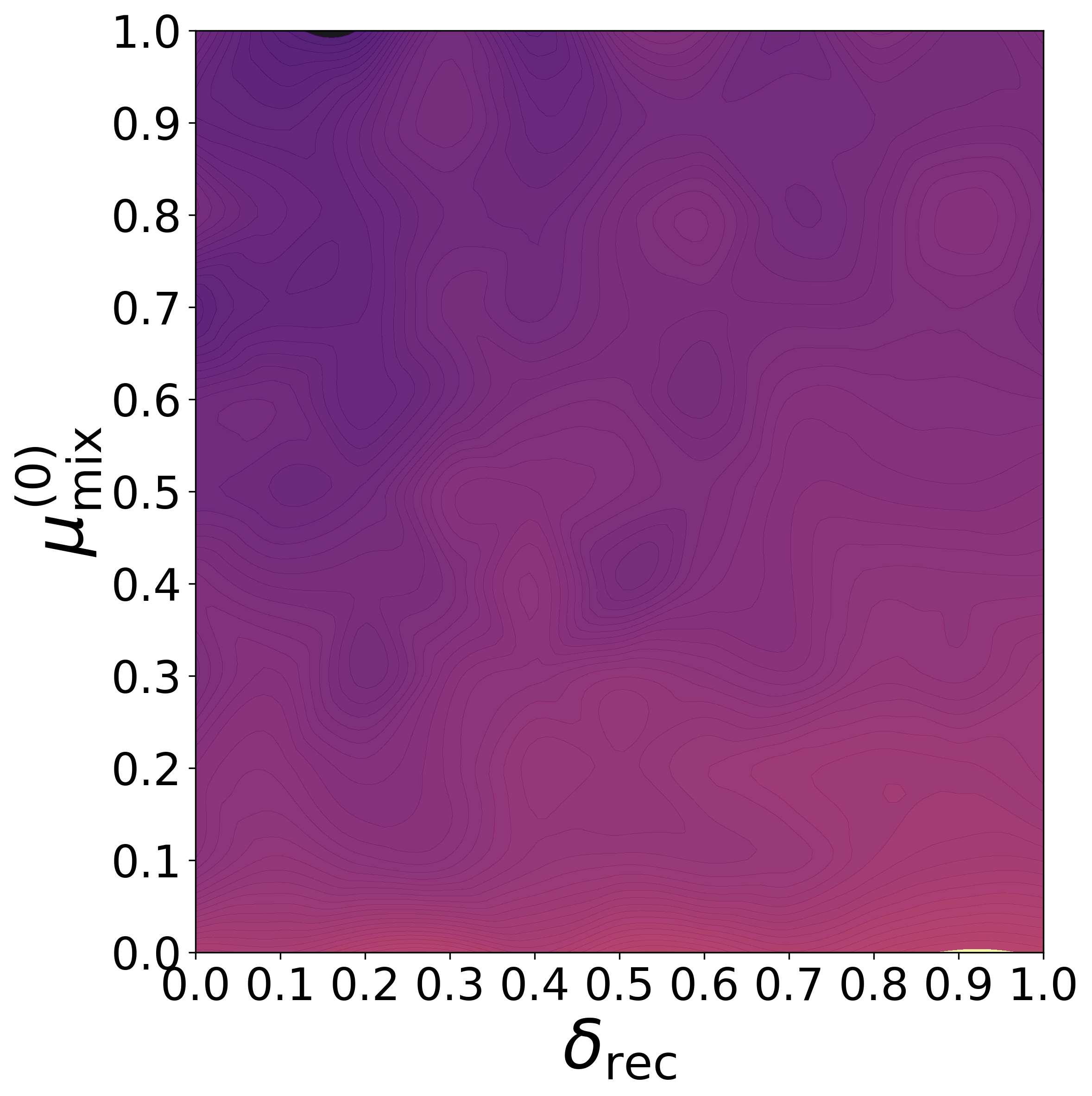}
        \centerline{(a) Total interactions: Unpolarized}
    \end{minipage}%
    \hfill
    \begin{minipage}[b]{\plotwidthfour}
        \centering
        \includegraphics[width=\linewidth]{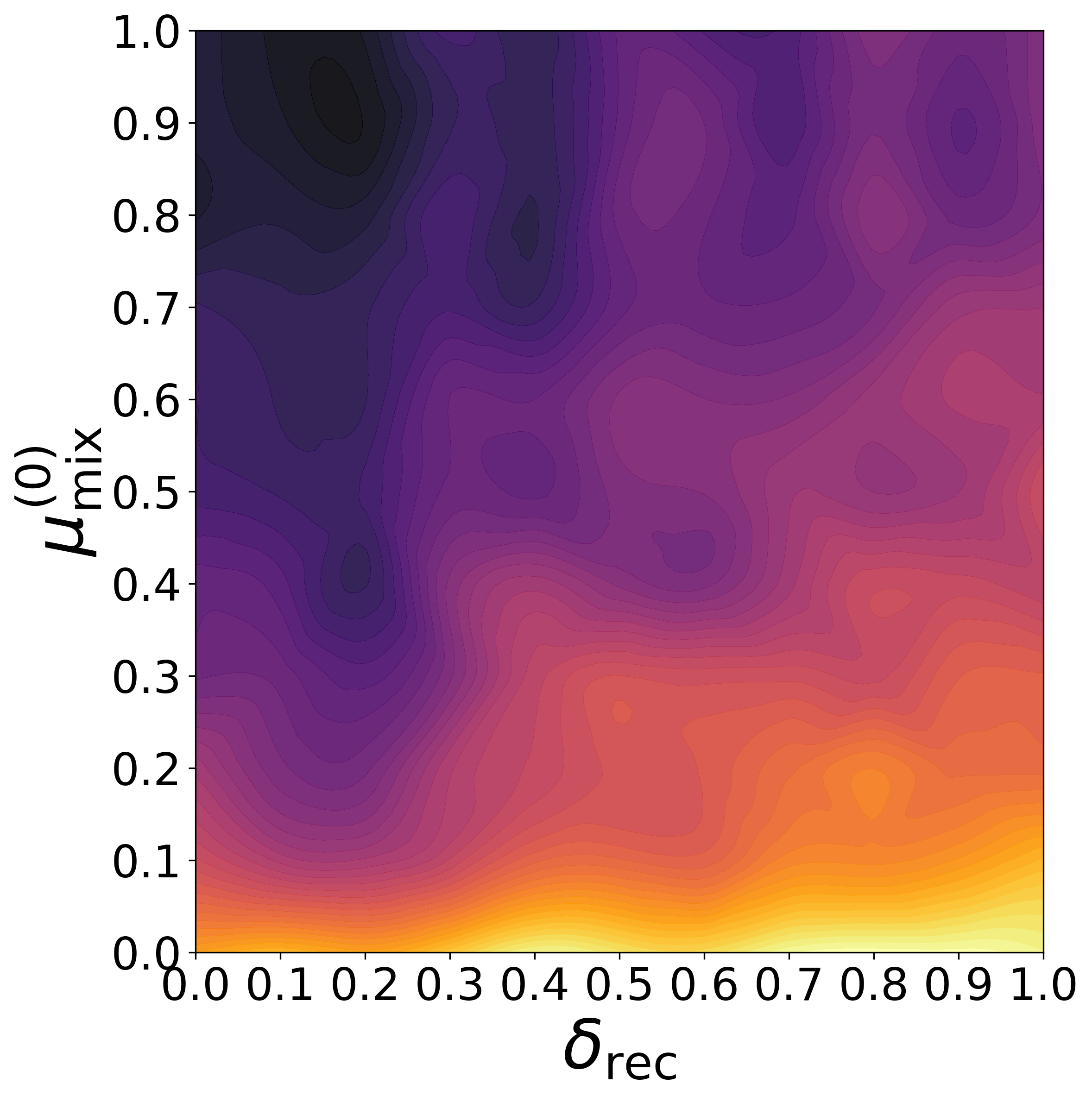}
        \centerline{(b) Total interactions: Polarized}
    \end{minipage}%
    \hfill
    \begin{minipage}[b]{\cbarwidthfour}
        \centering
        \includegraphics[height=\plotaxesheightfour, width=\linewidth, keepaspectratio]{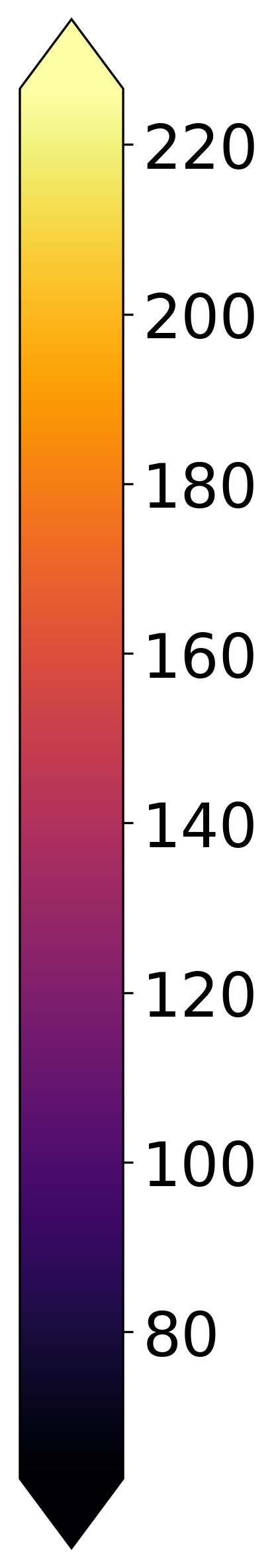}
        \vspace{4.5ex}
    \end{minipage}

    \vspace{1em}

    % Second row - Interaction ratios
    \begin{minipage}[b]{\plotwidthfour}
        \centering
        \includegraphics[width=\linewidth]{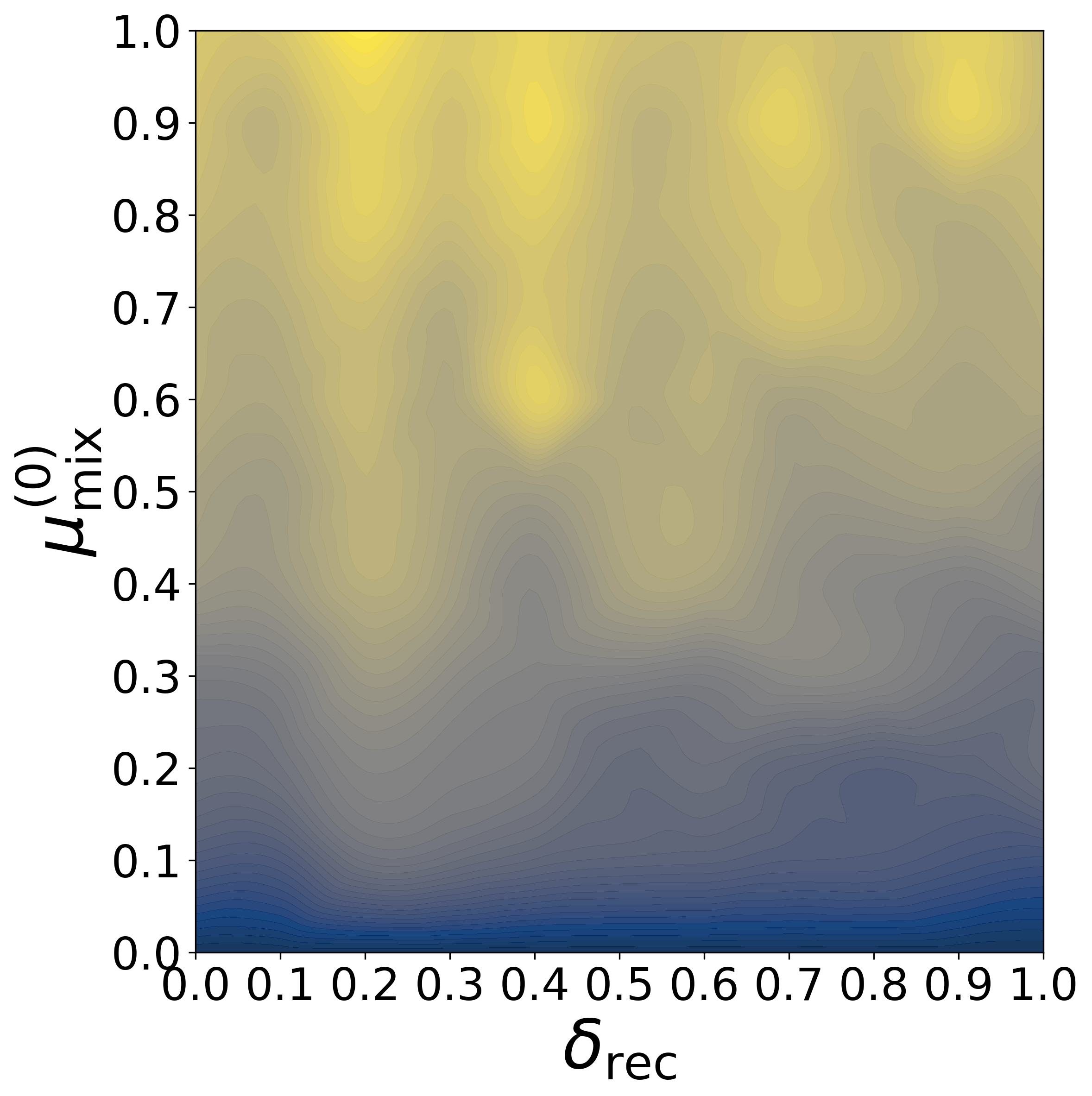}
        \centerline{(c) Cross-stance ratio: Unpolarized}
    \end{minipage}%
    \hfill
    \begin{minipage}[b]{\plotwidthfour}
        \centering
        \includegraphics[width=\linewidth]{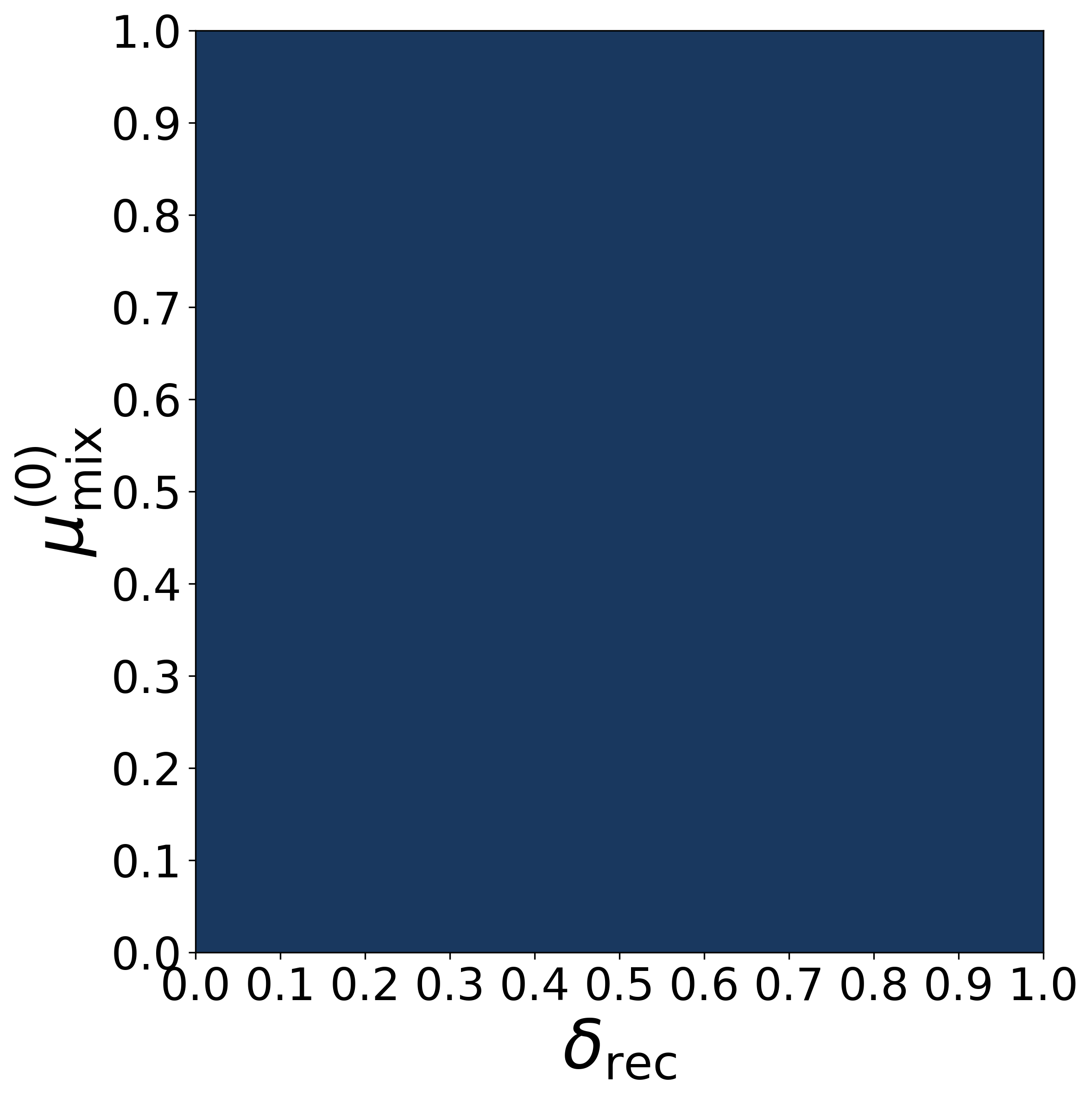}
        \centerline{(d) Cross-stance ratio: Polarized}
    \end{minipage}%
    \hfill
    \begin{minipage}[b]{\cbarwidthfour}
        \centering
        \includegraphics[height=\plotaxesheightfour, width=\linewidth, keepaspectratio]{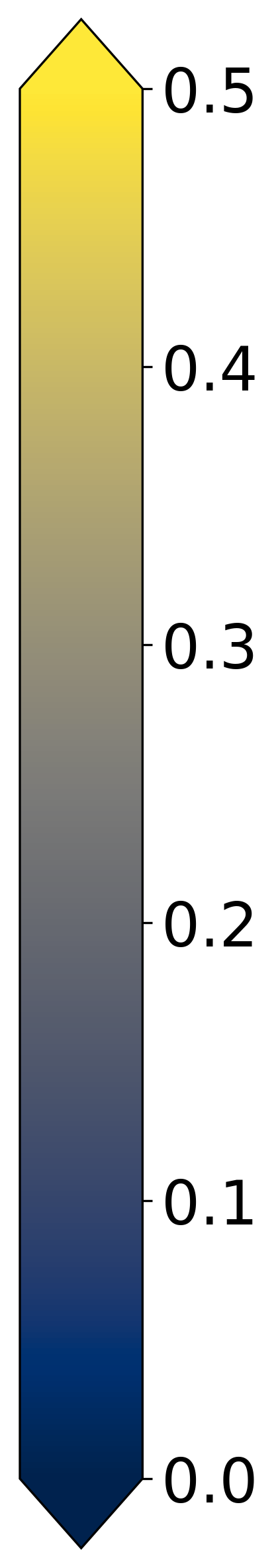}
        \vspace{4.5ex}
    \end{minipage}

    \caption{Agent engagement patterns under homophilic interaction model ($p_{\text{dis}}^{\text{(react)}} = 0$) as a function of discovery rate ($\delta_r$) and initial network mixing ($\mu_{\text{mix}}^{(0)}$).
    \textbf{Top row (a-b)}: Total interactions show moderate levels in unpolarized conditions irrespective of parameters, while polarized conditions exhibit maximum engagement with low mixing and high discovery.
    \textbf{Bottom row (c-d)}: Cross-stance interaction ratio (0 = complete homophily, 0.5 = egalitarian) increases with network mixing in unpolarized conditions but remains at zero (complete segregation) in polarized conditions regardless of structural parameters.
    Unpolarized conditions use initial opinions $U(-0.1, 0.1)$; polarized conditions use $U(-0.8, -0.6) \cup U(0.6, 0.8)$.}
    \label{fig:engagement-homophilic}
\end{figure*}

High discovery rates ($\delta_{\text{rec}} \ge 0.8$) produce fundamentally different outcomes that depend on extremist presence levels. With moderate extremist numbers (3-6 influencers), high discovery reduces polarization outcomes. Here, moderate voices vastly outnumber extremist ones in diverse information streams, allowing moderate agents to efficiently discover like-minded others and resist extremist influence \cite{stray_designing_2022}. However, with high extremist presence (7+ influencers), high discovery provides only temporal mitigation, delaying rather than preventing polarization as extremist content constitutes a substantial portion of the algorithmic feed (see \cref{fig:polarization-timeline-discovery} for temporal dynamics).

This creates a pronounced asymmetry between opinion and structural outcomes. Most strikingly, low discovery with many influencers produces extreme opinion polarization while maintaining relatively low structural segregation (Regime 2), demonstrating that beliefs can radicalize even when network integration persists. Specifically, the Gaussian Mixture Model analysis (four components, see \cref{app:exp6_coevolution_discovery}) reveals four distinct regimes: low-influence consensus (Regime 0), intermediate polarization with low segregation (Regime 1), extreme polarization with low segregation under low discovery (Regime 2), and substantial polarization with high segregation under high discovery (Regime 3).

These results indicate that algorithmic discovery can delay and moderate extremist influence, but not eliminate it when extremist presence reaches critical thresholds (Regime 3). The effectiveness of increasing exposure diversity depends on the existing balance of voices, reducing polarization when extremist influence is limited, but providing only temporal mitigation once extremist presence becomes substantial. While high discovery slows polarization progression and reduces its ultimate severity compared to low discovery scenarios, it does not prevent significant opinion shift in populations facing sustained extremist influence. The temporal dynamics suggest that early algorithmic intervention shows greater effectiveness, while reactive measures applied to established extremist presence demonstrate limited long-term impact and cannot eliminate the underlying polarization drivers.

\subsection{Impact of Polarization on Communication Content and Perception}
\label{subsec:eval_communication_content_and_perception}

Beyond structural patterns, polarization changes communication patterns substantially. Having observed the co-evolution of opinions and networks, we now analyze the content of messages themselves, examining how polarized environments shape the emotional tone, group identity salience, and certainty expressed in agent communications. These dimensions align with key theoretical constructs: emotionality connects to affective polarization research \cite{iyengar_affect_2012, mason_uncivil_2018}, group identity salience reflects social identity theory predictions \cite{tajfel_integrative_1979, huddy_social_2001}, and uncertainty expression relates to epistemic closure in polarized contexts \cite{nguyen_echo_2020}.

Our LLM-based content analysis shows that polarized environments, compared to unpolarized ones, generate discourse with significantly more extreme opinions, substantially higher group identity salience, increased emotionality, and reduced uncertainty (\cref{fig:message-analysis-violin-plots}; see \cref{subsec:offline-message-content-analysis} for detailed methodology and statistical validation). This analytical approach leverages recent advances demonstrating LLMs' reliability in assessing discourse dimensions \cite{ziems_can_2024, tornberg_chatgpt_2023}. Content generation and analysis were conducted as separate computational steps using distinct prompts to maintain analytical independence.

These patterns reveal polarization as a multi-dimensional communicative transformation involving cognitive, affective, and identity shifts \cite{iyengar_affect_2012, mason_uncivil_2018}, aligning with patterns observed in our human-subject study (\cref{sec:validation}). Notably, the overall polarization level appeared more influential in shaping these content characteristics than specific recommendation biases (except for opinion extremity).

\begin{figure*}[htbp]
    \centering
    \setlength{\plotaxesheightfour}{7.2cm} % Adjusted for 4-panel layout

    % First row - Total interactions
    \begin{minipage}[b]{\plotwidthfour}
        \centering
        \includegraphics[width=\linewidth]{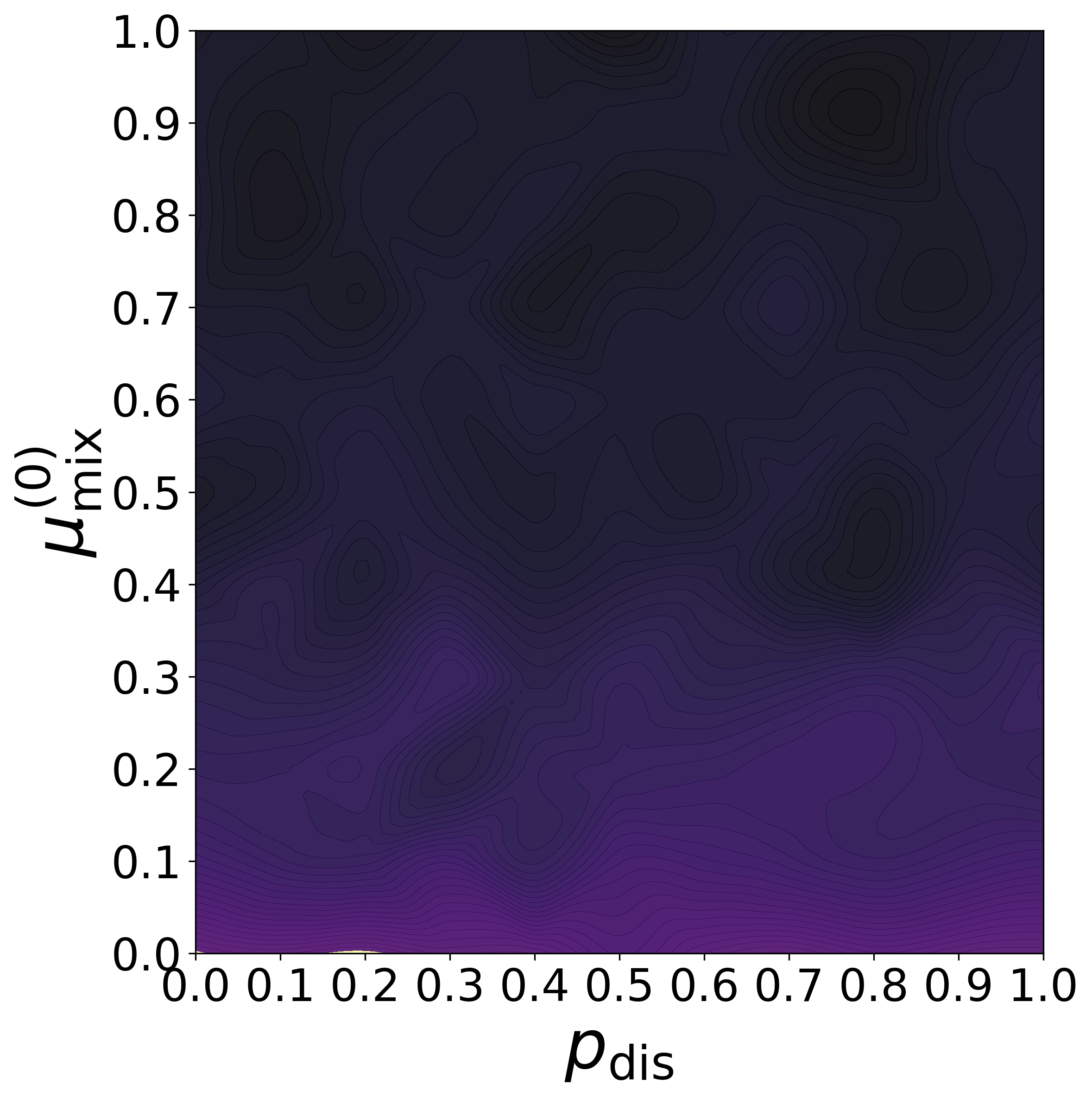}
        \centerline{(a) Total interactions: Unpolarized}
    \end{minipage}%
    \hfill
    \begin{minipage}[b]{\plotwidthfour}
        \centering
        \includegraphics[width=\linewidth]{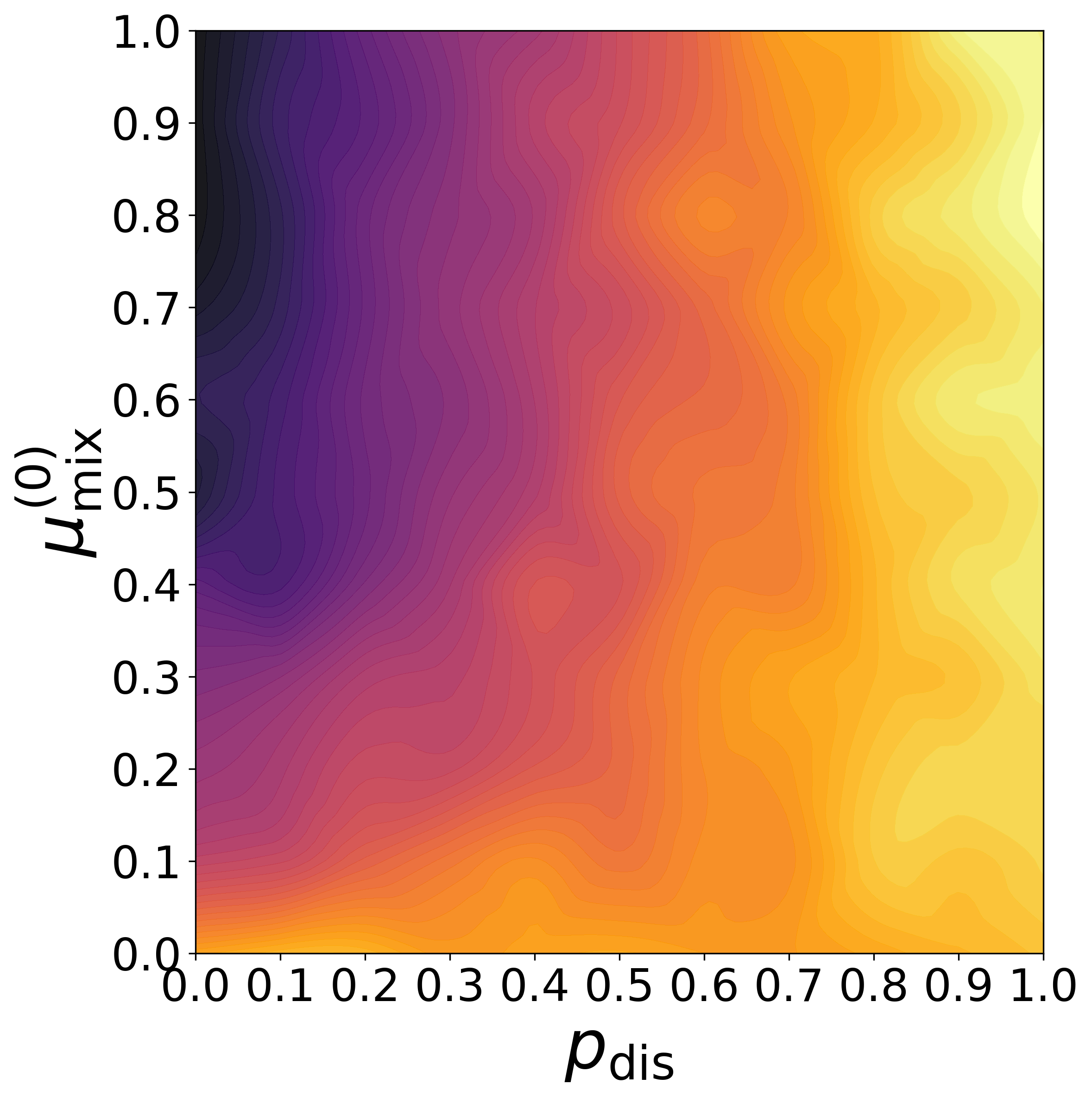}
        \centerline{(b) Total interactions: Polarized}
    \end{minipage}%
    \hfill
    \begin{minipage}[b]{\cbarwidthfour}
        \centering
        \includegraphics[height=\plotaxesheightfour, width=\linewidth, keepaspectratio]{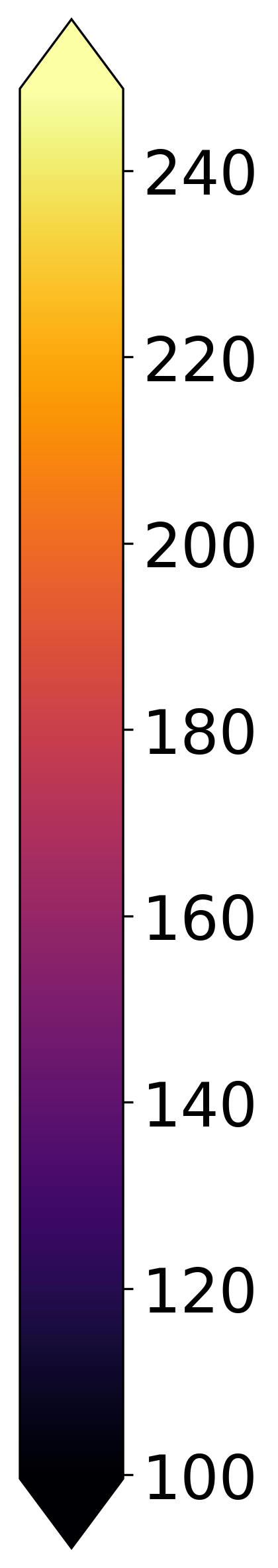}
        \vspace{4.5ex}
    \end{minipage}

    \vspace{1em}

    % Second row - Interaction ratios
    \begin{minipage}[b]{\plotwidthfour}
        \centering
        \includegraphics[width=\linewidth]{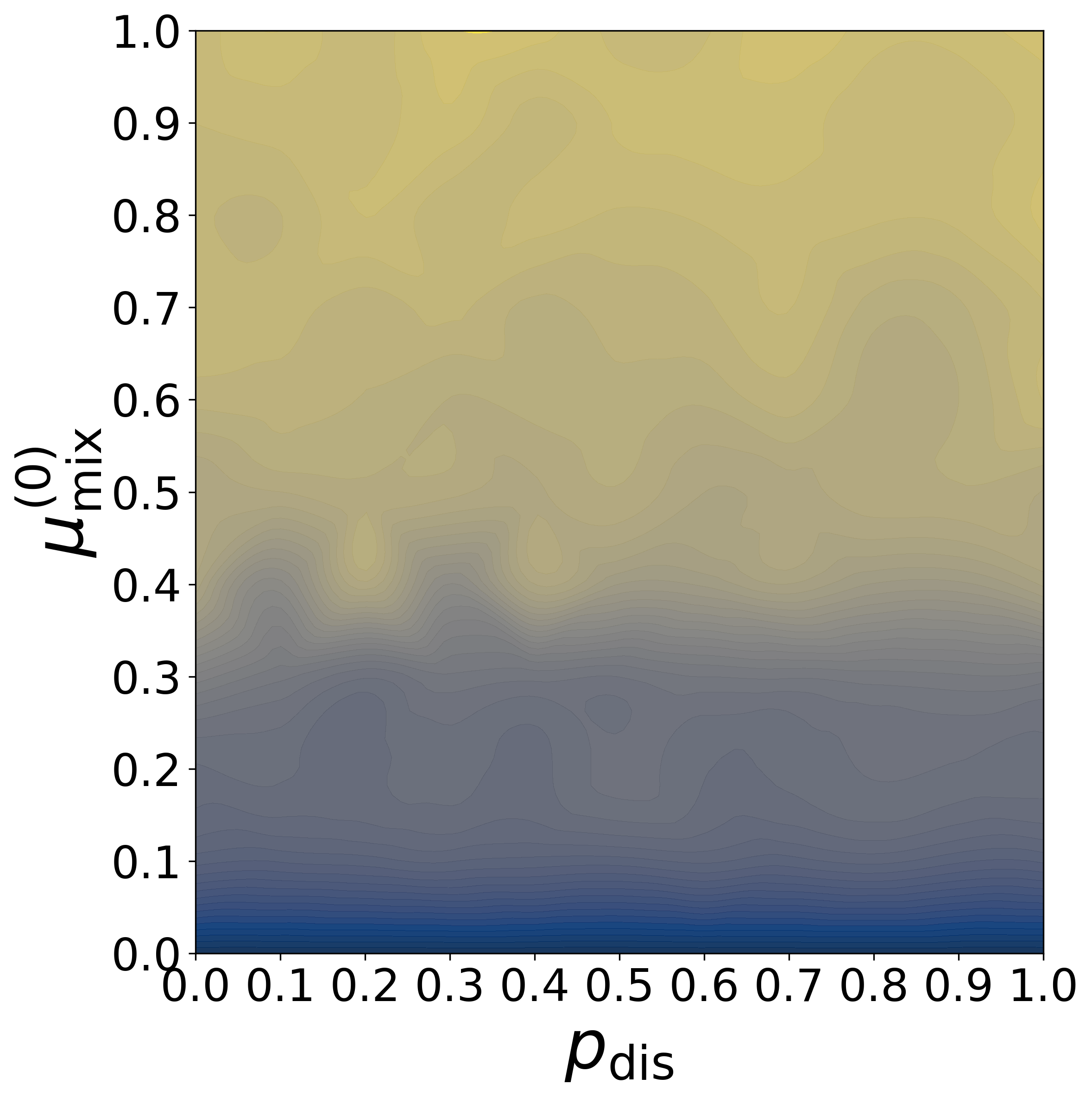}
        \centerline{(c) Cross-stance ratio: Unpolarized}
    \end{minipage}%
    \hfill
    \begin{minipage}[b]{\plotwidthfour}
        \centering
        \includegraphics[width=\linewidth]{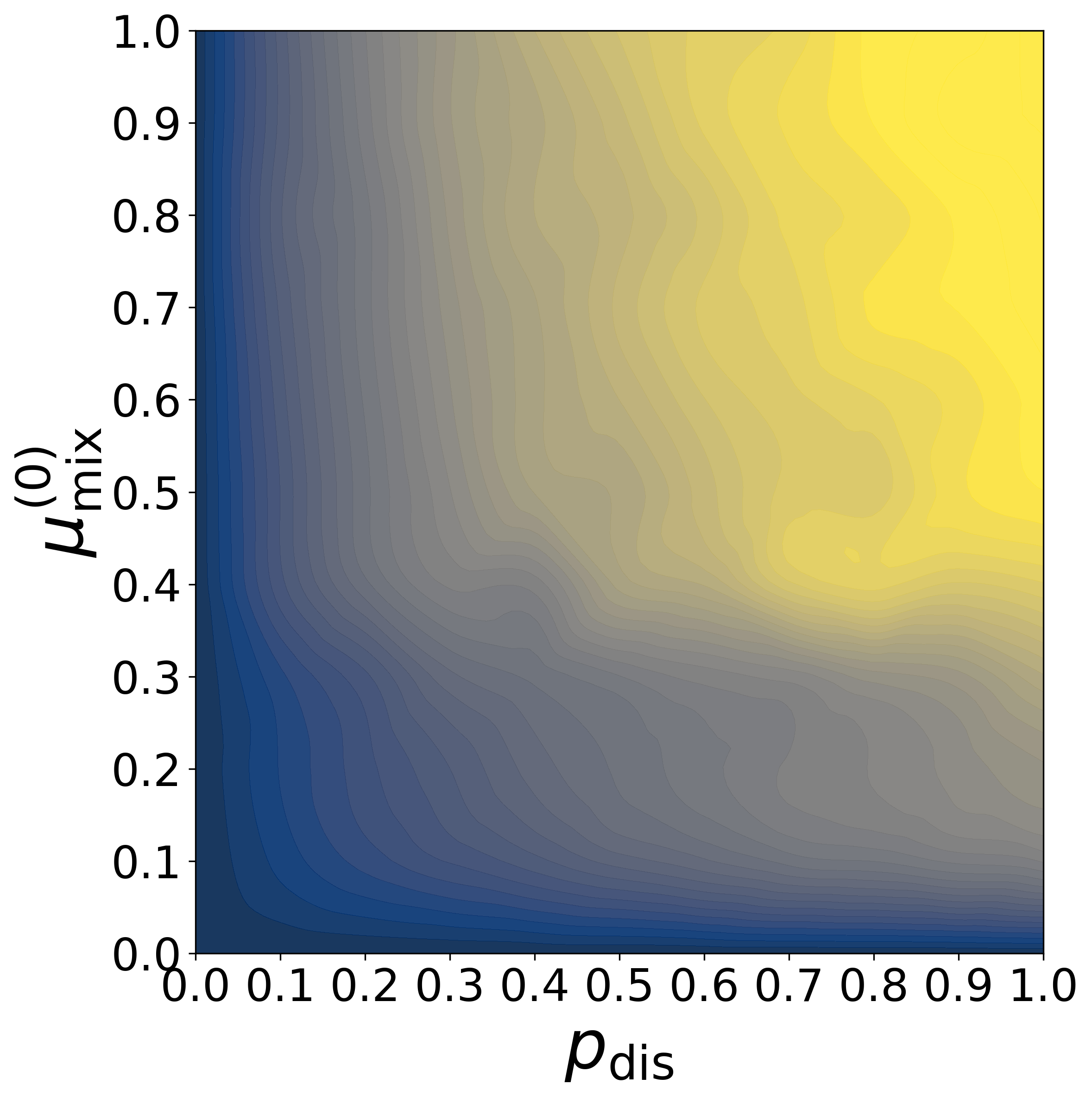}
        \centerline{(d) Cross-stance ratio: Polarized}
    \end{minipage}%
    \hfill
    \begin{minipage}[b]{\cbarwidthfour}
        \centering
        \includegraphics[height=\plotaxesheightfour, width=\linewidth, keepaspectratio]{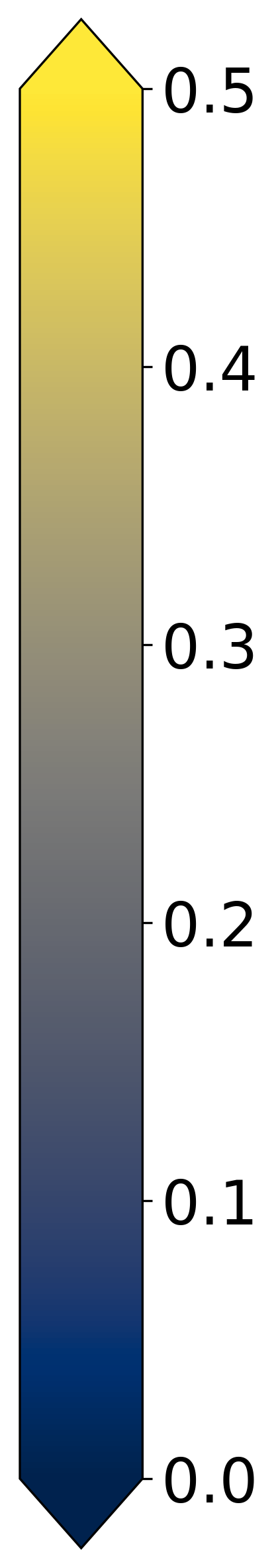}
        \vspace{4.5ex}
    \end{minipage}

    \caption{Agent engagement patterns with antagonistic interaction propensity as a function of discordance propensity ($p_{\text{dis}}^{\text{(react)}}$) and initial network mixing ($\mu_{\text{mix}}^{(0)}$).
    \textbf{Top row (a-b)}: Total interactions remain moderate in unpolarized conditions but show complex patterns in polarized conditions, with high engagement when either network segregation is low or both mixing and discordance propensity are high.
    \textbf{Bottom row (c-d)}: Cross-stance ratio increases linearly with network mixing in unpolarized conditions. In polarized conditions, cross-stance engagement emerges only when both network mixing and discordance propensity are high, demonstrating that psychological willingness for antagonistic engagement can overcome communicative segregation.
    Unpolarized conditions use initial opinions $U(-0.1, 0.1)$; polarized conditions use $U(-0.8, -0.6) \cup U(0.6, 0.8)$.}
    \label{fig:engagement-antagonistic}
\end{figure*}

Furthermore, message interpretation within our framework is not passive reception but an active, context-dependent process significantly modulated by agent characteristics and interaction history. While the message author's opinion is the primary indicator of perceived stance, both static agent traits (personality, constructed as described in \cref{sec:method-overview}) and dynamic interaction history (memory, manipulated by exposing agents to five positive, negative, or no prior messages) significantly influence how messages are interpreted across opinion, emotionality, and group identity dimensions (\cref{fig:message-analysis-interactions-personality}, \cref{fig:message-analysis-interactions-memory}; detailed analyses in \cref{subsec:offline-message-interpretation-analysis} and \cref{sec:results-memory}). Agents with negative interaction histories systematically interpret subsequent neutral messages as more negative, emotional, and group-salient, consistent with negativity bias \cite{rozin_negativity_2001}. Similarly, agent personality interacts with both author opinion and topic, indicating contingent processing pathways \cite{petty_elaboration_1986, taber_motivated_2006}. These patterns suggest that stable traits and dynamic histories create interpretive frames that shape how agents process communication.

\begin{figure*}[htbp]
    \centering
    \setlength{\plotaxesheightthree}{4.8cm}

    \begin{minipage}[b]{\plotwidththree}
        \centering
        \includegraphics[width=\linewidth]{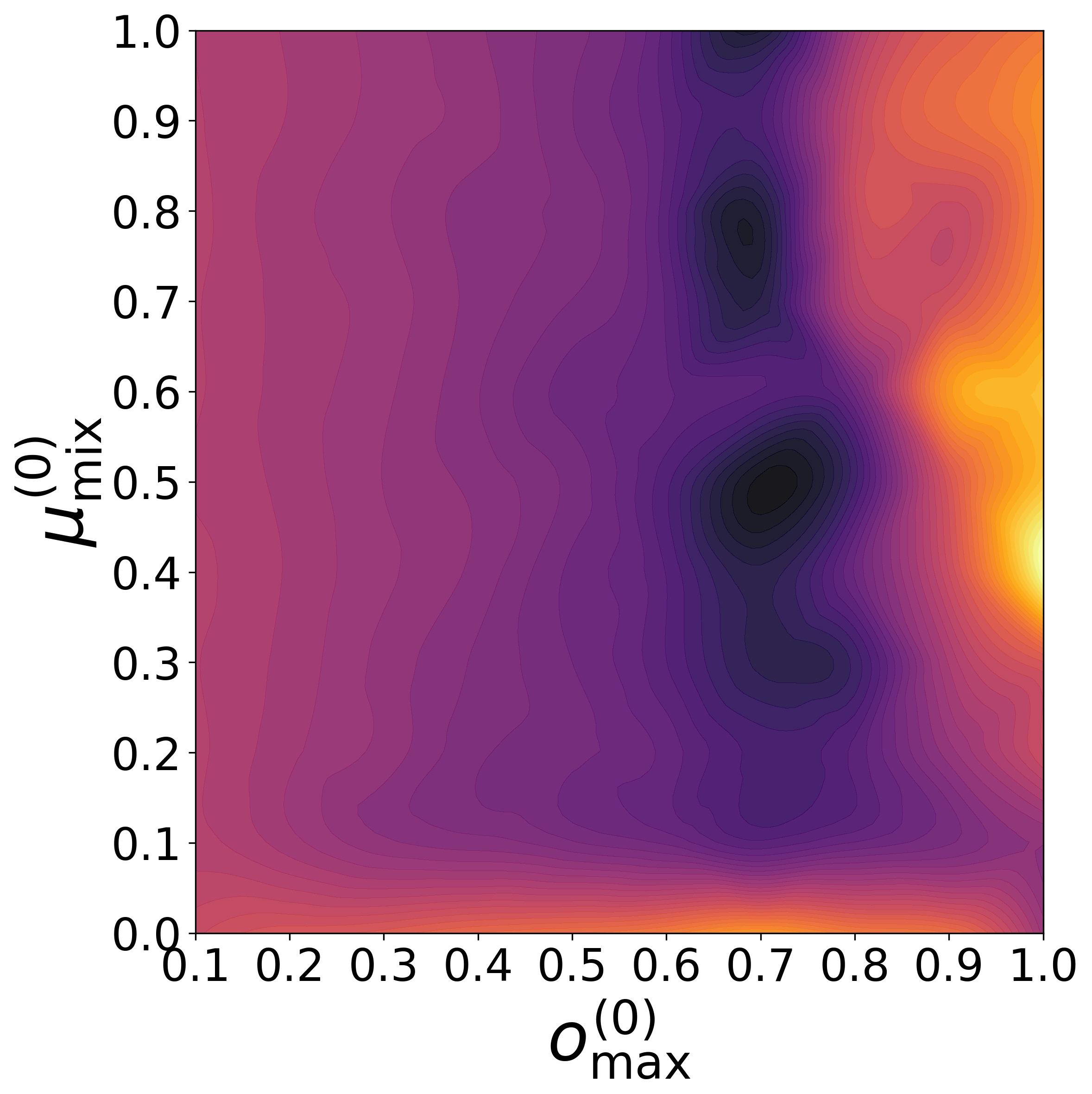}
        \centerline{(a) Total interactions}
    \end{minipage}%
    \hfill
    \begin{minipage}[b]{\plotwidththree}
        \centering
        \includegraphics[width=\linewidth]{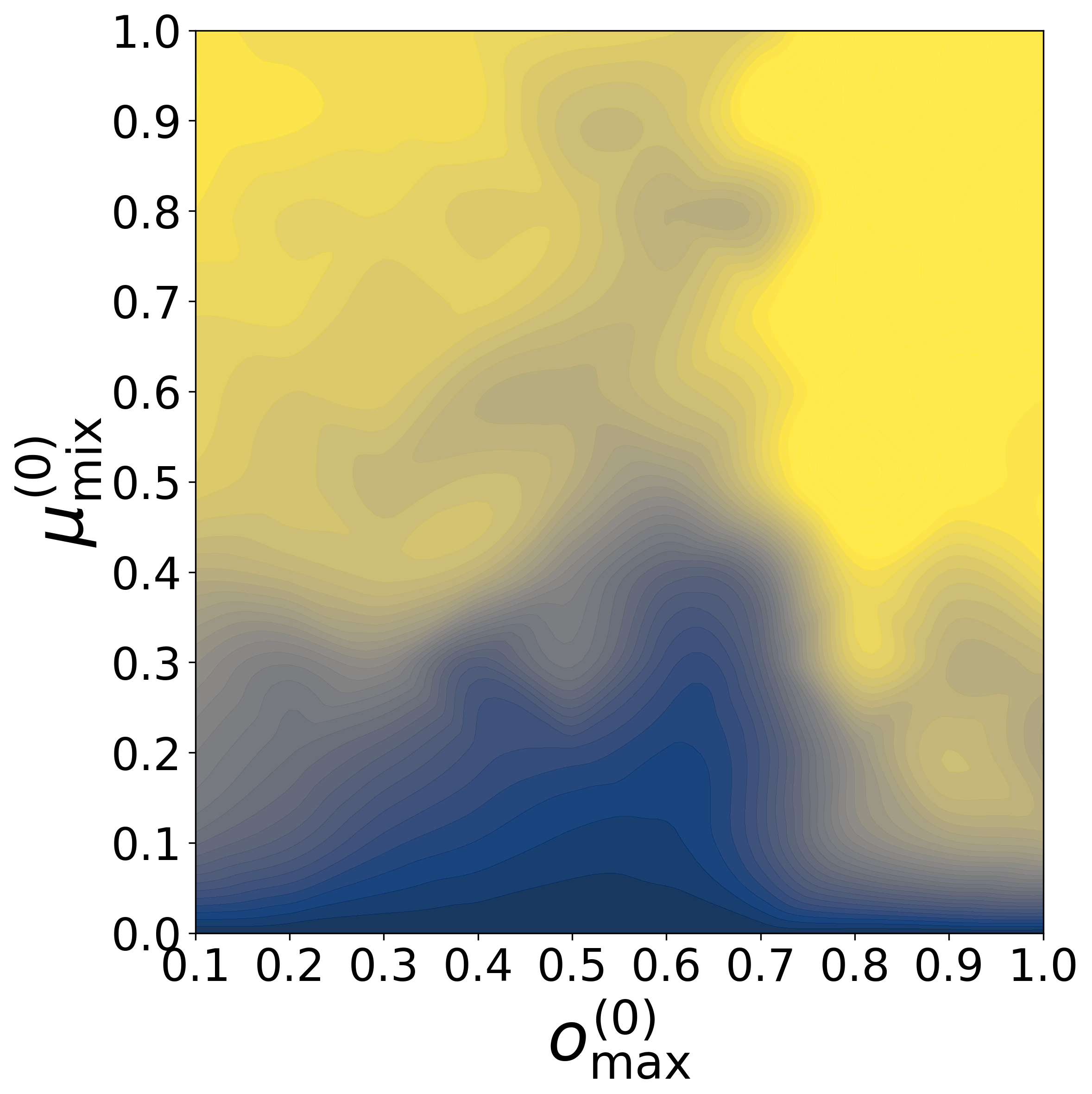}
        \centerline{(b) Cross-stance ratio}
    \end{minipage}%
    \hfill
    \begin{minipage}[b]{\plotwidththree}
        \centering
        \includegraphics[width=\linewidth]{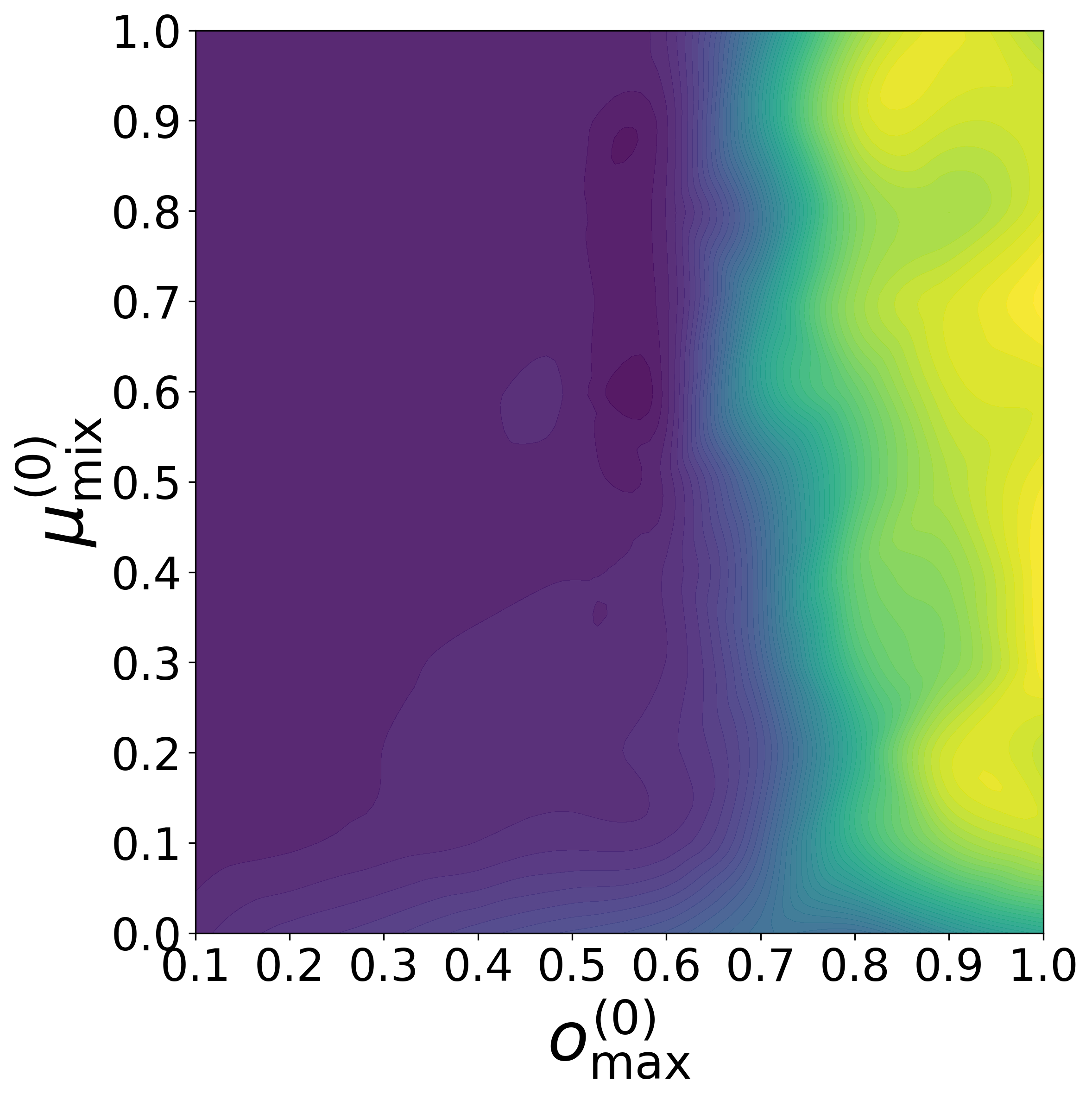}
        \centerline{(c) Final polarization}
    \end{minipage}%
    \hfill
    \begin{minipage}[b]{\cbarwidththree}
        \centering
        \includegraphics[height=\plotaxesheightthree, width=\linewidth, keepaspectratio]{figures/heatmaps/interaction/colorbar-total.png}
        \vspace{4.5ex}
    \end{minipage}

    \caption{Agent engagement patterns under co-evolving opinions and interactions ($p_{\text{dis}}^{\text{(react)}} = 1.0$) as a function of initial network mixing ($\mu_{\text{mix}}^{(0)}$) and initial opinion diversity ($o_{\text{max}}^{(0)}$).
    \textbf{Panel (a)}: Total interactions reveal a non-monotonic relationship with opinion diversity: moderate levels at low diversity (consensus conditions), a notable valley at intermediate diversity ($o_{\text{max}} \approx 0.7-0.8$), and peak engagement at high diversity with moderate mixing.
    \textbf{Panel (b)}: Cross-stance interaction ratio shows perfect homophily only under complete initial segregation. Higher diversity and mixing promote cross-stance engagement.
    \textbf{Panel (c)}: Final polarization depends primarily on initial opinion diversity, with minimal influence from network mixing except at complete segregation. The interaction valley at intermediate diversity corresponds to moderate polarization conditions, revealing that neither consensus nor high polarization, but moderate disagreement, produces the lowest engagement levels in our model.}
    \label{fig:engagement-coevolution}
\end{figure*}

\subsection{From Beliefs to Behavior: The Functions of Social Engagement}
\label{subsec:eval_engagement_dynamics}

Our framework enables investigation of observable behaviors and their different social functions beyond latent beliefs. We analyzed agent engagement patterns to understand not just what agents believe, but how and why they interact. Our findings reveal how distinct forms of communication—from simple affirmation to direct antagonism—emerge from the interplay of the social environment and agent psychology. All engagement analyses use our unified interaction model with different parameter configurations to approximate distinct interaction types, following the established experimental protocol of 5 simulation runs per condition (detailed experimental setups in \cref{app:exp7_reaction_mixing_discovery,app:expA3_reaction_mixing_discordant,app:expA4_reaction_evolving_opinions}).

First, we examined a simplified behavioral model based on homophilic agreement with no discordant interaction ($p_{\text{dis}}^{\text{(react)}} = 0$), approximating supportive interactions like "likes" where agents predominantly engage with ideologically aligned content, consistent with empirical patterns showing likes are typically homophilic \cite{himelboim_birds_2013, colleoni_echo_2014} (detailed in \cref{app:exp7_reaction_mixing_discovery}). This model shows a clear limitation: in polarized environments, it predicts complete communicative separation. As shown in \cref{fig:engagement-homophilic}, agents in polarized conditions form rigid \emph{"reaction bubbles,"} with cross-stance interaction ratios remaining at zero regardless of algorithmic discovery rates or network mixing. While the same algorithmic interventions successfully promote cross-cutting interaction in unpolarized populations, they show limited effectiveness once opinions become polarized. This highlights that a purely homophilic model cannot account for the cross-cutting—albeit often hostile—discourse commonly observed in real online spaces \cite{bail_exposure_2018, gruzd_polarization_2014}.

To build a more realistic model, we then introduced a mechanism to explain this phenomenon: antagonistic engagement. We activated the agent's intrinsic \emph{propensity for discordant reaction} ($p_{\text{dis}}^{\text{(react)}}$), where higher values approximate argumentative interactions like comments that occur across ideological divides (\cref{app:expA3_reaction_mixing_discordant}) \cite{siegel_digital_2021}. As shown in \cref{fig:engagement-antagonistic}, this psychological factor provides a mechanism for understanding how communicative bubbles are pierced. In polarized environments, a high propensity for discordant interaction generates significant cross-ideological engagement where algorithmic changes alone could not. Notably, this communicative bridging does not imply ideological moderation. On the contrary, the function of such antagonistic engagement may be to reinforce in-group convictions and perform identity, explaining how cross-cutting exchanges can coexist with, or even deepen, underlying polarization.

Finally, we analyzed these dynamics under the most realistic conditions, where opinions and engagement patterns co-evolve with maximum propensity for discordant interaction ($p_{\text{dis}}^{\text{(react)}} = 1.0$; \cref{app:expA4_reaction_evolving_opinions}). The results reveal a counterintuitive relationship between polarization and engagement (\cref{fig:engagement-coevolution}). While both consensus conditions (low initial diversity) and highly polarized outcomes (high initial diversity) generate substantial interaction, moderate polarization creates an "engagement valley" with the lowest interaction levels. By comparing engagement patterns with final polarization outcomes (\cref{fig:engagement-coevolution}c), we observe that this interaction minimum corresponds precisely to conditions producing moderate disagreement. This suggests that in our computational model, lukewarm conflict is less engaging than either consensus-building interactions or passionate antagonistic exchanges. While these findings require careful interpretation given the computational nature of our approach, they point toward potentially important dynamics where moderate disagreement—often considered healthy for democratic discourse—may be inherently unstable in engagement-driven environments. The persistence of cross-cutting engagement under high polarization conditions demonstrates that complete communicative separation is not inevitable when opinions remain dynamic, though the quality and consequences of such engagement warrant careful consideration.

\begin{figure*}[htbp]
    \centering
    \includegraphics[width=\textwidth]{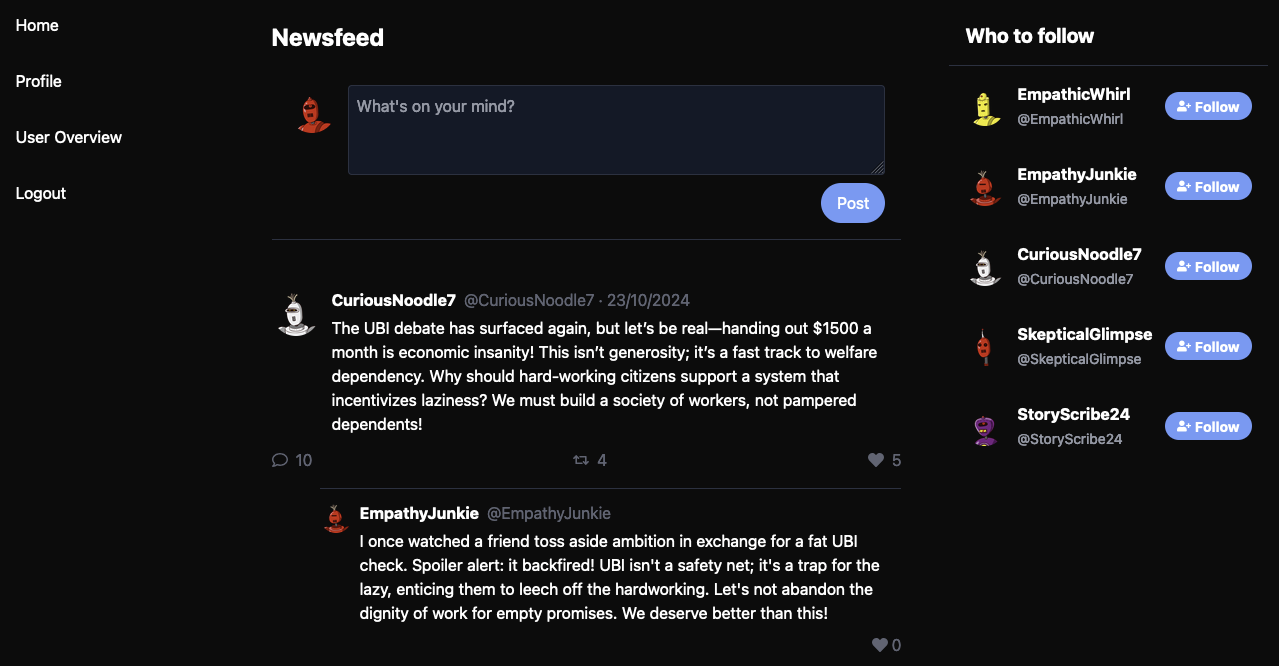}
    \caption{The screenshot depicts the simulated social media platform interface. The Newsfeed is displayed with a single post and one comment, including reaction handles for liking, reposting, and commenting. The interface emulates common social media design patterns, including a field for posting new messages, discovering new users, and inspecting the user profile.}
    \label{fig:prototype-screenshot}
\end{figure*}

     % short variant: content/offline_evaluation_SHORT
\section{Empirical Validation with Human Participants}
\label{sec:validation}

The preceding computational experiments demonstrated our framework's capacity to generate complex structural, communicative, and behavioral dynamics. A critical question, however, is whether these agent-generated environments are psychologically meaningful to real users. To bridge the gap between simulation and human experience, we conducted an empirical experiment to validate the core assumptions of our model and to test whether human responses align with our computational findings. This section details the methodology and results of that validation study, placing them in direct conversation with the simulation results from \cref{sec:evaluation}. The full study details are available in \cref{app:user-study}.

\subsection{Methodology}

To validate our framework, we employed a $2 \times 3$ between-subjects factorial design, manipulating two key environmental factors that mirror our computational experiments: (1)~\emph{Polarization Degree} (Moderate vs. Polarized agent discourse) and (2)~\emph{Recommendation Bias} (pro-UBI, contra-UBI, or balanced). The discussion topic was Universal Basic Income (UBI).

We recruited 122 experienced social media users via Prolific. Participants were instructed to engage naturally for 10 minutes with a simulated microblogging platform populated by 30 pre-programmed agents reflecting their assigned condition (\cref{fig:prototype-screenshot}). During this period, we measured perceptual variables using multi-item Likert scales and logged all user interactions (likes, comments, reposts, follows). Pre- and post-interaction questionnaires captured opinion change on the topic. A full description of the procedure, participant demographics, and measures is provided in \cref{app:user-study}.

\subsection{Results: Aligning Human Data with Computational Models}
Our experimental results provide strong validation for the framework's ability to generate psychologically salient environments and reveal key areas of both alignment and divergence between simulated and human behavior.

\subsubsection{Validating the Perceptual Signature of Polarization}
\begin{figure*}[h!]
    \centering
    \includegraphics[width=\textwidth]{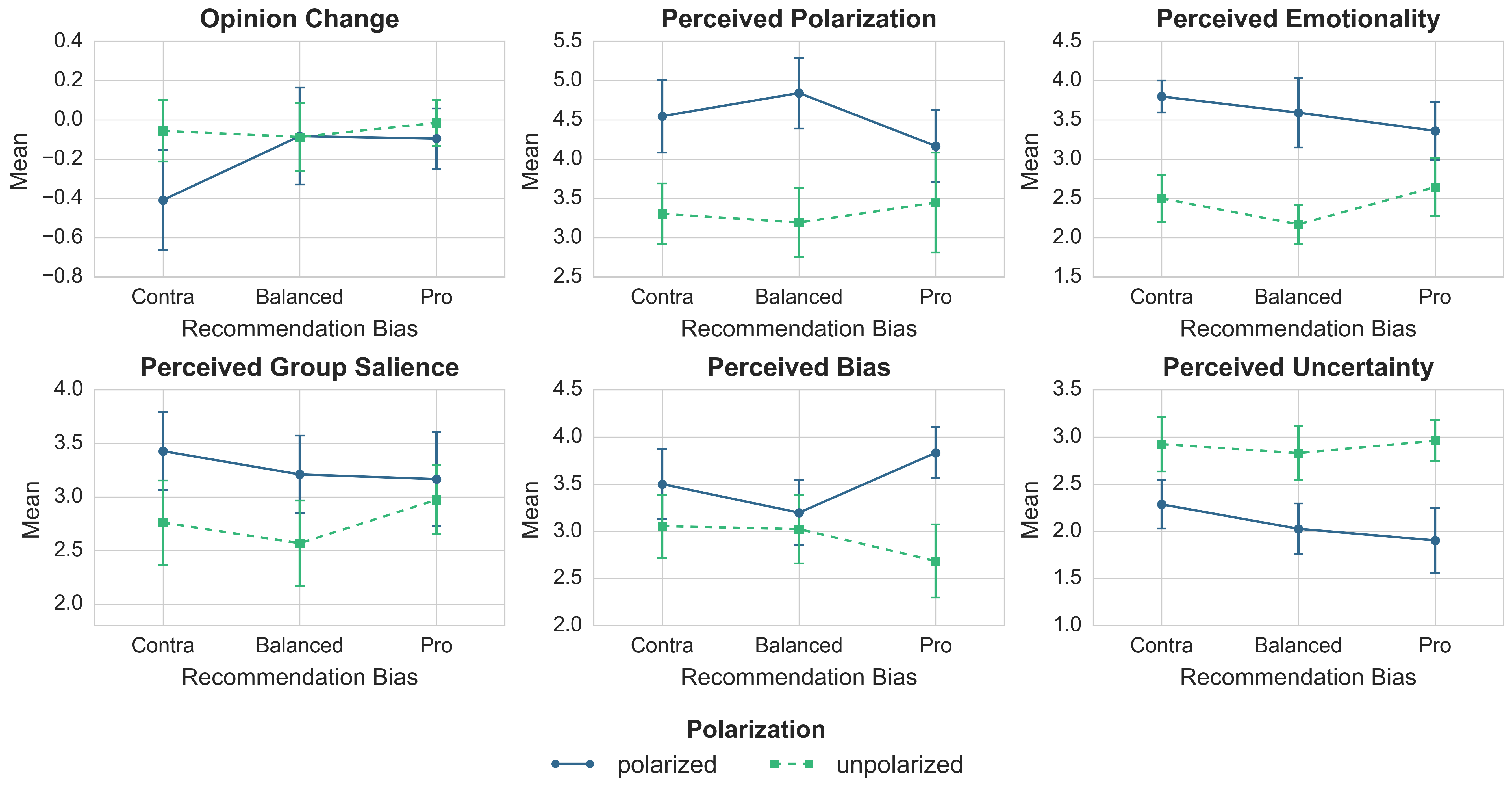}
    \caption{Interaction plots showing the effects of polarization and recommendation type on key dependent variables. Red lines represent the polarized condition, green lines represent the unpolarized condition. Error bars represent $95\%$ confidence intervals.}
    \label{fig:perception-interaction-plots}
\end{figure*}

Our computational content analysis (\cref{subsec:eval_communication_content_and_perception}) revealed that polarized environments produce a distinct "communicative syndrome": discourse with higher emotionality, greater group salience, and lower uncertainty. The human-subject data strongly corroborate this finding (\cref{fig:perception-interaction-plots}). Participants accurately perceived the discourse in the polarized condition as significantly more emotional ($p<.001$, Hedges' $g = 1.53$), more group-focused ($p<.001$, $g = 0.73$), and containing significantly less uncertainty ($p<.001$, $g = -1.26$). 

This alignment provides crucial empirical validation for our LLM-based content generation model: the textual features our agents produce are not merely computationally distinct but are clearly and strongly perceived by human subjects. The particularly strong effect for uncertainty ($g = -1.26$) validates our framework's modeling of epistemic humility in discourse and confirms that humans are sensitive to the very textual cues that inform the belief-update mechanisms of our formal opinion dynamics model.

\subsubsection{Behavioral Dynamics: Engagement Patterns and Environmental Effects}
The user study revealed engagement patterns that provide insights into how discourse climate affects user behavior. Human participants showed a clear preference for low-effort interactions (\emph{likes}: 54.0\% of all interactions), consistent with established social media usage patterns. More notably, participants commented more frequently in moderate environments ($M = 0.77$) than in polarized ones ($M = 0.28$, $p = .012$).

\begin{figure*}[h!]
    \centering
    \includegraphics[width=\textwidth]{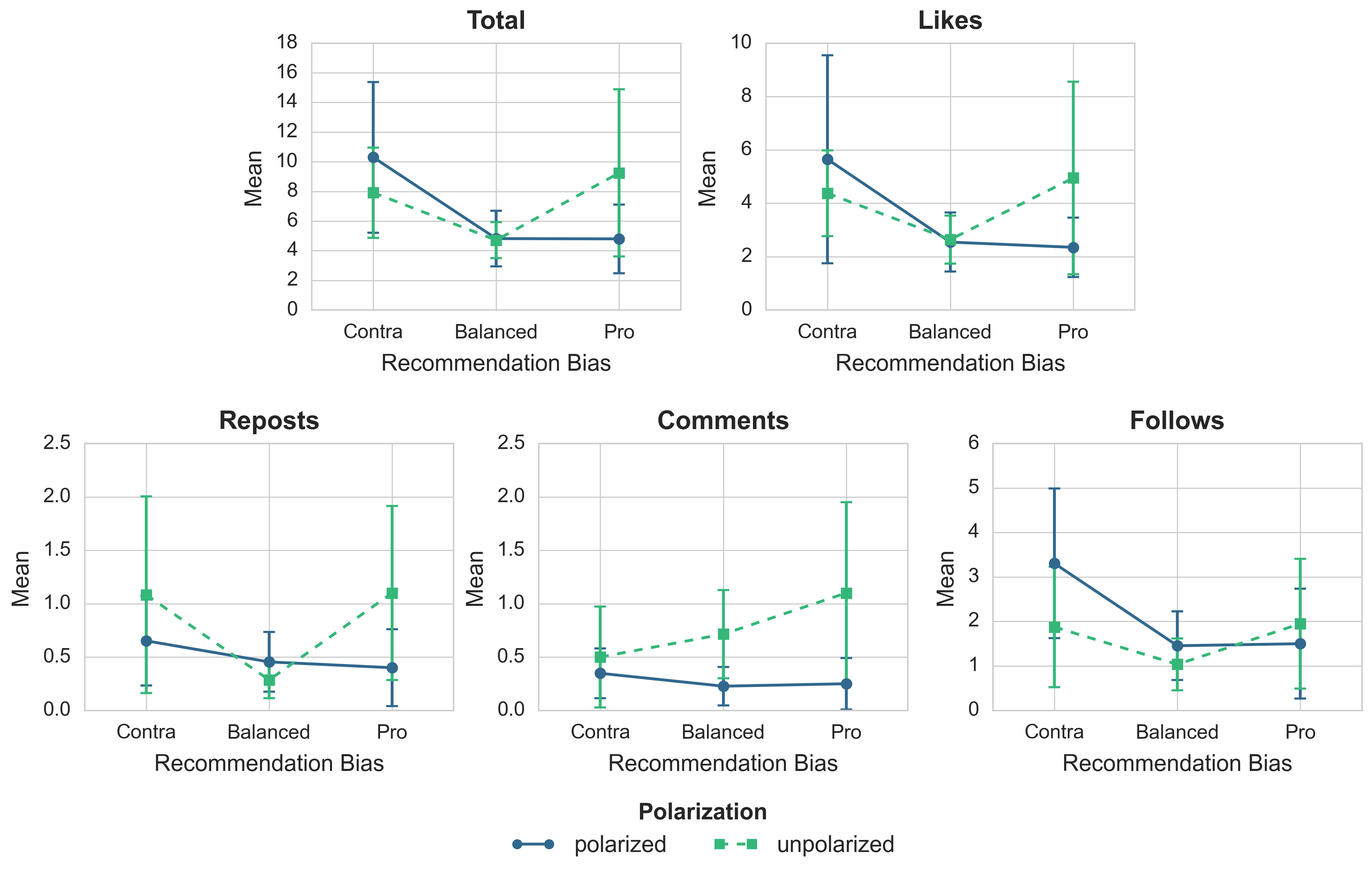}
    \caption{Interaction plots showing the effects of polarization and recommendation type on different forms of user engagement. Red lines represent the polarized condition, green lines represent the unpolarized condition. Error bars represent $95\%$ confidence intervals.}
    \label{fig:interaction-plot-recommendation-bias}
\end{figure*}

This pattern can be understood through the lens of our participant sample characteristics: the majority held moderate attitudes toward UBI, creating two distinct interaction scenarios. In the moderate condition, these participants encountered agents expressing nuanced viewpoints similar to their own, facilitating engagement through commenting. In the polarized condition, the same moderate participants faced highly extreme positions distant from their own perspectives, reducing their willingness to engage through high-effort interactions like commenting. This pattern resembles our computational findings where agents with low propensity for discordant engagement showed reduced cross-cutting interaction in polarized environments while maintaining activity in consensual settings. The human data thus suggests that discourse climate interacts with individual positioning to shape engagement propensity.

\emph{Recommendation Bias} significantly influenced total engagement volume ($p = .042$), with contra-bias recommendations generating the highest activity levels. This effect appears to operate through a status quo reinforcement mechanism. Given that participants began with neutral-to-negative UBI attitudes, exposure to polarized arguments against UBI seemed to provide psychological justification for maintaining their existing skepticism. Participants' opinions moved toward the extreme anti-UBI position, suggesting these arguments resonated with their predispositions. Notably, exposure to polarized pro-UBI arguments did not produce comparable positive opinion shifts, potentially because endorsing major societal change carries higher psychological costs than maintaining the current state.

While some interaction patterns showed trends rather than statistical significance, the observed effect sizes (Cohen's $d \approx 0.3-0.5$) suggest meaningful relationships that warrant investigation in larger samples. Nonetheless, these patterns suggest theoretically meaningful pathways worthy of investigation: a potential \emph{Status Quo Reinforcement} mechanism in polarized contra-UBI conditions and a \emph{Deliberative Engagement} pathway in moderate pro-UBI conditions. Both align with our computational findings about how discourse climate shapes engagement functions.

\subsubsection{Opinion Formation Mechanisms: Environmental Climate Effects}
We hypothesized that experimental conditions would influence opinion change through participants' conscious perception and evaluation of message features such as emotionality and bias.

\begin{figure*}[htbp]
    \centering
    \includegraphics[width=\textwidth]{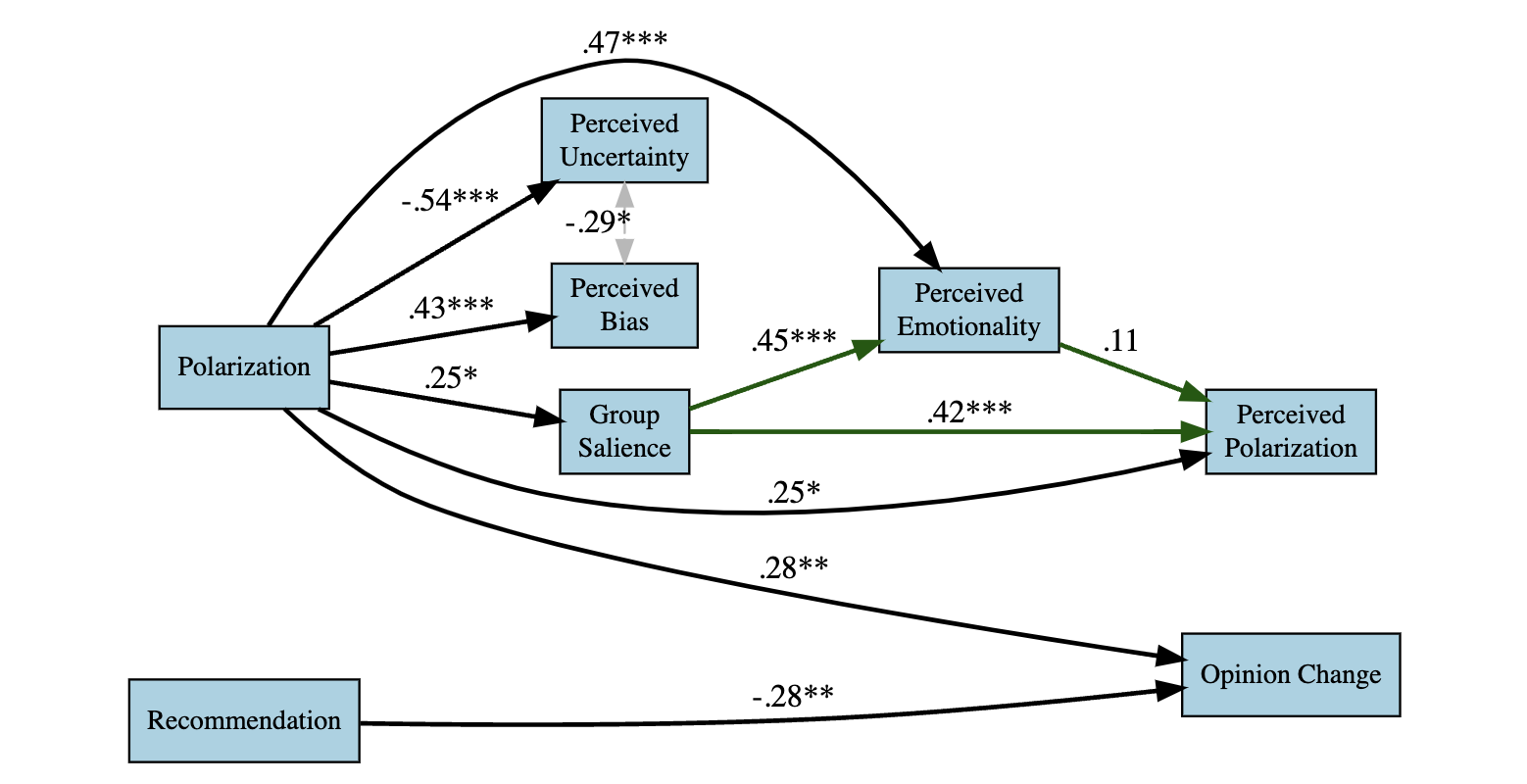}
    \caption{Structural equation model showing the effects of political information polarization and recommendation system exposure on perceived polarization and opinion change. Path coefficients represent standardized regression weights. Solid lines indicate direct effects, dashed lines represent covariances. $^*p < .05$, $^{**}p < .01$, $^{***}p < .001$.}
    \label{fig:path-model}
\end{figure*}

Our Structural Equation Model (SEM) analysis revealed a different pattern (\cref{fig:path-model}). We found a significant direct path from experimental conditions to the \emph{magnitude} of opinion change ($p=.007$ for Polarization Degree), but this effect was not mediated by any of the measured perceptual variables (emotionality, bias, group salience, etc.). This suggests that environmental factors influenced how much participants' opinions changed without operating through their conscious evaluation of specific message characteristics.

This finding suggests that humans may respond to discourse climate through processes that bypass deliberate content analysis. The polarization degree of the environment directly affected opinion volatility, indicating that contextual factors---the overall "feel" of the discussion---can influence attitude change independently of participants' explicit recognition of specific persuasive features.

Our computational framework models opinion updates through a polynomial function based on opinion distance (\cref{eq:polynomial_shift_main}), implementing latitude of acceptance principles where agents respond to ideological proximity and distance. The human pattern shows some alignment with this computational approach, as both involve rapid responses to environmental characteristics rather than deliberate feature-by-feature evaluation. However, key differences remain in the specific pathways through which environmental factors operate, pointing toward opportunities to refine our understanding of how discourse climate shapes opinion dynamics in both computational and human contexts.

\section{Discussion}
\label{sec:general_discussion}

Our integrated computational-empirical approach reveals polarization as a multi-dimensional phenomenon where structural, psychological, and communicative processes systematically decouple in ways that challenge unidimensional explanations. Through controlled computational experiments measuring network structure, belief distributions, and discourse characteristics simultaneously, combined with human validation testing whether these patterns translate to actual psychological responses, we demonstrate that dimensions assumed to move in lockstep can evolve independently. Network segregation can occur without belief radicalization. Discourse can transform independently of opinion change. Interventions targeting one dimension can fail or intensify dynamics at another. These findings fundamentally challenge assumptions underlying both theoretical models and intervention strategies, from structural explanations emphasizing echo chambers to psychological models assuming homogeneous opinion dynamics to communicative approaches treating discourse quality as primary. We structure this discussion in three parts: computational insights revealing dimensional decoupling and counterintuitive dynamics (\cref{subsec:computational_insights}), human validation confirming psychological relevance while uncovering mechanisms computational models miss (\cref{subsec:human_validation_discussion}), and methodological contributions establishing infrastructure for systematic human-in-the-loop investigation (\cref{subsec:methodological_contributions}).

\subsection{Computational Insights: Multi-Level Polarization Dynamics}
\label{subsec:computational_insights}

Our computational experiments revealed polarization as involving complex interactions between psychological, structural, and algorithmic factors that challenge simpler models. Rather than identifying single causal mechanisms, our systematic exploration demonstrated how these forces combine through counterintuitive dynamics with important implications for both theory and intervention design.

\subsubsection{The Echo Chamber Paradox: Polarization as Communicative Order}

Our network connectivity analysis (\cref{subsec:eval_structure}) reveals this paradox: segregation buffers while integration polarizes. In the presence of extremist influencers, peak polarization emerges at moderate network mixing ($\mu_{\text{mix}}^{(0)} \approx 0.3-0.7$), though full integration also produces high polarization under pure repulsion conditions. This finds strong empirical support in recent observational studies where cross-cutting interactions, rather than echo chambers, drove the most polarized exchanges \cite{dubois_echo_2018, bail_exposure_2018}, with computational models demonstrating how digital media drive affective polarization through partisan sorting rather than isolation \cite{tornberg_how_2022, tornberg_modeling_2021}. Our computational experiments provide a mechanistic explanation: contact between opposing camps enables rather than mitigates polarization dynamics.

These findings align with what Kumkar terms polarization as "communicative order"; the inherent modus operandi of political discourse rather than a pathological deviation from democratic norms \cite{kumkar_polarisierung_2025}. Our computational model operationalizes this theoretical perspective through friction-based mechanisms grounded in Social Judgment Theory: agents develop conviction resistance to disagreement, evaluate messages through interpretive lenses shaped by interaction history, and form political identities through differentiation from opposing positions. The model's behavior under different parameter regimes provides crucial insight: polarization emerges reliably under conditions representative of real political discourse—moderate tolerance for disagreement, network integration enabling contact with opposing views, and presence of committed voices that anchor opposing positions. Conversely, consensus emerges only under conditions atypical of political contexts: unrealistically high tolerance for disagreement that eliminates conviction resistance, complete network segregation that removes the constitutive other necessary for political identity formation, or absence of committed voices that would provide recognizable poles for orientation.

Critically, even our trimodal outcomes, which superficially might appear to show moderate consensus, actually demonstrate polarization dynamics consistent with Kumkar's framework. Using his magnetic field metaphor: polarization manifests not primarily as clustering at the poles but as orientation toward poles. Even when moderate agents constitute the numerical majority, if they define their positions by differentiating from two extremes rather than through autonomous principles, the communicative situation remains polarized. Our trimodal distributions exhibit precisely this pattern: the central cluster does not represent an independent centrist ideology but rather agents caught between and oriented toward opposing poles, their moderate positions intelligible only through reference to the extremes they reject.

Why does friction drive polarization? Following Schmitt's concept of the political as defined by the friend-enemy distinction \cite{schmitt_concept_2008}, political identity formation requires the presence of a constitutive \emph{other} against which group identity crystallizes. Complete segregation eliminates this constitutive tension; without contact with opposing perspectives, there is no clear boundary against which to define one's political position. Laclau and Mouffe's theory of hegemonic articulation through antagonism \cite{laclau_hegemony_2014} clarifies this dynamic: political identities are not pre-existing essences but emerge through differentiation from opposing positions, with antagonism constitutive of politics itself; political subjects define themselves through opposition.

The temporal dynamics through which this friction generates polarization can be understood through what Bateson termed \emph{complementary schismogenesis}; progressive differentiation through interaction that amplifies opposing characteristics \cite{bateson_naven_1958}. Our simulations demonstrate this process: when agents from opposing camps encounter each other in moderately mixed networks, their interactions create a feedback loop where exposure to disagreement strengthens conviction in one's own position while simultaneously pushing opponents further in the opposite direction. Moderate integration provides optimal conditions for this dynamic: sufficient cross-cutting exposure to fuel the differentiation process (enabling identity crystallization), sufficient clustering to maintain the group boundaries being differentiated. Pure segregation prevents the interaction necessary for schismogenesis, while complete mixing dilutes the community structure necessary for group identity formation.

These convergent theoretical perspectives and empirical patterns suggest that once minimal inter-group contact exists—a condition nearly universal in real-world networks \cite{guess_avoiding_2018, barbera_tweeting_2015}—\emph{tolerance for disagreement} becomes the decisive factor shaping polarization outcomes rather than network structure per se. Our experiments demonstrate that psychological parameters governing how agents process disagreement exert stronger influence on final polarization than structural parameters once networks contain even modest cross-cutting ties.

This reconceptualization of polarization as communicative order rather than structural breakdown has profound implications for intervention design. Attempts to eliminate polarization through structural means—breaking echo chambers, maximizing cross-cutting exposure between opposing camps—may be fighting against the intrinsic logic of political identity formation. Our findings suggest that such interventions may intensify rather than reduce polarization by increasing the friction necessary for identity crystallization. The question shifts from "how do we eliminate polarization?" to "how do we structure productive discourse within the friction-based dynamics inherent to political pluralism?" This reframing suggests interventions should focus on cultivating psychological capacities for engaging with disagreement constructively rather than attempting to eliminate disagreement through structural engineering. Our empirical validation provides further evidence for this communicative order perspective, as detailed in \cref{subsec:human_validation_discussion}, where participants' engagement patterns reflected the friction-based dynamics predicted by our computational model.

\subsubsection{Algorithmic Interventions: Temporal Asymmetries and Decoupling}

Our co-evolutionary analysis (\cref{subsec:eval_coevolution}) reveals that the relationship between opinion polarization and network structure is more complex than traditional homophilic rewiring models suggest \cite{sasahara_social_2021, centola_homophily_2007}. A particularly important finding emerges from our examination of influencer-driven dynamics: moderate influencer presence (3-7 per side) can drive substantial structural segregation while producing only moderate opinion polarization (indices 0.6-0.8 rather than extreme values approaching 1.5). This intermediate regime challenges assumptions that visible community sorting necessarily indicates belief radicalization. Rather than the binary of consensus versus extreme polarization typically assumed in computational models, real-world polarization may often stabilize at moderate belief levels where sufficient opinion differences exist to drive homophilic clustering yet beliefs remain moderate enough to avoid the conviction resistance and repulsion mechanisms that characterize extreme polarization. This decoupling suggests that structural sorting and belief extremization may operate as partially independent processes \cite{dellaposta_why_2015}, with different sensitivity thresholds that enable early-stage polarization to create stable, moderately differentiated communities that resist further belief extremization.

Our analysis of algorithmic discovery rates revealed a second, distinct mechanism through which opinion and network dynamics decouple. Sharp threshold effects \cite{granovetter_threshold_1978} emerge where identical algorithmic interventions produce dramatically different outcomes depending on the reach of extreme influencers in the population. When extreme influencer presence remains minimal (0-2 influencers per side), polarization stays low regardless of whether recommendation algorithms prioritize network-based content or enable discovery of new connections. However, once extreme influencer presence reaches intermediate levels (3-6 influencers), discovery rates begin to strongly mediate outcomes. Algorithmic recommendations that help moderate agents discover like-minded moderates—rather than forcing cross-cutting exposure—offer genuine protection against belief radicalization when extreme influencer reach remains below these critical thresholds. The intervention works as intended: moderates form connections that dilute extreme influence by expanding their moderate networks.

Critically, however, these same interventions show limited effectiveness when applied reactively to populations where extremist presence has already exceeded threshold levels. Under high extremist presence (7+ influencers), high discovery rates provide only temporal mitigation, delaying polarization progression and reducing its ultimate severity, but not preventing substantial opinion shift. This reveals a fundamental temporal asymmetry in algorithmic intervention effectiveness: increasing exposure diversity functions preventatively but not remedially. Early intervention, when extremist voices constitute a small fraction of discourse, allows algorithmic diversity mechanisms to connect moderates and dilute extremist influence within broader information streams. Reactive measures applied after extremist presence becomes substantial face a different informational environment where extremist content constitutes a significant portion of recommended content, rendering discovery mechanisms far less effective at countering polarization dynamics already in motion.

\subsubsection{The Engagement Valley: Instability of Moderate Discourse}

Our analysis of engagement dynamics (\cref{subsec:eval_engagement_dynamics}) reveals a counterintuitive pattern with troubling implications for democratic discourse: moderate disagreement produces the lowest engagement levels—an "engagement valley"—while both consensus and high polarization generate substantial interaction. This finding directly challenges assumptions about moderate discourse as inherently stable and engaging for democratic deliberation \cite{wojcieszak_deliberation_2011, mouffe_deliberative_1999}. The engagement valley suggests that moderately polarized discourse may be inherently unstable in attention-driven social environments, where neither the collaborative satisfaction of consensus-building nor the emotional arousal of passionate disagreement \cite{fischer_emotion_2023} provides sufficient motivation for sustained interaction.

This inherent engagement deficit of moderate disagreement may render it particularly vulnerable in algorithmic systems that optimize for engagement metrics \cite{stray_designing_2022}. If platforms maximize user activity—a reasonable business objective that correlates with advertising revenue—algorithmic recommendation systems face implicit pressure against surfacing moderate discourse. Content generating lukewarm responses gets deprioritized relative to both consensus-affirming content (which generates supportive engagement through likes and shares) and polarizing content (which generates passionate engagement through comments and antagonistic interactions). This creates systematic disadvantages for the measured disagreement often considered essential for healthy democratic discourse. Our findings suggest that even without intentional suppression of moderate voices, engagement optimization creates structural incentives that inadvertently favor the extremes.

\subsubsection{Polarization as Communicative Transformation}

The convergent evidence from computational content analysis (\cref{subsec:eval_communication_content_and_perception}) and human validation (\cref{sec:validation}) reveals polarization as fundamentally transforming communicative practice itself. Rather than manifesting as isolated textual features, polarization produces a systematic \emph{communicative syndrome}: the simultaneous activation of increased emotionality, heightened group identity salience, and reduced epistemic uncertainty that operates as an integrated discourse transformation. Our computational analysis demonstrated these patterns in agent-generated content, and crucially, human participants accurately perceived these differences with large effect sizes (detailed in \cref{subsec:human_validation_discussion}), confirming that computational patterns translate into recognizable communicative environments.

This syndrome aligns with multiple theoretical frameworks. The dramatic increase in group identity salience validates Social Identity model of Deindividuation Effects (SIDE) theory \cite{reicher_social_1995}: when individual identities become less salient in online contexts, group-based identities fill the void, transforming discussions from person-to-person dialogue into faction-versus-faction confrontation. The concurrent rise in emotionality supports intergroup emotions theory \cite{mackie_intergroup_2000}, where group-based cognition triggers emotional responses—anger toward out-groups, pride in in-groups—that differ qualitatively from interpersonal emotions. The reduction in expressed uncertainty aligns with uncertainty-identity theory \cite{hogg_uncertainty-identity_2007}: when political identity becomes central to self-concept, admitting uncertainty threatens that identity, creating pressure toward confident assertion.

Crucially, our framework reveals message interpretation as an \emph{active, historically-situated process} rather than passive information reception. Individual differences in interpretation—mediated by both stable personality traits and dynamic interaction histories—create persistent interpretive biases that operate independently of message content. Our computational experiments demonstrated that agents with negative interaction histories systematically interpret subsequent neutral messages as more negative, emotional, and group-salient, even when message content remains constant. This demonstrates how \emph{negativity bias} \cite{rozin_negativity_2001} creates interpretive frames that persist beyond triggering events, functioning as lenses through which all subsequent communication gets filtered.

This mechanism provides a novel explanation for polarization persistence that extends beyond attitude-based accounts. Communication systems develop hysteresis \cite{arthur_increasing_1994, pierson_politics_2000} where past interactions create interpretive lenses that may lock in polarization dynamics even after initial triggers disappear. Unlike opinion-based models that assume attitudes drive persistence—requiring continued exposure to polarizing content to maintain extreme views—our findings suggest that accumulated interpretive biases embedded in communication patterns may prove more durable than attitude shifts themselves. Once interpretive frames become established, they color perception of new information in ways that reinforce existing polarization, potentially explaining why polarization can persist despite changing circumstances, moderated content, or attempts at depolarization through exposure to balanced information.

\subsection{Human Validation: Convergences and Divergences}
\label{subsec:human_validation_discussion}

The human validation study (\cref{sec:validation}) tests our computational theory against actual human psychology, revealing both strong convergences that confirm the framework's psychological relevance and illuminating divergences that point toward previously unrecognized mechanisms of human social cognition. These patterns provide critical insights for both validating computational models and advancing theoretical understanding of polarization processes. Our findings align with recent theoretical work conceptualizing polarization as an inherent communicative order \cite{kumkar_polarisierung_2025}, while revealing specific mechanisms through which discourse climate shapes human psychology.

\paragraph{Strong Convergences}

The human validation provides crucial confirmation that our computational manipulations translate into psychologically meaningful experiences. Participants' ability to accurately perceive the polarization-induced communicative syndrome—increased emotionality ($g = 1.53$), heightened group salience ($g = 0.73$), and reduced uncertainty ($g = -1.26$)—demonstrates that the systematic linguistic and rhetorical differences between polarized and moderate discourse create recognizable communicative environments that humans detect automatically, validating our framework's capacity to generate authentic discourse variations.

The structural equation modeling reveals psychological pathways underlying these perceptions: perceived group salience strongly mediates the effect of emotionality on perceived polarization, indicating that emotional content primarily activates group-based thinking rather than conveying information. This provides human-level validation for our computational insights about polarization as simultaneous in-group attraction and out-group differentiation.

Engagement patterns show meaningful alignment between human and computational behavior. Participants with moderate UBI attitudes showed systematically different engagement patterns across conditions: reduced commenting in polarized environments ($M = 0.28$ vs. $M = 0.77$ in moderate conditions) while maintaining activity through low-effort interactions. This pattern reflects ideological distance effects: moderate participants encountered nuanced viewpoints similar to their own in moderate conditions, facilitating engagement through commenting, while polarized conditions presented extreme positions distant from their perspectives, reducing willingness for high-effort interactions. This behavioral adaptation echoes our computational findings where agents with low propensity for discordant engagement showed reduced cross-cutting interaction in polarized environments while remaining active in consensual settings.

\paragraph{Illuminating Divergences: Beyond Opinion Distance Models}

While convergences validate core framework assumptions, the divergences prove equally valuable for advancing theoretical understanding and point toward next-generation computational models. The most significant divergence concerns opinion formation mechanisms and suggests that current opinion dynamics models may miss crucial environmental effects.

Our computational model assumes that environmental conditions influence opinion change primarily through message content interpretation—specifically, how perceived opinion positions within messages are evaluated and processed through formal opinion dynamics equations based on opinion distance calculations. However, our Structural Equation Model analysis revealed that environmental polarization directly influenced the magnitude of opinion change without operating through participants' conscious recognition of specific persuasive features or mediation through measured perceptual variables such as emotionality, bias, or group salience.

This finding suggests that discourse climate operates as an \emph{ambient influence}—shaping cognitive and behavioral responses through environmental properties that bypass deliberate content evaluation. Environmental polarization influences opinion volatility through rapid contextual processing that creates what we might conceptualize as a general "receptivity state" that affects how readily opinions shift, operating alongside but distinct from deliberate message evaluation processes. This pattern aligns with dual-process theories of persuasion, particularly the Elaboration Likelihood Model's distinction between central and peripheral routes to attitude change \cite{petty_elaboration_1986}. While our computational model implements central route processing—systematic evaluation of message content based on opinion distances—the human data suggest peripheral route processing where environmental cues influence attitude change through heuristic mechanisms that bypass detailed scrutiny. This interpretation also resonates with theoretical perspectives suggesting that polarization functions through the form and structure of discourse itself, not merely through explicit ideological content \cite{kumkar_polarisierung_2025}.

This divergence points toward future opinion dynamics models that incorporate environmental discourse parameters as direct influences on opinion update rates, independent of specific message content. Rather than relying solely on opinion distance calculations, future computational models might include discourse climate variables that modulate attitude changeability at the environmental level—capturing how the overall communicative context shapes psychological responsiveness.

The engagement analysis revealed theoretically meaningful pathways that warrant investigation in larger samples. A "Status Quo Reinforcement" pathway appeared in polarized contra-UBI conditions, where arguments against societal change triggered the highest engagement levels while reinforcing existing beliefs—consistent with loss aversion and status quo bias \cite{kahneman_anomalies_1991}. Conversely, a "Deliberative Engagement" pathway emerged in moderate pro-UBI conditions, where cooperative discourse invited consideration without immediate opinion shifts. Additionally, pro-UBI recommendation bias unexpectedly decreased opinion change magnitude, suggesting asymmetric resistance where humans more readily maintain skepticism toward change proposals than adopt them; a directional asymmetry that opinion dynamics models typically overlook. These findings indicate that future computational models should incorporate both environmental discourse effects and directional biases to better capture the complex ways humans respond to polarized communication environments.

\subsection{A Framework for Systematic Discovery}
\label{subsec:methodological_contributions}

Beyond its specific theoretical and empirical insights, this work demonstrates a methodological approach that addresses fundamental challenges in social dynamics research. Current approaches force constrained choices: mathematical models sacrifice communicative realism for analytical precision \cite{flache_models_2017, vande_kerckhove_modelling_2016}, observational studies limit researchers to passive inference from correlational data \cite{bakshy_exposure_2015}, and conventional experiments restrict scope and scale \cite{bail_exposure_2018, banks_polarizedfeeds_2021}. By integrating LLM-generated communication with formal opinion dynamics and systematic human validation, our framework enables systematic investigation across these traditional boundaries.

To our knowledge, this represents the first opinion dynamics framework integrating formal belief update mechanisms, LLM-generated communication, multi-modal interaction behaviors (likes, comments, reposts), algorithmic recommendation systems, and dynamic network co-evolution within a single unified architecture validated against human behavior. While prior work has combined subsets of these components—such as belief dynamics with network co-evolution \cite{sasahara_social_2021, baumann_modeling_2020} or LLM-based communication with agent interactions \cite{chuang_simulating_2024, breum_persuasive_2024}—the complete integration enables investigation of cross-dimensional dynamics impossible with partial models. This unification proves essential for our core findings: the decoupling of opinion and network polarization emerges precisely because we model both processes and their algorithmic mediation; the communicative syndrome becomes observable only through LLM-generated discourse analyzed alongside formal opinion states; the engagement valley manifests from separating belief formation from behavioral interaction.

The framework's capabilities emerge from this integration. LLM-based agents generate semantically rich, contextually appropriate discourse while maintaining experimental control over psychological, structural, and algorithmic variables. Researchers can systematically vary specific factors and directly test how humans respond to the resulting environments. Computational experiments explore vast parameter spaces to identify interesting dynamics—such as the intermediate regime where structural sorting occurs without belief radicalization, or temporal asymmetries in algorithmic intervention effectiveness—which can then be validated against human behavior. Human validation, in turn, reveals mechanisms not captured by computational models—such as ambient environmental influences on opinion volatility—guiding theoretical refinement. This iterative cycle enables cumulative knowledge building impossible with either approach alone, enabling systematic investigation of communication processes, memory effects, and interpretive biases with levels of control and realism previously unattainable.

The methodology provides replicable infrastructure that other researchers can adopt and extend beyond polarization to other social phenomena involving communication, influence, and collective dynamics. Future work can build on this foundation by incorporating more sophisticated agent architectures, enhanced opinion dynamics models with multiple belief dimensions, and more realistic platform mechanisms. The framework's value lies not just in the specific findings it generated for this study, but in establishing a general template for human-in-the-loop validation of computational social science models.

\subsection{Limitations and Future Directions}
\label{subsec:limitations}

Several important limitations constrain current findings and suggest development priorities. First, it is essential to distinguish the function of LLM-based agents from human cognition. Despite their impressive capabilities, these agents lack genuine beliefs, emotions, and deep reasoning processes. Their primary value lies in generating controlled, stylistically appropriate textual stimuli that constitute a manipulable social environment for human participants, rather than perfectly replicating human actors. Future work should continue validating agent behavior against more fine-grained human interaction data and exploring more sophisticated agent architectures with explicitly modeled memory decay, adaptive strategies, or heterogeneous personality traits.

Second, our simulation necessarily simplifies complex real-world dynamics. The network structures, recommendation algorithms, and interaction mechanisms represent abstracted versions of the sophisticated systems governing actual platforms. The single recommendation algorithm we implement, while providing controlled experimental conditions, cannot capture the full complexity of deep learning-based recommendation systems used by major platforms. Further refinement of these components—particularly incorporating more realistic recommendation algorithms and platform mechanisms—would enhance ecological validity.

Third, our validation study focused on short-term exposure effects without a zero-treatment control group. The 10-minute interaction period serves as an initial validation of the framework's core principles rather than a comprehensive examination of long-term polarization. This brief exposure captures immediate heuristic processing and reactions to content but may not reflect the dynamics of identity entrenchment, social proof \cite{cialdini_influence_2021, festinger_social_1954}, and parasocial relationship development \cite{horton_parasocial_1956, rubin_parasocial_2000} that unfold over extended periods. While this design effectively compares the relative impact of different polarized environments, it cannot establish absolute effects against a non-exposed baseline. For example, while moderate environments initially fostered more commenting, extended exposure to polarized content might overcome initial inhibitions through habituation effects.

Fourth, the single-topic focus (Universal Basic Income) and predominantly moderate participant sample limit generalizability across different issues and populations. Political polarization manifests differently across topics, with some issues generating more intense affective responses than others. Similarly, our sample's demographic characteristics may not represent populations most susceptible to polarization or those with stronger prior political commitments.

Fifth, it is important to distinguish between \emph{model-specific quantitative predictions} and \emph{generalizable qualitative insights}. Precise parameter thresholds (e.g., $\sigma_{\text{base}} \approx 0.8$-$1.2$ for peak polarization) depend on our specific mathematical formulations and should be interpreted as demonstrating the existence of threshold effects rather than exact values applicable across contexts. Similarly, the specific numbers of influencers required for regime transitions reflect our model parameterization, though the ranges were selected to align with empirical estimates of extremist prevalence in real-world discourse \cite{pew_political_typology_2021, enders_american_2020}. However, the \emph{qualitative patterns}—that segregation can buffer against polarization, that extremist influence peaks at moderate network mixing, that opinion and structural polarization can decouple, that polarization produces a multi-dimensional communicative syndrome—represent theoretical insights that emerge from the model's structure rather than arbitrary parameter choices.

These limitations point toward several productive research directions. Longitudinal studies tracking attitude evolution over weeks or months would illuminate how polarization dynamics develop and potentially crystallize over time. Cross-topic validation studies would establish whether our findings generalize beyond economic policy to other domains such as social issues, science topics, or international conflicts. Investigations of how different demographic groups, including those with stronger initial political commitments, respond to polarized environments would enhance external validity. Technical extensions should focus on more sophisticated agent architectures, enhanced opinion dynamics models incorporating multiple belief dimensions, and more realistic platform mechanisms that better approximate real-world algorithmic systems.     % short variant: content/overall_discussion_SHORT
\section{Conclusion}

This study reveals online political polarization as a multi-dimensional phenomenon where opinion radicalization, network segregation, and discourse transformation systematically decouple in ways that challenge unidimensional theories and interventions. Through integrated computational experiments and human validation with 122 participants, we demonstrate that dimensions assumed to move in lockstep can evolve independently: structural sorting can occur without belief extremization, algorithmic interventions can reduce opinion polarization while increasing network segregation, and discourse can transform through ambient environmental effects that bypass deliberate message evaluation. These findings fundamentally reframe polarization from a problem of isolated dimensions—whether echo chambers, extreme beliefs, or toxic discourse—into a friction-based communicative order requiring multi-dimensional understanding and intervention.

Our computational experiments revealed counterintuitive dynamics that challenge conventional assumptions. Complete network segregation buffers against extreme polarization while moderate integration creates optimal conditions for extremism, contradicting echo chamber theories. Algorithmic interventions show sharp threshold effects, succeeding preventatively but failing remedially once extremist presence exceeds critical levels. Moderate disagreement occupies an unstable "engagement valley" in attention-driven environments. Polarization manifests as a systematic communicative syndrome—simultaneously increased emotionality, heightened group identity salience, and reduced uncertainty—operating through friction inherent to political discourse itself. Human validation confirmed that participants accurately perceived these computational patterns, yet their opinion changes operated through ambient environmental influences rather than deliberate evaluation, revealing mechanisms current models miss.

The framework's broader contribution lies in establishing replicable infrastructure for investigating social phenomena at the intersection of communication, influence, and collective behavior. By enabling systematic variation of psychological, structural, and algorithmic factors while maintaining communicative realism, this methodology creates what we term a "digital laboratory" for causal inquiry into dynamics that traditional approaches cannot access—from threshold effects in algorithmic intervention to memory effects in interpretive bias.

Important limitations constrain current findings. The 10-minute exposure captures immediate reactions but not long-term identity crystallization. The single-topic focus (Universal Basic Income) and predominantly moderate sample limit generalizability. The LLM-based agents generate realistic stimuli but lack genuine cognitive processes. Future research can extend these findings through longitudinal studies, cross-topic validation, investigation of populations with stronger political commitments, and technical enhancements incorporating more sophisticated agent architectures and platform mechanisms. By demonstrating how computational rigor and human realism can be productively integrated, this work provides both specific theoretical insights and methodological infrastructure for advancing understanding of our evolving digital public sphere.

%% =========================== BIBLIOGRAPHY ========================
\bibliography{bib}

%% =========================== APPENDICES ==========================
%% "Appendix for <title>" divider page (reconstructed -- sn-jnl has no built-in
%% command for it). Full-width page, then resume two-column appendix content.
\clearpage
\onecolumn
\thispagestyle{empty}
\makeatletter
\vspace*{0.30\textheight}
\begin{center}
  {\fontsize{22}{26}\selectfont \appendixname}\\[2.5em]
  for\\[2.5em]
  {\Large \@title\par}
\end{center}
\makeatother
\clearpage
\twocolumn
\pagenumbering{roman}  % appendix: restart page numbering with Roman numerals
\begin{appendices}
\section{Simulation Model}
\label{app:simulation-model}

Our computational framework integrates opinion dynamics principles with LLM-based agents to investigate polarization in social networks. Unlike traditional opinion dynamics models that often focus on simplified representations, our approach makes four key contributions:

First, we deeply integrate LLM-based message generation with an opinion dynamics model, enabling agents to adapt content based on individual opinion strengths, personal backgrounds, and interaction histories (Contribution A). Second, we develop a sophisticated opinion dynamics pipeline that captures both homophily and backfire effects, accounting for various initial opinion distributions, community structures, and the influence of extremist actors (Contribution B). Third, we model communication processes (liking, reposting, commenting) as independent from opinion evolution, allowing for more realistic representation of social media interactions (Contribution C). Fourth, we implement dynamic network evolution through following/unfollowing behaviors and personalized recommendation systems that extend beyond simple neighborhood-based approaches (Contribution D).

The system comprises three primary components: (1) an agent-based architecture where individuals possess distinct opinion values, personality traits, and interaction histories; (2) an LLM infrastructure that enables content generation, stance analysis, and interaction evaluation; and (3) a social network structure that governs information dissemination through a recommendation mechanism. Operating through discrete time steps, each iteration encompasses content generation, agent interactions, and network evolution processes.

\subsection{Adaptive Agent Architecture and Communication (Contribution A)}
\label{app:adaptive-agents}

Our simulation framework is built upon an agent-based model where each agent $A_i$ represents an individual participant in the online social network. The core innovation highlighted in this section (Contribution A) lies in the agent architecture's deep integration of dynamic internal states with sophisticated, LLM-driven communication capabilities. This allows agents to generate and interpret messages in a manner that is adaptively conditioned on their evolving characteristics and interaction context, moving beyond simpler communication models often employed in opinion dynamics studies.

\subsubsection{Agent State Representation}
\label{app:agent-state}

Each agent $A_i$ is characterized by a set of evolving state variables representing its identity, cognitive state, and social embedding. Conceptually, the agent state can be summarized as:
\begin{align}
    A_i(t) = \{ o_i(t), P_i, H_i(t), \mathcal{N}_i(t) \} \label{eq:agent_state}
\end{align}
where $o_i(t)$ is the agent's opinion at time $t$, $P_i = \{d_i, b_i\}$ represents the static persona components (personality traits $d_i$, biography $b_i$), $H_i(t)$ denotes the agent's dynamic interaction memory or history, and $\mathcal{N}_i(t)$ represents the agent's local network neighborhood (e.g., the set of agents $A_i$ follows) at time $t$.

The primary variable governing ideological stance is the opinion $o_i(t)$ concerning a specific topic, defined by a term and description. This opinion is represented as a continuous value within a normalized interval:
\begin{align}
    o_i(t) \in [-1, 1] \label{eq:opinion_range}
\end{align}
where $o_i = 1$ denotes maximal support, $o_i = -1$ maximal opposition, and $o_i = 0$ neutrality. This continuous representation captures both opinion polarity and strength, $|o_i|$, and agents maintain a history of their opinion trajectory. The persona components $P_i$ are algorithmically generated at initialization using an LLM to condition the agent's communicative style and contribute to population heterogeneity. The memory $H_i(t)$ stores representations of recent interactions or internal reflections, informing subsequent actions and communication, thus simulating cognitive constraints and path-dependency. The network neighborhood $\mathcal{N}_i(t)$ evolves based on following/unfollowing dynamics (detailed in \cref{app:connection-dynamics}).

Derived properties, such as influencer status and posting probability ($p_{\text{post},i}$), are assigned during initialization based on these core attributes or network structure.

\subsubsection{Initialization of Agent Population and States}
\label{app:agent-initialization}

The simulation environment is initialized with a population $V$ of $N$ agents. A key step is setting the initial opinion configuration $O(t=0) = \{o_1(t=0), ..., o_N(t=0)\}$. This configuration is generated based on distributional assumptions selected to reflect the specific scenario under investigation, such as modeling societal consensus, bipolarization, or fragmentation. The precise functional forms of these initial distributions are detailed in the Experimental Setup (\cref{app:computational_experiments}).

Following opinion initialization, specific agents may be designated as influencers based on predefined criteria or initial network properties (e.g., high connectivity). This designation can involve adjusting their initial opinions towards more extreme values to represent polarized figures.

Subsequently, the initial posting probability $p_{\text{post},i}$ is assigned to each agent $A_i$. This probability dictates the agent's likelihood of generating content in a given time step and reflects their expected activity level. It is modeled as a function combining the agent's structural position within the nascent social network and a stochastic component representing individual idiosyncrasies. First, a raw connectivity score $c_i$ (e.g., in-degree) is calculated for each agent based on the initial network topology. This score is normalized to the [0, 1] range:
\begin{align}
    \hat{c}_i = \frac{c_i - \min_j(c_j)}{\max(1, \max_j(c_j) - \min_j(c_j))} \label{eq:norm_connectivity}
\end{align}
where the denominator includes a safeguard against division by zero if all scores are identical. A random factor $r_i$ is drawn independently for each agent from a uniform distribution, $r_i \sim U(0, 1)$. The normalized connectivity $\hat{c}_i$ and the random factor $r_i$ are then combined using a connectivity weight parameter $w_c \in [0, 1]$:
\begin{align}
    f_i = w_c \cdot \hat{c}_i + (1 - w_c) \cdot r_i \label{eq:combined_factor}
\end{align}
This combined factor $f_i$ is used to linearly scale the posting probability between a predefined minimum $p_{\text{min}}$ and maximum $p_{\text{max}}$:
\begin{align}
    p_{\text{post},i} = p_{\text{min}} + f_i \cdot (p_{\text{max}} - p_{\text{min}}) \label{eq:posting_prob}
\end{align}
This formulation allows the model to capture the tendency for more central or connected individuals to be more active, while also incorporating inherent randomness in user behavior, with the balance controlled by $w_c$.

\subsubsection{LLM-Powered Adaptive Communication}
\label{app:llm-communication}

A central element of Contribution A is the mechanism by which agents generate communicative content. Rather than using predefined messages, agents leverage an LLM to produce text dynamically. This generation process is adaptive, being conditioned on multiple facets of the agent's current state and context. An agent's internal opinion state $o_i$ directly informs the generated message's stance and intensity, which in turn influences linguistic style and argument strength, guided by the specifications detailed in \cref{tab:prompt-components} and \cref{tab:message-intensity}. Simultaneously, the agent's unique persona $P_i$ guides the LLM in adopting a consistent and distinct communicative style. The process also incorporates the agent's interaction history $H_i$, allowing the LLM to generate contextually relevant messages that reflect past interactions or experiences, contributing to more coherent conversational behavior. The generated content's relevance to the ongoing discussion is maintained by conditioning on the simulation's defined topic term and description. In reactive communication, such as comments or reposts, the generation is additionally conditioned on the specific content and stance of the message being responded to, facilitating direct engagement. This multi-faceted conditioning allows the simulation to capture a richer, more realistic form of online communication where messages are dynamic linguistic acts sensitive to the sender's state, identity, history, and the immediate situation. This adaptive capability is fundamental to exploring the interplay between nuanced communication, opinion dynamics (detailed in \cref{app:opinion-dynamics}), and interaction patterns (\cref{app:communicative-actions}).

\subsection{Opinion Dynamics: Homophily and Reactance (Contribution B)}
\label{app:opinion-dynamics}

Building upon the adaptive agent architecture, our framework incorporates a sophisticated opinion dynamics pipeline (Contribution B) designed to capture nuanced psychological phenomena such as homophilous assimilation and reactance (backfire effects). This mechanism governs how agent opinions evolve in response to information exposure, aiming for greater realism compared to simpler linear or bounded confidence models.

\subsubsection{Modeling Opinion Shifts}
\label{app:opinion-shifts}

An agent $A_i$'s opinion $o_i(t)$ is updated at each time step $t$ based on the set of messages $\mathcal{M}_{\text{rec}, i}(t)$ it is exposed to through the recommendation system (detailed in \cref{app:information-propagation}). A critical prerequisite for calculating the opinion shift induced by a message $m \in \mathcal{M}_{\text{rec}, i}(t)$ is determining the perceived opinion expressed within that message. Our model achieves this through a dedicated message evaluation function, $\pi$. This function takes the agent $A_i$ and the message $m$ as input and returns a perceived opinion value, $o_m$, situated on the standard opinion scale:
\begin{align}
    o_m = \pi(A_i, m) \mapsto [-1, 1] \label{eq:message_evaluation}
\end{align}
This function $\pi$ leverages an LLM tasked with analyzing the textual content of message $m$ within the context of the simulation's defined topic. This evaluation ensures that the subsequent opinion update mechanism responds directly to the semantic and pragmatic content of the communication encountered, rather than solely relying on the generating agent's internal opinion state.

Once the perceived message opinion $o_m = \pi(A_i, m)$ has been determined, the calculation of the potential opinion shift $\Delta o_i(m)$ induced by message $m$ for agent $A_i$ proceeds using a polynomial function modulated by agent conviction and opinion alignment. This approach captures both assimilation towards similar opinions and potential repulsion from dissimilar ones, with a specific mechanism for resistance among highly opinionated agents.

First, the difference $d$ between the agent's current opinion $o_i(t)$ and the perceived message opinion $o_m$ is calculated:
\begin{align}
    d = o_i(t) - o_m \label{eq:opinion_difference}
\end{align}
The sign of $d$ indicates whether the agent's opinion is more positive ($d>0$) or negative ($d<0$) than the message's perceived opinion. If $d=0$, no shift occurs for this message.

The core of the shift mechanism relies on comparing this difference $d$ to an effective attraction zone, characterized by a squared width $\sigma_{\text{eff}}^2$. This squared width dynamically adapts based on the agent's conviction $|o_i|$ and whether the agent and message opinions share the same sign ($o_i \cdot o_m > 0$). The calculation begins with a fundamental parameter, the \emph{baseline attraction width}, $\sigma_{\text{base}}$. This parameter, $\sigma_{\text{base}} > 0$, defines the intrinsic scale governing the balance between assimilation and repulsion. It is modulated by factors incorporating agent conviction.

The resistance effect is primarily active when the agent holds a strong opinion ($|o_i|$ close to 1) and encounters a message on the same side of the opinion spectrum ($o_i \cdot o_m > 0$). The potential for resistance increases as the agent's opinion becomes more extreme. This is modeled via a \emph{conviction modulation factor}, $\chi(o_i)$:
\begin{align}
    \chi(o_i) = 1 - o_i^2 \label{eq:conviction_modulation_factor}
\end{align}
This factor, $\chi(o_i)$, quantifies an agent's susceptibility to moderating influences based on its current conviction. It approaches 0 for highly convicted agents ($|o_i| \to 1$), signifying reduced susceptibility, and is 1 for neutral agents ($o_i = 0$). This mechanism models the phenomenon where individuals with strong convictions are particularly resistant to information that might moderate their stance, even if that information originates from their own side of the opinion spectrum.

To determine how conviction modulates the attraction width, a smooth switch function, $w_{\text{con}}$, quantifies the concordance between the agent and message opinions based on their product:
\begin{align}
    w_{\text{con}}(o_i, o_m) = \frac{1}{2} \left( 1 + \tanh(o_i \cdot o_m) \right) \label{eq:concordance_switch}
\end{align}
This function transitions smoothly from $w_{\text{con}} \approx 0$ for discordant opinions (opposite signs) to $w_{\text{con}} \approx 1$ for concordant opinions (same sign).

An effective scaling factor for the attraction width, $f_{\text{scale}}$, is calculated based on the concordance and the conviction modulation factor $\chi(o_i)$:
\begin{align}
    f_{\text{scale}}(o_i, o_m) = 1 + w_{\text{con}}(o_i, o_m) \cdot (\chi(o_i) - 1) \label{eq:effective_scale_factor}
\end{align}
This factor is approximately 1 for discordant messages ($w_{\text{con}} \approx 0$) and approaches $\chi(o_i)$ for concordant messages ($w_{\text{con}} \approx 1$). Thus, for concordant messages, $f_{\text{scale}}$ decreases as agent conviction increases (as $\chi(o_i)$ decreases).

The dynamically adjusted squared attraction width $\sigma_{\text{eff}}^2$ is then computed:
\begin{align}
    \sigma_{\text{eff}}^2(o_i, o_m) = \left( \sigma_{\text{base}} \cdot f_{\text{scale}}(o_i, o_m) \right)^2 \label{eq:effective_width_sq}
\end{align}
This formulation ensures that if opinions are discordant, $\sigma_{\text{eff}}^2 \approx \sigma_{\text{base}}^2$. If opinions are concordant, $\sigma_{\text{eff}}^2$ is scaled down towards $(\sigma_{\text{base}} \cdot \chi(o_i))^2$. For agents with extreme opinions ($|o_i| \to 1$), $\chi(o_i) \to 0$, causing $\sigma_{\text{eff}}^2$ to shrink significantly when encountering concordant messages. This shrinking activates the resistance mechanism by narrowing the opinion range within which assimilation occurs, making repulsion more likely or reducing the magnitude of assimilation towards less extreme, same-sign opinions.

The raw opinion shift influence $\omega_i(m)$ is then computed using a cubic polynomial function involving the difference $d$ and the effective squared width $\sigma_{\text{eff}}^2$:
\begin{align}
    \omega_i(m) = d \cdot (d^2 - \sigma_{\text{eff}}^2(o_i, o_m)) \label{eq:polynomial_shift_raw}
\end{align}
The behavior of this function depends on the relationship between the opinion difference magnitude $|d|$ and the effective width $\sigma_{\text{eff}} = \sqrt{\sigma_{\text{eff}}^2}$:
\begin{itemize}
    \item \textbf{Attraction:} If $|d|^2 < \sigma_{\text{eff}}^2$, then $(d^2 - \sigma_{\text{eff}}^2)$ is negative. The overall shift $\omega_i(m)$ has the opposite sign to $d = o_i - o_m$, causing $o_i$ to shift \emph{towards} $o_m$.
    \item \textbf{Repulsion:} If $|d|^2 > \sigma_{\text{eff}}^2$, then $(d^2 - \sigma_{\text{eff}}^2)$ is positive. The overall shift $\omega_i(m)$ has the \emph{same} sign as $d = o_i - o_m$, causing $o_i$ to shift \emph{away} from $o_m$.
\end{itemize}
The resistance mechanism, by shrinking $\sigma_{\text{eff}}^2$ for extreme agents receiving same-sign messages, increases the likelihood that $|d|^2 > \sigma_{\text{eff}}^2$, thus promoting repulsion or significantly reducing attraction even for messages that are relatively close but represent a moderation of the agent's extreme stance.

To ensure the shift induced by a single message remains bounded and its magnitude controllable, the raw influence $\omega_i(m)$ is passed through the hyperbolic tangent function:
\begin{align}
    \Delta o_i(m) = \tanh \left( \omega_i(m) \right) \label{eq:single_message_shift_bounded}
\end{align}
This function maps the potentially large raw shift value to the interval $(-1, 1)$. 

The total opinion change $\Delta O_i$ for agent $A_i$ within a time step $t$ results from aggregating the individual shifts induced by all messages encountered $\mathcal{M}_{\text{rec}, i}(t)$, scaled by an overall learning rate $\lambda$:
\begin{align}
    \Delta O_i = \lambda \sum_{m \in \mathcal{M}_{\text{rec}, i}(t)} \Delta o_i(m) \label{eq:total_shift}
\end{align}
The learning rate $\lambda$ governs the overall pace of opinion evolution within the simulation. The agent's opinion $o_i(t+1)$ is then updated by adding this aggregated shift $\Delta O_i$ to the current opinion $o_i(t)$, and subsequently clipped to remain within the $[-1, 1]$ range:
\begin{align}
    o_i(t+1) = \text{clip}(o_i(t) + \Delta O_i, -1, 1) \label{eq:opinion_update_clipped}
\end{align}
This formulation provides a flexible mechanism for opinion dynamics. Agents are attracted to messages within an effective opinion distance determined by $\sigma_{\text{base}}$ and modulated by concordance and conviction (via $\sigma_{\text{eff}}^2$). They are repelled by messages outside this distance. Crucially, agents exhibit resistance to moderating influences from their own side when their convictions are strong. The key parameter $\sigma_{\text{base}}$ controls the baseline balance between attraction and repulsion, allowing the model to simulate various polarization phenomena depending on its configuration, driven by the semantic content of messages as evaluated by the $\pi$ function.

\subsection{Modeling Communicative Actions (Contribution C)}
\label{app:communicative-actions}

Distinct from the process of opinion evolution (\cref{app:opinion-dynamics}), our framework explicitly models the communicative actions agents perform in response to messages they encounter (Contribution C). These actions, such as liking, reposting, or commenting, are prevalent in online social networks and serve various functions beyond direct persuasion, including social signaling, identity expression, and engagement management. Modeling these actions as a separate probabilistic process allows for a richer simulation of online social behavior.

\subsubsection{Interaction Probability Model}
\label{app:interaction-probabilities}

When an agent $A_i$ encounters a message $m$ (with perceived stance $o_m = \pi(A_i, m)$), the probability $P_{\text{int}}(A_i, m)$ of performing a communicative action is modeled based on two potential triggers: one driven by opinion concordance (similarity) and another by opinion discordance (oppositeness).

\paragraph{Concordance Trigger Probability ($P_{\text{con}}$)} This component captures interaction driven by agreement or similarity. It depends on the direct opinion difference $d_{\text{con}} = o_i - o_m$. We model $P_{\text{con}}$ using a Gaussian-like function centered at zero difference, reflecting higher probability for interactions with ideologically closer messages:
\begin{align}
    P_{\text{con}}(o_i, o_m) = \exp\left( - \frac{d_{\text{con}}^2}{\sigma_{\text{con}}^2} \right) \label{eq:prob_concordant}
\end{align}
The parameter $\sigma_{\text{con}} > 0$ is the characteristic width controlling tolerance for difference; smaller $\sigma_{\text{con}}$ implies interactions are triggered only by very similar opinions (strong homophily).

\paragraph{Discordance Trigger Probability ($P_{\text{dis}}$)} This component captures interaction driven by disagreement or opposition. It depends first on the "oppositeness distance":
\begin{align}
    d_{\text{dis}}(o_i, o_m) = \max(0, - \text{sgn}(o_m) \cdot o_i) \label{eq:oppositeness_distance}
\end{align}
This measures how much the agent's opinion $o_i$ is on the opposite side of neutral (0) compared to the message $o_m$. The sensitivity to this distance is governed by a dynamically adapting width $\sigma_{\text{dis}}(o_m)$, which narrows for more extreme messages, making the trigger more sensitive to opposition. This dynamic width is defined relative to a \emph{baseline discordance width} $\sigma_{\text{dis,base}}$:
\begin{align}
    \sigma_{\text{dis}}(o_m) = \sigma_{\text{dis,base}} \cdot \exp(-|o_m|) \label{eq:dynamic_s_discordant}
\end{align}
The probability of a discordance trigger activating, which we denote as $P_{\text{dis}}$, increases with $d_{\text{dis}}$ relative to $\sigma_{\text{dis}}(o_m)$. We model this using a complementary exponential function:
\begin{align}
    P_{\text{dis}}(o_i, o_m) = 1 - \exp\left( - \frac{d_{\text{dis}}(o_i, o_m)^2}{\sigma_{\text{dis}}(o_m)^2} \right) \label{eq:prob_discordant_base}
\end{align}

\paragraph{Combined Interaction Probability} The agent interacts if \emph{either} the concordance or the discordance trigger is activated. The discordance trigger itself is scaled by a key psychological parameter, the \emph{propensity for discordant reaction ($p_{\text{dis}} \in [0, 1]$)}, which represents an agent's general tendency to engage antagonistically.

Assuming the triggers are independent events, the probability of at least one activating is given by the standard formula for the union of two events: $P(A \cup B) = P(A) + P(B) - P(A)P(B)$. In our model, this translates to:
\begin{align}
    P(\text{trigger}) = P_{\text{con}} + (p_{\text{dis}} \cdot P_{\text{dis}}) - P_{\text{con}} \cdot (p_{\text{dis}} \cdot P_{\text{dis}})
\end{align}
This reflects that the total probability of a discordance-driven interaction is the base probability $P_{\text{dis}}$ scaled by the agent's propensity $p_{\text{dis}}$.

Finally, this combined trigger probability is scaled by a \emph{base interaction probability ($p_{\text{base}} \in [0, 1]$)}, which represents the overall activity level for a given interaction type (e.g., likes are more frequent than comments). This yields the final interaction probability:
\begin{align}
    P_{\text{int}}(A_i, m) = p_{\text{base}} \left( P_{\text{con}} + p_{\text{dis}} \cdot P_{\text{dis}} - p_{\text{dis}} \cdot P_{\text{con}} \cdot P_{\text{dis}} \right)
    \label{eq:interaction_probability_final_appendix}
\end{align}
where $P_{\text{con}}$ and $P_{\text{dis}}$ are shorthand for their full functional forms dependent on $o_i$ and $o_m$. This model produces interaction likelihoods that peak for highly similar opinions (via $P_{\text{con}}$) but can also increase for highly opposing opinions if $p_{\text{dis}} > 0$. It thus captures both homophilic and antagonistic engagement within a single, unified formula.

\paragraph{Modeling Different Interaction Types} The flexibility of the parameters ($p_{\text{base}}$, $\sigma_{\text{con}}$, $p_{\text{dis}}$, $\sigma_{\text{dis,base}}$) allows this model to represent different types of communicative actions:
\begin{itemize}
    \item \textbf{Likes:} Often represent low-effort agreement or endorsement. One might model this with a relatively high base probability ($p_{\text{base}}$), a strong concordance trigger (small $\sigma_{\text{con}}$ to primarily like similar content), and a very low or zero discordance propensity ($p_{\text{dis}} \approx 0$).
    \item \textbf{Comments:} Can serve diverse functions, including agreement, clarification, or disagreement. Modeling comments might involve a lower base probability ($p_{\text{base}}$, higher effort than liking), potentially a wider concordance width ($\sigma_{\text{con}}$, allowing comments on slightly different views), and crucially, a non-zero discordance propensity ($p_{\text{dis}} > 0$) to capture argumentative replies. The sensitivity parameters ($\sigma_{\text{con}}, \sigma_{\text{dis, base}}$) would tune whether comments are primarily supportive or critical.
    \item \textbf{Reposts (without comment):} Typically strong endorsements. Could be modeled similarly to likes but perhaps with an even stronger concordance trigger (very small $\sigma_{\text{con}}$) and/or a lower base probability ($p_{\text{base}}$) reflecting lower frequency.
\end{itemize}
By setting distinct parameter sets for each action type, the simulation can capture differentiated engagement patterns reflecting the varied functions of interactions on social platforms.

\subsubsection{Contextual Response Generation}
\label{app:contextual-response-gen}

If the probabilistic determination results in an agent deciding to perform a communicative action that involves generating new content (specifically, commenting or reposting with commentary), the simulation leverages the LLM-based message generation capability described earlier (\cref{app:llm-communication}). Crucially, for these reactive message types, the generation process is provided with additional context: the content and perceived stance ($o_m$) of the original message $m$ being reacted to. The LLM is instructed to generate a response that directly addresses, agrees with, or disagrees with the original message, ensuring conversational coherence and relevance. This context-aware generation allows the simulation to produce more realistic dialogue sequences compared to models where reactions are merely abstract events.

\subsection{Co-evolving Social Network Structure (Contribution D)}
\label{app:network-evolution}

A key aspect of our simulation framework is the explicit modeling of the social network structure and its dynamic evolution alongside agent opinions (Contribution D). This allows us to investigate the interplay between network topology and polarization processes. The network is represented as a directed graph $G(t) = (V, E(t))$, where $V$ is the set of $N$ agents, and $E(t)$ is the set of directed edges $(A_i, A_j)$ at time $t$, signifying that agent $A_i$ follows agent $A_j$. This section details the initialization of the network at $t=0$ and its subsequent evolution.

\subsubsection{Network Representation and Initialization}
\label{app:network-representation}

The simulation begins by constructing an initial social graph $G(0)$ that reflects characteristics often observed in online social networks, namely community structure based on shared attributes (here, opinion) and heterogeneous degree distributions (presence of hubs/influencers). The generation process aims to create a network with a specified overall edge density $\rho_e$, tunable community mixing $\mu_{\text{mix}}^{(0)}$, and a degree distribution influenced by preferential attachment, shaped by an exponent parameter $\gamma_{\text{pl}}$.

The process initializes an empty directed graph with all $N$ agents as nodes. It then iteratively adds directed edges until a target number of edges, $M \approx \rho_e N(N-1)$, is reached. The formation of each potential edge $(A_i, A_j)$ involves several considerations.

First, the model incorporates \emph{Community Structure and Mixing}. Agents are implicitly grouped into communities based on their initial opinion sign (e.g., $o_i > 0$ vs. $o_i \leq 0$). The parameter $\mu_{\text{mix}}^{(0)} \in [0, 1]$ controls the probability of forming cross-community ties. When selecting a potential followee $A_j$ for a follower $A_i$, if $A_i$ and $A_j$ belong to different opinion communities, the connection is considered only with probability $\mu_{\text{mix}}^{(0)}$. A value of $\mu_{\text{mix}}^{(0)} = 0$ yields completely segregated communities in the initial graph, while $\mu_{\text{mix}}^{(0)} = 1$ makes the connection choice blind to community membership. Intermediate values allow for varying degrees of homophilous structure.

Second, connection formation is driven by \emph{Preferential Attachment}. The likelihood of an agent $A_j$ being chosen as a followee is proportional to their current "attractiveness" or popularity, typically related to their in-degree (number of followers) accumulated during the generation process. Let $\alpha_j(k)$ represent the attractiveness of agent $A_j$ after $k$ edges have been added (initialized with a small random value and incremented each time $A_j$ gains a follower). The probability of agent $A_i$ choosing to follow an \emph{allowed} (by community mixing rules) agent $A_j$ is proportional to $\alpha_j(k)$ raised to a power related to the preferential attachment exponent $\gamma_{\text{pl}}$:
\begin{align}
    P(A_i \text{ follows } A_j | A_j \text{ allowed}) \propto \alpha_j(k)^{1/\gamma_{\text{pl}}} \label{eq:pref_attachment}
\end{align}
The parameter $\gamma_{\text{pl}}$ (typically between 1 and 3) controls the strength of the preferential attachment. Lower values lead to a stronger "rich-get-richer" effect, resulting in more pronounced hubs and a heavier tail in the degree distribution, characteristic of scale-free networks.

In summary, the edge generation process iteratively selects a potential follower $A_i$, identifies potential followees $A_j$, filters these candidates based on the community mixing rule (probability $\mu_{\text{mix}}^{(0)}$ for cross-community ties), and then selects the final followee from the allowed candidates based on the preferential attachment mechanism (\cref{eq:pref_attachment}). Following edge addition $(A_i, A_j)$, the attractiveness $\alpha_j$ is increased. A final step ensures network connectivity by adding minimal outgoing edges for any potentially isolated nodes, guided by homophily and attractiveness.

This initialization procedure can be understood as generating a network structure that blends principles from \emph{Stochastic Block Models (SBMs)} and \emph{Preferential Attachment Models}. Like SBMs, it explicitly models community structures and controls the density of connections within and between these communities (via $\mu_{\text{mix}}^{(0)}$). However, unlike standard SBMs which often assume uniform connection probabilities within blocks, our approach incorporates a preferential attachment mechanism (via $\alpha_j$ and $\gamma_{\text{pl}}$). This hybrid approach aims to generate initial networks that exhibit both plausible community structures based on agent attributes (opinion homophily) and the heterogeneous, scale-free-like degree distributions commonly observed in real-world social networks. The resulting graph $G(0)$ provides the initial substrate for information diffusion and subsequent network evolution.

\subsubsection{Dynamic Connection Formation and Dissolution}
\label{app:connection-dynamics}

Beyond the initial structure, the social network $G(t)$ evolves over time as agents dynamically adjust their connections, reflecting processes like discovering new users and ceasing to follow others. This dynamic connection process is a core component of Contribution D, enabling the study of the feedback loop between opinion formation and network structure.

In each time step, agents have the opportunity to revise their outgoing connections (i.e., who they follow) based on the other users they encounter, primarily through the messages recommended to them ($\mathcal{M}_{\text{rec}, i}(t)$). For an agent $A_i$, the set of potential new users to follow consists of the authors $A_j$ of these recommended messages, provided $A_i$ is not already following $A_j$ and $A_j \neq A_i$.

The decision for agent $A_i$ to initiate a follow relationship with a potential candidate $A_j$ is probabilistic. The probability, $P_{\text{follow}}(A_i, A_j)$, is determined using a mechanism structurally similar to the general interaction probability model described in \cref{app:interaction-probabilities} (\cref{eq:interaction_probability_final_appendix}). However, instead of evaluating alignment based on the \emph{message's} perceived opinion $\pi(A_i, m)$, the calculation relies directly on the difference between the agent's own opinion $o_i$ and the \emph{potential followee's} (author's) opinion $o_j$.

A key aspect of the implemented dynamic is the constraint imposed on network structure changes. To maintain stability or simulate user attention limits, the model may enforce that an agent's out-degree (number of users they follow) remains constant or changes slowly. In the current mechanism, for each \emph{new} follow connection $(A_i, A_j)$ that is successfully established based on the probability above, agent $A_i$ simultaneously \emph{removes} one randomly selected existing outgoing connection $(A_i, A_k)$, where $A_k$ was previously followed by $A_i$. This "rewiring" approach ensures that the total number of outgoing links for agent $A_i$ remains constant throughout the update step, effectively modeling a process where new follows replace older ones.

The outcome of this process across all agents is an updated set of edges $E(t+1)$, reflecting both the formation of new social ties influenced by opinion similarity and the dissolution of previous ones. This evolving network structure directly impacts subsequent information propagation and agent interactions in the following time steps.

\subsubsection{Recommendation-Driven Information Propagation}
\label{app:information-propagation}

The flow of information within the simulated social network is governed by a recommendation system that determines message visibility for each agent $A_i$ at each time step $t$. This system defines the set of messages $\mathcal{M}_{\text{rec}, i}(t)$ that agent $A_i$ is exposed to, forming the basis for potential interactions and opinion updates. The design aims to balance exposure to content from an agent's existing social ties with opportunities for discovering content from beyond their immediate neighborhood, while allowing control over the ideological diversity of discovered content.

The recommendation process for agent $A_i$ begins by identifying the pool of eligible messages $\mathcal{M}_{\text{elig}, i}(t)$. This pool consists of all messages currently available in the system, excluding those authored by agent $A_i$ itself and any messages with which $A_i$ has previously interacted, thus preventing redundant exposure. Messages within this eligible pool are then conceptually categorized based on their source relative to agent $A_i$: messages originating from the agent's direct network neighborhood (authors $A_j$ such that $(A_i, A_j) \in E(t)$) constitute the \textit{Network Source}, while messages from all other authors constitute the \textit{Discovery Source}.

The composition of the final recommended set $\mathcal{M}_{\text{rec}, i}(t)$, which typically aims for a fixed size $N$, is controlled by a key parameter, the \emph{discovery rate} $\delta_{\text{rec}} \in [0, 1]$. This parameter specifies the target proportion of recommendations intended to originate from the Discovery Source. Consequently, the target number of recommendations from the Discovery Source is $N_{\text{disc}} = N \cdot \delta_{\text{rec}}$, and the target from the Network Source is $N_{\text{net}} = N \cdot (1 - \delta_{\text{rec}})$.

Furthermore, the ideological diversity of content drawn from the Discovery Source is regulated by the \emph{discovery community mixing} parameter $\mu_{c, \text{disc}} \in [0, 1]$. This parameter controls the balance between recommending discovery messages from authors sharing the same opinion sign as agent $A_i$ versus authors with the opposite opinion sign. Let $f_{\text{same}}(\mu_{c, \text{disc}})$ be the target fraction of discovery messages to be drawn from same-opinion authors (a function potentially decreasing from 1 to 0.5 as $\mu_{c, \text{disc}}$ goes from 0 to 1, depending on the desired interpretation). The number of targeted same-opinion discovery messages is then $N_{\text{disc, same}} = N_{\text{disc}} \cdot f_{\text{same}}(\mu_{c, \text{disc}})$, and the number of targeted different-opinion discovery messages is $N_{\text{disc, diff}} = N_{\text{disc}} - N_{\text{disc, same}}$. A value of $\mu_{c, \text{disc}} = 0$ (leading to $f_{\text{same}}=1$) creates an echo chamber within discovery, while higher values increase cross-ideological exposure.

The selection process involves randomly sampling the required number of messages from each category (Network, Discovery-Same-Opinion, Discovery-Different-Opinion) up to the calculated target for that category ($N_{\text{net}}$, $N_{\text{disc, same}}$, $N_{\text{disc, diff}}$) or the number of available eligible messages in that category, whichever is smaller. Configuration settings can determine whether the system strictly adheres to these targets, possibly providing fewer than $N$ recommendations if a source is exhausted (\textit{strict weights}), or whether it attempts to fill remaining slots using messages from other available sources (\textit{flexible weights}).

A significant feature is the potential for \emph{adaptive mixing}. Instead of using a fixed $\mu_{c, \text{disc}}$, this parameter can be dynamically computed each time step based on the current global network structure $G(t)$. This involves calculating the observed ratio of inter-community edges ($E_{\text{inter}}$) to intra-community edges ($E_{\text{intra}}$) in the graph, $\rho_{\text{obs}} = E_{\text{inter}} / E_{\text{intra}}$, and comparing it to the ratio expected in a random graph with equivalent community sizes, $\rho_{\text{exp}}$. The adaptive mixing parameter can then be set as a function of this comparison, for example, proportional to the normalized ratio $\min(1, \rho_{\text{obs}} / \rho_{\text{exp}})$, linking the diversity of discovered information to the emergent structural integration or segregation of the network.

Finally, the selected messages from all sources are combined and shuffled to form the final recommendation set $\mathcal{M}_{\text{rec}, i}(t)$ presented to agent $A_i$. This recommendation mechanism provides a flexible means to explore how different balances of network-based filtering versus controlled discovery influence opinion dynamics and polarization.

\subsection{Simulation Workflow}
\label{app:simulation-workflow}

The simulation progresses through two main phases: initialization and iterative execution over discrete time steps. This workflow orchestrates the interactions between the different model components described previously, as detailed in \cref{alg:social-simulation}.

The \emph{Initialization Phase} sets up the simulation environment. This involves creating the population $V$ of $N$ agents, assigning each agent $A_i$ its initial state including opinion $o_i(0)$ drawn from a selected distribution, and generating its unique persona $P_i$ using the LLM (\cref{app:agent-initialization}). Concurrently, the initial social network graph $G(0)$ is constructed based on specified parameters governing density, community mixing ($\mu_{\text{mix}}^{(0)}$), and preferential attachment ($\gamma_{\text{pl}}$), establishing the initial social structure (\cref{app:network-representation}). Initial posting probabilities $p_{\text{post},i}$ are also assigned (\cref{eq:posting_prob}).

The \emph{Iterative Simulation Phase} proceeds through a configured number of discrete time steps, $t=1, 2, ...$. Each time step represents a period of activity on the simulated social network. Within each step, first, agents engage in \emph{Content Generation}. Each agent $A_i$ stochastically decides whether to create a new message based on its posting probability $p_{\text{post},i}$. If creating content, the agent utilizes the LLM-powered adaptive communication mechanism (\cref{app:llm-communication}) reflecting its current state ($o_i(t)$, $P_i$, $H_i(t)$) and ideological context. These new messages augment the system's content pool. Following evaluation, the \emph{Information Recommendation} system generates a personalized set of messages $\mathcal{M}_{\text{rec}, i}(t)$ for each agent $A_i$. This selection balances content from the agent's network neighborhood and discovered content, governed by parameters $\delta_{\text{rec}}$ and $\mu_{c, \text{disc}}$ (\cref{app:information-propagation}). Next, agents undertake \emph{Interaction} with their recommended messages. For each message $m \in \mathcal{M}_{\text{rec}, i}(t)$, agent $A_i$ probabilistically decides whether to perform actions like liking, commenting, or reposting, according to the interaction probability model $P_{\text{react}}(A_i, m, \text{type})$ (\cref{eq:interaction_probability_final_appendix}). Generation of comments or reposts involves the contextual LLM mechanism (\cref{app:contextual-response-gen}), and all interactions are recorded. After interactions, \emph{Opinion Updates} are calculated. The total opinion shift $\Delta O_i$ for each agent is determined by aggregating individual shifts $\Delta o_i(m)$ (\cref{eq:single_message_shift_bounded}) from encountered messages, scaled by the learning rate $\lambda$, leading to the updated opinion $o_i(t+1)$ (\cref{eq:total_shift}). Parallel to or following opinion updates, the \emph{Network Update} occurs, where agents adjust their follow relationships based on recent encounters, guided by the connection dynamics model (\cref{app:connection-dynamics}), resulting in the updated graph $G(t+1)$. Finally, relevant simulation \emph{Metrics} are calculated and recorded for the completed time step.

This sequence creates interconnected feedback loops. Opinions influence communication (message generation, reactions) and network choices (following). Communication influences opinions (updates) and network choices. Network structure shapes information exposure (recommendations), which in turn influences opinions and future network changes. Executing this workflow over multiple time steps allows for the study of emergent phenomena like polarization, echo chamber formation, and the co-evolution of opinions and network structure.

\begin{algorithm*}[t]
    \caption{Social Network Simulation}
    \label{alg:social-simulation}
    \begin{algorithmic}[1] % [1] enables line numbering
    
        \State \textbf{Input:} $n_{agents}$, $n_{iterations}$, $n_{recs}$, $p_{reg}$, $p_{inf}$, $topic$, $\lambda$, $\sigma_{base}$
        \Statex
        
        \State $\mathcal{A} \gets \text{\Call{InitializeAgentPopulation}{$n_{agents}$, $topic$}}$
        \State $\mathcal{G} = (\mathcal{V}, \mathcal{E}) \gets \text{\Call{InitializeNetwork}{$\mathcal{A}$}}$
        \State $\mathcal{M} \gets \emptyset$
        \Statex
        
        \For{$t = 1$ \textbf{to} $n_{iterations}$}
            \State $\mathcal{M}_t \gets \emptyset$
            \Statex \Comment{Content Generation Phase}
            \ForAll{agent $A_i \in \mathcal{A}$}
                \State $p_{post} \gets (\text{\Call{IsInfluencer}{$A_i$}}) \text{ ? } p_{inf} : p_{reg}$
                \If{Random() < $p_{post}$}
                    \State $m \gets \text{\Call{CreateMessage}{$A_i, t$}}$ \Comment{Uses $o_i(t), P_i, H_i(t), topic$}
                    \State $\mathcal{M}_t \gets \mathcal{M}_t \cup \{m\}$
                \EndIf
            \EndFor
            \State $\mathcal{M} \gets \mathcal{M} \cup \mathcal{M}_t$
            \Statex \Comment{Interaction, Opinion Update, and Network Update Phase}
            \ForAll{agent $A_i \in \mathcal{A}$}
                \State $\mathcal{R}_{A_i} \gets \text{\Call{RecommendMessages}{$A_i, \mathcal{M}, \mathcal{G}, n_{recs}$}}$ \Comment{Uses $\delta_{rec}, \mu_{c,disc}$}
                \State $\mathcal{I}_{A_i} \gets \text{\Call{ModelInteractions}{$A_i, \mathcal{R}_{A_i}$}}$ \Comment{Generates likes, comments, reposts}
                \Statex 
                \State $\Delta O_i \gets 0$ \Comment{Initialize total shift for agent $A_i$}
                \ForAll{message $m_j \in \mathcal{R}_{A_i}$}
                    \State $o_m \gets \text{\Call{EvaluateMessageOpinion}{$A_i, m_j$}}$ \Comment{Corresponds to $\pi(A_i, m_j)$}
                    \State $\Delta o_{i}(m_j) \gets \text{\Call{CalculateIndividualMessageShift}{$o_i(t), o_m, \sigma_{base}$}}$
                \EndFor
                \State $\Delta O_i \gets \lambda \cdot \sum_{m_j \in \mathcal{R}_{A_i}} \Delta o_{i}(m_j)$ \Comment{Aggregate and scale}
                \State $o_i(t+1) \gets \text{\Call{ClipOpinion}{$(o_i(t) + \Delta O_i), -1, 1$}}$ \Comment{Update and clip}
                \State \text{\Call{UpdateAgentMemory}{$A_i, \mathcal{R}_{A_i}, \mathcal{I}_{A_i}$}} \Comment{Update memory}
                \Statex
                \State $\mathcal{G} \gets \text{\Call{UpdateNetwork}{$A_i, \mathcal{G}$}}$ \Comment{Network evolution}
            \EndFor
        \EndFor
        
    \end{algorithmic}
\end{algorithm*}              % app:simulation-model
\section{Detailed Computational Experiments}
\label{app:computational_experiments} % Label for the entire appendix section

This appendix provides detailed descriptions of the computational experiments conducted to assess the simulation framework's capabilities, explore the mechanisms driving opinion and network dynamics related to social polarization, and provide empirical support for the analyses presented in the main text. The general methodology involved systematic variation of key model parameters while observing the resultant changes in emergent phenomena such as agent opinion distributions, social network structure, and/or agent interaction patterns over time.

To isolate core mechanisms governing opinion shifts (Contribution B) and network evolution (Contribution D), many of these experiments employed a simplified agent logic. In these simulations, advanced functionalities leveraging LLMs for nuanced agent behavior (Contributions A and C aspects like persona-driven communication, complex message evaluation, and reactive content generation) were typically deactivated. Unless stated otherwise, agents in these runs utilized streamlined logic where message content followed templates, perceived message opinion ($o_m$) equaled the author's opinion ($o_j$), and explicit reactive interactions (i.e., likes, reposts, and comments) were inactive. Experiments focusing specifically on reaction dynamics (B.7-B.9) activated the relevant interaction models (Contribution C) as described in their respective setups.

The experiments detailed herein cover a range of investigations designed to build understanding from fundamental mechanisms to more complex, coupled dynamics. These include: the fundamental behavior of the opinion dynamics model under varying base attraction width ($\sigma_{\text{base}}$) and initial opinion fragmentation ($o_{\text{max}}^{(0)}$) (Experiment B.1); base attraction width ($\sigma_{\text{base}}$) and the influence of initial community mixing ($\mu_{\text{mix}}^{(0)}$) (Experiment B.2); the impact of extremist influencers ($N_{\text{inf}}$) and initial community mixing ($\mu_{\text{mix}}^{(0)}$) (Experiment B.3); the isolated dynamics of network evolution driven by agent opinions when opinions themselves are fixed (Experiment B.4); the co-evolutionary feedback between opinions and network topology, considering the roles of influencers ($N_{\text{inf}}$), initial mixing ($\mu_{\text{mix}}^{(0)}$), and algorithmic discovery rates ($\delta_{\text{rec}}$) (Experiments B.5, B.6); and the patterns of agent reaction dynamics under various structural, algorithmic, and psychological parameters, with both fixed and evolving opinions (Experiments B.7, B.8, B.9).

Each experiment description below follows a consistent structure: a brief introduction outlining its specific objective, a description of the experimental setup detailing the varied parameters and constant settings, and a presentation and interpretation of the results, often including visualizations like heatmaps, line graphs, or kernel density estimates. Parameter configurations generally adhere to a common baseline (defined in Table~\ref{tab:parameters-offline}), with specific parameter variations detailed within each experiment description.

% This table is for the appendix and should span both columns.

\begin{table*}[htbp]
    \centering
    \begin{threeparttable} % Use the threeparttable environment for notes
    \caption{Consolidated Model Parameters, Baseline Configuration, and Experimental Variations}
    \label{tab:parameters-offline}
    \small % Make font slightly smaller if needed
    
    \begin{tabular}{@{}llcl@{}}
    \toprule
    Symbol & Description & Baseline Value & Varied In Exp. \\
    \midrule
    \multicolumn{4}{l}{\textbf{General Simulation Setup}} \\
     $N$ & Number of agents & 50 & - \\
     $T$ & Number of simulation iterations & 100 & - \\
    \midrule
    \multicolumn{4}{l}{\textbf{Agent Initialization}} \\
     $o_{\text{max}}^{(0)}$ & Initial opinion range limit ($U[-o_{\text{max}}^{(0)}, o_{\text{max}}^{(0)}]$) & 0.7 & B.1, B.9 \\
     $N_{\text{inf}}$ & Number of influencers per side & 0 & B.3, B.5, B.6 \\
     $p_{\text{min}}$ & Minimum posting probability & 0.2 & - \\
     $p_{\text{max}}$ & Maximum posting probability & 0.5 & - \\
     $\omega_c$ & Connectivity weight for posting probability & 0.7 & - \\
    \midrule
    \multicolumn{4}{l}{\textbf{Network Initialization}} \\
     $\rho_e$ & Initial edge density & 0.15 & - \\
     $\gamma_{\text{pl}}$ & Preferential attachment exponent & 1.0 & - \\
     $\mu_{\text{mix}}^{(0)}$ & Initial community mixing probability & 0.5 & B.2 - B.9 \\
    \midrule
    \multicolumn{4}{l}{\textbf{Opinion Dynamics (\cref{app:opinion-dynamics})}} \\ % Use \cref for automatic linking
     $\lambda$ & Learning rate for opinion updates & 0.01 & - \\
     $\sigma_{\text{base}}$ & Base attraction width & 1.0 & B.1, B.2 \\
    \midrule
    \multicolumn{4}{l}{\textbf{Connection Dynamics (Following - \cref{app:network-evolution})}} \\ % Use \cref
     $p_{\text{base}}^{\text{(follow)}}$ & Base probability for \emph{follow} decision & 0.5 & - \\
     $\sigma_{\text{con}}^{\text{(follow)}}$ & Concordant \emph{follow} width & 0.1 & - \\
     $\sigma_{\text{dis, base}}^{\text{(follow)}}$ & Base discordant \emph{follow} width & 1.0 & - \\
     $p_{\text{dis}}^{\text{(follow)}}$ & Propensity for discordant \emph{follow} decision & 0.0 & B.4 \\
    \midrule
    \multicolumn{4}{l}{\textbf{Interaction Dynamics (Reactions - \cref{app:communicative-actions})}} \\ % Use \cref
     $p_{\text{base}}^{\text{(react)}}$ & Base probability for \emph{reaction} & 0.5 & - \\
     $\sigma_{\text{con}}^{\text{(react)}}$ & Concordant \emph{reaction} width & 0.1 & - \\
     $\sigma_{\text{dis, base}}^{\text{(react)}}$ & Base discordant \emph{reaction} width & 1.0 & - \\
     $p_{\text{dis}}^{\text{(react)}}$ & Propensity for discordant \emph{reaction} trigger & 0.0 & B.8 \\
    \midrule
    \multicolumn{4}{l}{\textbf{Recommendations (\cref{app:network-evolution})}} \\ % Use \cref
     $N_{\text{rec}}$ & Number of recommendations per agent per step & 8 & - \\
     $\delta_{\text{rec}}$ & Discovery rate (proportion from non-network) & 0.5 & B.6, B.7 \\
     $\mu_{\text{mix}, \text{rec}}$ & Discovery community mixing & Aligned with $\mu_{\text{mix}}$ & - \\
    \bottomrule
    \end{tabular}
    
    \begin{tablenotes}[para,flushleft] % [para] makes it a paragraph, flushleft aligns it
        \footnotesize
        \item[\textbf{Note:}] This table lists all core parameters and their baseline values used across the computational experiments (B.1-B.9) detailed in this appendix. Baseline values apply unless the parameter was explicitly varied in a specific experiment, as indicated in the final column. The notation `(follow)` or `(react)` distinguishes parameters specific to connection dynamics versus reactive interaction dynamics where symbols might overlap. Opinion dynamics were deactivated in Exp B.4 (isolated network evolution). Reaction dynamics were generally deactivated in Exp B.1-B.6, except where explicitly studied or enabled, but were active in Exp B.7-B.9. Network evolution was active only in Exp B.4, B.5, B.6, B.9 unless otherwise noted. Appendix A provides detailed mathematical definitions.
    \end{tablenotes}
    
    \end{threeparttable}
\end{table*} % Assumes your main parameter table is here

\subsection{Experiment 1: Initial Opinion Landscape and Base Attraction Width}
\label{app:exp1_landscape_attraction}

This first experiment investigates the fundamental interplay between the initial configuration of opinions within the population ($o_{\text{max}}^{(0)}$) and the core parameter of the opinion update mechanism, the base attraction width ($\sigma_{\text{base}}$). The primary objective is to understand how initial opinion diversity influences the system's trajectory towards consensus or polarization, depending on the baseline range of opinion differences agents tolerate before shifting from assimilation (attraction) to repulsion (backfire). This establishes the fundamental behavior of the opinion dynamics model (Contribution B) under varying initial conditions and assimilation/repulsion thresholds, without network evolution or specific structural features.

\subsubsection{Experimental Setup}
We systematically varied two key parameters:

\begin{enumerate}
    \item \textbf{Initial Opinion Range Limit ($o_{\text{max}}^{(0)}$):} Agent opinions initialized from $U[-o_{\text{max}}^{(0)}, o_{\text{max}}^{(0)}]$, varied over $[0.1, ..., 1.0]$. Higher values signify greater initial diversity.
    \item \textbf{Base Attraction Width ($\sigma_{\text{base}}$):} This parameter (\ref{app:opinion-shifts}) sets the baseline threshold for opinion difference ($|d|$) separating assimilation from repulsion. Agents assimilate opinions where $|d|^2 < \sigma_{\text{eff}}^2$ (and $\sigma_{\text{eff}}^2$ is primarily determined by $\sigma_{\text{base}}^2$ when conviction/concordance effects are minimal). We varied $\sigma_{\text{base}}$ over $[0.0, 0.2, ..., 2.0]$. A \emph{larger} $\sigma_{\text{base}}$ defines a wider range for assimilation, meaning agents are attracted towards more dissimilar opinions before repulsion occurs. A \emph{smaller} $\sigma_{\text{base}}$ defines a narrower assimilation range, triggering repulsion for smaller opinion differences. $\sigma_{\text{base}}=0$ implies repulsion for any non-zero difference.
\end{enumerate}
All other parameters followed the baseline configuration (Table~\ref{tab:parameters-offline}), notably a static network ($\mu_{\text{mix}}^{(0)}=0.5$, network evolution off) and no influencers ($N_{\text{inf}}=0$). Five simulation runs were performed over $T=100$ steps for each ($o_{\text{max}}^{(0)}$, $\sigma_{\text{base}}$) combination. The final mean opinion polarization (Esteban-Ray index) was the primary outcome measure.

\subsubsection{Results}

The combined influence of initial opinion diversity ($o_{\text{max}}^{(0)}$) and base attraction width ($\sigma_{\text{base}}$) on final polarization is summarized in Figure~\ref{fig:polarization-heatmap-center-limit}. The results reveal a complex relationship where polarization does not simply increase or decrease monotonically with $\sigma_{\text{base}}$.

Notably, strong polarization emerges under conditions of \emph{minimal or zero attraction width} ($\sigma_{\text{base}} \approx 0.0 - 0.4$, top rows). When $\sigma_{\text{base}}=0.0$ (pure repulsion), polarization is high across most initial diversity levels ($o_{\text{max}}^{(0)} > 0.1$), peaking around $o_{\text{max}}^{(0)} \approx 0.5 - 0.6$ (index $\approx 1.1-1.2$). This indicates that consistent repulsion from differing views strongly drives the system away from consensus. The \emph{highest polarization levels} (index $\approx 1.5$) occur for \emph{intermediate attraction widths} ($\sigma_{\text{base}} \approx 0.8 - 1.2$) combined with high initial diversity ($o_{\text{max}}^{(0)} \ge 0.9$). Conversely, polarization is consistently low, indicating \emph{consensus}, when the \emph{attraction width is wide} ($\sigma_{\text{base}} \ge 1.4$), regardless of the initial diversity. The impact of $o_{\text{max}}^{(0)}$ is significant primarily when the attraction width is not excessively large; high initial diversity is a prerequisite for substantial polarization.

\begin{figure*}[h!]
    \centering
    \includegraphics[width=\textwidth]{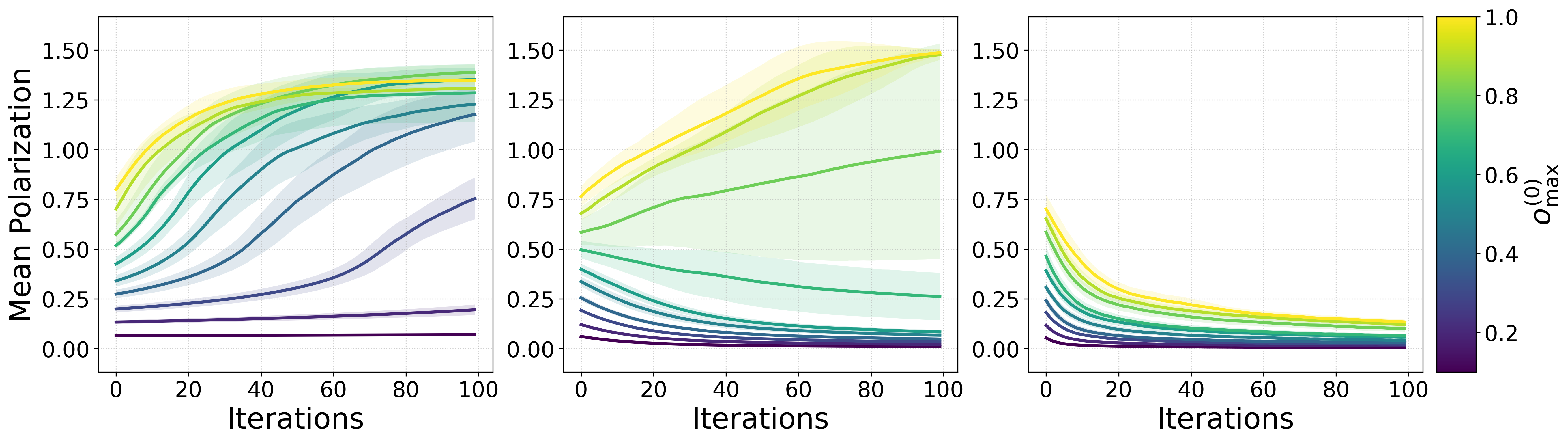}
    \caption{Mean polarization over time for varying baseline attraction widths ($\sigma_{\text{base}}$) and initial opinion spread ($o_{\text{max}}^{(0)}$). Panels from left to right: $\sigma_{\text{base}}=0.0$ (pure repulsion), $\sigma_{\text{base}}=1.0$ (intermediate), $\sigma_{\text{base}}=2.0$ (wide attraction). Line colors indicate $o_{\text{max}}^{(0)}$ values (see color bar); shaded areas are 95\% CIs (5 runs). Pure repulsion consistently polarizes except for low initial spread; intermediate $\sigma_{\text{base}}$ yields highest polarization if initial diversity is high, while wide attraction leads to consensus.}
    \label{fig:line-graphs-center-limit}
\end{figure*}

The temporal evolution of mean polarization (Figure~\ref{fig:line-graphs-center-limit}), confirms these regimes. The left panel ($\sigma_{\text{base}}=0.0$, pure repulsion) shows that for all but the lowest two diversity values ($o_{\text{max}}^{(0)} \le 0.2$), polarization \emph{increases} significantly over time, saturating at high levels (index $\approx 1.3$). This shows pure repulsion fosters polarization. The middle panel ($\sigma_{\text{base}}=1.0$, intermediate width) displays the previously noted bifurcation: low initial diversity ($o_{\text{max}}^{(0)} \le 0.7$) leads to decreasing polarization, while high initial diversity ($o_{\text{max}}^{(0)} \ge 0.8$) triggers rapid divergence to the \emph{highest} observed polarization levels (index $\approx 1.5$). The right panel ($\sigma_{\text{base}}=2.0$, wide attraction) shows that regardless of initial diversity, polarization consistently \emph{decreases} over time, converging towards strong consensus (index $< 0.25$). This confirms that a very wide attraction range dampens differences and prevents polarization.

Visualizing the underlying opinion distribution dynamics (Kernel Density Estimates, Figure~\ref{fig:kdes-center-limit}) provides crucial mechanistic insights. With a \emph{wide attraction width} ($\sigma_{\text{base}}=2.0$, right KDE plot), the initial uniform distribution rapidly collapses into a single, central peak ($o=0$), clearly demonstrating convergence to consensus. With \emph{pure repulsion} ($\sigma_{\text{base}}=0.0$, left KDE plot), the dynamic is different: agents are pushed away from dissimilar opinions. This forms two distinct peaks near the extremes ($o \approx \pm 1$). However, agents initially near the center are repelled by \emph{both} extremes, leading to the formation of a persistent \emph{third, central peak} around $o=0$. This results in a \emph{trimodal-like distribution}, explaining the high polarization index (capturing the distance between extreme clusters) despite significant density remaining at neutrality. With an \emph{intermediate attraction width} ($\sigma_{\text{base}}=1.0$, middle KDE plot), the system undergoes strong \emph{bipolarization}. The central region is effectively emptied as agents are strongly attracted towards the extremes they are closer to, forming two sharp, dominant peaks near $o=\pm 1$ with minimal central density.

 % Left: s_base=0, Middle: s_base=1, Right: s_base=2

In conclusion, the base attraction width $\sigma_{\text{base}}$ fundamentally determines the system's outcome, interacting strongly with initial diversity. Pure repulsion ($\sigma_{\text{base}}=0$) drives polarization, creating a trimodal-like state with extreme clusters and a neutral cluster repelled from both sides. A sufficiently wide attraction width ($\sigma_{\text{base}} \ge 1.4$) robustly drives the system to consensus by pulling agents towards the mean. Crucially, the \emph{highest} levels of polarization and the \emph{clearest bipolarization} (two extreme peaks, empty center) emerge at intermediate attraction widths ($\sigma_{\text{base}} \approx 1.0$) when initial diversity is high. This suggests that a balance where agents assimilate moderately different views but are repelled by highly dissimilar ones, rather than pure repulsion or very broad assimilation, is most effective at generating a strongly divided, two-party state in this model.

\subsection{Experiment 2: Structural Segregation and Base Attraction Width}
\label{app:exp2_mixing_attraction}

This second experiment explores the interplay between the initial structural segregation of the social network and the sensitivity of the opinion update mechanism. Specifically, we investigate how the level of initial community mixing ($\mu_{\text{mix}}^{(0)}$), which determines the prevalence of cross-group ties, interacts with the base attraction width ($\sigma_{\text{base}}$) to shape the final polarization outcome. The analysis aims to understand whether structural features like echo chambers (low $\mu_{\text{mix}}^{(0)}$) invariably lead to higher polarization, or if this relationship is modulated by the agents' propensity for assimilation versus repulsion ($\sigma_{\text{base}}$), keeping opinions dynamic but the network static.

\subsubsection{Experimental Setup}
We systematically varied two parameters:
\begin{enumerate} % Enumeration is fine for listing parameters
    \item \textbf{Initial Community Mixing ($\mu_{\text{mix}}^{(0)}$):} Governing the probability of forming cross-community ties during initial network generation (\ref{app:network-representation}), $\mu_{\text{mix}}^{(0)}$ was varied across the full range $[0.0, 0.1, ..., 1.0]$. $\mu_{\text{mix}}^{(0)}=0.0$ represents completely segregated communities based on initial opinion sign, while $\mu_{\text{mix}}^{(0)}=1.0$ represents a network structure generated without regard to opinion (fully mixed).
    \item \textbf{Base Attraction Width ($\sigma_{\text{base}}$):} As in Experiment 1, this parameter was varied over the range $[0.0, 0.2, ..., 2.0]$.
\end{enumerate}

All other parameters were held constant according to the baseline configuration (Table~\ref{tab:parameters-offline}). Crucially, the initial opinion range limit was set to a rather high initial diversity ($o_{\text{max}}^{(0)}=0.7$) to ensure sufficient initial diversity for polarization dynamics to potentially unfold, based on the findings of Experiment 1. The network remained static throughout the simulation ($T=100$). The final mean polarization index (Esteban-Ray) averaged over five runs was the primary outcome measure.

\subsubsection{Results}

The relationship between initial community mixing ($\mu_{\text{mix}}^{(0)}$) and the base attraction width ($\sigma_{\text{base}}$) on final polarization is depicted in Figure~\ref{fig:polarization-heatmaps}b. The results reveal a complex interplay, significantly modulated by the network structure.

One central finding is that \emph{minimal or zero attraction width} ($\sigma_{\text{base}} \approx 0.0 - 0.4$) consistently leads to \emph{high polarization} across nearly all levels of community mixing where interaction is possible ($\mu_{\text{mix}}^{(0)} > 0.0$). For pure repulsion ($\sigma_{\text{base}}=0.0$), final polarization is high (index $\approx1.3-1.5$) and increases slightly as mixing $\mu_{\text{mix}}^{(0)}$ increases towards 1.0. This indicates that when agents only repel differing views, any amount of cross-group interaction fuels divergence. Similar high polarization is observed for slightly larger, but still narrow, attraction widths (e.g., $\sigma_{\text{base}}=0.2, 0.4$).

The \emph{highest polarization levels} (index often > 1.45) are generally found for \emph{low attraction widths} ($\sigma_{\text{base}} \le 0.1$) combined with moderate to high community mixing ($\mu_{\text{mix}}^{(0)} \approx 0.6 - 1.0$). This suggests significant repulsion from the opposing side is the main driver of polarization, with high mixing providing sufficient cross-group exposure to fuel this dynamic without excessive dilution.

Conversely, polarization is strongly suppressed, leading to \emph{consensus}, when the \emph{attraction width is wide} ($\sigma_{\text{base}} \ge 1.4$). For $\sigma_{\text{base}}=2.0$, polarization is near zero across all $\mu_{\text{mix}}^{(0)}$. As $\sigma_{\text{base}}$ decreases from 2.0, the threshold $\mu_{\text{mix}}^{(0)}$ value required to maintain consensus increases, but wide attraction zones generally favor convergence.

The case of \emph{complete segregation} ($\mu_{\text{mix}}^{(0)}=0.0$) stands apart. Here, polarization settles at a moderate level (index $\approx 0.4-0.5$) and is remarkably insensitive to the attraction width $\sigma_{\text{base}}$, \emph{except} when the width is very large ($\sigma_{\text{base}}=2.0$), which forces consensus even within isolated groups. This confirms that without cross-community interaction, the assimilation/repulsion threshold $\sigma_{\text{base}}$ becomes largely irrelevant for system-wide polarization dynamics.

Further insight is provided by the temporal evolution of polarization (Figure~\ref{fig:line-graphs-center-mixing}) for different initial mixing levels ($\mu_{\text{mix}}^{(0)}$, colors) at representative attraction widths. Under \emph{pure repulsion} ($\sigma_{\text{base}}=0.0$, left panel), the segregated case ($\mu_{\text{mix}}^{(0)}=0.0$, purple) remains stable at moderate polarization. However, for all networks with mixing ($\mu_{\text{mix}}^{(0)} > 0.0$), polarization \emph{increases} significantly over time, rapidly reaching high levels (~1.5). Notably, higher mixing (greener/yellow lines) leads to slightly faster and higher final polarization in this pure repulsion regime. With an \emph{intermediate attraction width} ($\sigma_{\text{base}}=1.0$, middle panel), the dynamics show more variation with $\mu_{\text{mix}}^{(0)}$. The segregated case is stable. Low mixing ($\mu_{\text{mix}}^{(0)}=0.1, 0.2$) leads to depolarization (consensus). Moderate to high mixing ($\mu_{\text{mix}}^{(0)} \ge 0.3$) drives increasing polarization, peaking for intermediate $\mu_{\text{mix}}^{(0)}$ values and saturating slightly below the levels seen with pure repulsion for high $\mu_{\text{mix}}^{(0)}$. Finally, with a \emph{wide attraction width} ($\sigma_{\text{base}}=2.0$, right panel), the system consistently converges towards consensus for \emph{all} initial mixing levels ($\mu_{\text{mix}}^{(0)} \ge 0.0$), with polarization decreasing rapidly over time.

\begin{figure*}[h!]
    \centering
    \includegraphics[width=\textwidth]{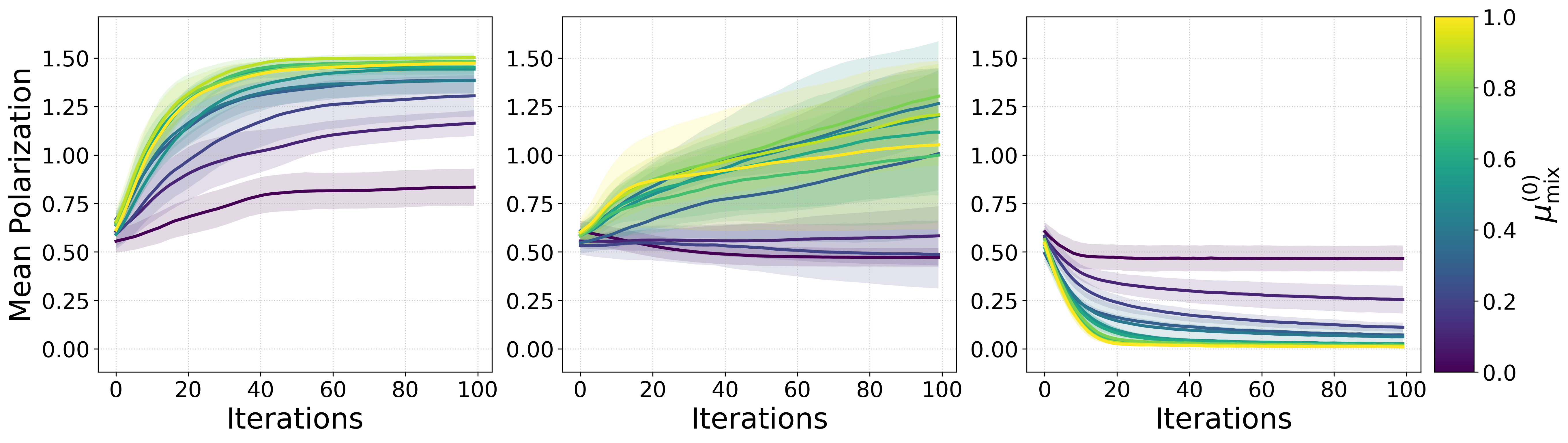}
    \caption{Mean polarization over time for varying baseline attraction widths ($\sigma_{\text{base}}$) and initial network mixing ($\mu_{\text{mix}}^{(0)}$). Panels from left to right: $\sigma_{\text{base}}=0.0$ (pure repulsion), $\sigma_{\text{base}}=1.0$ (intermediate), $\sigma_{\text{base}}=2.0$ (wide attraction). Line colors indicate $\mu_{\text{mix}}^{(0)}$ values (see color bar); shaded areas are 95\% CIs (5 runs). Pure repulsion consistently drives polarization; intermediate $\sigma_{\text{base}}$ leads to moderate polarization (higher for increased mixing), while wide attraction robustly yields consensus.}
    \label{fig:line-graphs-center-mixing}
\end{figure*}

These temporal plots confirm the heatmap findings: pure repulsion ($\sigma_{\text{base}}=0.0$) drives polarization when mixing occurs, wide attraction ($\sigma_{\text{base}}=2.0$) drives consensus universally, and intermediate attraction widths ($\sigma_{\text{base}}=1.0$) yield high polarization primarily under moderate mixing, while leading to consensus under low mixing. The sensitivity to network structure $\mu_{\text{mix}}^{(0)}$ is therefore highly dependent on the fundamental assimilation/repulsion balance set by $\sigma_{\text{base}}$.

The evolution of the opinion distributions (Kernel Density Estimates, Figure~\ref{fig:kdes-center-mixing}, using $o_{\text{max}}^{(0)}=1.0$) reveals the distinct population structures underlying these aggregate trends, depending on the base attraction width ($\sigma_{\text{base}}$) and initial mixing ($\mu_{\text{mix}}^{(0)}$). We focus on the contrasting outcomes under segregation ($\mu_{\text{mix}}^{(0)}=0$) versus full mixing ($\mu_{\text{mix}}^{(0)}=1$).

Under \emph{complete segregation} ($\mu_{\text{mix}}^{(0)}=0$), interaction is limited to within-group dynamics.
With pure repulsion ($\sigma_{\text{base}}=0$, KDE plot 1), a surprising trimodal-like structure emerges over time, with the highest density peak centered near neutrality ($o=0$). This suggests that within each isolated group, agents initially near the group's moderate edge are repelled by their more extreme counterparts, pushing them towards the global center, while the extremists remain near the poles. This internal dynamic limits extreme convergence within groups and contributes to the moderate overall polarization index observed.
With intermediate attraction ($\sigma_{\text{base}}=1$, KDE plot 3), segregation leads to two distinct clusters forming around moderate opinion values (approx. $\pm 0.5$), reflecting a balance of assimilation and repulsion within each isolated group.
With wide attraction ($\sigma_{\text{base}}=2$, KDE plot 5), strong assimilation within each group pulls agents towards their respective means, resulting in two moderate, relatively central peaks.
Common across all $\sigma_{\text{base}}$ values under segregation is the persistence of two separate distributions, confirming that the lack of interaction prevents system-wide convergence or divergence.

Under \emph{full mixing} ($\mu_{\text{mix}}^{(0)}=1$), agents interact across the spectrum, leading to system-wide dynamics dictated by $\sigma_{\text{base}}$.
Pure repulsion ($\sigma_{\text{base}}=0$, KDE plot 2) now drives strong \emph{bipolarization}. Unlike the segregated case, widespread repulsion pushes agents forcefully away from the center and the opposing side, leading to two dominant peaks near the extreme poles ($o \approx \pm 1$) and very low central density. This aligns with the high polarization index observed under these conditions.
Intermediate attraction ($\sigma_{\text{base}}=1$, KDE plot 4), however, results in a distinct \emph{trimodal distribution}. Two peaks form near the extremes, but a significant central peak around neutrality persists. This suggests that while attraction pulls agents towards nascent poles and repulsion pushes extremes apart, agents near the center experience conflicting signals or are repelled from both sides sufficiently to form a stable neutral cluster.
Wide attraction ($\sigma_{\text{base}}=2$, KDE plot 6) leads, as expected, to rapid convergence into a single \emph{unimodal distribution} centered at neutrality ($o=0$).

These distributional dynamics highlight the crucial role of both $\sigma_{\text{base}}$ and $\mu_{\text{mix}}^{(0)}$. Segregation ($\mu_{\text{mix}}^{(0)}=0$) prevents global dynamics, leading to varied within-group structures. Mixing ($\mu_{\text{mix}}^{(0)}=1$) allows $\sigma_{\text{base}}$ to determine the global outcome: pure repulsion drives bipolarization, wide attraction drives consensus, and intermediate attraction results in a trimodal state under these specific diversity initial conditions.

\subsubsection{Synthesis: Understanding Parameter Interactions Between Tolerance, Diversity, and Network Structure}
\label{app:synthesis-tolerance-structure}

The apparent discrepancies between Experiments B.1 and B.2 reveal crucial insights about how tolerance, initial diversity, and network structure interact to produce different polarization patterns. Understanding these interactions is essential for interpreting the main text results.

\paragraph{Why moderate tolerance produces different outcomes}
In Experiment B.1, moderate tolerance ($\sigma_{\text{base}} \approx 1.0$) combined with \emph{high initial diversity} ($o_{\text{max}}^{(0)} \ge 0.9$) produced the highest polarization levels through clear bipolarization. This occurs because sufficient extremist voices exist at the outset to drive dual attraction-repulsion dynamics: moderate agents are simultaneously attracted to their own side's extremists while being repelled by the opposing extreme, rapidly emptying the political center.

However, in Experiment B.2 with \emph{moderate initial diversity} ($o_{\text{max}}^{(0)} = 0.7$) and full integration, the same tolerance parameters produce trimodal distributions. The key difference is the initial extremist presence: without strong pre-existing extreme factions (as in the $o_{\text{max}}^{(0)} = 0.7$ case), there is insufficient "pull" to draw all moderate agents fully toward the poles. Instead, a persistent moderate center coexists with emerging extremes, creating the trimodal pattern.

\paragraph{Why pure repulsion produces different outcomes}
Pure repulsion ($\sigma_{\text{base}} = 0$) shows even more dramatic sensitivity to network structure:

In Experiment B.1 with \emph{moderate network integration} ($\mu_{\text{mix}}^{(0)} = 0.5$), pure repulsion produces trimodal distributions. Agents initially positioned near neutrality are exposed primarily to extremists from their own side (due to limited cross-cutting ties). This causes them to moderate \emph{inward}, away from their own side's extremists, creating a persistent central peak alongside the two extreme peaks.

In Experiment B.2 with \emph{full integration} ($\mu_{\text{mix}}^{(0)} = 1.0$), the same pure repulsion produces bipolarization. Here, near-neutral agents are exposed \emph{equally} to extremists from both sides. This symmetric exposure, combined with their slight initial lean toward one side, creates asymmetric repulsion: agents move further from the opposing camp than from their own faction. This neutralizes the inward moderation effect seen under limited integration, allowing cleaner bipolarization.

Under \emph{complete segregation} ($\mu_{\text{mix}}^{(0)} = 0.0$) in Experiment B.2, pure repulsion again produces trimodal distributions, but through a different mechanism: agents can only interact with their own community, so moderate agents are pushed toward the center by same-side extremists without any cross-cutting influence to counteract this effect.

\paragraph{Critical parameters for bipolarization}
These experiments reveal that strong bipolarization requires specific parameter combinations:
\begin{itemize}
\item \textbf{High initial diversity + moderate tolerance}: Works robustly across network structures (Experiment B.1)
\item \textbf{Moderate diversity + pure repulsion + full integration}: Creates bipolarization through asymmetric repulsion (Experiment B.2)
\item \textbf{Moderate diversity + moderate tolerance}: Produces trimodal distributions regardless of network structure due to insufficient extremist pull
\end{itemize}

The network structure thus acts as a critical mediator: it determines whether agents experience symmetric or asymmetric exposure to opposing views, which in turn determines whether psychological mechanisms (tolerance/repulsion) produce bipolar or trimodal outcomes.

In summary, this experiment demonstrates complex interactions between network structure and the assimilation/repulsion mechanism. While complete segregation stabilizes moderate polarization irrespective of $\sigma_{\text{base}}$ (except for $\sigma_{\text{base}}=2$), interaction across groups ($\mu_{\text{mix}}^{(0)} > 0$) allows $\sigma_{\text{base}}$ to shape the outcome. Pure repulsion ($\sigma_{\text{base}}=0$) or intermediate attraction ($\sigma_{\text{base}}=1$) can both lead to high polarization indices when mixing occurs, but via different underlying distributions (bipolar vs. trimodal under full mixing). Wide attraction ($\sigma_{\text{base}} \ge 1.4$) consistently promotes consensus.

\subsection{Experiment 3: Extremist Influence and Structural Segregation}
\label{app:exp3_influencer_mixing}

This experiment investigates the role of influential agents holding extreme initial opinions (influencers) in driving polarization within populations that initially exhibit high consensus. We examine how the polarizing impact of these influencers interacts with the initial structural segregation of the social network, controlled by the community mixing parameter $\mu_{\text{mix}}^{(0)}$. The objective is to understand whether the effectiveness of extremist actors in fostering societal division is contingent upon the network topology, specifically comparing their impact in segregated, integrated, or moderately mixed environments, under static network conditions.

\subsubsection{Experimental Setup}
For this analysis, we systematically varied two parameters:
\begin{enumerate} % Enumeration for parameters
    \item \textbf{Number of Influencers per Side ($N_{\text{inf}}$):} The number of agents designated as influencers on each side of the opinion spectrum (pro and contra) was varied from 0 to 10. These influencers were initialized with extreme opinions (drawn from a narrow distribution around $\pm 0.8$).
    \item \textbf{Initial Community Mixing ($\mu_{\text{mix}}^{(0)}$):} As in Experiment 2, this parameter governing initial network structure was varied across the full range $[0.0, 0.1, ..., 1.0]$.
\end{enumerate}

Crucially, the initial opinion distribution for the \emph{non-influencer} population was set to represent high consensus around neutrality ($o_{\text{max}}^{(0)}=0.1$). This isolates the influencers as the primary potential drivers of polarization. All other parameters followed the baseline configuration (Table~\ref{tab:parameters-offline}), including a static network and an assumed fixed agent resistance level conducive to polarization (based on prior experiments, $\sigma_{\text{base}} = 1.0$). The final mean polarization index after $T=100$ steps, averaged over five runs, served as the primary outcome measure.

\subsubsection{Results}

The impact of influencers ($N_{\text{inf}}$) on final polarization under varying initial community mixing ($\mu_{\text{mix}}^{(0)}$) is presented in Figure~\ref{fig:polarization-heatmaps}c. The results clearly demonstrate that the effectiveness of influencers is strongly contingent on the network's structure. In the baseline case without influencers ($N_{\text{inf}}=0$), polarization remains negligible across all mixing levels, confirming that the initial consensus state is stable on its own. As the number of influencers increases, polarization generally rises, indicating their potential to destabilize consensus. However, the extent of this effect critically depends on $\mu_{\text{mix}}^{(0)}$.

In completely segregated networks ($\mu_{\text{mix}}^{(0)}=0.0$), increasing $N_{\text{inf}}$ leads only to a moderate rise in polarization (saturating around 0.65). Here, the influencers' impact appears confined within their isolated communities, limiting system-wide division. The most potent polarizing effect emerges at intermediate levels of community mixing. Specifically, polarization reaches its highest values when a moderate-to-high number of influencers ($N_{\text{inf}} \approx 6-10$) operate within networks exhibiting moderate segregation ($\mu_{\text{mix}}^{(0)} \approx 0.2-0.7$). This suggests a scenario where influencers effectively pull their community members towards extremes, while the limited but existing cross-community links allow this polarization to generate friction or trigger reactance across the divide, thus amplifying overall polarization. Interestingly, as initial mixing increases further towards full integration ($\mu_{\text{mix}}^{(0)} \ge 0.8$), the polarizing effect of a given number of influencers diminishes slightly compared to the peak observed at moderate mixing. While influencers still induce significant polarization compared to the baseline, the well-mixed structure seemingly dilutes their impact, possibly due to non-influencers receiving more balanced exposure to countervailing moderate or opposing views.

\begin{figure*}[h!]
    \centering
    \includegraphics[width=\textwidth]{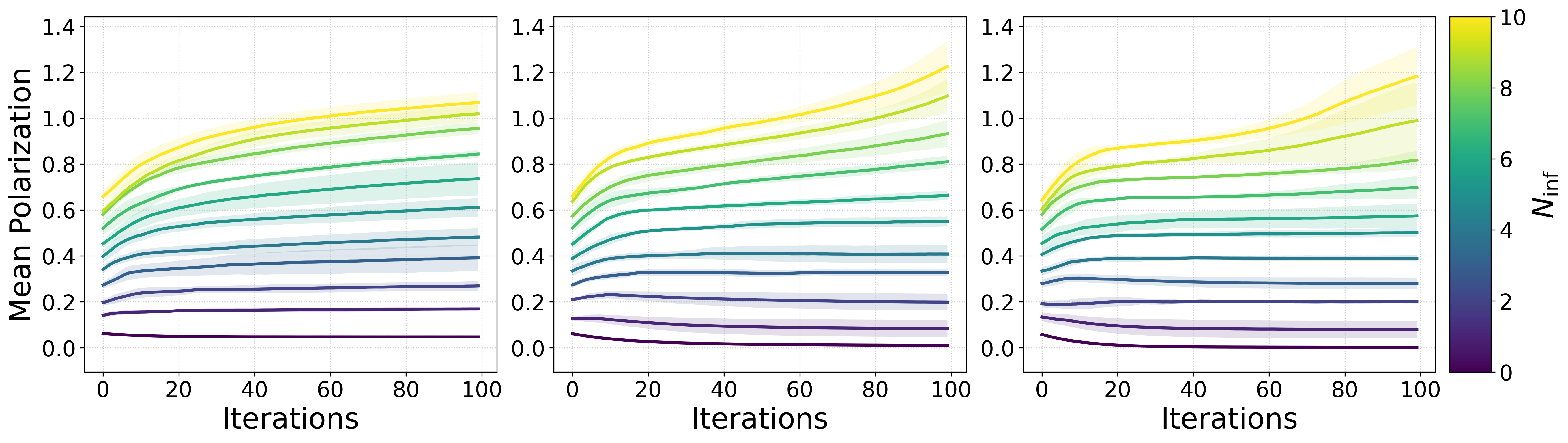}
    \caption{Mean polarization dynamics for baseline attraction widths $\sigma_{\text{base}} \in \{0.0, 1.0, 2.0\}$ (panels left to right) and varying number of influencers $N_{\text{inf}}$ (line colors; see color bar). Shaded areas: 95\% CIs (5 runs). A higher number of influencers consistently increases polarization; however, under pure repulsion ($\sigma_{\text{base}}=0.0$), the maximum polarization achieved is slightly less than with intermediate attraction and many influencers.}
    \label{fig:line-graphs-center-mixing}
\end{figure*}

Delving deeper into the underlying population dynamics, the Kernel Density Estimates in Figure~\ref{fig:kdes-mixing-influencers} illustrate the evolution of the opinion distribution for a high number of influencers ($N_{\text{inf}}=10$) under different mixing structures ($\mu_{\text{mix}}^{(0)} = 0.0, 0.5, 1.0$), starting from the initial consensus ($o_{\text{max}}^{(0)}=0.1$). Under complete segregation ($\mu_{\text{mix}}^{(0)}=0.0$, Panel a), the non-influencer population splits and converges towards moderate opinion values within their respective isolated groups, clearly influenced by but distinct from the extremists at the poles ($\pm 0.8$). This confirms the limited reach of influencers in fully segregated structures. With moderate mixing ($\mu_{\text{mix}}^{(0)}=0.5$, Panel b), the picture changes notably: the non-influencer population is strongly pulled away from the center towards the extremes, merging into broad clusters around the influencer positions. This demonstrates how moderate mixing enables influencers to effectively sway the broader population, leading to widespread bipolarization.

Perhaps most revealing is the outcome under high community mixing ($\mu_{\text{mix}}^{(0)}=1.0$, Panel c). Despite the persistent presence of extremists at the poles, the vast majority of the non-influencer population remains concentrated in a large peak around neutrality ($o=0$). This indicates that in a well-mixed environment, frequent exposure to diverse viewpoints effectively counteracts the polarizing pull of extremists for most agents. This finding provides crucial context for the aggregate heatmap results: high polarization scores observed in well-mixed networks with many influencers may largely reflect the fixed positions of the extremist minority, masking an underlying reality where the majority remains moderate.

In summary, this experiment highlights that the ability of extremist influencers to polarize a population is critically mediated by network structure. Neither complete isolation nor full integration provides the optimal conditions for maximizing societal division. Instead, moderate levels of community mixing appear most conducive, allowing influencers to leverage both in-group consolidation and cross-group friction. Furthermore, the distributional analysis reveals that aggregate polarization metrics can be misleading in well-mixed networks containing extremists, potentially overstating the extent of polarization within the general population.

\subsection{Parameter Realism and Model Applicability}
\label{app:parameter-realism}

While the experiments systematically explore parameter extremes to establish model boundaries, understanding real-world applicability requires mapping these parameters to observable social phenomena and assessing their empirical plausibility.

\emph{Psychological tolerance for disagreement} ($\sigma_{\text{base}}$) varies substantially across contexts and individuals. Pure repulsion ($\sigma_{\text{base}} = 0$) may characterize highly dogmatic individuals or discussions of deeply held moral convictions where any deviation triggers rejection. Moderate tolerance ($\sigma_{\text{base}} = 1.0$) likely reflects typical political discourse where individuals exhibit some openness to similar viewpoints while resisting strongly opposing positions. Full assimilation ($\sigma_{\text{base}} = 2.0$) represents idealized "rational discourse" where individuals remain genuinely open to persuasion from any position—potentially characterizing academic debates, deliberative forums, or discussions of technical issues where personal identity is less threatened. Most real-world scenarios likely fall in the moderate range ($\sigma_{\text{base}} = 0.5$–$1.5$), as complete openness or complete closure are psychologically and socially unsustainable.

\emph{Initial opinion diversity} ($o_{\text{max}}^{(0)}$) reflects how individuals enter discussions. High initial diversity ($o_{\text{max}}^{(0)} \ge 0.9$) may characterize highly polarized topics where strong factions already exist—such as abortion rights, gun control, or partisan political identities where prior socialization has created distinct camps. Moderate diversity ($o_{\text{max}}^{(0)} = 0.7$) may reflect emerging issues or topics where extreme positions exist but have not yet captured large populations—such as new technology policies, climate change solutions, or evolving social norms where most individuals hold intermediate positions. Low diversity scenarios are less theoretically interesting but may reflect topics with strong social consensus or limited prior exposure.

\emph{Network integration} ($\mu_{\text{mix}}$) depends on structural and algorithmic factors. Complete segregation ($\mu_{\text{mix}} = 0.0$) approximates echo chambers in highly curated social media environments or ideologically homogeneous communities. Full integration ($\mu_{\text{mix}} = 1.0$) may occur in diverse public forums, traditional mass media exposure, or algorithmic systems designed to maximize cross-cutting contact. Most realistic scenarios involve moderate integration ($\mu_{\text{mix}} = 0.3$–$0.7$), reflecting the permeable clustering typical of social networks where homophily creates tendencies toward like-minded connections while maintaining some cross-cutting ties through shared institutions, geographic proximity, or algorithmic recommendation.

\emph{Extremist influencer presence} reflects the number of highly committed actors attempting to shift moderate populations. Low influencer presence (0-2 per side) may characterize grassroots movements or emerging controversies where few dedicated activists exist. Moderate presence (3-6 per side) could reflect established political campaigns, advocacy organizations, or social movements with some but limited organizational capacity. High presence (7+ per side) might characterize highly mobilized contexts such as major political campaigns, culture wars, or topics where multiple well-funded organizations compete for influence. The "Goldilocks zone" effect observed in our results suggests that moderate influencer presence may be most effective for polarization, as insufficient numbers fail to create momentum while excessive numbers may trigger counter-mobilization or audience fatigue.

The model's predictive power emerges from identifying which parameter combinations characterize specific real-world contexts. Topics involving moral foundations or partisan identities may combine moderate-to-low tolerance with high initial diversity and low-to-moderate integration, predicting persistent polarization. Technical or policy debates in diverse forums may combine higher tolerance with moderate diversity and integration, suggesting more complex, potentially trimodal outcomes. The framework thus provides a systematic approach to anticipating polarization dynamics by mapping observable social and psychological characteristics onto specific parameter regimes.

These findings demonstrate that tolerance for disagreement becomes the decisive psychological factor once sufficient inter-group contact exists, but its effects are fundamentally mediated by both initial conditions and network structure. No single parameter determines polarization outcomes—rather, specific combinations create distinct regimes through different underlying mechanisms. This suggests that structural measures alone may be insufficient predictors of polarization without considering the psychological dispositions that govern how individuals process cross-cutting exposure and the initial distribution of extreme voices within the population.

\subsection{Experiment 4: Network Evolution under Fixed Opinion Landscapes}
\label{app:expA1_network_evolution}

This experiment isolates the dynamics of network structure evolution, investigating how the social graph changes based on agent interactions when opinions are held constant. Specifically, we examine the emergence of structural segregation (measured by network modularity) under different initial conditions. The goal is to understand the network rewiring mechanism driven by homophily/heterophily, disentangled from opinion change, and observe its dependence on the pre-existing opinion landscape (consensual vs. polarized) and the agents' propensity to connect with dissimilar others ($p_{\text{dis}}^{\text{(follow)}}$).

\subsubsection{Experimental Setup}
For this analysis, the core opinion update mechanism was deactivated; agent opinions remained fixed throughout the simulation ($T=100$). We explored two distinct initial opinion scenarios:
\begin{enumerate}
    \item \textbf{Unimodal:} Agent opinions initialized from $U(-0.1, 0.1)$ (high consensus).
    \item \textbf{Bimodal:} Agent opinions initialized from $U(-0.8, -0.6) \cup U(0.6, 0.8)$ (polarized).
\end{enumerate}
In both scenarios, the dynamic network evolution mechanism (\ref{app:connection-dynamics}) was active. Agents considered following others based on exposure, governed by the connection probability model. We systematically varied two parameters:

\begin{enumerate}
    \item \textbf{Initial Community Mixing ($\mu_{\text{mix}}^{(0)}$):} Varied across $[0.0, 0.1, ..., 1.0]$, controlling initial network segregation.
    \item \textbf{Propensity for Discordant Follow ($p_{\text{dis}}^{\text{(follow)}}$):} Varied over $[0.0, 0.1, ..., 1.0]$. $p_{\text{dis}}^{\text{(follow)}}=0.0$ implies strongly homophilous connection logic, while $p_{\text{dis}}^{\text{(follow)}}=1.0$ maximizes the chance of connecting despite discordance.
\end{enumerate}
All other parameters followed the baseline configuration (Table~\ref{tab:parameters-offline}). The primary metric tracked was the network modularity based on the fixed initial opinion groups.

\subsubsection{Results}

\begin{figure*}[htbp] % Use starred version for two-column span
    \centering

    \setlength{\plotaxesheighttwo}{7.0cm} % Using the command defined in main.tex

    \begin{minipage}[b]{\plotwidthtwo}
        \centering
        \includegraphics[width=\linewidth]
            {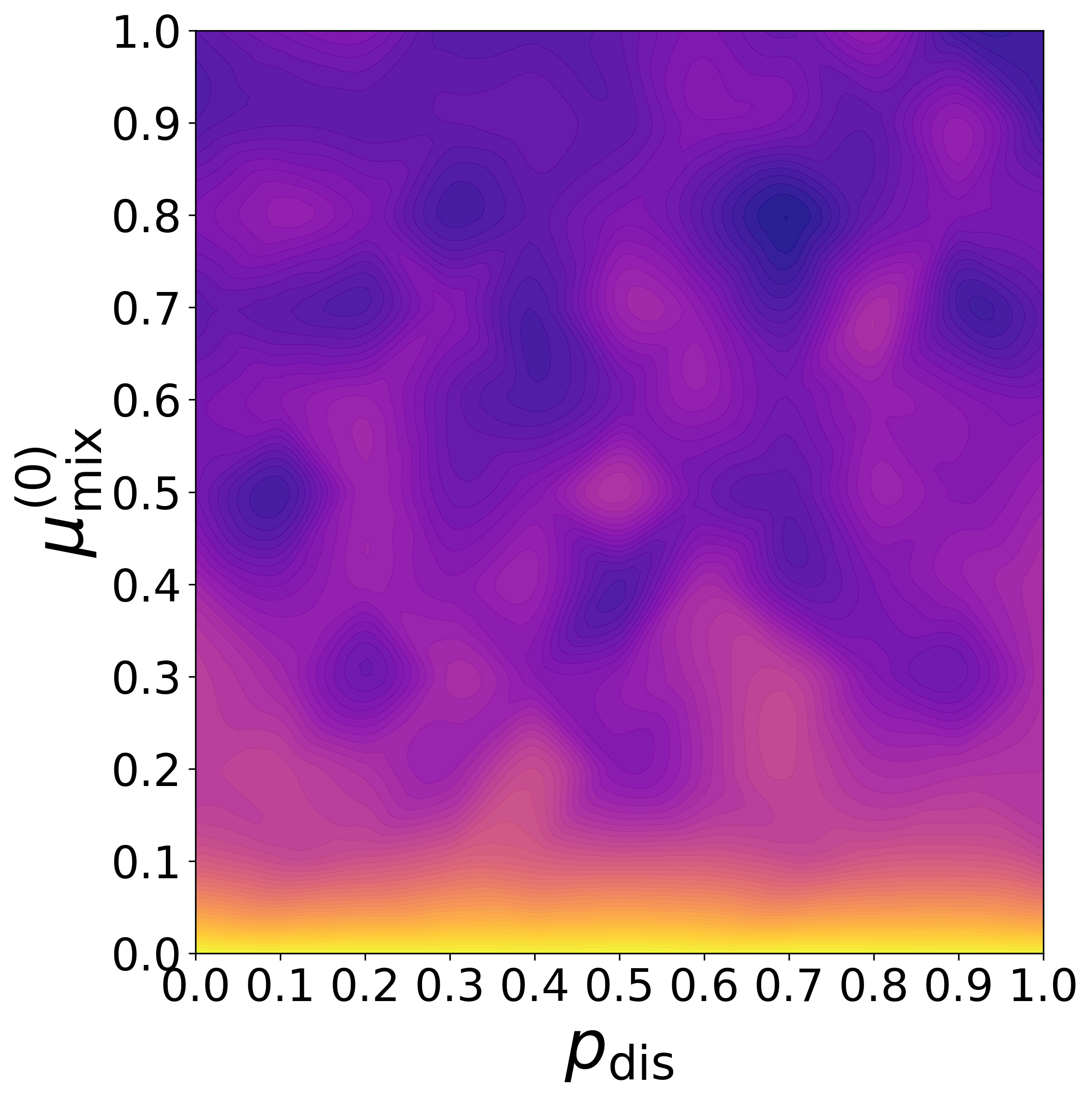}
        \centerline{(a) $U(-0.1, 0.1)$} % REPLACED \subcaption
    \end{minipage}%
    \hfill
    \begin{minipage}[b]{\plotwidthtwo}
        \centering
        \includegraphics[width=\linewidth]
             {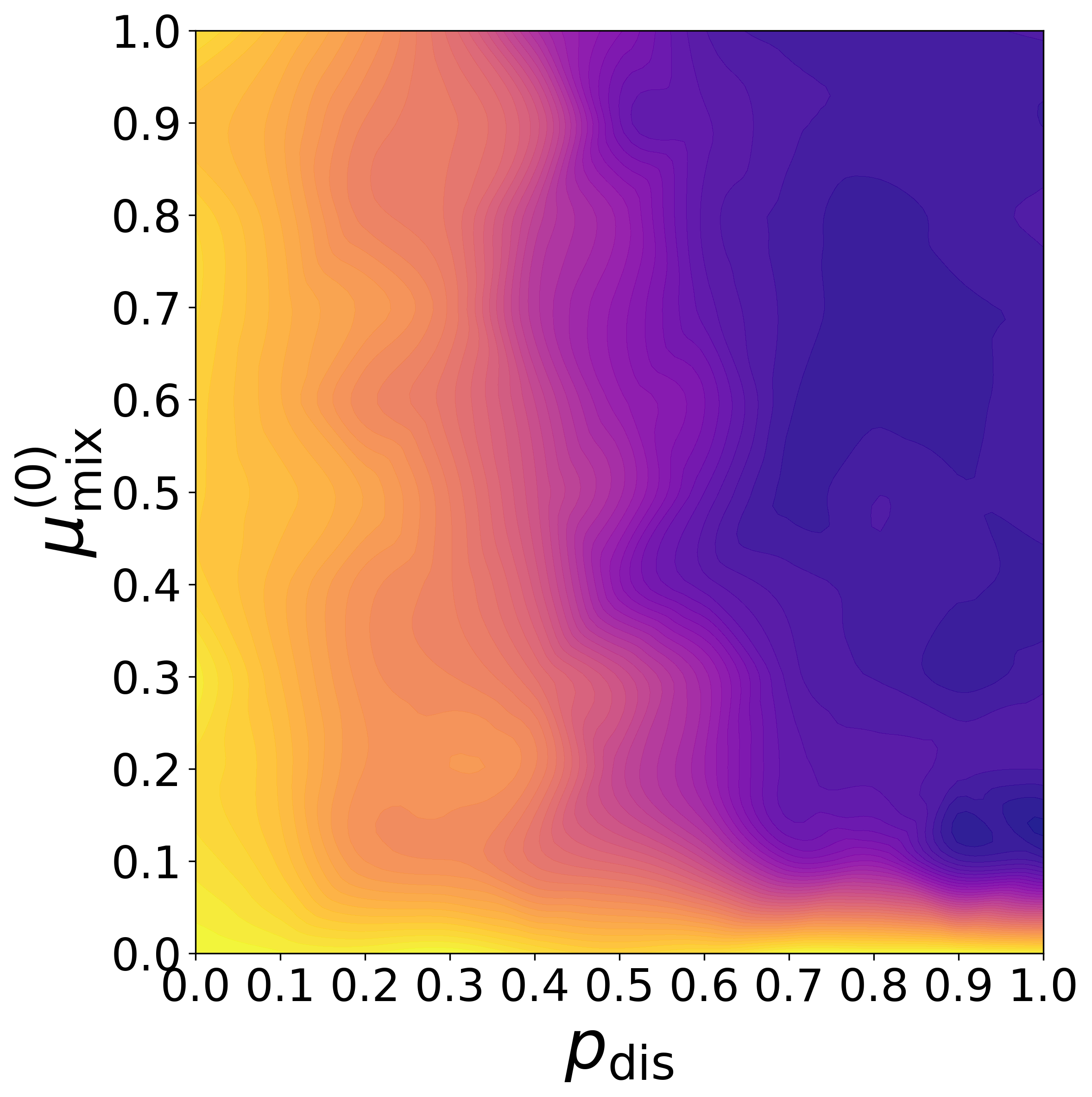}
        \centerline{(b) $U(-0.8, -0.6) \cup U(0.6, 0.8)$} % REPLACED \subcaption
    \end{minipage}%
    \hfill
    \begin{minipage}[b]{\cbarwidthtwo}
        \centering
        \includegraphics[height=\plotaxesheighttwo, width=\linewidth, keepaspectratio]
            {figures/heatmaps/modularity/colorbar.png}
        \vspace{4.5ex}
    \end{minipage}

    \caption{Network modularity as a function of discordance interaction propensity ($p_{\text{dis}}$) and initial network mixing ($\mu_{\text{mix}}^{(0)}$). 
    \textbf{(a)} Unimodal initial opinion distribution ($U(-0.1, 0.1)$, high consensus). 
    \textbf{(b)} Bimodal initial opinion distribution ($U(-0.8, -0.6) \cup U(0.6, 0.8)$, polarized). 
    Color intensity (brighter/yellower for higher values, darker/bluer for lower) represents the final network modularity. With unimodal initial opinions, modularity is generally low and increases mainly with decreasing initial mixing, peaking when $\mu_{\text{mix}}^{(0)} \approx 0$. With bimodal initial opinions, a non-linear relationship emerges: lower $p_{\text{dis}}$ (less discordant interaction) leads to higher modularity, and if $p_{\text{dis}}$ is high, high modularity is only achieved with minimal initial mixing.}
    \label{fig:bimodal-unimodal-modularity-heatmaps}
\end{figure*}

The evolution of network modularity under fixed opinions depends crucially on the underlying opinion landscape, as shown in Figure~\ref{fig:bimodal-unimodal-modularity-heatmaps}. In the unimodal case, where initial opinions exhibit high consensus, the final modularity is primarily determined by the initial network structure ($\mu_{\text{mix}}^{(0)}$). High initial segregation ($\mu_{\text{mix}}^{(0)}=0.0$) remains stable, resulting in high final modularity regardless of the discordant following propensity ($p_{\text{dis}}^{\text{(follow)}}$). As initial mixing increases ($\mu_{\text{mix}}^{(0)} > 0.1$), final modularity decreases significantly, largely reflecting the initial condition with minimal influence from $p_{\text{dis}}^{\text{(follow)}}$. The small opinion differences seemingly provide insufficient signal for substantial homophilic or heterophilic rewiring.

 % Original Figure: fig:exp4_heatmap_unimodal/bimodal

In stark contrast, when opinions are initially bimodal (polarized), the final network structure becomes strongly dependent on the discordant following propensity ($p_{\text{dis}}^{\text{(follow)}}$). A clear gradient across this parameter dominates the heatmap. Low values ($p_{\text{dis}}^{\text{(follow)}} \le 0.3$) lead to consistently high final modularity ($\approx 0.4-0.5$), indicating strong segregation driven by dominant homophilic rewiring. Conversely, high values ($p_{\text{dis}}^{\text{(follow)}} \ge 0.6$) result in consistently low final modularity ($\approx 0.18-0.22$), demonstrating that a propensity to connect with dissimilar others actively prevents structural segregation and promotes integration, even with polarized opinions. Initial mixing ($\mu_{\text{mix}}^{(0)}$) has a weaker effect, mainly modulating the outcome slightly at low $p_{\text{dis}}^{\text{(follow)}}$.

\begin{figure*}[htbp] % Use starred version for two-column span
    \centering

    \begin{minipage}[b]{\textwidth}
        \centering
        \includegraphics[width=\textwidth]{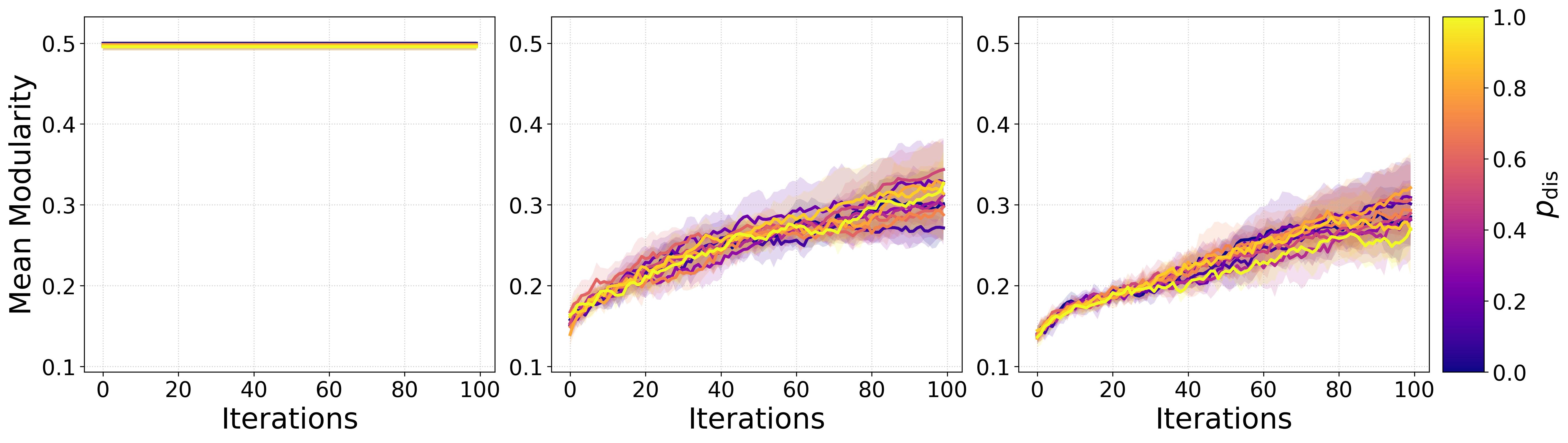}
        \centerline{(a) $U(-0.1, 0.1)$} % REPLACED \subcaption
    \end{minipage}
    
    \vspace{1em} % Vertical space between plots
    
    \begin{minipage}[b]{\textwidth}
        \centering
        \includegraphics[width=\textwidth]{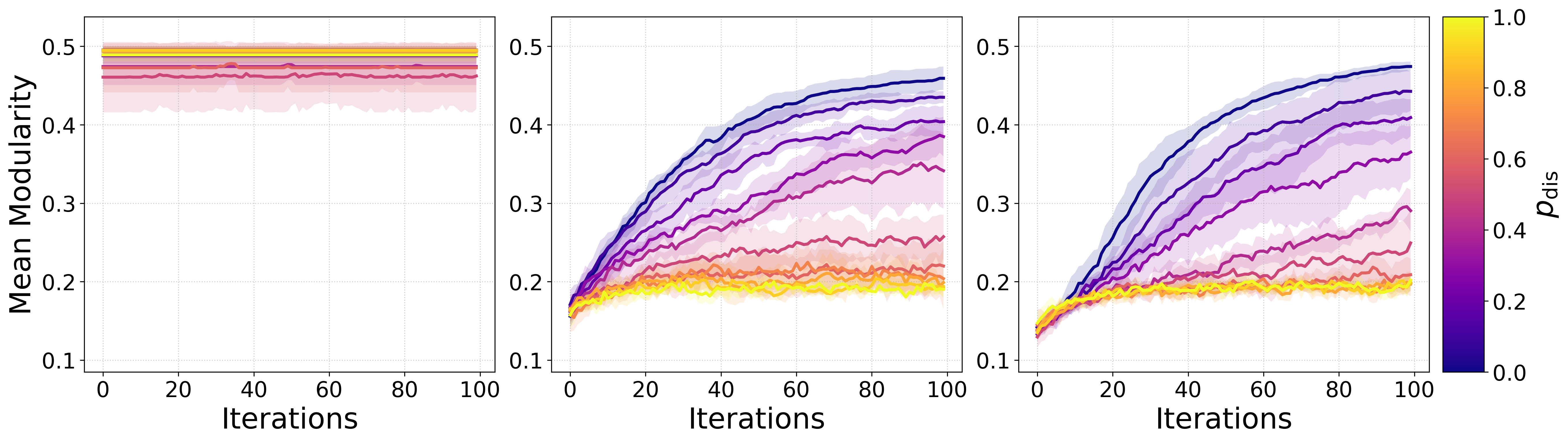}
        \centerline{(b) $U(-0.8, -0.6) \cup U(0.6, 0.8)$} % REPLACED \subcaption
    \end{minipage}

    \caption{Network modularity dynamics varying with initial network mixing ($\mu_{\text{mix}}^{(0)}$) and discordance propensity ($p_{\text{dis}}$). Panels show 
    \textbf{(a)} unimodal $U(-0.1, 0.1)$ and 
    \textbf{(b)} bimodal $U(-0.8, -0.6) \cup U(0.6, 0.8)$ initial opinions, for $\mu_{\text{mix}}^{(0)} \in \{0.0, 0.5, 1.0\}$ (columns L-R). Line colors: $p_{\text{dis}}$ (see color bar); shaded: 95\% CIs. Zero initial mixing ($\mu_{\text{mix}}^{(0)}=0.0$) consistently produces high modularity. For unimodal cases, higher mixing ($\mu_{\text{mix}}^{(0)} \ge 0.5$) leads to moderate modularity regardless of $p_{\text{dis}}$. For bimodal cases with higher mixing, lower $p_{\text{dis}}$ results in greater modularity.}
    \label{fig:bimodal-unimodal-modularity-line-graphs}
\end{figure*}

The temporal evolution plots (Figure~\ref{fig:bimodal-unimodal-modularity-line-graphs}) further illuminate these dynamics. The unimodal case confirms the negligible role of $p_{\text{dis}}^{\text{(follow)}}$ and the dominance of initial conditions, with modularity remaining relatively stable over time for any given $\mu_{\text{mix}}^{(0)}$. The bimodal case vividly illustrates the critical role of $p_{\text{dis}}^{\text{(follow)}}$. When the network starts segregated ($\mu_{\text{mix}}^{(0)}=0.0$, left panel), high $p_{\text{dis}}^{\text{(follow)}}$ actively dismantles this structure, decreasing modularity over time. When the network starts more mixed ($\mu_{\text{mix}}^{(0)}=0.5$ or $1.0$, middle and right panels), a clear bifurcation emerges: low $p_{\text{dis}}^{\text{(follow)}}$ drives a rapid increase in modularity (segregation), while high $p_{\text{dis}}^{\text{(follow)}}$ maintains or only marginally increases modularity (integration).

 % Original Figure: fig:exp4_lines_unimodal/bimodal

In summary, this experiment isolates the network evolution mechanism and shows its behavior is contingent on the opinion landscape. In consensual environments, structure is largely static and reflects initial conditions. In polarized environments, the network becomes highly dynamic, and the propensity for discordant following ($p_{\text{dis}}^{\text{(follow)}}$) acts as a switch determining whether the network evolves towards homophilic segregation or maintains integration.

\subsection{Experiment 5: Co-evolution of Polarization and Structure with Varying Initial Mixing}
\label{app:exp5_coevolution_mixing} % Updated label to reflect new numbering

This experiment investigates the coupled dynamics of opinion polarization and network structure evolution. Building upon Experiment 3, which examined influencer effects within static networks, we now activate the dynamic network evolution mechanism alongside opinion updates. The primary goal is to understand how the number of extremist influencers ($N_{\text{inf}}$) and the initial structural segregation of the network (community mixing, $\mu_{\text{mix}}^{(0)}$) jointly shape the emergent levels of both opinion polarization and structural segregation (measured as network modularity) when opinions and connections co-evolve. This directly addresses the feedback loop between opinion dynamics and network adaptation (Contribution D).

\subsubsection{Experimental Setup}
The simulation incorporated both active opinion updates (\ref{app:opinion-shifts}) and active dynamic network evolution (\ref{app:connection-dynamics}). We systematically varied two key parameters:
\begin{enumerate} % Enumeration for parameters
    \item \textbf{Number of Influencers per Side ($N_{\text{inf}}$):} Varied from 0 to 10, initialized with extreme opinions ($\pm 0.8$).
    \item \textbf{Initial Community Mixing ($\mu_{\text{mix}}^{(0)}$):} Varied from 0.1 to 1.0.
\end{enumerate}
The initial opinion landscape for non-influencers featured high consensus ($o_{\text{max}}^{(0)}=0.1$), isolating the impact of influencers. All other parameters adhered to the baseline configuration (Table~\ref{tab:parameters-offline}), including a fixed base attraction width ($\sigma_{\text{base}}=1.0$). The simulation ran for $T=100$ time steps. The analysis focused on the final opinion polarization (Esteban-Ray index) and the final network modularity, calculated based on the initial pro versus contra opinion groups and averaged over multiple runs.

\subsubsection{Results}

The co-evolutionary dynamics lead to distinct outcomes in both network structure and opinion distribution, strongly influenced by the number of influencers and initial mixing. Figure~\ref{fig:modularity-influencers-mixing} presents the final mean network modularity. A clear trend emerges where modularity generally increases with the number of influencers ($N_{\text{inf}}$). For any given initial mixing $\mu_{\text{mix}}^{(0)}$, adding influencers tends to enhance the final structural segregation. This suggests that influencers, by driving opinion divergence, create sharper divides that fuel subsequent homophilic network rewiring. Initial mixing $\mu_{\text{mix}}^{(0)}$ also plays a role; for a fixed $N_{\text{inf}} > 0$, higher initial mixing generally leads to lower final modularity. Consequently, the highest final modularity values (index > 0.55) arise from combining a large number of influencers ($N_{\text{inf}} \ge 6$) with low-to-moderate initial mixing ($\mu_{\text{mix}}^{(0)} \le 0.6$). However, even networks starting fully mixed ($\mu_{\text{mix}}^{(0)}=1.0$) can achieve considerable final modularity (index > 0.44) if enough influencers are present ($N_{\text{inf}} \ge 5$). Without influencers, modularity remains low regardless of $\mu_{\text{mix}}^{(0)}$.

Figure~\ref{fig:bimodal-unimodal-polarization-heatmaps}a displays the final mean polarization. As expected, polarization strongly increases with the number of influencers ($N_{\text{inf}}$). In scenarios with influencers, the effect of initial mixing $\mu_{\text{mix}}^{(0)}$ on polarization appears non-monotonic, consistent with the static network findings in Experiment 3. For a fixed, non-zero $N_{\text{inf}}$ (e.g., $N_{\text{inf}}=10$), polarization tends to peak at intermediate levels of initial mixing ($\mu_{\text{mix}}^{(0)} \approx 0.2-0.7$, potentially extending slightly higher) and is comparatively lower for both very low ($\mu_{\text{mix}}^{(0)}=0.1$) and very high ($\mu_{\text{mix}}^{(0)}=1.0$) initial mixing. This reinforces the idea of an optimal range of initial cross-group exposure for influencers to maximize system-wide polarization during co-evolution.

The relationship between the two outcome measures, final polarization and final modularity, is explored further in Figure~\ref{fig:mixing-influencers-scatter}. The scatter plot (left panel, colored by $\mu_{\text{mix}}^{(0)}$, sized by $N_{\text{inf}}$) reveals a generally positive correlation: higher modularity often accompanies higher polarization. Runs with few influencers cluster at low polarization and low modularity, while increasing $N_{\text{inf}}$ shifts outcomes towards the top-right. The coloring indicates that while high initial mixing (yellow, $\mu_{\text{mix}}^{(0)} \approx 1.0$) permits high polarization, it tends to result in slightly lower maximum modularity compared to intermediate mixing levels (green, $\mu_{\text{mix}}^{(0)} \approx 0.4-0.7$). The latter conditions appear capable of achieving both high polarization and high modularity simultaneously.

A Gaussian Mixture Model (GMM) analysis performed on the input parameters ($\mu_{\text{mix}}^{(0)}, N_{\text{inf}}$) identifies four distinct operational regimes (right panel), which map clearly onto the outcome space (middle panel). Regime 0 (Blue Circles, low $N_{\text{inf}}$) consistently yields low polarization and low modularity. Regime 1 (Orange Squares, intermediate $N_{\text{inf}}$) leads to intermediate levels of both outcomes. Regime 2 (Green Diamonds, high $N_{\text{inf}}$) corresponds to the highest observed levels of both polarization and modularity, signifying strongly polarized and structurally segregated states. Finally, Regime 3 (Red Crosses, low-intermediate $N_{\text{inf}}$ in moderate-to-high $\mu_{\text{mix}}^{(0)}$ networks) occupies a distinct region characterized by low-to-moderate polarization but moderate modularity. This suggests that under certain conditions, particularly with moderate influencer numbers (3-7 per side) in initially integrated networks, structural segregation and opinion polarization may decouple, with homophilic network formation occurring independently of extreme opinion convergence.

The mathematical basis for this decoupling lies in the different response functions governing network formation versus opinion updates, operating with different parameter sensitivities in the moderate influencer regime. Network formation decisions employ the interaction probability model (\cref{eq:interaction_probability_final_appendix}), where following probability depends on a Gaussian concordance function (\cref{eq:prob_concordant}) that responds smoothly to opinion similarity. Opinion updates, by contrast, use a polynomial function (\cref{eq:polynomial_shift_raw}) that creates sharp thresholds between attraction and repulsion, modulated by conviction-based resistance mechanisms (\cref{eq:effective_width_sq}). In the moderate influencer regime (3-7 per side), the system reaches a critical balance: influencers create sufficient directional signals to activate the Gaussian-based following decisions, leading to structural clustering around moderate opinion differences (polarization indices 0.6-0.8), but the same signals prove insufficient to consistently cross the polynomial threshold required for extreme opinion shifts. The discovery mechanism amplifies this effect by exposing agents to multiple weak influencer signals that accumulate to influence network structure without triggering strong opinion update mechanisms, creating a stable moderate differentiation state where agents sort into network communities based on detectable but non-extreme preference differences.

In summary, this experiment demonstrates a strong coupling between opinion polarization and network structure evolution when extremist influencers are present and the network can adapt. Influencers act as catalysts driving the system towards states of both higher polarization and increased structural segregation. The initial network mixing ($\mu_{\text{mix}}^{(0)}$) modulates this process, with intermediate mixing often facilitating the highest polarization, while lower initial mixing allows for higher final modularity given sufficient influencers. The GMM analysis effectively partitions the parameter space into regimes associated with distinct co-evolutionary outcomes, ranging from consensus/integration to polarization/segregation, and reveals potential intermediate states where structural segregation and opinion polarization can decouple.

\subsection{Experiment 6: Co-evolution with Algorithmic Discovery Rate}
\label{app:exp6_coevolution_discovery} % Updated label to reflect new numbering

This experiment further explores the co-evolution of opinion polarization and network structure, shifting the focus from initial network topology to the influence of the recommendation algorithm's information sourcing strategy. Specifically, we investigate how the interplay between extremist influencers ($N_{\text{inf}}$) and the recommendation system's discovery rate ($\delta_{\text{rec}}$) shapes the emergent levels of polarization and modularity when opinions and connections evolve dynamically. The discovery rate controls the balance between recommendations from within an agent's existing network versus those from outside (discovery). The objective is to understand how algorithmic curation of information flow interacts with polarizing actors to drive societal division, addressing Contribution D from an algorithmic perspective.

\subsubsection{Experimental Setup}
The simulation incorporated active opinion updates (\ref{app:opinion-shifts}) and dynamic network evolution (\cref{app:connection-dynamics}). The initial opinion landscape consisted of a consensus distribution ($o_{\text{max}}^{(0)}=0.1$) for non-influencers, supplemented by influencers with extreme opinions ($\pm 0.8$). The initial network structure was held constant, presumably representing a moderately or fully mixed state (e.g., baseline $\mu_{\text{mix}}^{(0)}=1.0$) to isolate the effects of $\delta_{\text{rec}}$ and rewiring. We systematically varied two key parameters:
\begin{enumerate} % Enumeration for parameters
    \item \textbf{Number of Influencers per Side ($N_{\text{inf}}$):} Varied from 0 to 10.
    \item \textbf{Discovery Rate ($\delta_{\text{rec}}$):} Varied from 0.0 (purely network-based recommendations) to 1.0 (purely discovery-based recommendations).
\end{enumerate}
All other parameters followed the baseline configuration (Table~\ref{tab:parameters-offline}), including a fixed base attraction width ($\sigma_{\text{base}}=1.0$). Final opinion polarization (Esteban-Ray index) and network modularity (based on initial opinion groups) after $T=100$ steps, averaged over five runs, were the primary outcome measures.

\subsubsection{Results}

The results demonstrate that the algorithmic discovery rate ($\delta_{\text{rec}}$) significantly mediates the impact of influencers on both opinion polarization and network structure, leading to distinct and sometimes opposing outcomes. Figure~\ref{fig:modularity-influencers-discovery} presents the final mean network modularity as a function of $N_{\text{inf}}$ and $\delta_{\text{rec}}$. Modularity generally increases with both parameters. For a fixed $\delta_{\text{rec}}$, increasing $N_{\text{inf}}$ enhances structural segregation. Similarly, for a fixed $N_{\text{inf}} > 0$, increasing $\delta_{\text{rec}}$ also tends to increase final modularity, particularly when influencers are numerous ($N_{\text{inf}} \ge 6$). Consequently, the highest modularity values (index > 0.5) are achieved with both a high number of influencers ($N_{\text{inf}} \ge 7$) and a high discovery rate ($\delta_{\text{rec}} \ge 0.8$). Without influencers, modularity remains low across all discovery rates.

However, the effect of the discovery rate on opinion polarization, shown in Figure~\ref{fig:bimodal-unimodal-polarization-heatmaps}b, is markedly different. While polarization again increases dramatically with the number of influencers ($N_{\text{inf}}$), its relationship with $\delta_{\text{rec}}$ is inverted compared to modularity. For a fixed, high number of influencers (e.g., $N_{\text{inf}}=10$), polarization reaches its peak values (index > 1.4) at \emph{low} discovery rates ($\delta_{\text{rec}} \le 0.1$). As the discovery rate increases, polarization tends to decrease significantly. Thus, high discovery rates, which promoted high structural modularity, simultaneously suppress maximum opinion polarization.

This reveals a striking \emph{asymmetrical relationship}: the conditions maximizing network segregation (high $N_{\text{inf}}$, high $\delta_{\text{rec}}$) are distinct from those maximizing opinion polarization (high $N_{\text{inf}}$, low $\delta_{\text{rec}}$). High modularity and high polarization do not necessarily peak together when co-evolution is driven by the interplay of algorithmic discovery and polarizing agents.

Further insight into this dissociation is provided by the scatter plot of final polarization versus final modularity in Figure~\ref{fig:discovery-influencers-scatter} (left panel, sized by $N_{\text{inf}}$, colored by $\delta_{\text{rec}}$). The plot confirms the separation: the highest polarization values (top region, large circles) occur across a range of modularity values but peak polarization is associated with relatively low modularity (approx. 0.15-0.25). Conversely, the highest modularity values (right region, > 0.5) correspond to only moderate polarization levels (approx. 0.8-1.1).

A Gaussian Mixture Model (GMM) analysis based on the input parameters ($N_{\text{inf}}, \delta_{\text{rec}}$) reinforces this asymmetry (right panel), identifying four distinct regimes mapped onto the outcome space (middle panel). Regime 0 (Blue Circles, low $N_{\text{inf}}$) results in low polarization and low modularity. Regime 1 (Orange Squares, intermediate $N_{\text{inf}}$, varied $\delta_{\text{rec}}$) yields intermediate polarization and low-to-moderate modularity. Crucially, Regime 2 (Green Diamonds, high $N_{\text{inf}}$, low $\delta_{\text{rec}}$) produces the highest levels of polarization but only low-to-moderate modularity. In contrast, Regime 3 (Red Crosses, mid/high $N_{\text{inf}}$, high $\delta_{\text{rec}}$) generates moderate-to-high modularity but yields significantly lower polarization compared to Regime 2. This GMM clustering effectively captures the observed divergence between conditions favoring opinion polarization versus structural segregation.

\begin{figure*}[htbp]
    \centering
    \includegraphics[width=0.95\textwidth]{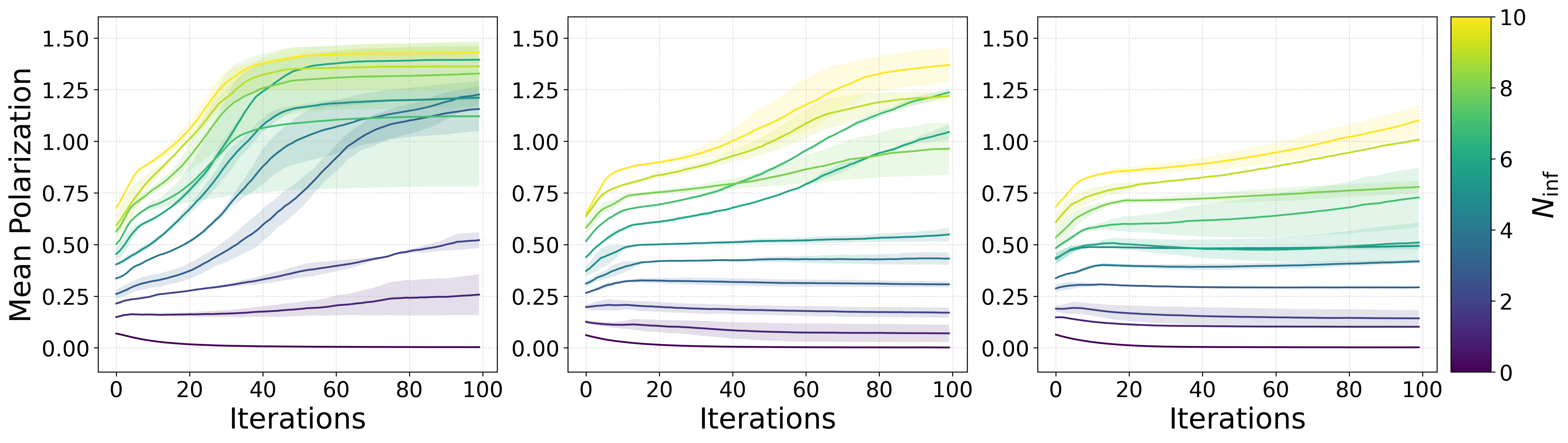}
    \caption{Temporal evolution of opinion polarization under co-evolutionary dynamics with varying algorithmic discovery rates and extremist influencer presence. Lines show mean polarization (Esteban-Ray index) over 100 time steps, with shaded regions indicating 95\% confidence intervals across 5 simulation runs. The figure demonstrates that under high extremist presence (7+ influencers), low discovery rates produce polarization trajectories that continue increasing throughout the simulation period, while high discovery rates show slower polarization progression. This temporal pattern supports the characterization of high discovery as providing temporal mitigation rather than elimination of polarization dynamics under sustained extremist influence.}
    \label{fig:polarization-timeline-discovery}
\end{figure*}

The temporal dynamics of polarization (\cref{fig:polarization-timeline-discovery}) provide crucial insights into the nature of algorithmic intervention effectiveness. Under high extremist presence (7+ influencers), polarization trajectories under low discovery rates show continued increase throughout the 100-step simulation period, while high discovery rates demonstrate slower progression. Critically, even under high discovery conditions with substantial extremist presence, polarization continues to increase rather than stabilizing, confirming that high discovery provides temporal mitigation—delaying polarization progression—rather than eliminating the underlying polarization dynamics. This temporal pattern validates the characterization of discovery-based interventions as effective for early prevention but offering only partial protection once extremist presence becomes substantial.

In summary, this experiment demonstrates that algorithmic discovery rates critically mediate the influence of extremist minorities on moderate populations. Low discovery rates enable disproportionate extremist influence by trapping moderate agents within their initial network connections, preventing them from discovering the broader moderate majority. This leads to high opinion polarization but limited structural segregation. High discovery rates facilitate "democratic sorting," where moderate agents can efficiently find like-minded others from the broader population, diluting extremist influence and enabling the formation of moderate coalitions. While this creates high structural segregation, it occurs through agents' connection choices rather than algorithmic content filtering, resulting in structurally segregated but informationally diverse environments. This suggests that discovery algorithms may serve as democratic protection mechanisms, preventing extremist capture while enabling informed homophilic sorting.

\subsection{Experiment 7: Reaction Dynamics under Varying Mixing and Discovery}
\label{app:exp7_reaction_mixing_discovery}

This experiment shifts the focus to agent interaction patterns, specifically reactive engagements like liking. It investigates how the overall volume of reactions and the tendency for agents to react homophilically (to content aligned with their own stance) are influenced by the interplay between initial network structure (community mixing, $\mu_{\text{mix}}^{(0)}$) and the algorithmic information discovery mechanism (discovery rate, $\delta_{\text{rec}}$). The analysis aims to provide insights into engagement dynamics (related to Contribution C) under different societal configurations (consensual vs. polarized) and information flow regimes, complementing the main text's focus on opinion and network structure.

\subsubsection{Experimental Setup}
The simulation framework was configured to track agent reactions ('likes'), activating the interaction probability model (\cref{app:interaction-probabilities}). Opinions and network connections were likely allowed to evolve alongside reactions (confirm if this was the case, or state if fixed). We explored two distinct initial opinion scenarios:

\begin{enumerate}
    \item \textbf{Unimodal:} Agent opinions initialized from $U(-0.1, 0.1)$ (high consensus).
    \item \textbf{Bimodal:} Agent opinions initialized from $U(-0.8, -0.6) \cup U(0.6, 0.8)$ (polarized).
\end{enumerate}

We systematically varied two parameters:

\begin{enumerate}
    \item \textbf{Initial Community Mixing ($\mu_{\text{mix}}^{(0)}$):} Varied across $[0.0, 0.1, ..., 1.0]$.
    \item \textbf{Discovery Rate ($\delta_{\text{rec}}$):} Varied across $[0.0, 0.1, ..., 1.0]$.
\end{enumerate}

All other parameters followed the baseline configuration (Table~\ref{tab:parameters-offline}). The primary metrics measured over $T=100$ steps were the mean total number of reactions per agent and the overall stance ratio (proportion of reactions given to same-stance content), averaged over five runs.

\subsubsection{Results}

The analysis reveals that reaction dynamics are fundamentally shaped by the initial opinion landscape, with distinct patterns emerging for consensual versus polarized populations, as shown in Figure~\ref{fig:engagement-homophilic}. 

Examining the total number of reactions per agent (left panels), the initial opinion state significantly alters overall engagement levels and their dependence on $\mu_{\text{mix}}^{(0)}$ and $\delta_{\text{rec}}$. In the unimodal (consensus) scenario, total reactions are moderately influenced by the parameters, peaking when the initial network is strongly segregated ($\mu_{\text{mix}}^{(0)} \le 0.1$) and the discovery rate is moderate to high ($\delta_{\text{rec}} \ge 0.4$). Engagement is lowest in initially mixed networks ($\mu_{\text{mix}}^{(0)} \ge 0.8$) with limited discovery ($\delta_{\text{rec}} \le 0.2$). In contrast, the bimodal (polarized) scenario exhibits considerably higher peak engagement and a different dependency pattern. Reaction volume is maximized under high initial segregation ($\mu_{\text{mix}}^{(0)}=0.0$) combined with high discovery rates ($\delta_{\text{rec}} \ge 0.6$). Engagement decreases markedly with increasing initial mixing ($\mu_{\text{mix}}^{(0)}$) or decreasing discovery ($\delta_{\text{rec}}$). Thus, while low $\mu_{\text{mix}}^{(0)}$ benefits engagement in both cases, high $\delta_{\text{rec}}$ is crucial for peak engagement only in the bimodal case.

\begin{figure*}[htbp] % Use starred version for two-column span
    \centering

    \setlength{\plotaxesheighttwo}{7.0cm} % Using the command defined in main.tex

    \begin{minipage}[b]{\plotwidthtwo}
        \centering
        \includegraphics[width=\linewidth]
            {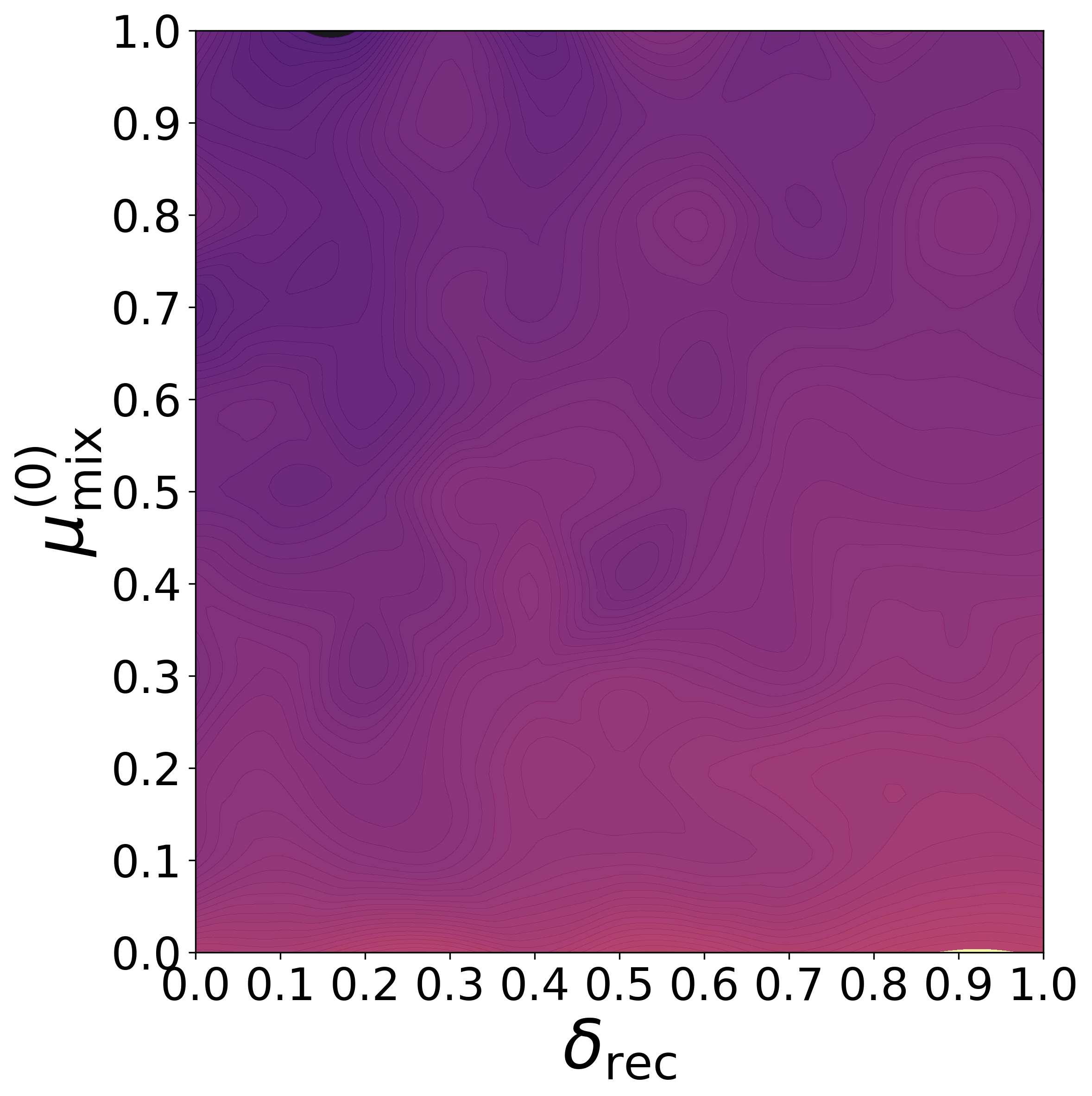}
        \centerline{(a) $U(-0.1, 0.1)$} % REPLACED \subcaption
    \end{minipage}%
    \hfill
    \begin{minipage}[b]{\plotwidthtwo}
        \centering
        \includegraphics[width=\linewidth]
             {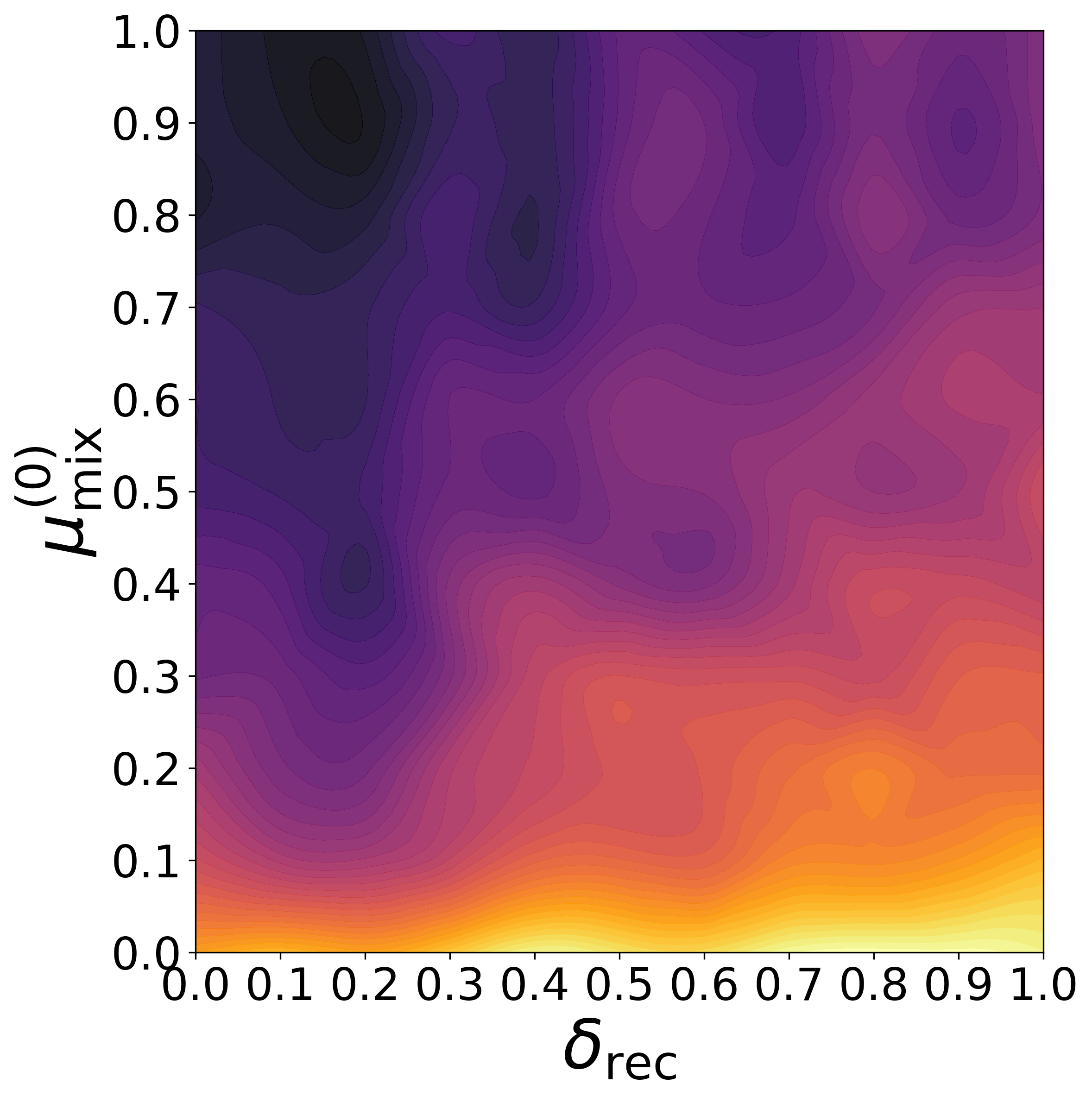}
        \centerline{(b) $U(-0.8, -0.6) \cup U(0.6, 0.8)$} % REPLACED \subcaption
    \end{minipage}%
    \hfill
    \begin{minipage}[b]{\cbarwidthtwo}
        \centering
        \includegraphics[height=\plotaxesheighttwo, width=\linewidth, keepaspectratio]
            {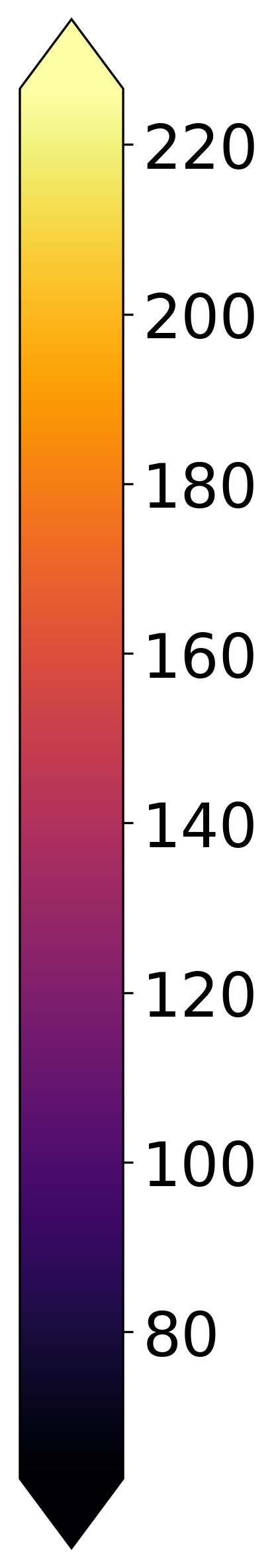}
        \vspace{4.5ex}
    \end{minipage}

    \caption{Total agent interaction contours based on discovery rate ($\delta_r$) and initial network mixing ($\mu_{\text{mix}}^{(0)}$). Panels show results for 
    \textbf{(a)} unimodal initial opinions ($U(-0.1, 0.1)$) and 
    \textbf{(b)} bimodal initial opinions ($U(-0.8, -0.6) \cup U(0.6, 0.8)$). 
    Brighter colors indicate more interactions. Unimodal initialization leads to moderate interaction counts irrespective of $\delta_r$ and $\mu_{\text{mix}}^{(0)}$. Bimodal initialization shows a more nuanced pattern: interactions are maximized with low initial mixing and high discovery (bottom-right region) and minimized with high initial mixing and low discovery (top-left region).}
    \label{fig:bimodal-unimodal-reaction-heatmaps-1}
\end{figure*} % Original Figure: fig:exp7_unimodal_metrics / fig:exp7_bimodal_metrics (Left Panels)

Turning to reaction homophily, measured by the overall stance ratio (Figure~\ref{fig:engagement-homophilic}, bottom panels), the difference between the scenarios is particularly stark. In the unimodal case, reaction selectivity is sensitive to both parameters. The stance ratio is minimal (0.0, perfect homophily) only under complete initial segregation ($\mu_{\text{mix}}^{(0)}=0.0$) and increases steadily as either initial mixing $\mu_{\text{mix}}^{(0)}$ or discovery rate $\delta_{\text{rec}}$ increases. In highly mixed networks with ample discovery ($\mu_{\text{mix}}^{(0)} \ge 0.8, \delta_{\text{rec}} \ge 0.5$), the ratio rises considerably (to approx. 0.25-0.3), indicating substantial cross-stance engagement facilitated by structural integration and algorithmic discovery. Conversely, in the bimodal case, the stance ratio remains uniformly minimal (0.0) across the entire parameter space. Regardless of the initial network mixing or the discovery rate, agents in the polarized scenario exclusively react to content aligning with their own opinion group. The underlying opinion polarization appears to completely dominate reaction selectivity, creating extreme homophily resistant to structural or algorithmic variations aimed at increasing exposure diversity.

In conclusion, the initial opinion landscape fundamentally determines agent reaction patterns. A consensual population exhibits moderate engagement levels and reaction selectivity that can be modulated by network structure and information discovery, with integration and discovery fostering cross-stance interaction. A polarized population, however, shows potentially higher peak engagement but drastically different selectivity: reaction homophily becomes extreme and immune to variations in initial network structure or algorithmic discovery. This suggests that once a society is polarized, interaction patterns might become rigidly segregated ("reaction bubbles"), even if the information environment technically allows cross-stance exposure, limiting the effectiveness of simply increasing diverse exposure to bridge divides in interaction patterns.

\subsection{Experiment 8: Reaction Dynamics under Varying Mixing and Discordant Engagement Propensity}
\label{app:expA3_reaction_mixing_discordant}

This experiment continues the investigation into agent reaction dynamics (e.g., liking), exploring how interaction patterns are shaped by the interplay between the initial network structure (community mixing, $\mu_{\text{mix}}^{(0)}$) and the agents' intrinsic propensity to engage with discordant information ($p_{\text{dis}}^{\text{(react)}}$). This parameter, part of the interaction probability model (\cref{app:interaction-probabilities}), modulates the likelihood of engagement driven specifically by opinion oppositeness. By varying $\mu_{\text{mix}}^{(0)}$ and $p_{\text{dis}}^{\text{(react)}}$, we aim to understand their combined influence on the volume and homophily of reactions in both consensual and polarized initial opinion landscapes.

\subsubsection{Experimental Setup}
The simulation tracked agent reactions with the interaction model active. We assumed opinions and network connections to be fixed and explored two initial opinion scenarios:

\begin{enumerate}
    \item \textbf{Unimodal:} Agent opinions initialized from $U(-0.1, 0.1)$ (high consensus).
    \item \textbf{Bimodal:} Agent opinions initialized from $U(-0.8, -0.6) \cup U(0.6, 0.8)$ (polarized).
\end{enumerate}

We systematically varied two parameters:

\begin{figure*}[htbp] % Use starred version for two-column span
    \centering

    \setlength{\plotaxesheighttwo}{7.0cm} % Using the command defined in main.tex

    \begin{minipage}[b]{\plotwidthtwo}
        \centering
        \includegraphics[width=\linewidth]
            {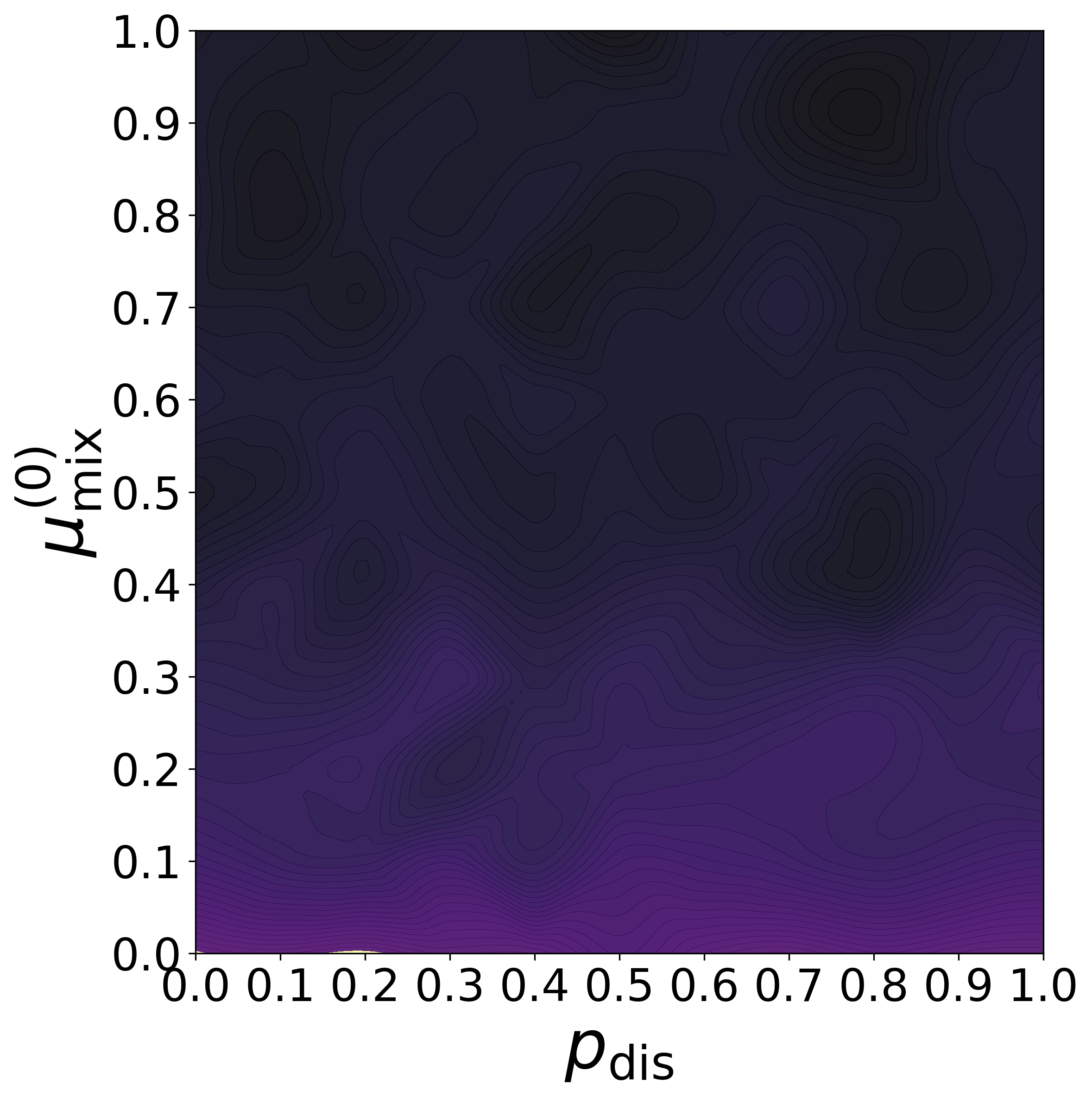}
        \centerline{(a) $U(-0.1, 0.1)$} % REPLACED \subcaption
    \end{minipage}%
    \hfill
    \begin{minipage}[b]{\plotwidthtwo}
        \centering
        \includegraphics[width=\linewidth]
             {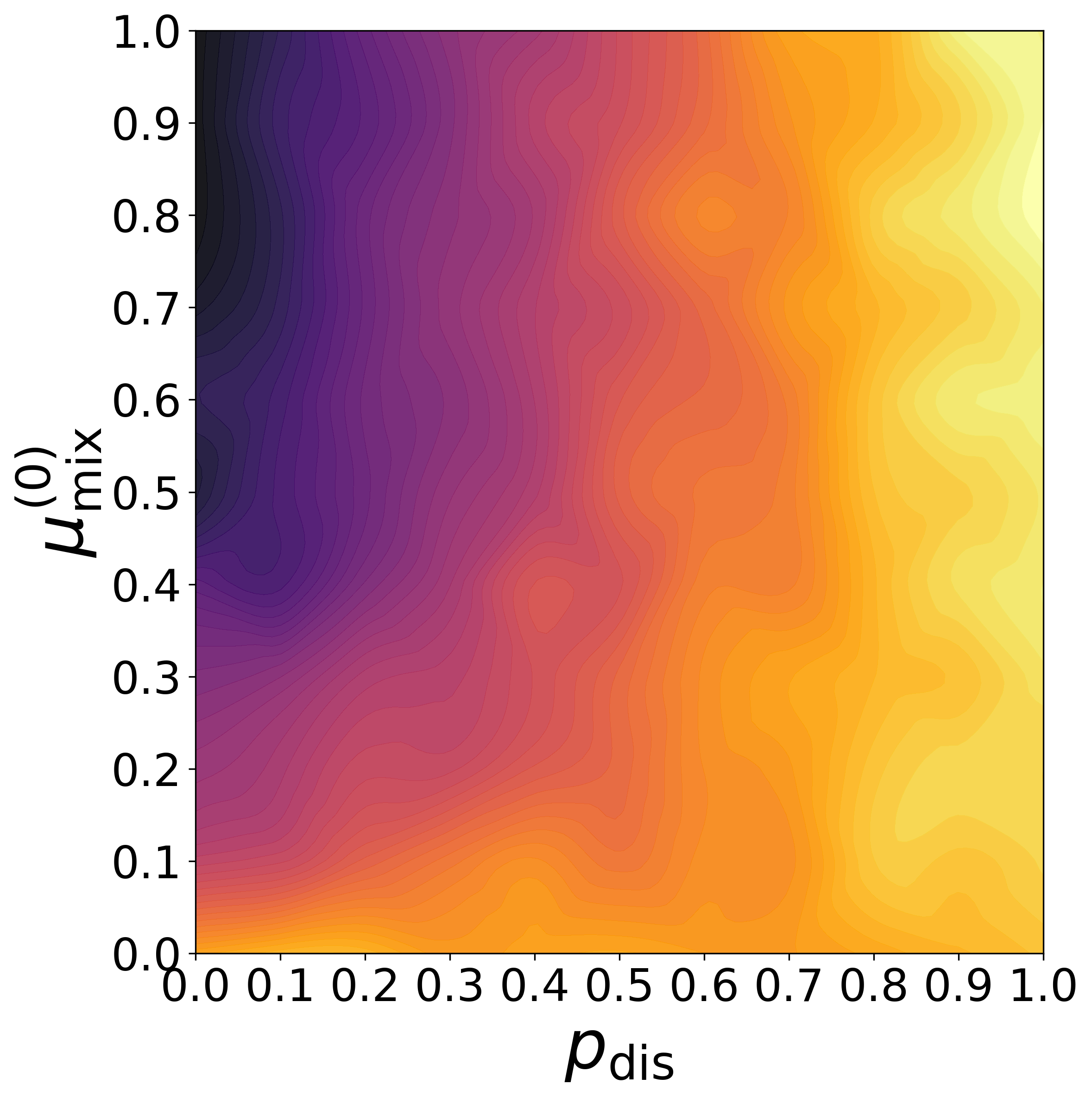}
        \centerline{(b) $U(-0.8, -0.6) \cup U(0.6, 0.8)$} % REPLACED \subcaption
    \end{minipage}%
    \hfill
    \begin{minipage}[b]{\cbarwidthtwo}
        \centering
        \includegraphics[height=\plotaxesheighttwo, width=\linewidth, keepaspectratio]
            {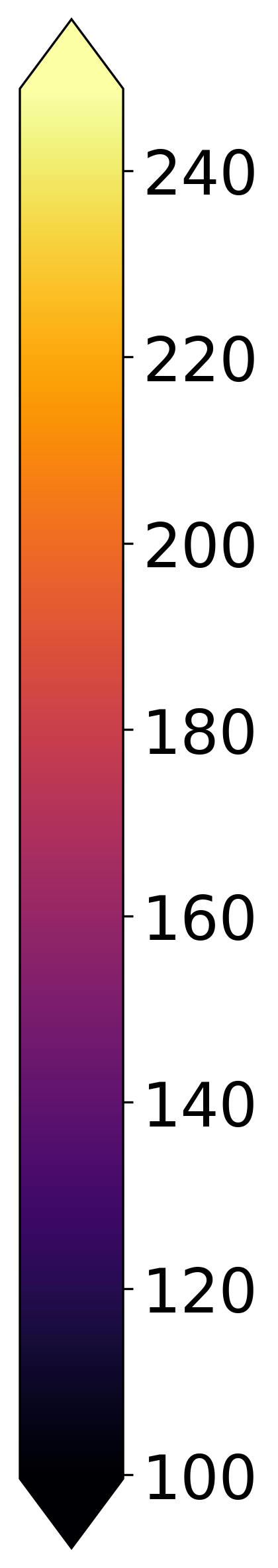}
        \vspace{4.5ex}
    \end{minipage}

    \caption{Total agent interaction contours based on discordance propensity ($p_{\text{dis}}$) and initial network mixing ($\mu_{\text{mix}}^{(0)}$). Panels show 
    \textbf{(a)} unimodal $U(-0.1, 0.1)$ and 
    \textbf{(b)} bimodal $U(-0.8, -0.6) \cup U(0.6, 0.8)$ initial opinions. 
    Brighter colors indicate more interactions. Unimodal initialization results in lower interaction counts overall, slightly increasing with lower $\mu_{\text{mix}}^{(0)}$ and largely independent of $p_{\text{dis}}$. Bimodal initialization shows a non-linear pattern: high interactions occur with low $\mu_{\text{mix}}^{(0)}$ (regardless of $p_{\text{dis}}$), or with moderate/high $\mu_{\text{mix}}^{(0)}$ when $p_{\text{dis}}$ is also high.}
    \label{fig:bimodal-unimodal-reaction-heatmaps-2}
\end{figure*} % Original Figure: fig:exp8_unimodal_metrics / fig:exp8_bimodal_metrics (Left Panels)

\begin{enumerate}
    \item \textbf{Initial Community Mixing ($\mu_{\text{mix}}^{(0)}$):} Varied across $[0.0, 0.1, ..., 1.0]$.
    \item \textbf{Propensity for Discordant Reaction ($p_{\text{dis}}^{\text{(react)}}$):} Varied over $[0.0, 0.1, ..., 1.0]$. Higher values increase the likelihood of reacting based on opinion oppositeness.
\end{enumerate}

All other parameters followed the baseline configuration (Table~\ref{tab:parameters-offline}). The primary metrics measured over $T=100$ steps were the mean total number of reactions per agent and the overall stance ratio, averaged over five runs.

\subsubsection{Results}

The initial opinion configuration significantly affects how the propensity for discordant engagement influences reaction patterns, as revealed by comparing the unimodal and bimodal scenarios in Figure~\ref{fig:engagement-antagonistic}.

Considering the total number of reactions (left panels), engagement levels in the unimodal (consensus) scenario are primarily determined by the initial mixing $\mu_{\text{mix}}^{(0)}$, with higher segregation leading to more reactions. The propensity for discordant interaction, $p_{\text{dis}}^{\text{(react)}}$, has only a marginal effect on the total reaction volume in this consensual environment. In sharp contrast, the bimodal (polarized) scenario shows much higher peak engagement and a strong dependence on both parameters. Reaction volume increases dramatically with $p_{\text{dis}}^{\text{(react)}}$, reaching its maximum when this propensity is high ($\ge 0.8$). Interestingly, under high $p_{\text{dis}}^{\text{(react)}}$, higher initial mixing $\mu_{\text{mix}}^{(0)}$ also tends to correlate with higher reaction counts. This suggests that in a polarized setting, a high tendency to engage with discordant views, especially when facilitated by structural mixing, drives significantly increased interaction, reflecting antagonistic engagement. Unlike the unimodal case, $p_{\text{dis}}^{\text{(react)}}$ emerges as a crucial driver of total engagement in polarized settings, particularly when initial mixing enables cross-group encounters.

Analyzing the stance ratio (Figure~\ref{fig:engagement-antagonistic}, bottom panels) again reveals significant differences contingent on the initial opinion landscape. In the unimodal case, the stance ratio is primarily governed by initial mixing $\mu_{\text{mix}}^{(0)}$, increasing as mixing increases, while $p_{\text{dis}}^{\text{(react)}}$ has very little impact. However, the bimodal case shows strong sensitivity to both parameters. While the ratio increases with increasing $\mu_{\text{mix}}^{(0)}$ (starting from perfect homophily at $\mu_{\text{mix}}^{(0)}=0.0$), it is also strongly influenced by $p_{\text{dis}}^{\text{(react)}}$. Increasing the propensity to react to discordant content significantly increases the stance ratio. Consequently, the highest ratios (indicating the least homophilic, or even heterophilic, reaction patterns) occur under high initial mixing ($\mu_{\text{mix}}^{(0)} \ge 0.8$) combined with a high propensity for discordant engagement ($p_{\text{dis}}^{\text{(react)}} \ge 0.8$).

This contrasts markedly with Experiment B.7, where varying the discovery rate $\delta_{\text{rec}}$ did not break the extreme reaction homophily observed in the bimodal case. Here, the agent's intrinsic tendency $p_{\text{dis}}^{\text{(react)}}$ actively promotes cross-stance reactions in the polarized setting when $\mu_{\text{mix}}^{(0)} > 0$. This suggests that reducing "reaction bubbles" in polarized environments may depend more on agents' psychological willingness to engage with disagreement rather than solely on algorithmically increasing exposure diversity.

In conclusion, the propensity to engage with discordant information ($p_{\text{dis}}^{\text{(react)}}$) interacts strongly with the opinion landscape and network structure. In consensual environments, its impact is minimal. In polarized environments, however, it becomes key: higher values dramatically increase total engagement (potentially antagonistic) and simultaneously decrease reaction homophily by fostering cross-stance interactions. This highlights the potential importance of individual psychological factors in mediating interaction patterns within polarized online spaces.

\subsection{Experiment 9: Reaction Dynamics with Evolving Opinions under Varying Mixing and Initial Diversity}
\label{app:expA4_reaction_evolving_opinions}

This final supplementary experiment examines reaction dynamics (e.g., liking) under conditions where agent opinions are allowed to evolve according to the opinion shift mechanism (\cref{app:opinion-shifts}). We investigate how the interplay between the initial network structure (community mixing, $\mu_{\text{mix}}^{(0)}$) and the initial diversity of opinions (range limit, $o_{\text{max}}^{(0)}$) influences the total volume of reactions and the degree of reaction homophily when opinions are dynamic. This setup allows observation of how reaction patterns emerge concurrently with the potential formation or dissolution of opinion clusters and network structures, offering a more integrated view compared to the fixed-opinion scenarios in Experiments B.7 and B.8.

\subsubsection{Experimental Setup}
The simulation tracked agent reactions while allowing both opinions and network connections to evolve dynamically. We systematically varied two initial condition parameters:

\begin{enumerate}
    \item \textbf{Initial Community Mixing ($\mu_{\text{mix}}^{(0)}$):} Varied across $[0.0, 0.1, ..., 1.0]$.
    \item \textbf{Initial Opinion Range Limit ($o_{\text{max}}^{(0)}$):} Varied over $[0.1, 0.2, ..., 1.0]$. Low values represent initial consensus, high values represent initial diversity.
\end{enumerate}

The propensity for discordant reaction was set to maximum ($p_{\text{dis}}^{\text{(react)}} = 1.0$) to enable antagonistic cross-cutting engagement. All other parameters followed the baseline configuration (Table~\ref{tab:parameters-offline}). The analysis focused on the mean total number of reactions per agent and the overall stance ratio (proportion of reactions to same-stance content) at the end of the simulation ($T=100$), averaged over five runs.

\subsubsection{Results}

Allowing opinions to evolve dynamically leads to intricate reaction patterns influenced by both initial network structure and initial opinion diversity, as shown in Figure~\ref{fig:engagement-coevolution}.

Analyzing the mean total number of reactions per agent (\cref{fig:engagement-coevolution}a), the results reveal a counterintuitive non-monotonic relationship with initial opinion diversity ($o_{\text{max}}^{(0)}$). Rather than engagement simply increasing with diversity, we observe three distinct regimes: moderate engagement levels under consensus conditions ($o_{\text{max}}^{(0)} \le 0.6$), a pronounced interaction valley at intermediate diversity ($o_{\text{max}}^{(0)} \approx 0.7-0.8$), and peak engagement under high diversity ($o_{\text{max}}^{(0)} \ge 0.9$) combined with moderate initial mixing ($\mu_{\text{mix}}^{(0)} \approx 0.4-0.6$). Understanding this pattern requires examining the relationship between engagement and final polarization outcomes (\cref{fig:engagement-coevolution}c). The final polarization depends primarily on initial opinion diversity: low diversity produces consensus (minimal polarization), intermediate diversity yields moderate polarization, and high diversity results in strong polarization. Initial network mixing shows minimal impact on polarization except under complete segregation ($\mu_{\text{mix}}^{(0)} = 0.0$). The engagement valley thus corresponds precisely to conditions producing moderate disagreement, suggesting that lukewarm conflict generates less interaction than either consensus-building exchanges or passionate antagonistic engagement. Compared to fixed-opinion scenarios (Experiments B.7, B.8), dynamic opinion evolution reveals that the relationship between polarization and engagement is fundamentally non-linear.

% Figure now integrated into main text as Figure engagement-coevolution with 3 panels including polarization

Examining the cross-stance interaction ratio (\cref{fig:engagement-coevolution}b), clearer and more systematic trends emerge. Consistent with previous experiments, complete initial segregation ($\mu_{\text{mix}}^{(0)} = 0.0$) results in perfect homophily (ratio = 0.0), regardless of initial opinion diversity. For networks with some initial mixing ($\mu_{\text{mix}}^{(0)} > 0$), both parameters systematically influence cross-stance engagement. Increasing initial mixing $\mu_{\text{mix}}^{(0)}$ consistently promotes cross-stance reactions, while increasing initial opinion diversity $o_{\text{max}}^{(0)}$ also generally leads to higher cross-stance ratios, especially when $\mu_{\text{mix}}^{(0)} > 0.1$. The highest cross-stance ratios (approx. 0.4-0.5) occur under conditions combining high initial mixing ($\mu_{\text{mix}}^{(0)} \ge 0.8$) with high initial diversity ($o_{\text{max}}^{(0)} \ge 0.8$). Notably, these high cross-stance interaction conditions correspond to scenarios that maintain strong polarization (\cref{fig:engagement-coevolution}c), demonstrating that cross-cutting engagement can persist alongside—or perhaps because of—sustained ideological disagreement when agents have maximum propensity for discordant interaction ($p_{\text{dis}}^{\text{(react)}} = 1.0$).

A key insight emerges from comparing these results to the fixed bimodal case in Experiment B.8, where extreme reaction homophily prevailed regardless of mixing or discovery rate. Here, allowing opinions to evolve dynamically from a diverse state within mixed networks leads to substantial cross-stance reaction, even under conditions that produce strong final polarization. This suggests that dynamic opinion adaptation can sustain cross-cutting engagement throughout the polarization process, preventing the complete communicative separation observed under fixed polarized opinions.

The most striking finding concerns the relationship between polarization level and engagement volume. Rather than high polarization universally suppressing interaction (as observed in fixed-opinion experiments), we find that moderate polarization produces the lowest engagement levels. This interaction valley suggests that neither consensus conditions nor highly polarized states, but lukewarm disagreement, may be least conducive to sustained engagement. Such moderate polarization appears to create a form of engagement paralysis—sufficient disagreement to prevent satisfying consensus but insufficient intensity to drive passionate exchange.

In conclusion, dynamic opinion evolution fundamentally alters the relationship between polarization and engagement. While total engagement shows complex non-monotonic dependencies on polarization levels, cross-stance interaction patterns demonstrate that structural integration can sustain significant cross-cutting engagement even under high polarization, provided agents maintain willingness to engage across ideological divides. These computational findings suggest that the timing and process of polarization formation may be as important as its final state for understanding engagement dynamics.

  % app:computational_experiments
\section{Detailed Message Analysis}
\label{app:message-analysis}

This appendix section details the methodologies and findings from our comprehensive message analyses. These analyses are twofold: first, we investigate how systemic factors (overall polarization level and algorithmic recommendation bias) influence the characteristics of message content generated within the simulation. Second, we explore how individual agent attributes (personality and interaction history) modulate the perception and interpretation of these messages.

\subsection{Content Analysis Under Polarization}
\label{subsec:offline-message-content-analysis}

This analysis investigates how the overall polarization level of the simulated environment and different algorithmic recommendation biases influence the characteristics of message content, specifically examining opinion extremity, group identity salience, emotionality, and uncertainty.

\subsubsection{Experimental Setup}

Using a $2\times3$ factorial design, we investigated how \emph{Polarization Level} and \emph{Recommendation Bias} influence content characteristics in social network discussions. We maintained similar network parameters as in the interaction analysis, but with a distinct approach to message recommendations.

For each condition (\emph{Polarized} and \emph{Unpolarized}), we implemented three \emph{Recommendation Bias} configurations: \emph{Pro-Biased} ($70$-$30$ ratio of pro to contra messages), \emph{Contra-Biased} ($30$-$70$ ratio), and \emph{Balanced} (equal proportions). Unlike the homophily-based recommendations in the polarization analysis, where recommendations were tailored to match each individual's stance, these biases were applied uniformly across all users regardless of their personal positions.

\begin{table*}[ht]
\centering
\small
\caption{Main Effects and Interactions Across Message Content Dimensions}
\label{tab:message-analysis-anova}
\begin{tabularx}{\textwidth}{>{\raggedright\arraybackslash}p{2.2cm}>{\raggedright\arraybackslash}p{1.8cm}*{3}{>{\centering\arraybackslash}X}>{\centering\arraybackslash}p{1.2cm}>{\centering\arraybackslash}p{1cm}}
\toprule
\multirow{2}{*}{\textbf{Metric}} & \multirow{2}{*}{\textbf{Condition}} & \multicolumn{3}{c}{\textbf{Position}} & \multicolumn{2}{c}{\textbf{ANOVA}} \\
\cmidrule(lr){3-5} \cmidrule(lr){6-7}
& & Contra & Balanced & Pro & $F$ & $p$ \\
\midrule
\multirow{2}{*}{Opinion} 
    & Polarized & \textbf{-0.43} (0.71) & -0.13 (0.81) & \textbf{0.52} (0.66) & \multirow{2}{*}{\textbf{901.86}$^a$} & \multirow{2}{*}{<.001***} \\
    & Unpolarized & -0.06 (0.21) & -0.12 (0.20) & -0.04 (0.18) & & \\
\midrule
\multirow{2}{*}{Group Identity} 
    & Polarized & \textbf{0.70} (0.10) & \textbf{0.74} (0.08) & \textbf{0.72} (0.09) & \multirow{2}{*}{\textbf{74.69}$^a$} & \multirow{2}{*}{<.001***} \\
    & Unpolarized & 0.15 (0.11) & 0.15 (0.11) & 0.14 (0.11) & & \\
\midrule
\multirow{2}{*}{Emotionality} 
    & Polarized & \textbf{0.71} (0.10) & \textbf{0.76} (0.08) & \textbf{0.76} (0.10) & \multirow{2}{*}{\textbf{156.85}$^a$} & \multirow{2}{*}{<.001***} \\
    & Unpolarized & 0.43 (0.09) & 0.42 (0.07) & 0.42 (0.07) & & \\
\midrule
\multirow{2}{*}{Uncertainty} 
    & Polarized & 0.19 (0.12) & 0.20 (0.08) & 0.21 (0.08) & \multirow{2}{*}{\textbf{17.58}$^a$} & \multirow{2}{*}{<.001***} \\
    & Unpolarized & \textbf{0.55} (0.07) & \textbf{0.56} (0.06) & \textbf{0.56} (0.07) & & \\
\bottomrule
\multicolumn{7}{p{.95\textwidth}}{\small \textbf{Note:} Values show means with standard deviations in parentheses. Bold values indicate significantly higher means between polarized and unpolarized conditions for each position. All F-statistics are significant at $p$ < .001.} \\
\multicolumn{7}{p{.95\textwidth}}{\small $^a$ F-statistic for Polarization × Position interaction (df = 2, 14095). ***$p$ < .001} \\
\end{tabularx}
\end{table*}

Our LLM-based content analysis measured four message dimensions: opinion (scale from $-1$ to $1$), group identity salience ($0$ to $1$), emotionality ($0$ to $1$), and uncertainty ($0$ to $1$). For each agent, the LLM evaluated all recommended messages across iterations, assessing the fundamental stance, group emphasis, emotional-rational balance, and expressed doubt in the discourse. The specific prompts and measurement scales used for this analysis are detailed in Table~\ref{tab:content-analysis-prompt}.

\begin{table*}[ht]
\centering
\caption{LLM Content Analysis Dimensions and Prompts}
\label{tab:content-analysis-prompt}
\begin{tabularx}{\textwidth}{>{\raggedright\arraybackslash}p{2.5cm}>{\raggedright\arraybackslash}p{1.5cm}>{\raggedright\arraybackslash}X}
\toprule
\textbf{Dimension} & \textbf{Scale} & \textbf{Analysis Prompt} \\
\midrule
Opinion Extremity & $-1$ to $1$ & Evaluate the fundamental stance expressed in the message regarding the topic. Consider: What position does the author take? How strongly is this position expressed? \\[0.5em]
Group Identity Salience & $0$ to $1$ & Assess the degree to which group membership and intergroup distinctions are emphasized. Consider: Does the message frame issues in terms of "us vs. them"? How prominent are references to group identities? \\[0.5em]
Emotionality & $0$ to $1$ & Measure the emotional versus rational tone of the message. Consider: Does the message appeal primarily to emotions or to logic? How charged is the language used? \\[0.5em]
Uncertainty & $0$ to $1$ & Evaluate expressions of doubt, qualification, or epistemic humility. Consider: Does the author acknowledge alternative viewpoints? Are claims presented as definitive or tentative? \\
\bottomrule
\multicolumn{3}{p{.95\textwidth}}{\small \textbf{Note:} The LLM evaluates each message along these dimensions using GPT-4. Content generation and analysis are conducted as separate computational steps with distinct prompts to maintain analytical independence. Full prompt specifications and implementation details are available in the supplementary materials.} \\
\end{tabularx}
\end{table*}

\subsubsection{Results}

Our analysis revealed significant effects of both \emph{Polarization Level} and \emph{Recommendation Bias} across all measured content dimensions (see Table~\ref{tab:message-analysis-anova} and Figure~\ref{fig:message-analysis-violin-plots}). The opinion analysis demonstrated strong main effects for \emph{Polarization Level} ($F(1, 14095) = 49.36$, $p < .001$) and \emph{Recommendation Homophily} ($F(2, 14095) = 1100.14$, $p < .001$), as well as a significant interaction between these factors ($F(2, 14095) = 901.86$, $p < .001$). In \emph{Polarized Conditions}, messages exhibited marginally negative average opinions ($M = -0.01$, $SD = 0.83$) compared to slightly more negative opinions in \emph{Unpolarized Conditions} ($M = -0.07$, $SD = 0.20$). The interaction manifested particularly strongly in opinion expression, where \emph{Polarized Conditions} showed marked differences between \emph{Pro-Biased} ($M = 0.52$, $SD = 0.66$), \emph{Contra-Biased} ($M = -0.43$, $SD = 0.71$), and \emph{Balanced} ($M = -0.13$, $SD = 0.81$) positions, while \emph{Unpolarized Conditions} exhibited substantially smaller variations between these positions (\emph{Pro-Biased}: $M = -0.04$, $SD = 0.18$; \emph{Contra-Biased}: $M = -0.06$, $SD = 0.21$; \emph{Balanced}: $M = -0.12$, $SD = 0.20$).

\emph{Group Identity Salience} showed a particularly pronounced main effect of \emph{Polarization Level} ($F(1, 14066) = 122007.50$, $p < .001$), with \emph{Polarized Conditions} eliciting substantially higher group identity expression ($M = 0.72$, $SD = 0.09$) compared to \emph{Unpolarized Conditions} ($M = 0.15$, $SD = 0.11$). While position effects were statistically significant ($F(2, 14066) = 52.47$, $p < .001$), the practical differences between positions were minimal, suggesting that polarization, rather than recommendation patterns, primarily drives group identity expression.

The \emph{Emotionality} analysis revealed strong effects of \emph{Polarization Level} ($F(1, 14066) = 50758.88$, $p < .001$), with messages in \emph{Polarized Conditions} showing markedly higher emotional content ($M = 0.75$, $SD = 0.10$) compared to \emph{Unpolarized Conditions} ($M = 0.42$, $SD = 0.07$). Though \emph{Recommendation Homophily} effects were significant ($F(2, 14066) = 104.63$, $p < .001$), the differences were relatively small in practical terms.

\emph{Uncertainty} levels displayed an inverse relationship with \emph{Polarization Level} ($F(1, 14066) = 66628.45$, $p < .001$), with \emph{Unpolarized Conditions} generating substantially higher uncertainty expression ($M = 0.56$, $SD = 0.06$) compared to \emph{Polarized Conditions} ($M = 0.20$, $SD = 0.10$). \emph{Recommendation Homophily} effects, while statistically significant ($F(2, 14066) = 41.64$, $p < .001$), showed only minor differences.

These findings collectively suggest that \emph{Polarization Level} plays a crucial role in shaping message content across all measured dimensions, with \emph{Polarized Conditions} generally amplifying opinion differences, increasing group identity salience and emotionality, while reducing uncertainty. \emph{Recommendation Homophily} effects, while significant, showed varying practical importance across different content dimensions, with the strongest impact observed in opinion expression.

\subsection{Evaluation of Message Interpretation considering Agent Personality}
\label{subsec:offline-message-interpretation-analysis}

This experiment explores how inherent agent personality traits (Skeptic, Neutral, Open-minded) modulate the perception of messages in terms of their opinion, emotionality, and group identity, depending on the message's authorial stance and discussion topic.

\subsubsection{Experimental Setup}

To evaluate how agent characteristics and message context influence message interpretation within the simulation, we conducted a series of 3 (Author Opinion: -0.7, 0.0, 0.7) $\times$ 3 (Agent Personality: Skeptic, Neutral, Open-minded) $\times$ 2 (Topic: Mandatory Vaccination, Universal Basic Income) between-subjects ANOVAs. The dependent variables were the receiving agent's perceived opinion, perceived emotionality, and perceived group identity of the message. Detailed results are presented in Table~\ref{tab:anova-results}.

\subsubsection{Results}

The analysis of perceived opinion revealed several significant effects. As expected, there was a very strong main effect of the message author's actual opinion ($F(2, 882) = 7971.75, p < .001$). Post-hoc Tukey HSD tests confirmed that messages from authors with negative opinions ($M = -0.56$) were perceived significantly more negatively than those from neutral authors ($M = 0.02$), which in turn were perceived significantly more negatively than messages from positive authors ($M = 0.53$, all $p < .001$). 
A significant main effect was also found for the receiving agent's personality ($F(2, 882) = 3.88, p = .021$), although mean differences were small (Skeptic: $M=-0.01$, Neutral: $M=0.01$, Open-minded: $M=0.01$) and post-hoc Tukey HSD tests did not reveal significant differences between any specific personality pairs ($p > .05$). The main effect of topic was not significant ($F(1, 882) = 1.75, p = .187$).

\begin{table*}[ht]
    \centering
    \small % Makes the font slightly smaller
    \caption{ANOVA Results for Message Perception Variables}
    \label{tab:anova-results}
    \sisetup{ table-format = 4.2, % Format for F-values (adjust numbers as needed)
              detect-weight,    % Use bold font if specified
              mode=text         % Ensure bold font works correctly in text mode parts
            } 
    \begin{tabular}{@{} l l c S[table-format=4.2] l @{}} 
    \toprule
    \textbf{Dependent Variable} & \textbf{Source of Variation} & \textbf{df} & {\textbf{F}} & \textbf{p} \\ 
    \midrule
    \multirow{7}{*}{\parbox{2.5cm}{\raggedright Perceived Opinion}} % Use parbox for line breaking if needed
        & Author Opinion & 2, 882 & \bfseries 7971.75 & {$<$ .001***} \\ % Use {$< .001^{***}$}
        & Agent Personality & 2, 882 & 3.88 & .021* \\
        & Topic & 1, 882 & 1.75 & .187 \\
        & Author Opinion $\times$ Agent Personality & 4, 882 & \bfseries 6.36 & {$<$ .001***} \\ % Use {$< .001^{***}$}
        & Author Opinion $\times$ Topic & 2, 882 & 2.41 & .090 \\
        & Agent Personality $\times$ Topic & 2, 882 & \bfseries 5.61 & .004** \\
        & Author Opinion $\times$ Agent Personality $\times$ Topic & 4, 882 & \bfseries 6.48 & {$<$ .001***} \\ % Use {$< .001^{***}$}
    \midrule
    \multirow{7}{*}{\parbox{2.5cm}{\raggedright Perceived Emotionality}} 
        & Author Opinion & 2, 882 & \bfseries 399.74 & {$<$ .001***} \\ % Use {$< .001^{***}$}
        & Agent Personality & 2, 882 & 1.37 & .255 \\
        & Topic & 1, 882 & \bfseries 134.05 & {$<$ .001***} \\ % Use {$< .001^{***}$}
        & Author Opinion $\times$ Agent Personality & 4, 882 & \bfseries 5.25 & {$<$ .001***} \\ % Use {$< .001^{***}$}
        & Author Opinion $\times$ Topic & 2, 882 & \bfseries 142.10 & {$<$ .001***} \\ % Use {$< .001^{***}$}
        & Agent Personality $\times$ Topic & 2, 882 & 0.72 & .487 \\
        & Author Opinion $\times$ Agent Personality $\times$ Topic & 4, 882 & 1.87 & .114 \\
    \midrule
    \multirow{7}{*}{\parbox{2.5cm}{\raggedright Perceived Group Identity}} 
        & Author Opinion & 2, 882 & \bfseries 199.43 & {$<$ .001***} \\ % Use {$< .001^{***}$}
        & Agent Personality & 2, 882 & 2.33 & .098 \\
        & Topic & 1, 882 & \bfseries 53.62 & {$<$ .001***} \\ % Use {$< .001^{***}$}
        & Author Opinion $\times$ Agent Personality & 4, 882 & \bfseries 6.58 & {$<$ .001***} \\ % Use {$< .001^{***}$}
        & Author Opinion $\times$ Topic & 2, 882 & \bfseries 102.09 & {$<$ .001***} \\ % Use {$< .001^{***}$}
        & Agent Personality $\times$ Topic & 2, 882 & \bfseries 3.39 & .034* \\
        & Author Opinion $\times$ Agent Personality $\times$ Topic & 4, 882 & \bfseries 3.50 & .008** \\
    \bottomrule
    \multicolumn{5}{@{}p{\dimexpr\linewidth-2\tabcolsep}@{}}{\footnotesize \textbf{Note:} Table displays results from three-way ANOVAs with Author Opinion (-0.7, 0.0, 0.7), Agent Personality (Skeptic, Neutral, Open-minded), and Topic (Mandatory Vaccination, Family Voting Rights) as independent variables. Degrees of freedom (df) are shown for the effect and the residual term (df = effect df, residual df). F-values and p-values are reported. Bold F-values indicate significance at p < .05. Significance levels: *p* < .05, **p* < .01, ***p* < .001.} \\
    \end{tabular}
    \end{table*}

Critically, these main effects were qualified by significant interactions. There was a significant interaction between author opinion and agent personality ($F(4, 882) = 6.36, p < .001$), and between agent personality and topic ($F(2, 882) = 5.61, p = .004$). Most importantly, a significant three-way interaction between author opinion, agent personality, and topic emerged ($F(4, 882) = 6.48, p < .001$). This indicates that the way an agent's personality influences their perception of a message's opinion depends on both the original author's stance and the specific topic being discussed. For instance, the difference in perceived opinion between messages discussing mandatory vaccination versus Universal Basic Income might be interpreted differently by skeptic versus open-minded agents, particularly when considering messages from authors with strong opinions.

For perceived emotionality, we observed significant main effects for author opinion ($F(2, 882) = 399.74, p < .001$) and topic ($F(1, 882) = 134.05, p < .001$). Post-hoc tests showed that messages from authors with negative opinions ($M = 0.45$) were perceived as most emotional, followed by positive opinions ($M = 0.36$), and lastly neutral opinions ($M = 0.23$, all pairwise $p < .001$). Messages concerning mandatory vaccination ($M = 0.38$) were perceived as significantly more emotional than those about UBI ($M = 0.31$, $p < .001$). The main effect of agent personality was not significant ($F(2, 882) = 1.37, p = .255$).

Significant two-way interactions were found between author opinion and agent personality ($F(4, 882) = 5.25, p < .001$) and between author opinion and topic ($F(2, 882) = 142.10, p < .001$). This suggests that personality type moderates how author opinion maps onto perceived emotionality, and that the relationship between author opinion and perceived emotionality differs depending on the topic. The three-way interaction was not significant ($F(4, 882) = 1.87, p = .114$).

Finally, the analysis of perceived group identity (ingroup vs. outgroup salience) showed significant main effects for author opinion ($F(2, 882) = 199.43, p < .001$) and topic ($F(1, 882) = 53.62, p < .001$). Post-hoc tests for author opinion revealed a non-linear pattern: messages from authors with both negative ($M = 0.36$) and positive ($M = 0.35$) opinions were perceived as having significantly higher group identity salience than messages from neutral authors ($M = 0.19$, $p < .001$). However, there was no significant difference in perceived group identity between negative and positive author opinions ($p = .465$). Messages about mandatory vaccination ($M = 0.33$) were perceived as having higher group identity salience than those about UBI ($M = 0.27$, $p < .001$). The main effect of agent personality was not significant ($F(2, 882) = 2.33, p = .098$).

Similar to perceived opinion, we found significant interactions involving all three factors. Significant two-way interactions emerged between author opinion and agent personality ($F(4, 882) = 6.58, p < .001$), author opinion and topic ($F(2, 882) = 102.09, p < .001$), and agent personality and topic ($F(2, 882) = 3.39, p = .034$). Furthermore, a significant three-way interaction was present ($F(4, 882) = 3.50, p = .008$). This complex interaction highlights that the perception of group identity cues in messages is shaped by a combination of the author's stance, the receiver's personality, and the discussion topic. For example, a skeptical agent might perceive high group salience in a negative message about mandatory vaccination, while an open-minded agent might perceive less group salience in the same message or topic combination.

\begin{figure*}[h!]
    \centering
    \includegraphics[width=0.9\textwidth]{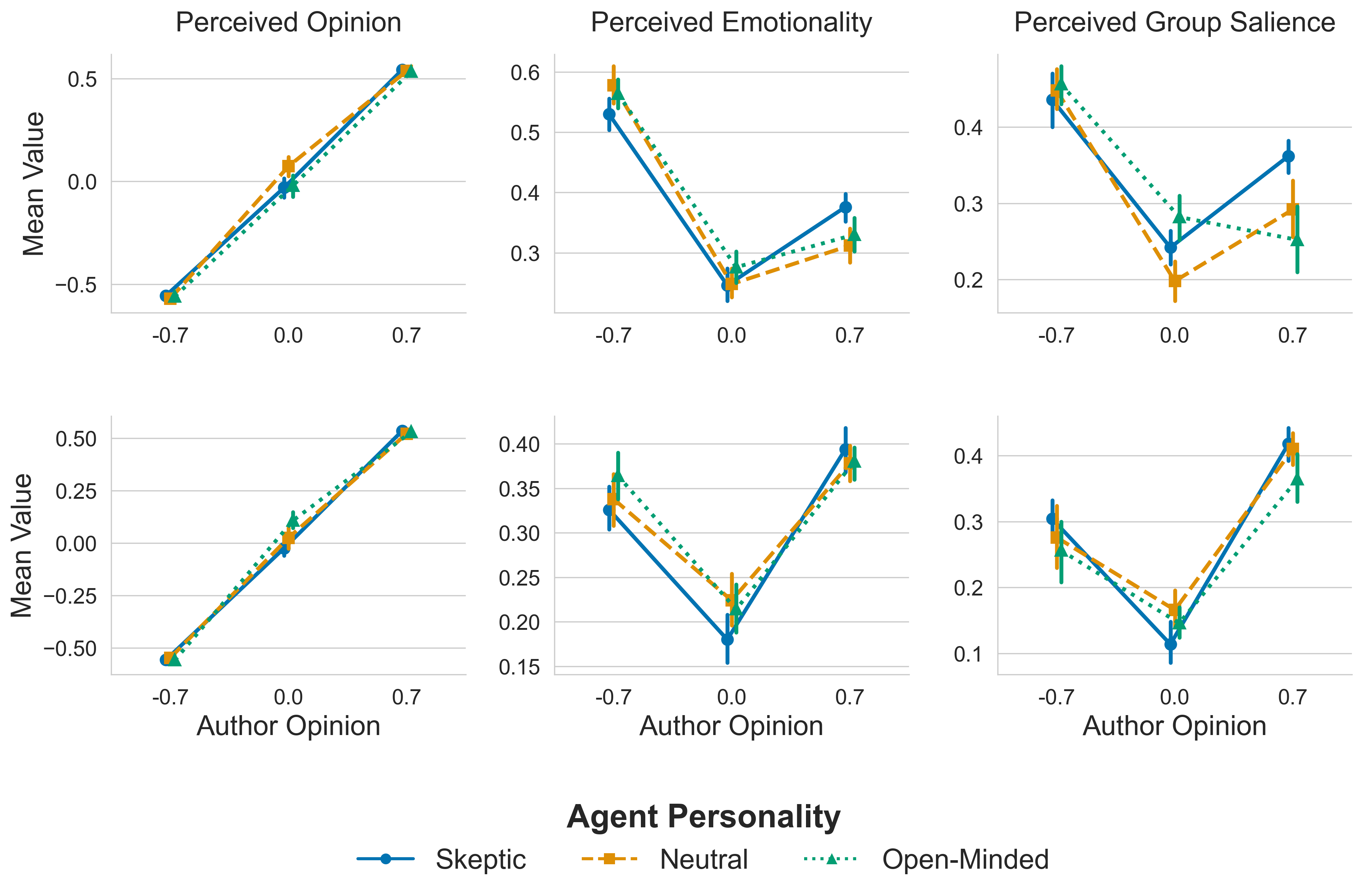}
    \caption{Mean perceived message characteristics (columns: opinion, emotionality, group salience) vs. actual author opinion (x-axis) for different agent personalities (lines) and topics (rows: Mandatory Vaccination, Universal Basic Income). Error bars show SEM. Perceived opinion accurately tracks author opinion across both topics. Perceived emotionality and group salience show V-shaped patterns (minimal for neutral authors, maximal for extremists), largely consistent across personality types and topics, indicating neutral messages are perceived as less emotional and group-focused. Notable personality differences emerge in specific topic-opinion combinations, demonstrating the contingent nature of message interpretation.}
    \label{fig:message-analysis-interactions-personality-appendix}
\end{figure*}

Overall, the analyses demonstrate that message interpretation in the simulation is a complex process influenced by multiple factors. The author's opinion consistently and strongly shaped perceptions across all dimensions. The discussion topic significantly influenced perceived emotionality and group identity. While the main effect of agent personality was only significant (though weakly) for perceived opinion, personality played a crucial role in moderating the effects of author opinion and topic, as evidenced by the significant two-way and particularly the three-way interactions for perceived opinion and perceived group identity. These complex interactions suggest the simulation successfully captures nuanced interpretation dynamics where individual differences (personality) interact with message content (author opinion) and context (topic). The complete interaction patterns across both topics are visualized in \cref{fig:message-analysis-interactions-personality-appendix}.

\subsection{Evaluation of Message Interpretation considering Agent Memory}
\label{sec:results-memory}

Parallel to the personality analysis, this experiment investigates how an agent's accumulated interaction history (memory: Mostly Negative, Mostly Positive, No Memory) influences their interpretation of messages across the same dimensions (perceived opinion, emotionality, group identity) and contextual factors (author opinion, topic).

\begin{table*}[ht]
    \centering
    \small % Makes the font slightly smaller
    \caption{ANOVA Results for Message Perception Variables with Agent Memory}
    \label{tab:anova-results-memory}
    \sisetup{ table-format = 4.2, % Format for F-values (adjust numbers as needed)
              detect-weight,    % Use bold font if specified
              mode=text         % Ensure bold font works correctly in text mode parts
            } 
    \begin{tabular}{@{} l l c S[table-format=4.2] l @{}} 
    \toprule
    \textbf{Dependent Variable} & \textbf{Source of Variation} & \textbf{df} & {\textbf{F}} & \textbf{p} \\ 
    \midrule
    \multirow{7}{*}{\parbox{2.5cm}{\raggedright Perceived Opinion}} % Use parbox for line breaking if needed
        & Author Opinion & 2, 882 & \bfseries 5891.48 & {$<$ .001***} \\ 
        & Agent Memory & 2, 882 & \bfseries 27.71 & {$<$ .001***} \\
        & Topic & 1, 882 & 1.23 & .268 \\
        & Author Opinion $\times$ Agent Memory & 4, 882 & \bfseries 13.87 & {$<$ .001***} \\ 
        & Author Opinion $\times$ Topic & 2, 882 & \bfseries 7.52 & {$<$ .001***} \\
        & Agent Memory $\times$ Topic & 2, 882 & \bfseries 3.75 & .024* \\
        & Author Opinion $\times$ Agent Memory $\times$ Topic & 4, 882 & \bfseries 3.48 & .008** \\
    \midrule
    \multirow{7}{*}{\parbox{2.5cm}{\raggedright Perceived Emotionality}} 
        & Author Opinion & 2, 882 & \bfseries 304.43 & {$<$ .001***} \\ 
        & Agent Memory & 2, 882 & \bfseries 4.33 & .013* \\
        & Topic & 1, 882 & \bfseries 101.21 & {$<$ .001***} \\ 
        & Author Opinion $\times$ Agent Memory & 4, 882 & \bfseries 3.21 & .012* \\ 
        & Author Opinion $\times$ Topic & 2, 882 & \bfseries 149.47 & {$<$ .001***} \\
        & Agent Memory $\times$ Topic & 2, 882 & \bfseries 3.66 & .026* \\
        & Author Opinion $\times$ Agent Memory $\times$ Topic & 4, 882 & 1.61 & .170 \\
    \midrule
    \multirow{7}{*}{\parbox{2.5cm}{\raggedright Perceived Group Identity}} 
        & Author Opinion & 2, 882 & \bfseries 179.08 & {$<$ .001***} \\
        & Agent Memory & 2, 882 & \bfseries 6.62 & .001** \\
        & Topic & 1, 882 & \bfseries 68.13 & {$<$ .001***} \\ 
        & Author Opinion $\times$ Agent Memory & 4, 882 & \bfseries 9.26 & {$<$ .001***} \\ 
        & Author Opinion $\times$ Topic & 2, 882 & \bfseries 103.50 & {$<$ .001***} \\
        & Agent Memory $\times$ Topic & 2, 882 & \bfseries 11.74 & {$<$ .001***} \\
        & Author Opinion $\times$ Agent Memory $\times$ Topic & 4, 882 & \bfseries 9.28 & {$<$ .001***} \\
    \bottomrule
    \multicolumn{5}{@{}p{\dimexpr\linewidth-2\tabcolsep}@{}}{\footnotesize \textbf{Note:} Table displays results from three-way ANOVAs with Author Opinion (-0.7, 0.0, 0.7), Agent Memory (Mostly Negative, Mostly Positive, No Memory), and Topic (Mandatory Vaccination, Family Voting Rights) as independent variables. Degrees of freedom (df) are shown for the effect and the residual term (df = effect df, residual df). F-values and p-values are reported. Bold F-values indicate significance at p < .05. Significance levels: *p* < .05, **p* < .01, ***p* < .001.} \\
    \end{tabular}
\end{table*}

\subsubsection{Experimental Setup}

Following the initial analysis with agent personality, we conducted a parallel set of 3 (Author Opinion: -0.7, 0.0, 0.7) $\times$ 3 (Agent Memory: Mostly Negative, Mostly Positive, No Memory) $\times$ 2 (Topic: Mandatory Vaccination, Universal Basic Income) between-subjects ANOVAs to examine the influence of prior interaction history (memory) on message interpretation. The dependent variables remained perceived opinion, emotionality, and group identity. Detailed results are presented in Table~\ref{tab:anova-results-memory}. Main effects of Author Opinion and Topic largely replicated the patterns observed in the personality analysis and are not reiterated in detail here.

\subsubsection{Results}

Consistent with the previous analysis, Author Opinion was the strongest predictor ($F(2, 882) = 5891.48, p < .001$), and the main effect of Topic remained non-significant ($F(1, 882) = 1.23, p = .268$). A significant main effect was found for Agent Memory ($F(2, 882) = 27.71, p < .001$). Agents with mostly negative prior interactions perceived messages slightly more negatively ($M = -0.05$) compared to those with mostly positive ($M = 0.02$) or no prior interactions ($M = 0.01$). However, post-hoc Tukey HSD tests indicated these pairwise differences were not statistically significant ($p > .05$), suggesting the overall effect, while significant, is diffuse across the memory conditions in this specific measure. 

Significant interactions involving memory were observed. Interactions between Author Opinion $\times$ Agent Memory ($F(4, 882) = 13.87, p < .001$), Agent Memory $\times$ Topic ($F(2, 882) = 3.75, p = .024$), and the three-way Author Opinion $\times$ Agent Memory $\times$ Topic interaction ($F(4, 882) = 3.48, p = .008$) were all significant. This indicates that an agent's interaction history significantly moderates how author opinion and topic influence the perception of a message's stance, mirroring the complex role previously observed for personality. For the first time in these analyses, we also observed a significant Author Opinion $\times$ Topic interaction ($F(2, 882) = 7.52, p < .001$), suggesting the mapping between author opinion and perceived opinion differed slightly across the two topics, independent of agent characteristics in this model.

The main effects of Author Opinion ($F(2, 882) = 304.43, p < .001$) and Topic ($F(1, 882) = 101.21, p < .001$) remained highly significant, following the same pattern as before (negative/positive opinions and mandatory vaccination topic perceived as more emotional). Agent Memory also showed a significant main effect ($F(2, 882) = 4.33, p = .013$), with agents having mostly negative memory perceiving messages as slightly more emotional ($M=0.38$) than those with positive ($M=0.36$) or no memory ($M=0.36$). Again, post-hoc Tukey HSD tests did not find significant differences between specific pairs ($p > .05$).

Significant interactions were found for Author Opinion $\times$ Agent Memory ($F(4, 882) = 3.21, p = .012$) and Agent Memory $\times$ Topic ($F(2, 882) = 3.66, p = .026$), as well as the previously observed Author Opinion $\times$ Topic interaction ($F(2, 882) = 149.47, p < .001$). Unlike the personality analysis for emotionality, the interactions involving agent memory suggest that prior interaction history modulates how author opinion translates to perceived emotionality and how topics influence emotional perception. The three-way interaction was not significant ($F(4, 882) = 1.61, p = .170$).

Similar patterns for Author Opinion ($F(2, 882) = 179.08, p < .001$) and Topic ($F(1, 882) = 68.13, p < .001$) main effects were observed, with extreme opinions and the mandatory vaccination topic eliciting higher perceived group identity. Agent Memory demonstrated a significant main effect ($F(2, 882) = 6.62, p = .001$). Post-hoc Tukey HSD tests revealed that agents with mostly negative prior interactions perceived significantly higher group identity ($M = 0.32$) compared to those with mostly positive interactions ($M = 0.28$, $p = .019$). Agents with no memory fell in between ($M = 0.30$) and did not differ significantly from the other two groups ($p > .05$).

Critically, all interactions were significant: Author Opinion $\times$ Agent Memory ($F(4, 882) = 9.26, p < .001$), Author Opinion $\times$ Topic ($F(2, 882) = 103.50, p < .001$), Agent Memory $\times$ Topic ($F(2, 882) = 11.74, p < .001$), and the three-way Author Opinion $\times$ Agent Memory $\times$ Topic interaction ($F(4, 882) = 9.28, p < .001$). This complex pattern strongly suggests that the perception of group cues is highly contingent on the interplay between the message's stance, the agent's specific interaction history related to the topic, and the topic itself.

\begin{figure*}[h!]
    \centering
    \includegraphics[width=0.9\textwidth]{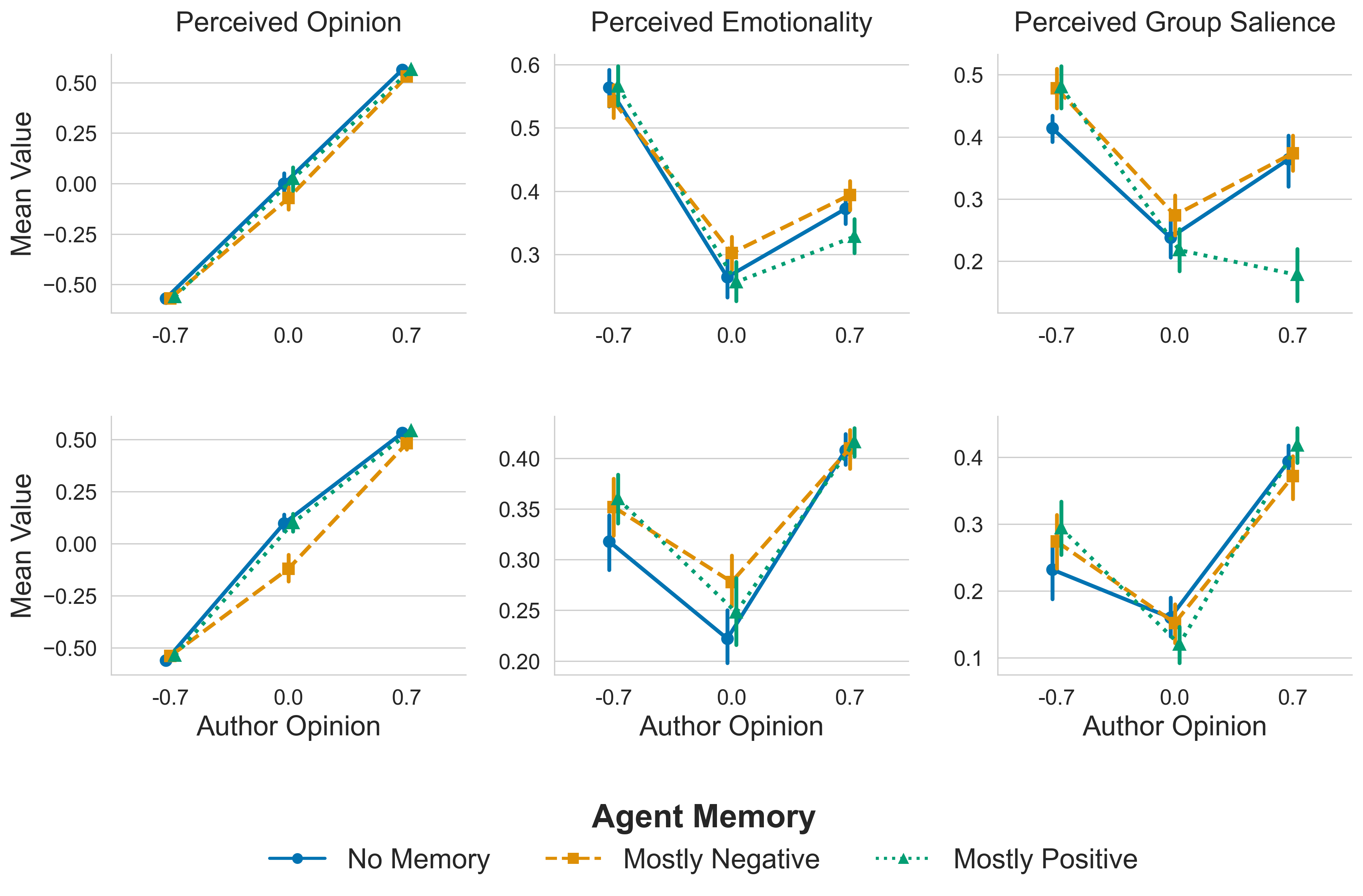}
    \caption{Mean perceived message characteristics (columns: opinion, emotionality, group salience) vs. actual author opinion (x-axis) for different agent memory states (lines: No Memory, Mostly Negative, Mostly Positive) and topics (rows: Mandatory Vaccination, Universal Basic Income). Perceived opinion consistently tracks author opinion across both topics. Perceived emotionality and group salience generally show V-shaped patterns (minimal for neutral authors). Agents with mostly positive memories perceive extreme messages (especially positive ones) as less emotional and group-salient than other agents, with topic-specific variations demonstrating how interaction history creates interpretive frames that persist across different discussion contexts.}
    \label{fig:message-analysis-interactions-memory-appendix}
\end{figure*}

Replacing agent personality with agent memory maintained the strong influence of author opinion and topic on message perception. Agent memory emerged as a significant factor influencing perceived opinion, emotionality, and group identity, although post-hoc tests often showed nuanced rather than large pairwise differences for its main effects. Crucially, agent memory consistently participated in significant two-way and three-way interactions (except for emotionality's three-way interaction). This highlights that an agent's accumulated interaction history, much like its inherent personality, plays a vital, moderating role in how messages are interpreted within the simulation, interacting dynamically with message content and context. The presence of significant interactions involving memory across all dependent variables underscores its importance in shaping the simulation's communication dynamics. The complete interaction patterns across both topics are visualized in \cref{fig:message-analysis-interactions-memory-appendix}.

    % app:message-analysis

\section{Detailed User Study Methodology and Findings}
\label{app:user-study} % Label for referencing

To supplement the discussion in \cref{sec:validation}, this appendix serves as a detailed account of the methodology and results from the user study, which was conducted to assess the framework's empirical utility. The study had human participants interact with a simulated social media environment where conditions of polarization and recommendation bias were experimentally manipulated. The following sections cover the complete details of the experimental design, the implementation of the prototype platform and its recommendation algorithm, the agent pre-simulation, participant recruitment and characteristics, the validation of measurement scales, the full experimental procedure, and in-depth statistical analyses, all of which provide the necessary background for the key results highlighted in the main text.

\subsection{Experimental Design}

A $2 \times 3$ between-subjects factorial design was utilized to investigate the dynamics of opinion formation and the perception of polarization within online discussions. In this design, two key dimensions were manipulated: the \emph{Polarization Degree} within the artificial agent population and a systematic \emph{Recommendation Bias}. Universal Basic Income (UBI) was held as the constant discussion topic throughout the experiment.

The first manipulated dimension placed highly polarized discussions in contrast with moderate ones by altering the behavior of the artificial agents. For the polarized condition, agents were set to express extreme viewpoints and use confrontational discourse, characterized by emotional language, forceful assertions, and minimal acknowledgment of opposing perspectives. In contrast, the moderate condition was defined by more nuanced discussions and cooperative interaction styles, with agents programmed to express uncertainty, recognize limitations in their knowledge, and engage with opposing views constructively.

The second dimension involved the introduction of a systematic bias into the platform's recommendation system. This was implemented across three distinct levels: a 'neutral' setting (featuring 50\% pro-UBI and 50\% contra-UBI content), a 'pro-bias' setting (70\% pro-UBI, 30\% contra-UBI), and a 'contra-bias' setting (30\% pro-UBI, 70\% contra-UBI). This manipulation was designed to allow for an investigation into how algorithmic content curation can influence both the formation of opinions and the perception of a debate's polarization.

The decision to select UBI as the focal topic was guided by several strategic factors. It represents an emerging policy proposal where public opinion is still relatively malleable, unlike heavily polarized topics where individuals' positions are often entrenched, making it an ideal context for studying opinion formation. Although it may evoke fewer preconceived notions, UBI is a sufficiently concrete and consequential subject to fuel meaningful discourse, with a complexity that touches upon economic, social, and technological domains. Support for its suitability is found in recent polling data, which indicates a balanced distribution of opinions among Europeans (approx. 51.2\% support, \cite{vlandas_politics_2019}) and Americans (approx. 48\% support, \cite{hamilton_people_2022}). Furthermore, because of UBI's limited real-world implementation, participants' opinions are more likely to be formed based on the arguments presented in the experiment rather than on direct experience or pre-existing partisan loyalties. This allows for a clearer examination of how social media dynamics can shape opinion before entrenched polarization develops.

\subsection{System Prototype}

The implementation of the prototype is a web application that simulates a social media platform, stylistically reminiscent of X (formerly Twitter), for the purpose of studying social polarization dynamics. The user interface, shown in Figure~\ref{fig:prototype-screenshot}, was designed with a familiar layout to promote intuitive user engagement and interaction.

\subsubsection{User Interface}
The application's main interface is structured into three primary components: a navigation sidebar on the left, a central Newsfeed, and a recommendation panel on the right. Essential functionalities such as the user's profile, a general user overview, and a logout option are accessible via the navigation sidebar. The primary space for interaction is the central Newsfeed, where users can view and engage with content posted by others. At the top of this feed, a text input area prompts users to share their own thoughts, mimicking the spontaneous nature of social media communication.

Displayed within the Newsfeed is a series of posts, each one accompanied by standard social media elements like user avatars, usernames, timestamps, and interaction metrics (\emph{likes}, \emph{comments}, and \emph{reposts}). The purpose of this design is to encourage user engagement by providing visual cues related to the popularity and impact of each post. On the right side of the interface, the recommendation panel suggests other users to follow, which can potentially influence the expansion of a user's network and their exposure to diverse viewpoints.

Dynamically generated user profiles display each user's posts, their follower relationships, and other relevant metadata such as their handle and biography (for details on the agent model, see \cref{app:adaptive-agents} and \cref{app:network-evolution}). The system also permits users to follow and unfollow the artificial agents.

\subsection{Newsfeed Recommendations}

An adaptive recommendation system for content presentation that evolves with user activity is implemented in the web application. This system makes use of two distinct algorithmic approaches: a 'default' variant for new users and a 'collaborative' variant that is activated once a user has established an interaction history.

The default variant utilizes a popularity-based scoring mechanism that takes into account multiple forms of engagement to establish content visibility. For any given message $m$, a composite popularity score is calculated by the system using the following formula:
\begin{align}
    S_p(m) = l_m + 2c_m + 2r_m
\end{align}
where $l_m, c_m, \text{ and } r_m$ correspond to the number of \emph{likes}, \emph{comments}, and \emph{reposts} for that message, respectively. The coefficients are weighted to reflect the higher importance assigned to more active forms of engagement.

After a user begins to interact with the platform, the system shifts to a collaborative variant that incorporates not only popularity metrics but also ideological proximity and a stochastic element to ensure recommendation diversity. This enhanced scoring function combines these elements into a composite score:
\begin{align}
    S_c(m) = \omega_p \cdot \frac{S_p(m)}{S_{max}} + \omega_i \cdot \frac{2 - |o_u - o_a|}{2} + \omega_r \cdot \epsilon
\end{align}
In this equation, $S_p(m)$ is the popularity score normalized by the maximum observed score $S_{max}$; $o_u$ and $o_a$ are the opinion scores of the active user and the message author; $\epsilon$ is a uniform random variable in $[0,1]$; and $\omega_p, \omega_i, \text{ and } \omega_r$ are weighting parameters that sum to one. For the current implementation, these weights are configured to $\omega_p = 0.6, \omega_i = 0.2, \text{ and } \omega_r = 0.2$, which balances the influence of popularity, ideological similarity, and randomization.

To maintain temporal relevance, both variants present a user's most recent contributions at the top of their feed upon page load. This design ensures users remain aware of their own contributions while still experiencing the broader content landscape delivered by the scoring-based recommendation system.

This dual-variant system design allows the platform to provide meaningful content recommendations even without initial user interaction data, while also enabling a smooth transition to more personalized recommendations as users become active. By incorporating popularity metrics, mild ideological factors, and controlled randomization, the system generates diverse recommendations that create the impression of a dynamic network environment, all while operating within the conditional \emph{Recommendation Bias} regime. The inclusion of a stochastic element is particularly helpful in preventing recommendation stagnation and ensuring dynamic content delivery.

\begin{table*}[htbp]
\centering
\small
\caption{Factor loadings and scale reliability for key measures. Note: [R] indicates reverse-coded items. Factor loadings are displayed for all items retained after cleaning (loading threshold $|.40|$).}
\begin{tabularx}{\textwidth}{>{\raggedright\arraybackslash}X>{\centering\arraybackslash}p{2cm}}
\toprule
\multicolumn{2}{l}{\textbf{Universal Basic Income (Pre)} ($\alpha = .893$)} \\
\midrule
A universal basic income would benefit society as a whole & .828 \\
Providing everyone with a basic income would do more harm than good [R] & .728 \\
Universal basic income is a fair way to ensure everyone's basic needs are met & .701 \\
Giving everyone a fixed monthly payment would reduce people's motivation to work [R] & .671 \\
Universal basic income would lead to a more stable and secure society & .867 \\
People should earn their income through work rather than receiving it unconditionally from the government [R] & .649 \\
A universal basic income would give people more freedom to make choices about their lives & .724 \\
\midrule
\multicolumn{2}{l}{\textbf{Universal Basic Income (Post)} ($\alpha = .890$)} \\
\midrule
A universal basic income would benefit society as a whole & .847 \\
Providing everyone with a basic income would do more harm than good [R] & .769 \\
Universal basic income is a fair way to ensure everyone's basic needs are met & .782 \\
Giving everyone a fixed monthly payment would reduce people's motivation to work [R] & .610 \\
Universal basic income would lead to a more stable and secure society & .921 \\
People should earn their income through work rather than receiving it unconditionally from the government [R] & .522 \\
A universal basic income would give people more freedom to make choices about their lives & .698 \\
\midrule
\multicolumn{2}{l}{\textbf{Perceived Polarization} ($\alpha = .776$)} \\
\midrule
The discussions on the platform were highly polarized & .674 \\
Users on the platform expressed extreme views & .755 \\
Users appeared to be firmly entrenched in their positions & .644 \\
There were frequent hostile interactions between users with differing views & .644 \\
\midrule
\multicolumn{2}{l}{\textbf{Perceived Emotionality} ($\alpha = .842$)} \\
\midrule
The discussions were highly charged with emotional content & .864 \\
Users frequently expressed strong feelings in their messages & .691 \\
The debate maintained a predominantly calm and neutral tone [R] & .692 \\
Participants typically communicated in an unemotional manner [R] & .779 \\
\midrule
\multicolumn{2}{l}{\textbf{Perceived Group Salience} ($\alpha = .639$)} \\
\midrule
Messages frequently emphasized "us versus them" distinctions & .585 \\
Users often referred to their group membership when making arguments & .806 \\
\midrule
\multicolumn{2}{l}{\textbf{Perceived Uncertainty} ($\alpha = .715$)} \\
\midrule
The agents frequently acknowledged limitations in their knowledge & .541 \\
Users often expressed doubt about their own positions & .772 \\
Messages typically contained absolute statements without room for doubt [R] & .566 \\
The agents seemed very certain about their claims and positions [R] & .671 \\
\midrule
\multicolumn{2}{l}{\textbf{Perceived Bias} ($\alpha = .830$)} \\
\midrule
The discussion seemed to favor one particular viewpoint & .782 \\
Certain perspectives received more attention than others in the debate & .738 \\
The platform provided a balanced representation of different viewpoints [R] & .821 \\
Different perspectives were given equal consideration in the discussion [R] & .646 \\
\bottomrule
\end{tabularx}
\label{tab:factor-loadings}
\end{table*}

\subsection{Procedure}

The experiment was conducted across three phases: a pre-interaction phase, an interaction phase, and a post-interaction phase.

\paragraph{Pre-interaction Phase} In the first phase, participants completed a comprehensive questionnaire designed to assess various baseline measures. This included collecting demographic information, data on social media usage patterns, and initial attitudes toward UBI.

\begin{table*}[ht]
    \centering
    \small % Keep the small font size if needed
    \caption{Agent Platform Statistics at Final Iteration}
    \label{tab:agent_platform_stats}

    \sisetup{
        separate-uncertainty = true,
        table-align-uncertainty = true
    }
    \begin{tabular}{
        l % Condition (left aligned text)
        l % User Type (left aligned text)
        S[table-format=2.1(1)] % Followers
        S[table-format=1.1(1)] % Followees
        S[table-format=1.1(1)] % Posts
        S[table-format=1.1(1)] % Likes
        S[table-format=2.1(1)] % Comments
        S[table-format=1.1(1)] % Reposts
    }
    \toprule
    \textbf{Condition} & \textbf{User Type} & {\textbf{Followers}} & {\textbf{Followees}} & {\textbf{Posts}} & {\textbf{Likes}} & {\textbf{Comments}} & {\textbf{Reposts}} \\
    \midrule
    \multirow{3}{*}{Polarized}
        & Overall     & 7.3  \pm 0.5  & 7.3  \pm 0.5 & 2.2  \pm 0.0 & 8.1  \pm 1.4 & 5.5  \pm 0.4  & 2.9  \pm 0.1 \\
        & Influencers & 13.9 \pm 0.9  & 7.6  \pm 1.3 & 4.9  \pm 0.2 & 8.8  \pm 1.2 & 12.3 \pm 1.4 & 5.3  \pm 0.3 \\
        & Regular     & 5.6  \pm 0.5  & 7.2  \pm 0.4 & 1.5  \pm 0.1 & 7.9  \pm 1.6 & 3.8  \pm 0.2  & 2.3  \pm 0.1 \\
    \midrule
    \multirow{3}{*}{Moderate}
        & Overall     & 8.1  \pm 0.1  & 8.1  \pm 0.1 & 2.1  \pm 0.1 & 6.5  \pm 0.3 & 1.6  \pm 0.1 & 1.6  \pm 0.0 \\
        & Influencers & 17.4 \pm 0.3  & 7.2  \pm 0.3 & 4.9  \pm 0.6 & 7.8  \pm 1.6 & 4.1  \pm 0.6 & 3.4  \pm 0.2 \\
        & Regular     & 5.7  \pm 0.2  & 8.3  \pm 0.1 & 1.4  \pm 0.2 & 6.1  \pm 0.6 & 1.0  \pm 0.1 & 1.1  \pm 0.1 \\
    \bottomrule
    \multicolumn{8}{p{\dimexpr\linewidth-2\tabcolsep\relax}}{\footnotesize \textbf{Note:} Values represent mean $\pm$ standard deviation (SD) from final iteration values across three recommendation conditions (pro, contra, balanced). Statistics reflect the cumulative state of the agent-based platform presented to users before interaction.} \\
    \end{tabular}
\end{table*}

\paragraph{Interaction Phase} In the second phase, participants were introduced to the simulated social media platform. Every participant interacted individually within an isolated instance of the platform and did not communicate with other human users. The platform environment was populated by a set of 30 pre-programmed artificial agents, each specific to the participant's assigned experimental condition, a set that included three designated 'influencers' on each side of the UBI debate. The initial state of the platform, including a history of posts, comments, reposts, likes, and follow relationships, was generated by first running agent-only simulations for 10 iterations with agent opinions held fixed.

For example, in the polarized condition, agents generated an average of $2.16$ posts, $8.07$ likes, $5.47$ comments, and $2.90$ reposts each. In contrast, the moderate condition exhibited similar posting patterns ($2.07$ posts per agent) but reduced commenting ($1.60$) and reposting ($1.57$) activity. Complete statistics can be found in Appendix B, Table~\ref{tab:agent_platform_stats}.

Participants received instructions to engage with the platform naturally, as they would with their regular social media, based on the following text:
\begin{quote}
"You will now interact with a social media platform discussing \emph{Universal Basic Income}. Please use the platform as you normally would use social media. You can read posts, like them, comment on them, or create your own posts. Your goal is to form an opinion on the topic. You will have 10 minutes for this task."
\end{quote}

Interactions from this 10-minute phase (including likes, comments, reposts, and follows) were recorded for later analysis. A crucial aspect of the design was that while participants were interacting, the artificial agents did not generate any new content. This was a deliberate choice to guarantee exposure to a controlled initial environment. The participant's newsfeed, however, remained dynamic, as it was updated by the recommendation algorithm, which presented content from the pre-generated pool based on their individual interaction behavior.

\paragraph{Post-interaction Phase} In the final phase, after the interaction period concluded, participants were asked to complete a post-test questionnaire. This questionnaire included measures of their perception of the key constructs listed below, as well as an evaluation of the platform's realism and effectiveness.

\subsection{Measures}

All key constructs were quantified using four-item scales rated on 5-point Likert-type scales, ranging from 1 (Strongly Disagree) to 5 (Strongly Agree). The full item list and corresponding factor loadings can be found in this appendix, Table~\ref{tab:factor-loadings}. The key constructs measured in the study are enumerated below:

\begin{itemize}
    \item \emph{Opinion Change}: This captured the shift in a participant's opinion on UBI between the pre- and post-interaction phases. It was operationalized as the difference between the average scores of the opinion items before and after the interaction.
    
    \item \emph{Perceived Polarization}: This construct assessed how participants perceived the extremity and ideological division of the discourse, focusing on the perceived distance between opposing viewpoints.
    
    \item \emph{Perceived Group Salience}: This evaluated the degree to which participants saw the discussion as being driven by group identities as opposed to individual perspectives. It should be noted that, following psychometric analysis, this scale was refined to retain only two of its four initial items (details in Appendix \ref{app:user-study}).
    
    \item \emph{Perceived Emotionality}: This measure captured participants' judgment of the discussion's emotional intensity and affective tone, assessing the perceived level of emotional versus rational communication.
    
    \item \emph{Perceived Uncertainty}: This was designed to capture the extent to which participants noticed expressions of doubt or acknowledgments of knowledge limitations from the agents.
    
    \item \emph{Perceived Bias}: This measured participants' evaluation of the discussion's viewpoint balance and fairness of representation.
\end{itemize}

\subsection{Participants}

A total of $122$ participants were recruited for the study via the Prolific\footnote{https://www.prolific.com/} platform, with compensation set according to the platform's recommended rate of $\$12$/hour. These participants were distributed evenly among the six experimental conditions, resulting in approximately 20 participants per condition. The gender distribution of the sample skewed male (63.6\%) over female (35.7\%), with one participant opting not to disclose. The age distribution indicated a predominantly young to middle-aged cohort, as roughly 61.5\% of participants were between 20 and 39 years old. The modal age bracket was 25-29 years (18.6\%), with the 35-39 year bracket being the next most frequent (15.0\%).

In terms of educational background, the sample showed a high level of formal education, with nearly half (49.3\%) possessing university degrees. The other participants were distributed across a range of educational qualifications, including A-levels/IB, GCSE, and vocational certifications, each of which accounted for about 11\% of the sample.

Professionally, the majority of the sample was active, with 58.6\% employed and 10.7\% self-employed. A notable portion of the sample consisted of students (15.7\% from both university and school), indicating a diversity of occupational backgrounds.

The participants reported a high level of engagement with social media platforms, as 80\% indicated daily or near-constant usage. A majority (75\%) spent between one and four hours per day on these platforms. The most frequently used platforms were YouTube (25.0\%) and Facebook (24.3\%), followed by Instagram (17.9\%) and X (formerly Twitter, 10.7\%). This usage profile suggests that the participants were highly familiar with social media interfaces and interaction patterns, making them well-suited as subjects for the simulated environment in this study.

\subsection{Preliminary Analysis}

An evaluation of the platform's ecological validity was conducted. Participants provided ratings on several aspects using 7-point scales (1 = Not at all, 7 = Extremely). The platform garnered favorable ratings on multiple dimensions, with scores consistently above the scale's midpoint of 4. Interface usability was rated as particularly high ($M = 5.52, SD = 1.15$). The platform's similarity to existing social media ($M = 4.69, SD = 1.65$) and its capacity to facilitate meaningful discussion ($M = 4.59, SD = 1.41$) were also rated positively. A satisfactory rating for overall realism ($M = 4.47, SD = 1.61$) suggests that the experimental environment was perceived as sufficiently realistic and engaging. Regarding attitudes toward UBI, a slight decline was observed from the pre-interaction phase ($M = 3.12, SD = 0.90$) to the post-interaction phase ($M = 2.99, SD = 0.99$), though these attitudes remained near the scale's midpoint.

Additionally, an examination of the psychometric properties of our key measures was performed (see Table~\ref{tab:factor-loadings}). For each scale, principal component analyses confirmed that items loaded on their intended factors. High reliability was observed for most scales, with Cronbach's alpha ranging from $.715$ to $.893$ and factor loadings generally exceeding $|.40|$. The \emph{Perceived Group Salience} scale, however, required modification from its initial four-item structure. Two items ("The debate focused on ideas rather than group affiliations" and "Individual perspectives were more prominent...") were removed because of poor factor loadings (.049 and .051, respectively). The two remaining items exhibited modest to acceptable loadings ($.585$ and $.806$). Given the theoretical importance of this construct, this modified measure was retained for further analysis despite these acknowledged psychometric limitations.

\subsection{Analysis of Debate Perception}
\label{subsec:debate-perception}

\begin{table*}[ht]
\small
\centering
\caption{Effects of Polarization and Recommendation Bias on Key Dependent Variables}
\label{tab:perception-anova}
\begin{tabularx}{\textwidth}{>{\raggedright\arraybackslash}p{2.5cm}>{\raggedright\arraybackslash}p{1.5cm}*{3}{>{\centering\arraybackslash}X}>{\centering\arraybackslash}p{1.2cm}>{\centering\arraybackslash}p{1.2cm}>{\centering\arraybackslash}p{1.2cm}}
\toprule
\multirow{2}{*}{\textbf{Metric}} & \multirow{2}{*}{\textbf{Condition}} & \multicolumn{3}{c}{\textbf{Recommendation Bias}} & \multicolumn{3}{c}{\textbf{ANOVA $\boldsymbol{F}$($\boldsymbol{p}$)}} \\
\cmidrule(lr){3-5} \cmidrule(lr){6-8}
& & Contra & Balanced & Pro & Pol & Rec & Pol×Rec \\
\midrule
\multirow{2}{*}{Opinion Change} 
   & Polarized & \textbf{-0.41} (0.60) & -0.08 (0.55) & -0.10 (0.33) & 3.48 & 1.74 & 1.90 \\
   & Unpolarized & -0.06 (0.38) & -0.09 (0.43) & -0.02 (0.26) & (.065) & (.180) & (.154) \\
\midrule
\multirow{2}{*}{|Opinion Change|} 
   & Polarized & \textbf{0.56} (0.45) & 0.35 (0.42) & 0.22 (0.26) & \textbf{5.21*} & \textbf{3.61*} & 2.54 \\
   & Unpolarized & 0.24 (0.30) & 0.32 (0.28) & 0.18 (0.18) & (.024) & (.030) & (.083) \\
\midrule
\multirow{2}{*}{Polarization} 
   & Polarized & \textbf{4.47} (0.82) & \textbf{4.61} (0.97) & \textbf{4.29} (0.75) & \textbf{56.48***} & 0.02 & 1.50 \\
   & Unpolarized & 3.30 (0.69) & 3.23 (0.83) & 3.54 (0.84) & (.000) & (.979) & (.228) \\
\midrule
\multirow{2}{*}{Emotionality} 
   & Polarized & \textbf{3.80} (0.48) & \textbf{3.59} (0.99) & \textbf{3.36} (0.80) & \textbf{73.54***} & 1.42 & 2.44 \\
   & Unpolarized & 2.50 (0.74) & 2.17 (0.60) & 2.65 (0.83) & (.000) & (.245) & (.092) \\
\midrule
\multirow{2}{*}{Group Salience} 
   & Polarized & \textbf{2.94} (0.60) & \textbf{2.86} (0.51) & \textbf{3.21} (0.42) & \textbf{17.18***} & \textbf{5.63**} & 0.07 \\
   & Unpolarized & 2.51 (0.59) & 2.42 (0.65) & 2.86 (0.40) & (.000) & (.005) & (.934) \\
\midrule
\multirow{2}{*}{Bias} 
   & Polarized & \textbf{3.50} (0.87) & 3.20 (0.76) & \textbf{3.83} (0.59) & \textbf{15.08***} & 0.44 & \textbf{3.73*} \\
   & Unpolarized & 3.05 (0.82) & 3.02 (0.87) & 2.68 (0.87) & (.000) & (.648) & (.027) \\
\midrule
\multirow{2}{*}{Uncertainty} 
   & Polarized & 2.29 (0.60) & 2.03 (0.60) & 1.90 (0.75) & \textbf{48.86***} & 0.97 & 1.06 \\
   & Unpolarized & \textbf{2.92} (0.71) & \textbf{2.83} (0.69) & \textbf{2.96} (0.48) & (.000) & (.383) & (.350) \\
\bottomrule
\multicolumn{8}{p{.95\textwidth}}{\small \textbf{Note:} Values show means with standard deviations in parentheses. Bold values indicate significantly higher means between polarized and unpolarized conditions. F-statistics in bold are significant. Pol = Polarization main effect, Rec = Recommendation main effect, Pol×Rec = Interaction effect. *$p$ < .05, **$p$ < .01, ***$p$ < .001} \\
\end{tabularx}
\end{table*}

The first analysis examines how users perceive and process the discussions they were exposed to under the varying conditions of \emph{Polarization Degree} and \emph{Recommendation Bias}. The specific focus is on investigating whether participants could recognize the polarized discourse patterns, how they processed the emotional and group-based content, and how \emph{Recommendation Bias} may have moderated these perceptions. The goal of this analysis is to understand the psychological mechanisms by which the climate of a discussion and the curation of its content shape the user experience.

\subsubsection{Variance Analysis}

The effects of \emph{Polarization Degree} and \emph{Recommendation Bias} were analyzed with a series of $2 \times 3$ ANOVAs (see Table~\ref{tab:perception-anova} for full statistics and Figure~\ref{fig:perception-interaction-plots} for interaction plots). A consistent pattern emerged from the results, showing significant main effects for \emph{Polarization Degree} across most of the perceptual variables, while \emph{Recommendation Bias} exerted a more limited impact.

With respect to \emph{Opinion Change}, neither \emph{Polarization Degree} nor \emph{Recommendation Bias} yielded a statistically significant main effect (Table~\ref{tab:perception-anova}). Descriptive statistics, however, pointed to a trend of stronger negative opinion shifts, particularly within the polarized condition that was combined with contra-bias recommendations ($M = -0.41$), as compared to the minimal changes in other conditions.

The manipulation of \emph{Polarization Degree} had a strong influence on the perceived climate of the discussion. Ratings for \emph{Perceived Polarization} were significantly higher in the polarized condition ($M = 4.46$) than in the unpolarized condition ($M = 3.35$). This difference was confirmed by post-hoc tests ($p < .001$, Hedges' $g = 1.36$) and represented a substantial main effect (Table~\ref{tab:perception-anova}).

The strongest main effect of \emph{Polarization Degree} in the study was observed for \emph{Perceived Emotionality} ($\eta_p^2 = 0.388, p < .001$). Polarized discussions were perceived as being significantly more emotional ($M = 3.58$ vs. $M = 2.44$ for unpolarized; Hedges' $g = 1.53$). Although the main effect for \emph{Recommendation Bias} was not significant, a marginal interaction with \emph{Polarization Degree} was noted ($p = .092$).

Similarly, a strong main effect of \emph{Polarization Degree} was found for \emph{Perceived Uncertainty} ($\eta_p^2 = 0.296, p < .001$; Hedges' $g = -1.26$), which was rated as significantly lower in polarized discussions ($M = 2.07$) compared to unpolarized ones ($M = 2.90$).

Both \emph{Polarization Degree} ($p < .001$) and \emph{Recommendation Bias} ($p = .005$) significantly affected \emph{Perceived Group Salience}. Perceptions of group-based discourse were higher in polarized conditions ($M = 3.00$ vs. $M = 2.59$ for unpolarized; Hedges' $g = 0.73$). Post-hoc tests for \emph{Recommendation Bias} showed a significantly higher salience in the pro-bias condition compared to both the contra-bias ($p=.011$) and neutral ($p=.001$) conditions (see Table~\ref{tab:perception-anova} for means).

Finally, for \emph{Perceived Bias}, a significant main effect for \emph{Polarization Degree} ($p < .001$) was found, as was a significant interaction effect ($p = .027$). Participants perceived a higher level of bias in the polarized condition ($M = 3.51$) compared to the unpolarized condition ($M = 2.92$; Hedges' $g = 0.69$), with the interaction indicating this effect was especially pronounced under pro-bias recommendations (Table~\ref{tab:perception-anova}).

\subsubsection{Path Analysis}

Structural equation modeling (SEM) was employed to better understand the mechanisms through which the experimental conditions influence both perceptual dimensions and opinion change. This method makes it possible to simultaneously estimate multiple interdependent relationships while also accounting for measurement error and the covariation between constructs.

A theoretical model was developed to examine how \emph{Polarization Degree} and \emph{Recommendation Bias} affect \emph{Opinion Change Magnitude} and \emph{Perceived Polarization}, both directly and via various perceptual pathways (see Figure~\ref{fig:path-model}). The model was estimated with maximum likelihood estimation using standardized variables, and the results indicated an excellent fit to the data ($\chi^2(6) = 2.335, p = .886$, CFI $= 1.000$, TLI $= 1.109$, RMSEA $= .000$ [90\% CI: .000, .068], SRMR $= .023$). It also explained a substantial amount of variance in key outcome variables, such as \emph{Perceived Emotionality} (53.3\%) and \emph{Perceived Polarization} (38.0\%).

The SEM analysis brought several significant pathways to light. First, significant direct effects on \emph{Opinion Change Magnitude} were observed for both experimental conditions. A higher \emph{Polarization Degree} led to an increased \emph{Opinion Change Magnitude} ($\beta = .275, p = .007$), whereas the pro-UBI \emph{Recommendation Bias} was associated with a decreased magnitude ($\beta = -.280, p = .006$). It is noteworthy that these effects emerged even though none of the perceptual variables showed a significant direct effect on opinion change magnitude.

\begin{table*}[ht]
\centering
\small
\caption{Main Effects of Experimental Conditions on User Engagement}
\label{tab:interactions-anova}
\begin{tabularx}{\textwidth}{>{\raggedright\arraybackslash}p{2cm}>{\raggedright\arraybackslash}p{1.8cm}*{3}{>{\centering\arraybackslash}X}>{\centering\arraybackslash}p{1.2cm}>{\centering\arraybackslash}p{1cm}}
\toprule
\multirow{2}{*}{\textbf{Metric}} & \multirow{2}{*}{\textbf{Condition}} & \multicolumn{3}{c}{\textbf{Recommendation Bias}} & \multicolumn{2}{c}{\textbf{ANOVA}} \\
\cmidrule(lr){3-5} \cmidrule(lr){6-7}
& & Contra & Balanced & Pro & $F$ & $p$ \\
\midrule
\multirow{2}{*}{Interactions} 
   & Polarized & \textbf{10.30} (12.43) & 4.82 (4.49) & 4.80 (5.29) & \multirow{2}{*}{\textbf{3.25}$^a$} & \multirow{2}{*}{.042*} \\
   & Unpolarized & 7.92 (7.62) & 4.71 (3.26) & \textbf{9.25} (12.85) & & \\
\midrule
\multirow{2}{*}{Likes} 
   & Polarized & \textbf{5.65} (9.53) & 2.55 (2.65) & 2.35 (2.54) & \multirow{2}{*}{2.25$^a$} & \multirow{2}{*}{.110} \\
   & Unpolarized & 4.38 (4.03) & 2.64 (2.44) & \textbf{4.95} (8.22) & & \\
\midrule
\multirow{2}{*}{Reposts} 
   & Polarized & 0.65 (1.03) & 0.45 (0.67) & 0.40 (0.82) & \multirow{2}{*}{1.89$^a$} & \multirow{2}{*}{.156} \\
   & Unpolarized & \textbf{1.08} (2.30) & 0.29 (0.46) & \textbf{1.10} (1.86) & & \\
\midrule
\multirow{2}{*}{Comments} 
   & Polarized & 0.35 (0.57) & 0.23 (0.43) & 0.25 (0.55) & \multirow{2}{*}{\textbf{6.51}$^b$} & \multirow{2}{*}{.012*} \\
   & Unpolarized & 0.50 (1.18) & 0.71 (1.12) & \textbf{1.10} (1.94) & & \\
\midrule
\multirow{2}{*}{Follows} 
   & Polarized & \textbf{3.30} (4.12) & 1.45 (1.84) & 1.50 (2.82) & \multirow{2}{*}{2.60$^a$} & \multirow{2}{*}{.078} \\
   & Unpolarized & 1.88 (3.38) & 1.04 (1.57) & 1.95 (3.33) & & \\
\bottomrule
\multicolumn{7}{p{.95\textwidth}}{\small \textbf{Note:} Values show means with standard deviations in parentheses. Bold values indicate highest means within recommendation bias conditions. $^a$F-statistic for main effect of Recommendation (df = 2, 134). $^b$F-statistic for main effect of Polarization (df = 1, 135). *$p$ < .05} \\
\end{tabularx}
\end{table*}

Regarding the perceptual pathways, \emph{Polarization Degree} exerted strong direct effects on several dimensions: \emph{Perceived Emotionality} ($\beta = .472, p < .001$), \emph{Perceived Uncertainty} ($\beta = -.541, p < .001$), \emph{Perceived Bias} ($\beta = .433, p < .001$), and \emph{Perceived Group Salience} ($\beta = .251, p = .019$). Further analysis revealed that \emph{Polarization Degree} shaped \emph{Perceived Polarization} through multiple paths. In addition to its direct effect ($\beta = .247, p = .022$), it also operated through two indirect routes: one via \emph{Perceived Group Salience} ($IE = .105, p = .044$) and another via \emph{Perceived Emotionality} ($IE = .050, p = .406$). The combined direct and indirect effects resulted in a substantial total effect of \emph{Polarization Degree} on \emph{Perceived Polarization} ($\beta = .402, p < .001$).

The model also revealed interesting relationships among the mediating variables. There were significant positive effects of \emph{Perceived Group Salience} on both \emph{Perceived Emotionality} ($\beta = .447, p < .001$) and \emph{Perceived Polarization} ($\beta = .417, p < .001$). A significant negative covariance was also found between \emph{Perceived Uncertainty} and \emph{Perceived Bias} ($\beta = -.291, p = .012$), which suggests these two perceptions tend to operate in opposition to one another.

These findings point to a complex pattern wherein the experimental conditions shape both behavioral outcomes (\emph{Opinion Change Magnitude}) and perceptual experiences, although these two paths appear to function independently rather than sequentially. The substantial total effect that \emph{Polarization Degree} has on \emph{Perceived Polarization}, when decomposed into its direct and indirect pathways, serves to highlight how features of the environment can influence user perceptions through multiple, complementary mechanisms.

\subsection{Analysis of User Engagement}

\begin{figure*}[h!]
    \centering
    \includegraphics[width=\textwidth]{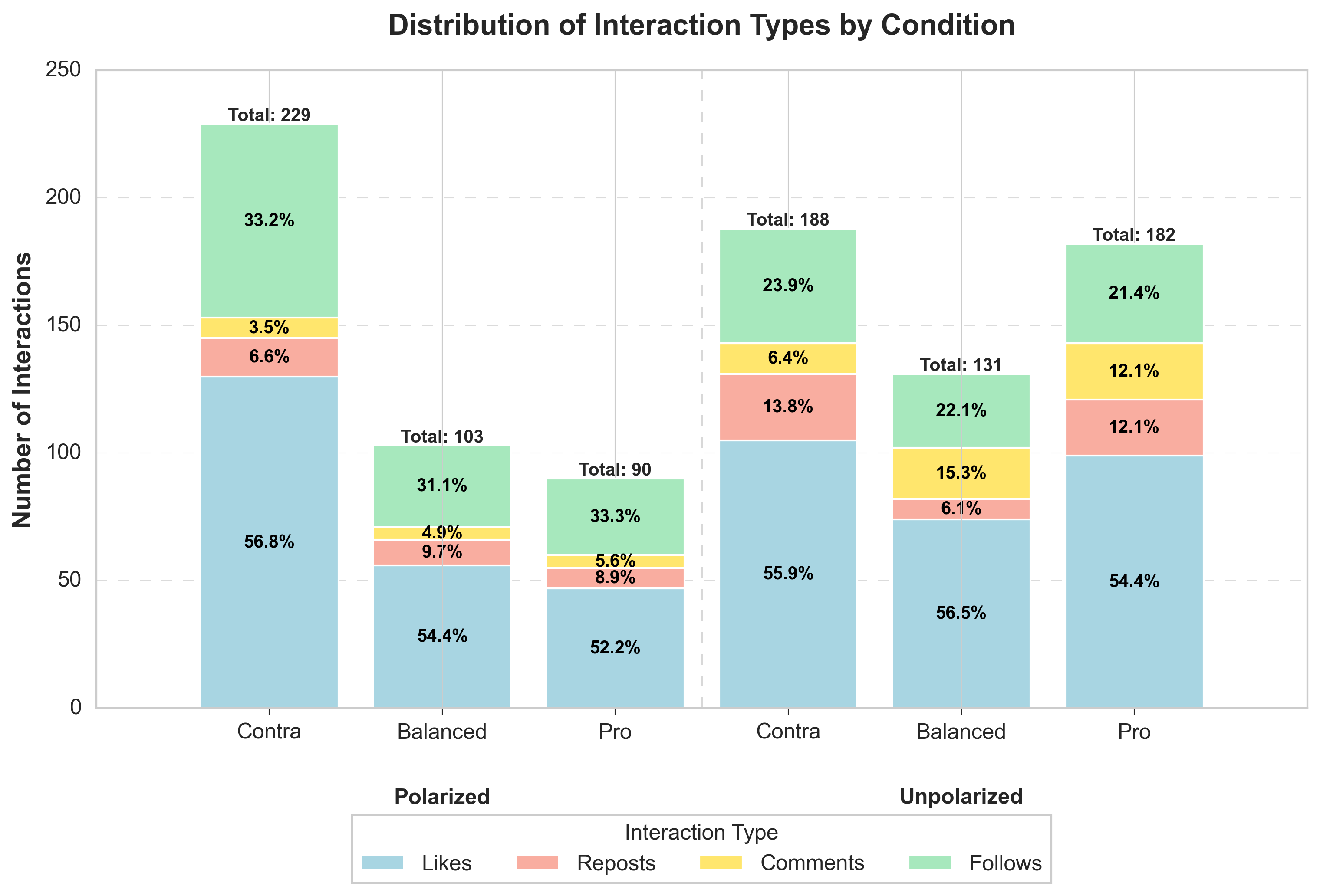}
    \caption{Distribution of interaction types across experimental conditions. The stacked bars show the relative proportion of different interaction types (\emph{likes}, \emph{reposts}, \emph{comments}, and \emph{follows}) for each combination of polarization level and recommendation bias. Total interaction counts are displayed above each bar.}
    \label{fig:stacked-interaction-distribution}
\end{figure*}

The second analysis is focused on how polarization and recommendation bias influence user engagement behaviors on the platform. It investigates both the quantity and the quality of interactions by analyzing how the different experimental conditions affect user preferences for specific types of engagement (\emph{likes}, \emph{comments}, \emph{reposts}, and \emph{follows}). The aim of this analysis is to determine whether polarized environments and algorithmic bias affect not only how much users engage, but also the manner in which they choose to participate in discussions.

\subsubsection{Descriptive and Variance Analysis}

The analysis of user engagement revealed distinct patterns across the experimental conditions. A total of 946 interactions were generated by participants ($M = 6.91, SD = 8.50$), showing substantial individual variation. The majority of participants (81.02\%) engaged at least once. For these active users, engagement levels could be categorized as low (1-5 interactions; 36.5\%), moderate (6-10; 24.82\%), or high ($>10$; 19.71\%).

A clear hierarchy among interaction types is shown in Figure~\ref{fig:stacked-interaction-distribution}. \emph{Likes} were the dominant form of interaction ($54.02\%, M = 3.73$), followed by \emph{follows} ($26.53\%, M = 1.83$), \emph{reposts} ($9.41\%, M = 0.65$), and finally \emph{comments} ($7.61\%, M = 0.53$). This hierarchy indicates a general preference for lower-effort forms of engagement.

Correlation analysis showed strong positive associations between \emph{total interactions} and both \emph{likes} ($r = .93, p < .001$) and \emph{follows} ($r = .73, p < .001$). The analysis also revealed a moderate correlation with \emph{comments} ($r = .45, p < .001$) and a weaker one with \emph{reposts} ($r = .37, p < .001$). Notably, there was minimal correlation between \emph{comments} and \emph{reposts} ($r = .01, p = .899$), which suggests they fulfill distinct user purposes.

Results from the ANOVA (Table~\ref{tab:interactions-anova}) indicated a significant main effect for \emph{Recommendation Bias} on \emph{total interactions} ($\eta^2_p = .046$). Post-hoc tests showed that there were significantly more interactions in the contra-bias condition compared to the balanced condition ($p = .008$, Hedges' $g = 0.56$), while other comparisons were not significant (see Table~\ref{tab:interactions-anova} for means).

The analysis of specific interaction types revealed a significant main effect of \emph{Polarization Degree} on the act of \emph{commenting} ($\eta^2_p = .046$). Participants in unpolarized conditions commented more ($M = 0.77$) than those in polarized conditions ($M = 0.28$), which suggests that moderate environments may be more conducive to this form of engagement.

For other interaction types, no significant main effects were found, although some trends did emerge. For example, \emph{following} behavior showed a marginal effect of \emph{Recommendation Bias} ($p = .078, \eta^2_p = .037$), with a tendency for users to follow more people when exposed to opposing views (contra-bias) versus balanced content. The number of \emph{likes} followed a non-significant trend that was consistent with that of \emph{total interactions} ($p = .110$).

An examination of the interaction patterns (Table~\ref{tab:interactions-anova}, Figure~\ref{fig:interaction-plot-recommendation-bias}) showed that engagement within polarized conditions reached its peak under contra-bias recommendations, while being notably lower in the balanced and pro-bias conditions. For unpolarized conditions, engagement was more evenly distributed, but it remained at an elevated level in both the contra- and pro-bias conditions compared to the balanced one.

\subsection{Extended Discussion: Aligning Human-Subject Data with Computational Models}
\label{subsec:user_study_discussion_extended}

This discussion provides a detailed interpretation of the validation study's findings, placing them in direct conversation with the computational experiments presented in the main text. By synthesizing human-subject data with agent-based modeling results, we can more deeply inform the framework's ecological validity, identify areas of alignment and divergence, and chart paths for future research.

A central achievement of the validation study is its confirmation of the framework's ability to generate psychologically meaningful discourse. Participants proved highly attuned to the manipulated \emph{Polarization Degree}, perceiving the polarized environment as more emotional, more biased, and higher in group salience (Hedges' $g = 1.53$ for emotionality). This provides crucial empirical validation for our LLM-based content analysis (Section \ref{subsec:eval_communication_content}), which demonstrated the model's capacity to generate text with these specific communicative features. Critically, the strong inverse relationship observed between polarization and perceived uncertainty ($g = -1.26$) resonates with theoretical work on epistemic humility in discourse \citep{holtz_intergroup_2001, holtz_relative_2008, winter_toward_2019} and confirms that humans are sensitive to the very same textual cues that inform the belief-update mechanisms of our formal opinion dynamics model.

The SEM analysis of human data reveals a more nuanced perceptual pathway than our computational models assume. It suggests that the effect of emotionality on perceived polarization is primarily mediated by group salience ($\beta = .417$), a finding consistent with theories of affective polarization \citep{iyengar_affect_2012}. This implies that for humans, emotional content may function as a key trigger for activating group-based antagonisms, echoing political theories that frame conflict in terms of friend-enemy distinctions \citep{schmitt_concept_2008, laclau_hegemony_2014}. This insight complements our computational findings, which identified structural factors like network mixing as primary drivers of polarization. The user study thus suggests an interplay where structural conditions may determine the potential for division, while affective communication acts as the proximate trigger for its perception.

Regarding opinion formation, the study highlights a divergence between the formal model and human behavior. While the experimental conditions directly influenced the \emph{magnitude} of participants' opinion change, the SEM found this effect was not mediated by any of the measured perceptual variables. This contrasts with our agent update rule (Eq. \ref{eq:opinion_update_clipped}), which models opinion shifts as a direct function of perceived content. This divergence strongly suggests that human opinion adjustment in these contexts may involve non-deliberative processing routes \citep{petty_elaboration_1986, chaiken_the_2014} or a direct response to the "environmental press" \citep{lewin_principles_2013}, pathways that our current formal model does not capture. This finding is not a limitation but a key insight, providing a clear directive for future work to enhance the psychological realism of agent-based opinion dynamics models.

Turning to user engagement, participants' preference for low-effort interactions like \emph{likes} over \emph{comments} mirrors typical social media behavior. Interestingly, the finding that participants commented more in moderate environments \citep{koudenburg_polarized_2022, yousafzai_political_2022} aligns with our extended agent reaction simulations (\cref{app:computational_experiments}), where agents in consensual environments also exhibited more variable, less homophilic engagement patterns. The cautious interpretation of this finding due to potential participant-environment mismatch \citep{simchon_troll_2022} remains valid, but the alignment with agent behavior suggests the framework is capturing a genuine dynamic. The strong effect of contra-bias recommendations on engagement, especially in polarized conditions, further underscores the power of exposing users to challenging arguments, a finding that connects to literature on loss aversion and status quo bias \citep{kahneman_anomalies_1991}.

In conclusion, the validation study confirms that our hybrid framework can produce controlled, manipulable environments that elicit complex and psychologically meaningful responses from human users. The strong alignment on perceptual dimensions validates the core of our content generation model. The divergences observed in behavioral and opinion-update pathways do not invalidate the framework; rather, they highlight its utility as a tool for discovery, pointing toward specific, data-driven avenues for refining the next generation of opinion dynamics models. The study thus confirms the value of an integrated approach that places computational modeling and human-subject experimentation in a continuous, productive dialogue.
% app:user-study
\end{appendices}

\end{document}